\documentclass[pra,aps,twocolumn,eqsecnum,showpacs,letterpaper,10pt]{revtex4-2}
\usepackage{amsmath}
\usepackage{amssymb}
\usepackage{amsfonts}
\usepackage{bm} 
\usepackage{hyperref}
\usepackage{color}

\begin{document}
\title{Microscopic density-functional approach to nonlinear elasticity theory}

\author{Rudolf Haussmann}
\affiliation{Fachbereich Physik, Universit\"at Konstanz, D-78457 Konstanz, Germany}

\email[email:\enskip]{rudolf.haussmann@uni-konstanz.de}
\homepage[\newline homepage:\enskip]{https://www.kormoran-tech.com}

\date{May 31, 2022; \enskip published in \ Journal of Statistical Mechanics: Theory and Experiment, 053210 (2022)}

\begin{abstract}
Starting from a general classical model of many interacting particles we present a well defined step by step 
procedure to derive the continuum-mechanics equations of nonlinear elasticity theory with fluctuations which 
describe the macroscopic phenomena of a solid crystal. As the relevant variables we specify the coarse-grained 
densities of the conserved quantities and a properly defined displacement field which describes the local translations, 
rotations, and deformations. In order to stay within the framework of the conventional density-functional theory we 
first and mainly consider the isothermal case and omit the effects of heat transport and warming by friction where 
later we extend our theory to the general case and include these effects. We proceed in two steps. First, we apply 
the concept of local thermodynamic equilibrium and minimize the free energy functional under the constraints that 
the macroscopic relevant variables are fixed. As results we obtain the local free energy density and we derive 
explicit formulas for the elastic constants which are exact within the framework of density-functional theory. 
Second, we apply the methods of nonequilibrium statistical mechanics with projection-operator techniques. We extend 
the projection operators in order to include the effects of coarse-graining and the displacement field. As a result 
we obtain the time-evolution equations for the relevant variables with three kinds of terms on the right-hand 
sides: reversible, dissipative, and fluctuating terms. We find explicit formulas for the transport coefficients 
which are exact in the limit of continuum mechanics if the projection operators are properly defined. By construction 
the theory allows the diffusion of particles in terms of point defects where, however, in a normal crystal this 
diffusion is suppressed.
\end{abstract}


\maketitle

\tableofcontents

\section{Introduction}
\label{section::01}
A solid crystal is a densely packed system of many atoms or molecules which interact with each other by forces of 
electromagnetic origin. Whenever the temperature is below the critical value of the melting transition the particles 
are confined on the sites of a regular lattice structure. In a perfect crystal the particles are completely arrested 
so that no diffusive motion of the particles over larger distances is possible. Only one particle may be on each site. 
The strong repulsive forces between the particles prohibit that two or more particles stay on one site. 

However, there may be a \emph{vacancy} which means that a lattice site is empty. On the other hand, an additional 
atom may be on an \emph{interstitial} place. In this way two kinds of \emph{point defects} are possible which are 
vacancies and interstitial particles, respectively. These two point defects may move around so that an effective 
diffusion of particles within the crystal lattice is possible. However, large energy barriers compared to the 
temperature strongly reduce the density and the diffusivity of the point defects. Thus, the point defects and the 
diffusion of particles are minor effects in a \emph{normal crystal}.

The situation is completely different if the interaction potentials of any two particles have a finite maximum 
at zero distance and decay to zero with increasing distance after a certain range. In this case the lattice is 
determined by the interaction range as the characteristic length scale. For short distances the repulsive forces 
are small and tend to zero. On the other hand, the repulsive forces are maximum for intermediate distances in 
the middle to the next neighboring lattice site. As a result a \emph{cluster crystal} may form where more than one 
or even many particles may be on each site. The hopping of a particle from one site to the next is no more blocked 
so that the diffusion of particles is possible over large distances through the crystal. Cluster crystals can be 
modeled and investigated in Computer simulations.

In the conventional elasticity theory \cite{LL07} normal crystals are considered where the particles are confined 
to the lattice sites. Martin, Parodi, and Pershan \cite{MPP72} were the first who extended the theory by including 
the diffusion of the particles. They derived rather general time-evolution equations for the relevant variables 
from rather general phenomenological principles. Besides the densities of the conserved quantities they considered 
the \emph{displacement field} as a further relevant variable. Fleming and Cohen \cite{FC76} extended and improved 
the time-evolution equations. In place of the displacement field they considered the \emph{strain tensor} as the 
relevant variable for local deformations which is obtained from gradients of the displacement field. Later 
Grabert and Michel \cite{GM83} provided the full set of completely nonlinear time-evolution equations which 
describe the elastic properties and the macroscopic behavior of a solid crystal in the nonlinear regime for 
large deformations. Many years later Temmen \emph{et al.}\ \cite{Te00} investigated the nonlinear properties of 
non-Newtonian fluids and solid crystals for large deformations where they presented essential parts of the nonlinear 
theory but not a complete and elaborated set of equations.

Starting from a microscopic theory of many interacting classical particles and using projection operators for the 
macroscopic relevant variables a microscopic approach was first developed by Szamel and Ernst \cite{SE93,Sz97}. 
For small deformations and small deviations from the thermal equilibrium a microscopic expression for the displacement 
field was derived and defined via a scalar product in terms of the microscopic density. Microscopic formulas for 
the elastic constants \cite{SE93} and for the transport coefficients \cite{Sz97} were obtained in linear response. 
The theory was extended by Walz and Fuchs \cite{WF10} by defining the change of the particle density in terms of 
the displacement field and the change of the vacancy density. First explicit numerical calculations were done by 
H\"aring \emph{et al.}\ \cite{HF15} to obtain some elastic properties of a cluster crystal. A further extension 
is due to Ras, Szafarczyk, and Fuchs \cite{RSF20} by including several species of particles. Miserez \cite{Mi21} 
extended the theory to include heat transport and warming by friction. An alternative microscopic approach was 
recently presented by Mabillard and Gaspard \cite{MG20,MG21} applying and extending the method of Sasa \cite{Sa14} 
which avoids the use of projection operators. In all these approaches a decoupling of the motion of the lattice 
structure from the motion of the material was found due to the diffusion of point defects. However, all the 
microscopic considerations and calculations were done for small deformations in the linear-response regime and 
close to thermal equilibrium. Moreover, stochastic forces and fluctuations were not taken into account.

In a previous publication \cite{Ha16} we have presented a complete step by step procedure to derive the 
hydrodynamic equations of a simple liquid from microscopic physics including all \emph{nonlinear} and 
\emph{fluctuating} effects. In the present paper we extend this approach to the continuum mechanics of a solid 
crystal in order to describe the elastic properties for strong deformations in the nonlinear regime including 
the fluctuations. Our main motivation is to combine and to extend the previous approaches into a complete theory 
of nonlinear elasticity. We start with a very general microscopic theory of many interacting particles and want 
to end up with a macroscopic theory of continuum mechanics where any approximations in between are reduced to 
a minimum. We claim that all necessary approximations are related to the limit of large scales in space and time
so that our macroscopic theory is \emph{exact} in this limit. We want to cover all nonlinear effects for large 
deformations and far from global equilibrium, the diffusion of particles or point defects through the lattice 
structure, and the fluctuations in the nonlinear regime.

Our theory is developed in two stages. First, we develop a density-functional theory for the nonequilibrium 
states of continuum mechanics by using the concept of local thermodynamic equilibrium. We minimize the free 
energy functional under the constraints that the macroscopic relevant variables are fixed. Second, we derive 
macroscopic time-evolution equations for the relevant variables by applying the concept of projection operators 
in nonequilibrium statistical mechanics. The success of our approach will depend on a careful definition of the 
coarse-grained densities and the displacement field as the relevant variables and a careful definition of the 
related projection operators.

In order to stay within the formalism of conventional density-functional theory where many explicit formulas 
are available and many explicit calculations can be done we first omit the energy density as a relevant variable. 
Thus, in the first and main part of the paper we develop an \emph{isothermal} nonlinear elasticity theory where 
the temperature is constant while the transport of heat by convection and diffusion and the warming by friction 
and dissipation are neglected. Usually, in crystals the thermal or kinetic energy of the particles is much smaller 
than the potential energy which implies the nonlinear elastic effects. Thus, a constant temperature may be a 
reasonable approximation. Nevertheless, in a later section we extend our theory to the \emph{general} case where 
we include the energy density as a relevant variable and take into account all the heat and entropy effects. 
Thus, in the later section we derive and present the complete theory of nonlinear elasticity for strong 
deformations, with diffusion of particles or point defects, and with all nonlinear fluctuations.

In Sec.\ \ref{section::02} we describe the classical interacting many-particle system which we use as a microscopic 
foundation of our procedure to derive the macroscopic theory of continuum mechanics. We apply the methods of 
statistical physics in order to define the partition function and the grand canonical thermodynamic potential 
for nonequilibrium states in local thermodynamic equilibrium. In Sec.\ \ref{section::03} via a Legendre transformation 
we define the Helmholtz free energy as a functional of the densities of the conserved quantities which will 
serve as the functional for the density-functional theory in the later sections. We present the results of the 
Mayer-Ursell cluster expansion \cite{MM41,Ur27} which in principle provides \emph{complete} formulas for the 
thermodynamic potential and the free energy in perturbation theory if the infinite perturbation series is summed 
up \emph{exactly}. Alternatively, one can use any sophisticated ansatz of the density-functional theory as a 
starting point for the later calculations.

In Sec.\ \ref{section::04} the free energy functional is minimized under constraints. From the related necessary 
conditions a linear response equation is derived which is solved exactly. In this way the change of the microscopic 
density is calculated depending on the displacement field and the vacancy density. Our exact solution extends 
the formula of Walz and Fuchs \cite{WF10} by a correction term. In Sec.\ \ref{section::05} this solution is used to 
calculate the free energy explicitly as a Taylor series expansion in terms of the relevant variables which are the 
coarse grained particle densities, the coarse grained momentum density, and the displacement field. We obtain the 
free energy as an integral of a local free energy density. In this way the concept of the local thermodynamic 
equilibrium is established. We present exact results for the elastic constants which again extend the formulas 
of Walz and Fuchs \cite{WF10} by correction terms.

In Sec.\ \ref{section::06} we develop a geometric approach to describe strong and nonlinear deformations by considering 
the coordinate transformation between the Cartesian laboratory frame and the curvilinear coordinates attached to the 
material. Here we apply some methods of General Relativity and Riemannian Geometry. The transformation matrices or 
the metric tensors are taken as measures for the strains in the crystal. In this way several conjugated pairs of 
strain and stress tensors are found which allow a simple and straightforward classification of the several strain 
and stress tensors of the conventional elasticity theory.

In Secs.\ \ref{section::07} and \ref{section::08} we turn to dynamic phenomena. While in Sec.\ \ref{section::07} 
we stay within the isothermal case, in Sec.\ \ref{section::08} we extend our theory to the general case. First, we 
extend the Grabert projection operator \cite{Gr82} for a correct treatment of the coarse-grained densities and the 
displacement field as the relevant variables. Then, we use this projection operator to derive the time-evolution 
equations within the framwork of the GENERIC formalism of \"Ottinger and Grmela \cite{GO97A,GO97B,Ot05}. As results 
we obtain explicit time-evolution equations for the coarse grained densities and the displacement field which include 
all nonlinear effects and which agree with the phenomenological equations of Grabert and Michel \cite{GM83}. In 
Sec.\ \ref{section::09} we discuss the essential features and consequences of our theory and of our results. Finally, 
in Sec.\ \ref{section::10} we summarize our results and conclude with some comments and remarks.

\section{Interacting many-particle system}
\label{section::02}
As a physical system we consider many particles with masses $m_a$, coordinates $\mathbf{r}_{ai}$, and momenta $\mathbf{p}_{ai}$. 
The particles interact with each other by pair potentials $V_{ab}( \mathbf{r}_{ai} - \mathbf{r}_{bj} )$. The dynamics of the system 
is described by the classical Hamilton equations
\begin{equation}
  \frac{ d \mathbf{r}_{ai} }{ d t } = \{ \mathbf{r}_{ai} , \hat{H} \} \ , \qquad 
  \frac{ d \mathbf{p}_{ai} }{ d t } = \{ \mathbf{p}_{ai} , \hat{H} \}
  \label{equation::B_010}
\end{equation}
for the coordinates and momenta where
\begin{equation}
  \hat{H} = \sum_{ai} \frac{ \mathbf{p}_{ai}^2 }{ 2 m_a } 
  + \frac{ 1 }{ 2 } \sum_{ai \neq bj} V_{ab}( \mathbf{r}_{ai} - \mathbf{r}_{bj} )
  \label{equation::B_020}
\end{equation}
is the Hamilton function. As a convention we put a hat onto the symbol of a quantity in order to indicate that the quantity 
depends on the coordinates $\mathbf{r}_{ai}$ and momenta $\mathbf{p}_{ai}$ of the particles. We use two kinds of indices. 
The first indices
\begin{equation}
  a,b,c,\ldots = 1,2,3,\ldots,n
  \label{equation::B_030} 
\end{equation}
count the particle species while the second indices
\begin{equation}
  i,j,k,\ldots = i_a,j_a,k_a,\ldots = 1,2,3,\ldots,N_a
  \label{equation::B_040} 
\end{equation}
count the particles of a given species $a$. For maximum precision of the notation the second indices $i_a$ must have a species 
index $a$ by itself. However, for convenience we omit this species index, simply write $i$, and expect that it can be reconstructed 
easily from the context. The number of particles of a given species may be written in the form
\begin{equation}
  \hat{N}_a = \sum_i 1 \ .
  \label{equation::B_050} 
\end{equation}
A further important quantity is the total momentum of the system
\begin{equation}
  \hat{\mathbf{P}} = \sum_{ai} \mathbf{p}_{ai} \ .
  \label{equation::B_060} 
\end{equation}
Since the Hamiltonian does not depend explicitly on the time $t$ the energy $\hat{E} = \hat{H}$ is a conserved quantity. 
The pair potentials $V_{ab}( \mathbf{r}_{ai} - \mathbf{r}_{bj} )$ depend only on coordinate differences and hence are invariant 
under spatial translations. Consequently, the total momentum $\hat{\mathbf{P}}$ is a conserved quantity. Finally, we do not 
allow chemical reactions so that the species particle numbers $\hat{N}_a$ are conserved. We assume that the space has $d$ 
dimensions so that the particle coordinates $\mathbf{r}_{ai}$, the particle momenta $\mathbf{p}_{ai}$, and the total momentum 
$\hat{\mathbf{P}}$ are $d$-dimensional vectors.

We consider a statistical description of the system in the phase space of the coordinates and momenta 
$\Gamma = ( \mathbf{r}_{ai},\mathbf{p}_{ai} )$. Integrations are performed with the volume element of the phase space
\begin{equation}
  d \Gamma = \prod_a \frac{ 1 }{ N_a! } \prod_i \frac{ d^dr_{ai} \, d^dp_{ai} }{ h^d } \, .
  \label{equation::B_070} 
\end{equation}
In the denominator, the constant $h$ normalizes the volume of the phase space. We may use Planck's quantum constant $h$ here 
even though we consider a classical system. The additional prefactors $1 / N_a!$ avoid multiple counting for identical 
particles and originate from quantum theory.

In thermal equilibrium for a grand canonical ensemble the phase-space distribution function is given by
\begin{equation}
  \varrho( \Gamma ) = Z^{-1} \ \exp\Bigl( - \beta \Bigl[ \hat{H} - \mathbf{v} \cdot \hat{\mathbf{P}} 
  - \sum_a \mu_a \hat{N}_a \Bigr] \Bigr) \ .
  \label{equation::B_080} 
\end{equation}
Here, $Z$ is the normalization factor, $\beta = 1 / k_B T$ is the inverse temperature, $\mathbf{v}$ is the velocity,
and $\mu_a$ are the chemical potentials for the several species of the particles. From the normalization condition
\begin{equation}
  \int d\Gamma \ \varrho( \Gamma ) = 1
  \label{equation::B_090} 
\end{equation}
the normalization constant $Z = Z(T,\mathbf{v},\mu)$ is calculated which may be interpreted as the partition function
\begin{equation}
  Z(T,\mathbf{v},\mu) = \int d\Gamma \ \exp\Bigl( - \beta \Bigl[ \hat{H} - \mathbf{v} \cdot \hat{\mathbf{P}} 
  - \sum_a \mu_a \hat{N}_a \Bigr] \Bigr) \ .
  \label{equation::B_100} 
\end{equation}
It depends on the Lagrange parameters of the distribution function \eqref{equation::B_080} which are the temperature $T$, 
the velocity $\mathbf{v}$, and the chemical potentials $\mu_a$ of the several particle species. We note that in 
Eqs.\ \eqref{equation::B_090} and \eqref{equation::B_100} the integration over the phase space includes a summation over all 
possible values of the particles numbers $\{ N_a \}$ so that $Z(T,\mathbf{v},\mu)$ is the \emph{grand canonical} partition function.

In this paper we consider the many-particle system in a \emph{local} thermodynamic equilibrium which is the foundation 
for any macroscopic hydrodynamic or elasticity theory. This means, that the Lagrange parameters $T = T(\mathbf{r})$, 
$\mathbf{v} = \mathbf{v}(\mathbf{r})$, and $\mu_a = \mu_a(\mathbf{r})$ depend weakly on the space coordinate $\mathbf{r}$. 
A weak space dependence means that the Lagrange parameters vary on length scales larger than a mesoscopic length scale $\ell$ 
which is much larger than the mean distance between the atoms $a$ so that $\ell \gg a$. In order to establish the local 
thermodynamic equilibrium we need local densities of the conserved quantities. We define the local momentum density
\begin{equation}
  \hat{\mathbf{j}}(\mathbf{r}) = \mathbf{j}(\mathbf{r},\Gamma) = 
  \sum_{ai} \mathbf{p}_{ai} \ \delta( \mathbf{r} - \mathbf{r}_{ai} )
  \label{equation::B_110} 
\end{equation}
and the local particle densities of the different species
\begin{equation}
  \hat{n}_a(\mathbf{r}) = n_a(\mathbf{r},\Gamma) = \sum_i \delta( \mathbf{r} - \mathbf{r}_{ai} ) \ .
  \label{equation::B_120} 
\end{equation}
As a convention we put a hat onto the symbol of a quantity whenever we suppress the phase-space variables 
$\Gamma = ( \mathbf{r}_{ai},\mathbf{p}_{ai} )$ in the argument. Then, the phase-space distribution function is given 
by the more general form
\begin{equation}
  \begin{split}
    \varrho( \Gamma ) = Z^{-1} \ \exp\Bigl( - \beta \Bigl[ \hat{H} 
    - \int d^dr \ \mathbf{v}(\mathbf{r}) \cdot \hat{\mathbf{j}}(\mathbf{r}) \\ 
    - \sum_a \int d^dr \ \mu_a(\mathbf{r}) \, \hat{n}_a(\mathbf{r}) \Bigr] \Bigr) \ ,
  \end{split}
  \label{equation::B_130} 
\end{equation}
and the partition function is rewritten as
\begin{equation}
  \begin{split}
    Z[T,\mathbf{v},\mu] = \int d\Gamma \ \exp\Bigl( - \beta \Bigl[ \hat{H} 
    - \int d^dr \ \mathbf{v}(\mathbf{r}) \cdot \hat{\mathbf{j}}(\mathbf{r}) \\ 
    - \sum_a \int d^dr \ \mu_a(\mathbf{r}) \, \hat{n}_a(\mathbf{r}) \Bigr] \Bigr) \ .
  \end{split}
  \label{equation::B_140} 
\end{equation}
Now, the partition function $Z = Z[T,\mathbf{v},\mu]$ is a function of the constant temperature $T$ and a functional of 
the locally varying velocity $\mathbf{v}(\mathbf{r})$ and chemical potentials $\mu_a(\mathbf{r})$.

It is more difficult to introduce a local and space dependent temperature $T(\mathbf{r})$. The reason is the pair 
potential in the Hamiltonian \eqref{equation::B_020} because it is nonlocal. For the kinetic energy represented by the 
first term in \eqref{equation::B_020} a local density $\varepsilon_{\rm{kin}}(\mathbf{r})$ can be defined in analogy to 
the momentum density \eqref{equation::B_110}. However, the second term with the potential is nonlocal so that the definition 
of a local potential energy $\varepsilon_{\rm{pot}}(\mathbf{r})$ is difficult. In order to overcome this problem one 
would need a local Hamiltonian. For this purpose one should alternatively consider a local field theory which models 
the interaction by a dynamic field like the electromagnetic field. The field theory might be classical or of quantum 
nature. An approach of this kind was used in our previous paper \cite{Ha16} for the derivation of the macroscopic 
hydrodynamics of a simple fluid from a microscopic system. 

In the first and main part of this paper we restrict our considerations to a constant temperature $T$. We consider the 
\emph{isothermal} elastic deformation of a solid crystal with a periodic lattice structure. For a solid system the change of 
the kinetic energy due to heat production is much smaller than the change of the potential energy by the storage of elastic 
energy. Thus, in a leading approximation the change of temperature, heat, and entropy might be negligible. Hence, it does not 
matter if we either keep the temperature or the entropy constant. The differences in the results are expected to be small.

In the next step we define the grand canonical thermodynamic potential by taking the logarithm
\begin{equation}
  \Omega[T,\mathbf{v},\mu] = - k_B T \ \ln Z[T,\mathbf{v},\mu] \ .
  \label{equation::B_150} 
\end{equation}
Again, $\Omega = \Omega[T,\mathbf{v},\mu]$ is a function of the constant temperature $T$ and a functional of the locally varying 
velocity $\mathbf{v}(\mathbf{r})$ and chemical potentials $\mu_a(\mathbf{r})$. Taking the differential we obtain the well 
known thermodynamic relation which in our case reads
\begin{equation}
  d \Omega = -S \, dT - \int d^dr \ \mathbf{j}(\mathbf{r}) \cdot d \mathbf{v}(\mathbf{r}) 
  - \sum_a \int d^dr \ n_a(\mathbf{r}) \, d \mu_a(\mathbf{r})
  \label{equation::B_160} 
\end{equation}
where $S$ is the \emph{total} entropy of the system, $\mathbf{j}(\mathbf{r})$ is the \emph{local} average momentum density, and
$n_a(\mathbf{r})$ are the \emph{local} particle densities of the different species of particles. One should be aware that here 
in this context the \emph{average} densities $\mathbf{j}(\mathbf{r})$ and $n_a(\mathbf{r})$ differ from the \emph{microscopic} 
densities $\hat{\mathbf{j}}(\mathbf{r})$ and $\hat{n}_a(\mathbf{r})$ defined in Eqs.\ \eqref{equation::B_110} and 
\eqref{equation::B_120} which depend on the phase space variables $\Gamma = ( \mathbf{r}_{ai},\mathbf{p}_{ai} )$ and which 
occur in the microscopic formula \eqref{equation::B_140}. For distinction we put a hat onto the microscopic quantities as seen 
in the equations above. 

Now, we investigate to which extent we can evaluate the phase space integral of the partition function \eqref{equation::B_140}. 
In the Hamiltonian \eqref{equation::B_020} and in the total momentum \eqref{equation::B_060} the particle momenta $\mathbf{p}_{ai}$ 
appear quadratically and linearly, respectively. Consequently, the argument of the exponential in the partition function 
\eqref{equation::B_140} is a quadratic polynomial in $\mathbf{p}_{ai}$. Hence the integrals over $\mathbf{p}_{ai}$ are Gaussian 
integrals which can be evaluated exactly. The remaining integrals over the particle coordinates $\mathbf{r}_{ai}$ are evaluated by 
constructing the Mayer-Ursell cluster expansion \cite{MM41,Ur27}. For this purpose we define the fugacities
\begin{equation}
  z_a( \mathbf{r}_{ai} ) = \lambda_a^{-d} \, \exp( \beta [ \mu_a( \mathbf{r}_{ai} ) 
  + \textstyle{\frac{ 1 }{ 2 }} m_a (\mathbf{v}( \mathbf{r}_{ai} ))^2 ] )
  \label{equation::B_170} 
\end{equation}
and the Mayer functions
\begin{equation}
  f_{ab}( \mathbf{r}_{ai} - \mathbf{r}_{bj} ) = \exp( - \beta \, V_{ab}( \mathbf{r}_{ai} - \mathbf{r}_{bj} ) ) - 1
  \label{equation::B_180} 
\end{equation}
where 
\begin{equation}
  \lambda_a = h / \sqrt{ 2 \pi m_a k_B T }
  \label{equation::B_190} 
\end{equation}
is the thermal wave length of a particle of species $a$. The cluster expansion series is obtained by first drawing all diagrams 
made of \emph{nodes} and \emph{lines}, then identifying the elements
\begin{eqnarray}
  \mbox{node} &=& z_a( \mathbf{r}_{ai} ) \ ,
  \label{equation::B_200} \\
  \mbox{line} &=& f_{ab}( \mathbf{r}_{ai} - \mathbf{r}_{bj} ) \ ,
  \label{equation::B_210} 
\end{eqnarray}
integrating over all particle coordinates $\mathbf{r}_{ai}$, and finally summing up all terms. Thus, for the partition function 
we obtain
\begin{equation}
  Z[T,\mathbf{v},\mu] = 1 + \begin{Bmatrix} \mbox{sum of all Mayer diagrams} \end{Bmatrix} \, .
  \label{equation::B_220} 
\end{equation}
Taking the logarithm only the connected diagrams survive. Consequently, for the grand canonical thermodynamic potential 
\eqref{equation::B_150} we find 
\begin{equation}
  \Omega[T,\mathbf{v},\mu] = - k_B T \, \begin{Bmatrix} \mbox{sum of all connected} \\ \mbox{Mayer diagrams} \end{Bmatrix} \, .
  \label{equation::B_230} 
\end{equation}
The cluster expansion is well known. The details can be found in standard textbooks like Refs.\ \onlinecite{Re80} and \onlinecite{HM86}.

We note that the chemical potentials $\mu_a( \mathbf{r}_{ai} )$ and the velocity field $\mathbf{v}( \mathbf{r}_{ai} )$ enter only 
via the fugacities \eqref{equation::B_170} in the combinations shown in the argument of the exponential function. These combinations 
express the \emph{Galilean invariance} of our interacting many-particle system.

\section{Density-functional theory}
\label{section::03}
In a further step we define the \emph{Helmholtz free energy} by the Legendre transformation
\begin{equation}
  \begin{split}
    F[T,\mathbf{j},n] = &\, \Omega[T,\mathbf{v},\mu] 
    + \int d^dr \ \mathbf{v}(\mathbf{r}) \cdot \mathbf{j}(\mathbf{r}) \\
    &+ \sum_a \int d^dr \ \mu_a(\mathbf{r}) \, n_a(\mathbf{r})
  \end{split}
  \label{equation::C_010} 
\end{equation}
which is a function of the temperature $T$ and a functional of the space dependent momentum density $\mathbf{j}(\mathbf{r})$ and the 
space dependent particle densities $n_a(\mathbf{r})$. Taking the differential in analogy to Eq.\ \eqref{equation::B_160} we obtain 
the thermodynamic relation
\begin{equation}
  d F = -S \, dT + \int d^dr \ \mathbf{v}(\mathbf{r}) \cdot d \mathbf{j}(\mathbf{r}) 
  + \sum_a \int d^dr \ \mu_a(\mathbf{r}) \, d n_a(\mathbf{r}) \, .
  \label{equation::C_020} 
\end{equation}
The Legendre transformation can be performed explicitly for the Mayer-Ursell expansion. The free energy splits into three terms 
according to
\begin{equation}
  F[T,\mathbf{j},n] = F_0[T,n] + F_\mathrm{corr}[T,n] + F_\mathrm{kin}[\mathbf{j},n] \ .
  \label{equation::C_030} 
\end{equation}
The three contributions are the free energy of the noninteracting particle system
\begin{equation}
  F_0[T,n] = k_B T \sum_a \int d^dr \ n_a(\mathbf{r}) \, [ \ln( ( \lambda_a )^d \, n_a(\mathbf{r}) ) - 1 ] \, ,
  \label{equation::C_040} 
\end{equation}
the correlation energy
\begin{equation}
  F_\mathrm{corr}[T,n] = - k_B T \, \begin{Bmatrix} \mbox{sum of all irreducible} \\ \mbox{Mayer diagrams} \end{Bmatrix} \, ,
  \label{equation::C_050} 
\end{equation}
and the kinetic energy
\begin{equation}
  F_\mathrm{kin}[\mathbf{j},n] = \int d^dr \ \frac{ [ \mathbf{j}(\mathbf{r}) ]^2 }{ 2 \sum_a m_a \, n_a(\mathbf{r}) } \, .
  \label{equation::C_060} 
\end{equation}
In the noninteracting part \eqref{equation::C_040} the temperature dependence is partially implicit via the thermal wave length 
\eqref{equation::B_190}. The correlation energy \eqref{equation::C_050} represents all contributions caused by the interaction 
potential via the Mayer function \eqref{equation::B_180}. It is given by an infinite sum of Mayer diagrams which are irreducible 
in the sense that they do not fall into pieces if a node is removed. Here in the diagrams the nodes and lines are identified 
with the densities and the Mayer functions according to
\begin{eqnarray}
  \mbox{node} &=& n_a( \mathbf{r}_{ai} ) \ ,
  \label{equation::C_070} \\
  \mbox{line} &=& f_{ab}( \mathbf{r}_{ai} - \mathbf{r}_{bj} ) \ .
  \label{equation::C_080} 
\end{eqnarray}
For the nodes the rule has changed where for the lines it remains the same as before.

The kinetic energy is the only contribution which depends on the local average motion of the condensed matter system via 
the momentum density $\mathbf{j}(\mathbf{r})$. This contribution is given by the exact formula \eqref{equation::C_060} 
which is a consequence of the Galilean invariance of the physical system. In the special case where only elastic properties 
are considered and the physical system is at rest the momentum density $\mathbf{j}(\mathbf{r})$ is zero so that the kinetic 
energy $F_\mathrm{kin}[\mathbf{j},n]$ is zero. In this case the free energy reduces to the non moving contribution
\begin{equation}
  F[T,n] = F_0[T,n] + F_\mathrm{corr}[T,n] \ .
  \label{equation::C_090} 
\end{equation}

A theory which is build upon the Helmholtz free energy $F[T,n]$ is called a \emph{density-functional theory}. While the 
temperature $T$ is assumed to be constant, this free energy is a functional of the particle densities $n_a(\mathbf{r})$. 
If all Mayer diagrams of Eq.\ \eqref{equation::C_050} are taken into account and the complete perturbation series is summed 
up, then our Helmholtz free energy $F[T,n]$ is exact and hence our density-functional theory is exact. This fact might be 
true in principle. However, in practical calculations it can never be realized because the perturbation series cannot be 
summed up exactly. Rather, in practical calculations for the correlation energy $F_\mathrm{corr}[T,n]$ a simplified ansatz 
is used which is called e.g.\ fundamental measure theory (FMT) \cite{Ro89} or White Bear II \cite{Ta00,Ro10}. These ansatzes 
are not exact but approximations. However, they work rather well for interacting many particle systems which form crystalline 
solids.

Our following calculations of the stress tensor and of the elastic constants will be exact within the framework of the 
density-functional theory. This means whenever the free energy $F[T,n]$ is given, the following calculations are exact and 
provide no further approximations. If the free energy is exact, then also the results of the stress tensor and of the elastic 
constants are exact. Vice versa, if the free energy is an approximate density functional, then the results are approximate, too.

\section{Crystalline solids in states of local thermodynamic equilibrium}
\label{section::04}
\subsection{Minimization of the free energy}
\label{section::04a}
From the differential thermodynamic relation \eqref{equation::C_020} we obtain the two functional derivatives
\begin{equation}
  \frac{ \delta F }{ \delta \mathbf{j}(\mathbf{r}) } = \mathbf{v}(\mathbf{r}) \, , \qquad
  \frac{ \delta F }{ \delta n_a(\mathbf{r}) } = \mu_a(\mathbf{r}) \, .
  \label{equation::D_010} 
\end{equation}
If the momentum density $\mathbf{j}(\mathbf{r})$ and the particle densities $n_a(\mathbf{r})$ are given then the velocity 
$\mathbf{v}(\mathbf{r})$ and the chemical potentials $\mu_a(\mathbf{r})$ can be calculated explicitly by the functional 
derivatives.

Alternatively, the two equations \eqref{equation::D_010} may be interpreted as the necessary conditions for the minimization 
of the Helmholtz free energy $F[T,\mathbf{j},n]$ under certain constraint conditions. If on the right-hand sides the 
velocity $\mathbf{v}(\mathbf{r})$ and the chemical potentials $\mu_a(\mathbf{r})$ are given then the momentum density 
$\mathbf{j}(\mathbf{r})$ and the particle densities $n_a(\mathbf{r})$ are determined by solving the equations.

Here the \emph{function class} is important to which these functions belong, whether a function varies slowly on a macroscopic 
scale or whether it varies fast on a microscopic scale. For a crystal the particle densities $n_a(\mathbf{r})$ and also 
the momentum density $\mathbf{j}(\mathbf{r})$ have a periodic structure on a microscopic scale $a$ which is the mean 
particle spacing. On the other hand, the chemical potentials $\mu_a(\mathbf{r})$ and the velocity $\mathbf{v}(\mathbf{r})$ 
must be either constant in thermal equilibrium or slowly varying on a macroscopic scale in order to describe a local 
thermodynamic equilibrium.

We denote by $f(\mathbf{r})$, $g(\mathbf{r})$, etc.\ some arbitrary functions which depend on the space position $\mathbf{r}$. 
As a minimum property we require that these functions are sufficiently smooth or at least continuous in space. We denote 
this class of functions by $\mathcal{C}$. We define a scalar product 
\begin{equation}
  \langle f \vert g \rangle = \int d^dr \, [ f(\mathbf{r}) ]^* \, g(\mathbf{r}) 
  \label{equation::D_020} 
\end{equation}
where the asterisk means taking the complex conjugate value of the first factor. In some cases as here in the scalar 
product we must assume that the functions decay sufficiently fast at infinity for $\mathbf{r} \to \infty$ so that the 
integrals are well defined and converge.

We use the mesoscopic scale $\ell$ much larger than the mean particle distance $a$ in order to split the function class 
$\mathcal{C}$ into two subclasses $\tilde{\mathcal{C}}(\ell)$ and $\mathcal{C}^\prime(\ell)$. We define $\tilde{\mathcal{C}}(\ell)$ 
as the subclass of all functions $\tilde{f}(\mathbf{r})$, $\tilde{g}(\mathbf{r})$, etc.\ which vary only on macroscopic 
scales which are larger than $\ell$. On the other hand, we define $\mathcal{C}^\prime(\ell)$ as the complementary 
orthogonal class of functions $f^\prime(\mathbf{r})$, $g^\prime(\mathbf{r})$, etc.\ which vary on microscopic scales.
Then, the total function class $\mathcal{C}$ may be written as the direct sum of the two subclasses as
\begin{equation}
  \mathcal{C} = \tilde{\mathcal{C}}(\ell) \oplus \mathcal{C}^\prime(\ell) \, .
  \label{equation::D_030} 
\end{equation}
We note that the function classes can be interpreted as vector spaces of infinite dimensions.

We may define the function $\tilde{f}(\mathbf{r}) \in \tilde{\mathcal{C}}(\ell)$ by the \emph{coarse-graining} procedure
\begin{equation}
  \tilde{f}(\mathbf{r}) = \int d^dr^\prime \, w( \mathbf{r} - \mathbf{r}^\prime ) \, f(\mathbf{r}^\prime)
  \label{equation::D_040} 
\end{equation}
where $w( \mathbf{r} - \mathbf{r}^\prime )$ is an integral kernel which smears out the original function 
$f(\mathbf{r}) \in \mathcal{C}$ over the mesoscopic scale $\ell$ so that all microscopic variations are smoothed. 
We may assume that the integration kernel satisfies the projection property
\begin{equation}
  w( \mathbf{r} - \mathbf{r}^{\prime\prime} ) = \int d^dr^\prime \, w( \mathbf{r} - \mathbf{r}^\prime ) 
  \, w( \mathbf{r}^\prime - \mathbf{r}^{\prime\prime} ) \, .
  \label{equation::D_050} 
\end{equation}
A possible choice for this integration kernel is the Fourier integral
\begin{equation}
  w( \mathbf{r} - \mathbf{r}^\prime ) = \int \frac{ d^dk }{ ( 2 \pi )^d } \, \theta( \Lambda^2 - \mathbf{k}^2 ) \, 
  \exp( i \mathbf{k} \cdot ( \mathbf{r} - \mathbf{r}^\prime ) ) 
  \label{equation::D_060} 
\end{equation}
where the Heaviside theta function restricts the wave vectors $\mathbf{k}$ to the maximum absolute value $\Lambda = 2\pi / \ell$ 
which is related to the mesoscopic length $\ell$. Then, the coarse-graining procedure \eqref{equation::D_040} may be 
interpreted as a projection operation. On the other hand, the related complement function 
$f^\prime(\mathbf{r}) \in \mathcal{C}^\prime(\ell)$ is defined in order to enable the decomposition
\begin{equation}
  f(\mathbf{r}) = \tilde{f}(\mathbf{r}) + f^\prime(\mathbf{r}) \, .
  \label{equation::D_070} 
\end{equation}

The chemical potentials $\mu_a(\mathbf{r})$ and the velocity $\mathbf{v}(\mathbf{r})$ are slowly varying in space so 
that they belong to the coarse grained function class $\tilde{\mathcal{C}}(\ell)$ so that 
$\mu_a(\mathbf{r}), \mathbf{v}(\mathbf{r}) \in \tilde{\mathcal{C}}(\ell)$. Consequently, if we apply the coarse-graining 
procedure \eqref{equation::D_040} we find the relations
\begin{equation}
  \mu_a(\mathbf{r}) = \tilde{\mu}_a(\mathbf{r}) \, , \qquad
  \mathbf{v}(\mathbf{r}) = \tilde{\mathbf{v}}(\mathbf{r}) \, .
  \label{equation::D_080} 
\end{equation}
The related complement functions are just zero so that $\mu^\prime_a(\mathbf{r}) = 0$ and $\mathbf{v}^\prime(\mathbf{r}) = \mathbf{0}$.

On the other hand, the particle densities $n_a(\mathbf{r})$ and the momentum density $\mathbf{j}(\mathbf{r})$ vary spatially 
on macroscopic and microscopic scales so that they belong to the full function class 
$\mathcal{C} = \tilde{\mathcal{C}}(\ell) \oplus \mathcal{C}^\prime(\ell)$. Consequently we obtain the decomposition into 
a macroscopic coarse grained function and a microscopic complement function according to
\begin{equation}
  n_a(\mathbf{r}) = \tilde{n}_a(\mathbf{r}) + n^\prime_a(\mathbf{r}) \, , \qquad
  \mathbf{j}(\mathbf{r}) = \tilde{\mathbf{j}}(\mathbf{r}) + \mathbf{j}^\prime(\mathbf{r}) \, .
  \label{equation::D_090} 
\end{equation}
For a crystal, the complement functions $n^\prime_a(\mathbf{r})$ and $\mathbf{j}^\prime(\mathbf{r})$ are nonzero. 

Now, the minimization of the free energy must be formulated as
\begin{equation}
  \tilde{F}[T,\tilde{\mathbf{j}},\tilde{n}] = \min_{\mathbf{j}^\prime, n^\prime} 
  F[T,\mathbf{j} = \tilde{\mathbf{j}} + \mathbf{j}^\prime,n = \tilde{n} + n^\prime] \, .
  \label{equation::D_100} 
\end{equation}
For the minimization the complement functions $n^\prime_a(\mathbf{r})$ and $\mathbf{j}^\prime(\mathbf{r})$ 
are varied where the coarse grained functions $\tilde{n}_a(\mathbf{r})$ and $\tilde{\mathbf{j}}(\mathbf{r})$ 
are kept constant and work as constraints. The necessary conditions for the minimization remain unchanged 
and are given by Eqs.\ \eqref{equation::D_010} as before. However, now on the right-hand sides the coarse 
grained functions \eqref{equation::D_080} are used where on the left-hand sides the decompositions 
\eqref{equation::D_090} are inserted. The complement functions $n^\prime_a(\mathbf{r})$ and 
$\mathbf{j}^\prime(\mathbf{r})$ are determined by solving these equations where the coarse grained functions 
$\tilde{n}_a(\mathbf{r})$ and $\tilde{\mathbf{j}}(\mathbf{r})$ are given by the constraints.

\subsection{Galilean invariance}
\label{section::04b}
The Galilean invariance of the physical system considerably simplifies the structure of the free energy 
functional $F[T,\mathbf{j},n]$ defined in \eqref{equation::C_030}. Only the kinetic energy \eqref{equation::C_060} 
depends on the momentum density $\mathbf{j}$, and this in a simple quadratic form. As a consequence, for the minimization 
the first of the necessary conditions \eqref{equation::D_010} considerably simplifies into
\begin{equation}
  \frac{ \delta F }{ \delta \mathbf{j}(\mathbf{r}) } = \frac{ \delta F_\mathrm{kin} }{ \delta \mathbf{j}(\mathbf{r}) } 
  = \frac{ \mathbf{j}(\mathbf{r}) }{ \sum_a m_a \, n_a(\mathbf{r}) } = \mathbf{v}(\mathbf{r}) \, .
  \label{equation::D_110} 
\end{equation}
Thus, we obtain the well known equation
\begin{equation}
  \mathbf{j}(\mathbf{r}) = \rho(\mathbf{r}) \, \mathbf{v}(\mathbf{r})
  \label{equation::D_120} 
\end{equation}
where 
\begin{equation}
  \rho(\mathbf{r}) = \sum_a m_a \, n_a(\mathbf{r})
  \label{equation::D_130} 
\end{equation}
may be interpreted as the mass density. As required from Newtonian mechanics the momentum density is the mass density 
times the velocity. We note that all involved functions are elements of the general function space $\mathcal{C}$ which 
means that $n_a(\mathbf{r}), \rho(\mathbf{r}), \mathbf{v}(\mathbf{r}), \mathbf{j}(\mathbf{r}) \in \mathcal{C}$.

We may now ask the question to which extent we can split the equations \eqref{equation::D_120} and \eqref{equation::D_130} 
into the macroscopic and the microscopic part according to the decomposition of the function class \eqref{equation::D_030} 
into a coarse grained macroscopic class and a complement microscopic class. Function classes are vector spaces. For this 
reason, a linear equation can be split easily. Thus, if we insert Eq.\ \eqref{equation::D_090} and use the decomposition 
\begin{equation}
  \rho(\mathbf{r}) = \tilde{\rho}(\mathbf{r}) + \rho^\prime(\mathbf{r})
  \label{equation::D_140} 
\end{equation}
then from Eq.\ \eqref{equation::D_130} we obtain
\begin{eqnarray}
  \tilde{\rho}(\mathbf{r}) &=& \sum_a m_a \, \tilde{n}_a(\mathbf{r}) \, ,
  \label{equation::D_150} \\
  \rho^\prime(\mathbf{r}) &=& \sum_a m_a \, n^\prime_a(\mathbf{r}) \, .
  \label{equation::D_160} 
\end{eqnarray}
For the nonlinear equation \eqref{equation::D_120} the situation is more complicated since on the right-hand side 
two functions are multiplied. Nevertheless, if we insert Eqs.\ \eqref{equation::D_080}, \eqref{equation::D_090}, 
and \eqref{equation::D_140} we obtain approximately
\begin{eqnarray}
  \tilde{\mathbf{j}}(\mathbf{r}) &\approx& \tilde{\rho}(\mathbf{r}) \, \tilde{\mathbf{v}}(\mathbf{r}) \, ,
  \label{equation::D_170} \\
  \mathbf{j}^\prime(\mathbf{r}) &\approx& \rho^\prime(\mathbf{r}) \, \tilde{\mathbf{v}}(\mathbf{r}) \, .
  \label{equation::D_180} 
\end{eqnarray}
We note that the velocity field defined in \eqref{equation::D_080} has only a macroscopic contribution. This 
fact implies that the right-hand side of Eq.\ \eqref{equation::D_170} may have a small contribution in the 
complement space $\mathcal{C}^\prime(\ell)$ where the right-hand side of Eq.\ \eqref{equation::D_180} 
may have a small contribution in the macroscopic space $\tilde{\mathcal{C}}(\ell)$. These small contributions 
imply that the \emph{approximate equal} sign $\approx$ must be used here for mathematical precision. However, the 
small contributions will vary on length scales close to the mesoscopic scale $\ell$ which divides the function space 
$\mathcal{C}$ into the macroscopic space $\tilde{\mathcal{C}}(\ell)$ and the complement space $\mathcal{C}^\prime(\ell)$.
While from mathematical point of view this division is sharp and located precisely at $\ell$, for physical 
arguments a smeared length scale is sufficient which is known only up to an order of magnitude. For this reason, 
for our physical discussions we may replace the \emph{approximate equal} signs $\approx$ by \emph{equal} signs 
$=$ in Eqs.\ \eqref{equation::D_170} and \eqref{equation::D_180} and ignore the mathematical subtleties.

Thus, we conclude that the important relation of Galilean invariance $\mathbf{j} = \rho \mathbf{v}$ may be written 
in both versions, in the microscopic version \eqref{equation::D_120} together with \eqref{equation::D_130} and 
in the macroscopic coarse grained version \eqref{equation::D_170} together with \eqref{equation::D_150}.

\subsection{Normal and cluster crystals}
\label{section::04c}
For investigating the elastic properties of a crystal we assume that the motions and the deformations of the system 
are very slow. In this case we can assume that the momentum density $\mathbf{j}(\mathbf{r})$ and the 
velocity field $\mathbf{v}(\mathbf{r})$ are nearly zero. Thus, we simply set $\mathbf{j}(\mathbf{r}) = \mathbf{0}$ 
and $\mathbf{v}(\mathbf{r}) = \mathbf{0}$. As a consequence, the first of the necessary conditions for the minimization 
\eqref{equation::D_010} which explicitly is given by Eqs.\ \eqref{equation::D_110} or \eqref{equation::D_120} is 
satisfied identically.

As a further consequence the kinetic energy \eqref{equation::C_060} is zero so that the total free energy 
\eqref{equation::C_030} reduces to the first two terms which are written in Eq.\ \eqref{equation::C_090}. There 
remains the second of the necessary conditions for the minimization \eqref{equation::D_010} which reads
\begin{equation}
  \begin{split}
    \frac{ \delta F }{ \delta n_a(\mathbf{r}) } &= \frac{ \delta F_0 }{ \delta n_a(\mathbf{r}) } 
    + \frac{ \delta F_\mathrm{corr} }{ \delta n_a(\mathbf{r}) } \\ 
    &= k_B T \, \Bigl[ \ln[ ( \lambda_a )^d \, n_a(\mathbf{r}) ] - c^{(1)}_a(\mathbf{r}) \Bigr] = \mu_a(\mathbf{r}) \, .
  \end{split}
  \label{equation::D_190} 
\end{equation}
Here $c^{(1)}_a(\mathbf{r})$ is the one-point direct correlation function which arises from the functional derivatives 
of the correlation energy $F_\mathrm{corr}[T,n]$.

In a first step we may insert $\mu_a(\mathbf{r}) = \tilde{\mu}_a(\mathbf{r})$ on the right-hand side and calculate 
the microscopic particle densities $n_a(\mathbf{r}) = \tilde{n}_a(\mathbf{r}) + n^\prime_a(\mathbf{r})$ by solving 
Eq.\ \eqref{equation::D_190}. In this way we obtain the \emph{microscopic} particle densities for given \emph{macroscopic} 
chemical potentials. In a second step we may vary the macroscopic functions $\tilde{\mu}_a(\mathbf{r})$, solve 
Eq.\ \eqref{equation::D_190} again and observe how the macroscopic densities $\tilde{n}_a(\mathbf{r})$ change. We may 
adjust the macroscopic chemical potentials $\tilde{\mu}_a(\mathbf{r})$ in such a way so that the macroscopic densities 
$\tilde{n}_a(\mathbf{r})$ tend to the given constraint functions. In this way we obtain the \emph{microscopic} densities 
for given \emph{macroscopic} particle densities. 

Within the framework of thermodynamics we may interpret the macroscopic functions $\tilde{n}_a(\mathbf{r})$ and 
$\tilde{\mu}_a(\mathbf{r})$ as \emph{conjugate} variables. The densities $\tilde{n}_a(\mathbf{r})$ are \emph{extensive} 
variables divided by volume while the chemical potentials $\tilde{\mu}_a(\mathbf{r})$ are \emph{intensive} variables. 
As a consequence the macroscopic functions $\tilde{n}_a(\mathbf{r})$ and $\tilde{\mu}_a(\mathbf{r})$ are not independent. 
Rather they are related to each other by a Legendre transformation.

However, for given functions $\tilde{\mu}_a(\mathbf{r})$ or $\tilde{n}_a(\mathbf{r})$ the solution of 
Eq.\ \eqref{equation::D_190} is not unique. The microscopic densities 
$n_a(\mathbf{r}) = \tilde{n}_a(\mathbf{r}) + n^\prime_a(\mathbf{r})$ will have a crystal structure which is 
represented by the complement function $n^\prime_a(\mathbf{r})$. It is well known, that many different lattice structures 
with different symmetry properties exist for crystals. Furthermore, the crystal may be deformed, especially in the case 
if the given functions are not constant but vary slowly within the space. Consequently, we need some more variables to specify 
the physical state and the related microscopic densities $n_a(\mathbf{r}) = \tilde{n}_a(\mathbf{r}) + n^\prime_a(\mathbf{r})$.

A special case is a crystal in thermal equilibrium with a uniform regular lattice in space. In this case besides the 
temperature $T$ all the other macroscopic functions are constant in space so that
\begin{equation}
  \tilde{\mu}_{0,a}(\mathbf{r}) = \tilde{\mu}_{0,a} \, , \qquad
  \tilde{n}_{0,a}(\mathbf{r}) = \tilde{n}_{0,a} \, .
  \label{equation::D_200} 
\end{equation}
Then the microscopic densities $n_a(\mathbf{r}) = \tilde{n}_a(\mathbf{r}) + n^\prime_a(\mathbf{r})$ show a perfect 
periodic structure and satisfy
\begin{equation}
  n_{0,a}(\mathbf{r}) = n_{0,a}(\mathbf{r} + \mathbf{a}) \quad \mbox{for all $\mathbf{a} \in \mathcal{A}$}
  \label{equation::D_210} 
\end{equation}
where we define the set of all lattice vectors by 
\begin{equation}
  \mathcal{A} = \Bigl\{ \mathbf{a} \, \Big| \, \mathbf{a} = \sum_{i=1}^d x_i \, \mathbf{a}_i , \, x_i \in \mathcal{Z} \Bigr\}
  \label{equation::D_220} 
\end{equation}
with basis vectors $\mathbf{a}_i$ and integer numbers $x_i \in \mathcal{Z}$. Thus, as an additional variable to 
describe the lattice of a crystal we have the lattice-vector set $\mathcal{A}$.

In a normal crystal there is usually one particle on each lattice site. The reason is that in most cases the interaction 
potential $V_{ab}( \mathbf{r}_{ai} - \mathbf{r}_{bj} )$ has a repulsive hard core so that
\begin{equation}
  V_{ab}( \mathbf{r}_{ai} - \mathbf{r}_{bj} ) \to + \infty \quad \mbox{for $\mathbf{r}_{ai} - \mathbf{r}_{bj} \to \mathbf{0}$ } \, .
  \label{equation::D_230} 
\end{equation}
This fact does not allow that two particles come close to each other. It is observed for hard spheres, Lennard-Jones potentials 
and also repulsive Coulomb potentials.

On the other hand, there might be potentials with a finite core as e.g.
\begin{equation}
  V_{ab}( \mathbf{r}_{ai} - \mathbf{r}_{bj} ) = V_{ab} \, \exp( - \lambda_{ab} \, | \mathbf{r}_{ai} - \mathbf{r}_{bj} |^4 ) 
  \label{equation::D_240} 
\end{equation}
so that the related repulsive forces 
$\mathbf{F}_{ab}( \mathbf{r}_{ai} - \mathbf{r}_{bj} ) = - \nabla V_{ab}( \mathbf{r}_{ai} - \mathbf{r}_{bj} )$ decay to zero 
rapidly for $\mathbf{r}_{ai} - \mathbf{r}_{bj} \to \mathbf{0}$.
In this case two or more particles may be close to each other on one lattice site and form clusters. This type of crystals 
are called \emph{cluster crystals}. They can be realized experimentally in colloidal systems of small spheres in solvents 
and theoretically in computer simulations. For finite temperatures, the average numbers of particles on a lattice site even 
need not be \emph{integer}. Rather they may have some continuous value in between some integers.

Our approach using the free energy $F[T,\mathbf{j},n]$ as a density functional never requires that only one particle is 
on each lattice site. The number of particles per lattice site is completely free and will be determined by the minimization 
procedure \eqref{equation::D_100} or by the related necessary conditions \eqref{equation::D_010}.

\subsection{Deformation of the crystal and the displacement field}
\label{section::04d}
As a reference state we consider a crystal with a perfect homogeneous structure and a periodic lattice. The densities 
are given by the functions $n_{0,a}(\mathbf{r}_0)$ which we have considered before. We assume that these densities 
minimize the free energy because we calculate them by solving the necessary conditions \eqref{equation::D_190}. The 
positions in space are denoted by the vector
\begin{equation}
  \mathbf{r}_0 = r_0^i \, \mathbf{e}_i \, .
  \label{equation::D_250} 
\end{equation}
In order to indicate the \emph{reference state} we use a subscript zero for the densities and for the coordinate vectors.
Here $\mathbf{e}_i$ are the basis vectors of a Cartesian coordinate system which are orthonormalized and satisfy the 
scalar products $\mathbf{e}_i \cdot \mathbf{e}_j = \delta_{ij}$ where $\delta_{ij}$ are Kronecker symbols. $r_0^i$ 
denote the coordinates of the positions. We use the notations and conventions of General Relativity. This means that 
we use \emph{upper} indices $i$ for the coordinates $r_0^i$ and lower indices $i$ for the basis vectors $\mathbf{e}_i$ 
and for partial derivatives $\partial_{0,i} = \partial / \partial r_0^i$. Furthermore, we use the \emph{sum convention} 
for vector and tensor indices. If there are two equal indices, one upper and one lower, then we automatically sum over 
the index according to $r_0^i \, \mathbf{e}_i = \sum_{i=1}^d r_0^i \, \mathbf{e}_i$.

On the other hand we consider the crystal in a deformed state which is described by the density functions $n_a(\mathbf{r})$.
The positions in space are denoted by the vector
\begin{equation}
  \mathbf{r} = r^i \, \mathbf{e}_i \, .
  \label{equation::D_260} 
\end{equation}
In this case the densities and the coordinate vectors have no subscript.

Now, we are looking for a nonlinear coordinate transformation which describes the deformation process and which connects 
the two different states of the crystal. In the reference state, we consider a point with the coordinate $\mathbf{r}_0$. 
We perform a deformation and shift this point by a \emph{displacement} vector $\mathbf{u} = \mathbf{u}(\mathbf{r})$. 
Afterwards the point is located at the new coordinate $\mathbf{r}$ which is given by
\begin{equation}
  \mathbf{r} = \mathbf{r}_0 + \mathbf{u}(\mathbf{r}) \, .
  \label{equation::D_270} 
\end{equation}
Since $\mathbf{u} = \mathbf{u}(\mathbf{r})$ is a function of the space variable $\mathbf{r}$ Eq.\ \eqref{equation::D_270} 
is a \emph{nonlinear} coordinate transformation. The inverse transformation is given by the related equation
\begin{equation}
  \mathbf{r}_0 = \mathbf{r} - \mathbf{u}(\mathbf{r}) \, .
  \label{equation::D_280} 
\end{equation}
Here we may insert the basis-vector representations \eqref{equation::D_250}, \eqref{equation::D_260}, and 
$\mathbf{u}(\mathbf{r}) = u^i(\mathbf{r}) \, \mathbf{e}_i$. Then we find the nonlinear transformation in coordinate 
representation
\begin{equation}
  r_0^i = r^i - u^i(\mathbf{r}) \, .
  \label{equation::D_290} 
\end{equation}

In the next step we construct the density functions $n_a(\mathbf{r})$ from the reference functions $n_{0,a}(\mathbf{r}_0)$ 
by the nonlinear coordinate transformation and the formula
\begin{equation}
  n_a(\mathbf{r}) \, d^dr = n_{0,a}(\mathbf{r}_0) \, d^dr_0 \, .
  \label{equation::D_300} 
\end{equation}
This formula guarantees that the particle numbers of the two states are equal if we integrate over the whole space 
which includes the crystal so that $N_a = N_{0,a}$. The volume elements $d^dr_0$ and $d^dr$ are transformed by the 
formula
\begin{equation}
  d^dr_0 = \frac{ \partial r_0 }{ \partial r } \, d^dr 
  \label{equation::D_310} 
\end{equation}
where 
\begin{equation}
  \frac{ \partial r_0 }{ \partial r } = \det \Bigl\{ \frac{ \partial r_0^i }{ \partial r^j } \Bigr\}
  \label{equation::D_320} 
\end{equation}
is the \emph{Jacobi determinant} and
\begin{equation}
  \frac{ \partial r_0^i }{ \partial r^j } = \delta^i_{\ j} - \partial_j u^i
  \label{equation::D_330} 
\end{equation}
is the \emph{Jacobi matrix} of the nonlinear transformation. If we insert Eq.\ \eqref{equation::D_310} into 
Eq.\ \eqref{equation::D_300} and omit the volume element $d^dr$ on both sides, then we obtain the densities 
of the deformed state in terms of the densities of the reference state 
\begin{equation}
  n_a(\mathbf{r}) = n_{0,a}(\mathbf{r}_0) \, \frac{ \partial r_0 }{ \partial r } 
  = n_{0,a}( \mathbf{r} - \mathbf{u}(\mathbf{r}) ) \, \frac{ \partial r_0 }{ \partial r } \, .
  \label{equation::D_340} 
\end{equation}
These densities are \emph{constructed} by the deformation. However, while the reference densities $n_{0,a}(\mathbf{r}_0)$ 
minimize the free energy $F[T,n]$ the deformed densities $n_a(\mathbf{r})$ as defined in Eq.\ \eqref{equation::D_340} in 
general do not. Consequently, we must add some correction terms $\Delta n^\prime_a(\mathbf{r})$ in order to improve the 
densities. Thus, we write
\begin{equation}
  n_a(\mathbf{r}) = n_{0,a}( \mathbf{r} - \mathbf{u}(\mathbf{r}) ) \, \frac{ \partial r_0 }{ \partial r } 
  + \Delta n^\prime_a(\mathbf{r}) \, .
  \label{equation::D_350} 
\end{equation}
The correction terms are determined by inserting the densities \eqref{equation::D_350} into the necessary conditions 
\eqref{equation::D_190} of the minimization procedure and solving the resulting equations with respect to 
$\Delta n^\prime_a(\mathbf{r})$. In this way the deformed densities $n_a(\mathbf{r})$ minimize the free energy $F[T,n]$ 
either.

For crystals the densities vary on the microscopic scale. Hence, we expect that all terms in Eq.\ \eqref{equation::D_350} 
are functions in the function class $\mathcal{C} = \tilde{\mathcal{C}}(\ell) \oplus \mathcal{C}^\prime(\ell)$. 
However, the last term is an exception. The correction term $\Delta n^\prime_a(\mathbf{r})$ represents the densities 
which are adjusted in the minimization procedure \eqref{equation::D_100}. Here only functions in the complement space 
$\mathcal{C}^\prime(\ell)$ are adjusted while the contributions in the macroscopic space $\tilde{\mathcal{C}}(\ell)$ 
are kept fixed as constraints. Consequently, the correction term $\Delta n^\prime_a(\mathbf{r})$ is restricted to 
the complement space $\mathcal{C}^\prime(\ell)$ where we have added a prime to the symbol in order to indicate this fact.

The deformation is a \emph{macroscopic} phenomenon. It is parameterized by the displacement field 
$\mathbf{u}(\mathbf{r}) = u^i(\mathbf{r}) \, \mathbf{e}_i$ with functions $u^i(\mathbf{r})$ in the \emph{macroscopic} 
function space $\tilde{\mathcal{C}}(\ell)$. For this reason also the Jacobi matrix \eqref{equation::D_330} and the 
Jacobi determinant \eqref{equation::D_320} are macroscopic functions of $\tilde{\mathcal{C}}(\ell)$. We consider 
the first term on the right-hand side of Eq.\ \eqref{equation::D_350}. If we add a macroscopic function to the 
argument of a microscopic function as done in $n_{0,a}( \mathbf{r} - \mathbf{u}(\mathbf{r}) )$ or if we multiply 
by a macroscopic function $\partial r_0 / \partial r$ the resulting function stays in the full function class 
$\mathcal{C}$. Thus, the first term on the right-hand side of Eq.\ \eqref{equation::D_350} does not make troubles.

Now, we apply the coarse-graining procedure \eqref{equation::D_040} and project the microscopic densities 
\eqref{equation::D_350} to the macroscopic function space $\tilde{\mathcal{C}}(\ell)$ and obtain
\begin{equation}
  \tilde{n}_a(\mathbf{r}) = \tilde{n}_{0,a}( \mathbf{r} - \mathbf{u}(\mathbf{r}) ) \, \frac{ \partial r_0 }{ \partial r } \, .
  \label{equation::D_360} 
\end{equation}
The last term on the right-hand side is gone because the correction term is restricted to the complement space 
and hence has no contribution in the macroscopic function space. In the first term on the right-hand side we 
have performed some approximations since for simplicity we have applied the projection to each function separately. 
This simplification implies that this term is not perfectly restricted to the macroscopic function space 
$\tilde{\mathcal{C}}(\ell)$ but may have some small contributions in the complement space $\mathcal{C}^\prime(\ell)$ 
because we have a macroscopic function in the argument and we have a product of two macroscopic functions. 
This fact might be a problem for strict mathematics. However, for our physical argumentations and conclusions 
it can be ignored.

If we consider the terms on the right-hand sides of Eqs.\ \eqref{equation::D_350} and \eqref{equation::D_360} 
we clearly see that the \emph{microscopic} and the \emph{macroscopic} densities are affected by the deformation 
in the same functional way. In both cases we insert the argument $\mathbf{r}_0 = \mathbf{r} - \mathbf{u}(\mathbf{r})$ 
into the reference densities and we multiply by the Jacobi determinant $\partial r_0 / \partial r$. In order to 
obtain an alternative expression for the deformed density we may eliminate the Jacobi determinant from the two 
equations. For this reason, we resolve the macroscopic Eq.\ \eqref{equation::D_360} with respect to 
$\partial r_0 / \partial r$ and insert the resulting expression into the microscopic Eq.\ \eqref{equation::D_350}. 
Thus, we obtain
\begin{equation}
  n_a(\mathbf{r}) = \frac{ n_{0,a}( \mathbf{r} - \mathbf{u}(\mathbf{r}) ) }{ \tilde{n}_{0,a}( \mathbf{r} - \mathbf{u}(\mathbf{r}) ) } 
  \, \tilde{n}_a(\mathbf{r}) + \Delta n^\prime_a(\mathbf{r}) \, .
  \label{equation::D_370} 
\end{equation}
We may subtract a reference density $n_{0,a}(\mathbf{r})$ from both sides and split the first term on the right-hand 
side into two terms. Then, we obtain
\begin{equation}
  \begin{split}
    n_a(\mathbf{r}) - n_{0,a}(\mathbf{r}) =& 
    \, \frac{ \tilde{n}_{0,a}(\mathbf{r}) }{ \tilde{n}_{0,a}( \mathbf{r} - \mathbf{u}(\mathbf{r}) ) } 
    \, n_{0,a}( \mathbf{r} - \mathbf{u}(\mathbf{r}) ) - n_{0,a}(\mathbf{r}) \\
    &+ \frac{ n_{0,a}( \mathbf{r} - \mathbf{u}(\mathbf{r}) ) }{ \tilde{n}_{0,a}( \mathbf{r} - \mathbf{u}(\mathbf{r}) ) } 
    \, [ \tilde{n}_a(\mathbf{r}) - \tilde{n}_{0,a}(\mathbf{r}) ] \\
    &+ \Delta n^\prime_a(\mathbf{r}) \, .
  \end{split}
  \label{equation::D_380} 
\end{equation}
For the reference state we consider a perfect homogeneous crystal. While the microscopic reference density 
$n_{0,a}(\mathbf{r})$ is a perfect periodic function, the related macroscopic reference density is constant 
in space so that 
\begin{equation}
  \tilde{n}_{0,a}(\mathbf{r}) = \tilde{n}_{0,a} = \mbox{constant} \, .
  \label{equation::D_390} 
\end{equation}
As a consequence, the above formulas \eqref{equation::D_360}, \eqref{equation::D_370} and \eqref{equation::D_380} 
considerably simplify. For the latter formula we obtain
\begin{equation}
  \begin{split}
    n_a(\mathbf{r}) - n_{0,a}(\mathbf{r}) =& 
    \, n_{0,a}( \mathbf{r} - \mathbf{u}(\mathbf{r}) ) - n_{0,a}(\mathbf{r}) \\
    &+ \frac{ n_{0,a}( \mathbf{r} - \mathbf{u}(\mathbf{r}) ) }{ \tilde{n}_{0,a} } 
    \, [ \tilde{n}_a(\mathbf{r}) - \tilde{n}_{0,a} ] \\
    &+ \Delta n^\prime_a(\mathbf{r}) \, .
  \end{split}
  \label{equation::D_400} 
\end{equation}
Until now we have considered only deformations of the crystal. The only free macroscopic functions are the 
components of the displacement field $\mathbf{u}(\mathbf{r}) = u^i(\mathbf{r}) \, \mathbf{e}_i$. On the right-hand 
side the macroscopic density $\tilde{n}_a(\mathbf{r})$ is given by the formula \eqref{equation::D_360} in terms 
of the displacement field. However, we may generalize our formula \eqref{equation::D_370}, \eqref{equation::D_380} 
and \eqref{equation::D_400} by taking the macroscopic densities $\tilde{n}_a(\mathbf{r})$ as \emph{independent} 
functions. Then, our microscopic densities $n_a(\mathbf{r})$ are parameterized by two types of macroscopic functions, 
the displacement fields $\mathbf{u}(\mathbf{r})$ and the densities $\tilde{n}_a(\mathbf{r})$. This generalization 
is important for \emph{cluster crystals} because in this case the change of the densities is independent from the 
deformation.

Eq.\ \eqref{equation::D_400} shows that the microscopic density changes $n_a(\mathbf{r}) - n_{0,a}(\mathbf{r})$ 
depend \emph{nonlinearly} on the displacement field $\mathbf{u}(\mathbf{r})$ and \emph{linearly} on the macroscopic 
density changes $\tilde{n}_a(\mathbf{r}) - \tilde{n}_{0,a}$. We may expand in a Taylor series with respect to the 
displacement field. If we keep the linear terms only then we obtain the linearized formula
\begin{equation}
  \begin{split}
    n_a(\mathbf{r}) - n_{0,a}(\mathbf{r}) =& 
    - [ \nabla n_{0,a}(\mathbf{r}) ] \cdot \mathbf{u}(\mathbf{r}) \\
    &+ \frac{ n_{0,a}( \mathbf{r} ) }{ \tilde{n}_{0,a} } 
    \, [ \tilde{n}_a(\mathbf{r}) - \tilde{n}_{0,a} ] \\
    &+ \Delta n^\prime_a(\mathbf{r}) \, .
  \end{split}
  \label{equation::D_410} 
\end{equation}
Using the notation with deltas $\delta$ for the changes the linear formula can be rewritten in the more 
convenient form
\begin{equation}
  \delta n_a(\mathbf{r}) = - [ \nabla n_{0,a}(\mathbf{r}) ] \cdot \delta \mathbf{u}(\mathbf{r})
  + \frac{ n_{0,a}( \mathbf{r} ) }{ \tilde{n}_{0,a} } 
  \, \delta \tilde{n}_a(\mathbf{r}) + \Delta n^\prime_a(\mathbf{r}) \, .
  \label{equation::D_420} 
\end{equation}
Later we will insert the density changes into the Taylor expansion of the free energy functional $F[T,n]$ in order 
to calculate the stress tensor and the elastic constants. There will arise the question if the linear formulas 
are sufficient or if we must expand the nonlinear formulas up to higher order in the displacement field.

The density variation \eqref{equation::D_420} has been used in previous works in order to derive the macroscopic 
elasticity theory from microscopic approaches. Szamel and Ernst \cite{SE93,Sz97} have used the first term on the 
right-hand side in order to construct a microscopic expression for the displacement field. Walz and Fuchs \cite{WF10}, 
H\"aring \emph{et al.}\ \cite{HF15} and further consecutive works use the first two terms on the right-hand side for 
a microscopic approach to cluster crystals. However, until now in all cases the formula \eqref{equation::D_420} 
has been used as an \emph{ansatz} where the correction term is omitted and $\Delta n^\prime_a(\mathbf{r}) = 0$
has been set. In order to prepare an improvement of the theory, in the next subsection we determine and calculate 
the correction densities $\Delta n^\prime_a(\mathbf{r})$ explicitly for the linear formulas \eqref{equation::D_410} 
and \eqref{equation::D_420}. As a consequence the density changes are no more an ansatz but rather a \emph{solution} 
of the minimization procedure of the free energy functional $F[T,n]$.

\subsection{Correction to the density of a deformed crystal}
\label{section::04e}
Starting point for the calculation is the necessary condition \eqref{equation::D_190} for the minimization of the free 
energy. We assume that $n_{0,a}(\mathbf{r})$ are the densities of the reference state where the crystal is homogeneous 
in space and the lattice is perfectly periodic. These densities solve the necessary condition \eqref{equation::D_190} 
for a constant chemical potentials $\mu_{0,a}$. For the deformed crystal we use the ansatz 
\begin{eqnarray}
  n_a(\mathbf{r}) &=& n_{0,a}(\mathbf{r}) + \delta n_a(\mathbf{r}) \, ,
  \label{equation::D_430} \\
  \mu_a(\mathbf{r}) &=& \mu_{0,a} + \delta \mu_a(\mathbf{r}) 
  \label{equation::D_440} 
\end{eqnarray}
and linearize Eq.\ \eqref{equation::D_190} with respect to the variations. Thus, we obtain 
\begin{equation}
  \sum_b \int d^dr_2 \, \frac{ \delta^2 F[T,n] }{ \delta n_a(\mathbf{r}_1) \delta n_b(\mathbf{r}_2) } \bigg|_{n = n_0} 
  \delta n_b(\mathbf{r}_2) = \delta \mu_a(\mathbf{r}_1) \, .
  \label{equation::D_450} 
\end{equation}
Here the second variational derivative of the free energy is given by
\begin{equation}
  \begin{split}
    \frac{ \delta \mu_a(\mathbf{r}_1) }{ \delta n_b(\mathbf{r}_2) } &= 
    \frac{ \delta^2 F[T,n] }{ \delta n_a(\mathbf{r}_1) \delta n_b(\mathbf{r}_2) } \\
    &= k_B T \, \Bigl[ \frac{ 1 }{ n_a(\mathbf{r}_1) } \, \delta_{ab} \, \delta( \mathbf{r}_1 - \mathbf{r}_2 ) 
    - c^{(2)}_{ab}( \mathbf{r}_1, \mathbf{r}_2 ) \Bigr] 
  \end{split}
  \label{equation::D_460} 
\end{equation}
where $c^{(2)}_{ab}( \mathbf{r}_1, \mathbf{r}_2 )$ is the two-point \emph{direct} correlation function. It serves 
as an integral kernel in a linear inhomogeneous integral equation. The Legendre transformation between the 
thermodynamic potentials $F[T,n]$ and $\Omega[T,\mu]$ yields an explicit expression for the inverse integral kernel
\begin{equation}
  \begin{split}
    \frac{ \delta n_a(\mathbf{r}_1) }{ \delta \mu_b(\mathbf{r}_2) } &= 
    - \frac{ \delta^2 \Omega[T,\mu] }{ \delta \mu_a(\mathbf{r}_1) \delta \mu_b(\mathbf{r}_2) } \\
    &= \frac{ 1 }{ k_B T } \, \Bigl[ n_a(\mathbf{r}_1) \, \delta_{ab} \, \delta( \mathbf{r}_1 - \mathbf{r}_2 ) \\
    &\hspace{14mm} + n_a(\mathbf{r}_1) \, h^{(2)}_{ab}( \mathbf{r}_1, \mathbf{r}_2 ) \, n_b(\mathbf{r}_2) \Bigr] 
  \end{split}
  \label{equation::D_470} 
\end{equation}
where $h^{(2)}_{ab}( \mathbf{r}_1, \mathbf{r}_2 )$ is the two-point \emph{indirect} correlation function. The 
two integral kernels satisfy the equation
\begin{equation}
  \begin{split}
  &- \sum_b \int d^dr_2 \, \frac{ \delta^2 \Omega[T,\mu] }{ \delta \mu_a(\mathbf{r}_1) \delta \mu_b(\mathbf{r}_2) }
  \, \frac{ \delta^2 F[T,n] }{ \delta n_b(\mathbf{r}_2) \delta n_c(\mathbf{r}_3) } \\
  &= \sum_b \int d^dr_2 \, \frac{ \delta n_a(\mathbf{r}_1) }{ \delta \mu_b(\mathbf{r}_2) }
  \, \frac{ \delta \mu_b(\mathbf{r}_2) }{ \delta n_c(\mathbf{r}_3) } 
  = \delta_{ac} \, \delta( \mathbf{r}_1 - \mathbf{r}_3 ) 
  \end{split}
  \label{equation::D_480} 
\end{equation}
which transforms into the Ornstein-Zernike equation 
\begin{equation}
  \begin{split}
    h^{(2)}_{ac}( \mathbf{r}_1, \mathbf{r}_3 ) = \,& c^{(2)}_{ac}( \mathbf{r}_1, \mathbf{r}_3 ) \\ 
    &+ \sum_b \int d^dr_2 \, h^{(2)}_{ab}( \mathbf{r}_1, \mathbf{r}_2 ) \, n_b(\mathbf{r}_2) 
    \, c^{(2)}_{bc}( \mathbf{r}_2, \mathbf{r}_3 )
  \end{split}
  \label{equation::D_490} 
\end{equation}
if we insert the explicit expressions \eqref{equation::D_460} and \eqref{equation::D_470} and omit the delta 
functions on both sides. 

Thus, the necessary condition \eqref{equation::D_450} can be solved explicitly by
\begin{equation}
  \delta n_a(\mathbf{r}_1) = - \sum_b \int d^dr_2 
  \, \frac{ \delta^2 \Omega[T,\mu] }{ \delta \mu_a(\mathbf{r}_1) \delta \mu_b(\mathbf{r}_2) } \bigg|_{\mu = \mu_0} 
  \delta \mu_b(\mathbf{r}_2) \, .
  \label{equation::D_500} 
\end{equation}
If we insert the explicit expressions \eqref{equation::D_470} then we obtain
\begin{equation}
  \delta n_a(\mathbf{r}) = \sum_b \, a_{ab}(\mathbf{r}) \, \frac{ \delta \mu_b(\mathbf{r}) }{ k_B T }
  \label{equation::D_510} 
\end{equation}
where for convenience we define the matrix function
\begin{equation}
  a_{ab}(\mathbf{r}_1) = n_{0,a}(\mathbf{r}_1) \, \Bigl[ \delta_{ab} 
    + \int d^dr_2 \, h^{(2)}_{ab}( \mathbf{r}_1, \mathbf{r}_2 ) \, n_{0,b}(\mathbf{r}_2) \Bigr] \, .
  \label{equation::D_520} 
\end{equation}
In Eq.\ \eqref{equation::D_510} we have performed an approximation. We assume that the chemical potentials 
$\delta \mu_a(\mathbf{r}) = \delta \tilde{\mu}_a(\mathbf{r})$ are slowly varying functions defined in the 
macroscopic function space $\tilde{\mathcal{C}}(\ell)$. Thus, we may take them out off the integral of 
Eq.\ \eqref{equation::D_520}.

However, Eq.\ \eqref{equation::D_500} or Eq.\ \eqref{equation::D_510} is only a \emph{particular} solution. 
It depends on the chemical potentials $\delta \mu_a(\mathbf{r}) = \delta \tilde{\mu}_a(\mathbf{r})$ only 
where the effects of deformations described by a displacement field $\delta \mathbf{u}(\mathbf{r})$ are missing. 
The particular solution describes the changes of the density profiles $\delta n_a(\mathbf{r})$ implied by 
the changes of the chemical potentials $\delta \mu_a(\mathbf{r})$ where no deformations happen.

The integral kernel \eqref{equation::D_460} has zero eigenvalues which are related to the translation 
invariance of the physical system. Consequently, there exist homogeneous solutions of the necessary condition 
\eqref{equation::D_450} which are parameterized by the displacement field $\delta \mathbf{u}(\mathbf{r})$. 
Hence, we must solve the homogeneous equation
\begin{equation}
  \sum_b \int d^dr_2 \, \frac{ \delta^2 F[T,n] }{ \delta n_a(\mathbf{r}_1) \delta n_b(\mathbf{r}_2) } \bigg|_{n = n_0} 
  \delta n_b(\mathbf{r}_2) = 0 
  \label{equation::D_530} 
\end{equation}
where the right-hand side is zero. As an ansatz we use a formula similar like Eq.\ \eqref{equation::D_420} 
which we write in the form 
\begin{equation}
  \delta n_a(\mathbf{r}) = - [ \nabla n_{0,a}(\mathbf{r}) ] \cdot \delta \mathbf{u}(\mathbf{r})
  + \delta n_{1,a}(\mathbf{r})
  \label{equation::D_540} 
\end{equation}
where the deformation term has been separated explicitly. If we insert these densities $\delta n_a(\mathbf{r})$ 
into the homogeneous equation \eqref{equation::D_530} we obtain the inhomogeneous equation \eqref{equation::D_450} 
for the densities $\delta n_{1,a}(\mathbf{r})$ where on the right-hand side we have the chemical potentials
\begin{widetext}
\begin{equation}
  \delta \mu_{1,a}(\mathbf{r}_1) = \sum_b \int d^dr_2 
  \, \frac{ \delta^2 F[T,n] }{ \delta n_a(\mathbf{r}_1) \delta n_b(\mathbf{r}_2) } \bigg|_{n = n_0} 
  [ \nabla n_{0,b}(\mathbf{r}_2) ] \cdot \delta \mathbf{u}(\mathbf{r}_2)
  \label{equation::D_550} 
\end{equation}
as inhomogeneities. If we solve the inhomogeneous equation then we obtain densities
\begin{equation}
  \delta n_{1,a}(\mathbf{r}_1) = - \sum_b \int d^dr_2 
  \, \frac{ \delta^2 \Omega[T,\mu] }{ \delta \mu_a(\mathbf{r}_1) \delta \mu_b(\mathbf{r}_2) } \bigg|_{\mu = \mu_0} 
  \delta \mu_{1,b}(\mathbf{r}_2) \, .
  \label{equation::D_560} 
\end{equation}
Now, translation invariance implies that Eq.\ \eqref{equation::D_550} yields zero if the displacement field is 
constant under the integral. Thus, on the right-hand side we may replace 
$\delta \mathbf{u}(\mathbf{r}_2) \to \delta \mathbf{u}(\mathbf{r}_2) - \delta \mathbf{u}(\mathbf{r}_1)$ so that
\begin{equation}
  \delta \mu_{1,a}(\mathbf{r}_1) = \sum_b \int d^dr_2 
  \, \frac{ \delta^2 F[T,n] }{ \delta n_a(\mathbf{r}_1) \delta n_b(\mathbf{r}_2) } \bigg|_{n = n_0} 
  [ \nabla n_{0,b}(\mathbf{r}_2) ] \cdot [ \delta \mathbf{u}(\mathbf{r}_2) - \delta \mathbf{u}(\mathbf{r}_1) ] \, .
  \label{equation::D_570} 
\end{equation}
Next, we insert the explicit second functional derivative \eqref{equation::D_460} and use the equation
\begin{equation}
  \delta( \mathbf{r}_1 - \mathbf{r}_2 ) \, [ \delta \mathbf{u}(\mathbf{r}_1) - \delta \mathbf{u}(\mathbf{r}_2) ] = 0 \, .
  \label{equation::D_580} 
\end{equation}
Then, the chemical potentials $\delta \mu_{1,a}(\mathbf{r})$ can be expressed in terms of the two-point direct 
correlation function $c^{(2)}_{ab}( \mathbf{r}_1, \mathbf{r}_2 )$ according to
\begin{equation}
  \delta \mu_{1,a}(\mathbf{r}_1) = k_B T \, \sum_b \int d^dr_2 \, c^{(2)}_{ab}( \mathbf{r}_1, \mathbf{r}_2 )
  \, [ \nabla n_{0,b}(\mathbf{r}_2) ] \cdot [ \delta \mathbf{u}(\mathbf{r}_1) - \delta \mathbf{u}(\mathbf{r}_2) ] \, .
  \label{equation::D_590} 
\end{equation}
Since the direct correlation function is nonzero only for short distances $\mathbf{r}_1 - \mathbf{r}_2$ on 
microscopic length scales $a$ we may use the Taylor expansion up to first order and write 
\begin{equation}
  c^{(2)}_{ab}( \mathbf{r}_1, \mathbf{r}_2 ) \, [ \delta \mathbf{u}(\mathbf{r}_1) - \delta \mathbf{u}(\mathbf{r}_2) ] 
  \approx c^{(2)}_{ab}( \mathbf{r}_1, \mathbf{r}_2 ) [ ( \mathbf{r}_1 - \mathbf{r}_2 ) \cdot \nabla ] 
  \, \delta \mathbf{u}(\mathbf{r}_1) ) \, .
  \label{equation::D_600} 
\end{equation}
Then, we obtain the chemical potentials 
\begin{equation}
  \delta \mu_{1,a}(\mathbf{r}_1) = k_B T \, \sum_b \int d^dr_2 \, c^{(2)}_{ab}( \mathbf{r}_1, \mathbf{r}_2 )
  \, ( \partial_i n_{0,b}(\mathbf{r}_2) ) \, ( r_1^j - r_2^j ) \, ( \partial_j \delta u^i(\mathbf{r}_1) ) \, .
  \label{equation::D_610} 
\end{equation}
Finally, we calculate the densities $\delta n_{1,a}(\mathbf{r})$ from the particular solution \eqref{equation::D_500} 
by inserting the chemical potentials $\delta \mu_{1,a}(\mathbf{r})$ and the explicit second functional derivative 
\eqref{equation::D_470} on the right-hand side. Thus, we obtain
\begin{equation}
  \delta n_{1,a}(\mathbf{r}_1) = n_{0,a}(\mathbf{r}_1) \Bigl[ \delta \mu_{1,a}(\mathbf{r}_1)
  + \int d^dr_2 \, h^{(2)}_{ab}( \mathbf{r}_1, \mathbf{r}_2 ) \, n_b(\mathbf{r}_2) 
  \, \delta \mu_{1,b}(\mathbf{r}_2) \Bigr] / k_B T \, .
  \label{equation::D_620} 
\end{equation}
Eventually, the homogeneous solution is given by the formula \eqref{equation::D_540}.

For a more compact notation we define the basis functions
\begin{equation}
  b_{a,i}^{\ \ \ j}(\mathbf{r}_1) = n_{0,a}(\mathbf{r}_1) \Bigl[ d_{a,i}^{\ \ \ j}(\mathbf{r}_1)
  + \int d^dr_2 \, h^{(2)}_{ab}( \mathbf{r}_1, \mathbf{r}_2 ) \, n_b(\mathbf{r}_2) 
  \, d_{b,i}^{\ \ \ j}(\mathbf{r}_2) \Bigr] 
  \label{equation::D_630} 
\end{equation}
together with
\begin{equation}
  d_{a,i}^{\ \ \ j}(\mathbf{r}_1) = \sum_b \int d^dr_2 \, c^{(2)}_{ab}( \mathbf{r}_1, \mathbf{r}_2 )
  \, ( \partial_i n_{0,b}(\mathbf{r}_2) ) \, ( r_1^j - r_2^j ) \, .
  \label{equation::D_640} 
\end{equation}
\end{widetext}
Then, we write 
\begin{equation}
  \delta n_{1,a}(\mathbf{r}) = b_{a,i}^{\ \ \ j}(\mathbf{r}) \, ( \partial_j \delta u^i(\mathbf{r}) )
  \label{equation::D_650} 
\end{equation}
where again we have performed an approximation because we have taken the slowly varying macroscopic function 
$\partial_j \delta u^i(\mathbf{r})$ out of the integrals. Eventually, from Eq.\ \eqref{equation::D_540} we 
obtain the homogeneous solution
\begin{equation}
  \delta n_a(\mathbf{r}) = - [ \nabla n_{0,a}(\mathbf{r}) ] \cdot \delta \mathbf{u}(\mathbf{r})
  + b_{a,i}^{\ \ \ j}(\mathbf{r}) \, ( \partial_j \delta u^i(\mathbf{r}) ) \, .
  \label{equation::D_660} 
\end{equation}

In summary, we have calculated the particular solution \eqref{equation::D_510} and the homogeneous solution 
\eqref{equation::D_660}. If we add these two results we obtain the general solution 
\begin{equation}
  \begin{split}
    \delta n_a(\mathbf{r}) = &- [ \nabla n_{0,a}(\mathbf{r}) ] \cdot \delta \mathbf{u}(\mathbf{r})
    + b_{a,i}^{\ \ \ j}(\mathbf{r}) \, ( \partial_j \delta u^i(\mathbf{r}) ) \\
    &+ \sum_b \, a_{ab}(\mathbf{r}) \, \frac{ \delta \mu_b(\mathbf{r}) }{ k_B T }
  \end{split}
  \label{equation::D_670} 
\end{equation}
as a final result. It is parameterized by two kinds of slowly varying macroscopic functions, the displacement 
field $\delta \mathbf{u}(\mathbf{r}) = \delta \tilde{\mathbf{u}}(\mathbf{r})$ and the chemical potentials 
$\delta \mu_a(\mathbf{r}) = \delta \tilde{\mu}_a(\mathbf{r})$ which are defined in the macroscopic function 
space $\tilde{\mathcal{C}}(\ell)$. The densities $\delta n_a(\mathbf{r})$ itself are microscopic functions in 
the full function class $\mathcal{C} = \tilde{\mathcal{C}}(\ell) \oplus \mathcal{C}^\prime(\ell)$. On the 
right-hand side $n_{0,a}(\mathbf{r})$, $b_{a,i}^{\ \ \ j}(\mathbf{r})$, and $a_{ab}(\mathbf{r})$ are microscopic 
functions in $\mathcal{C}$.

Once again we apply the coarse-graining procedure \eqref{equation::D_040} and project the microscopic densities 
\eqref{equation::D_670} into the macroscopic function space $\tilde{\mathcal{C}}(\ell)$. We obtain
\begin{equation}
  \delta \tilde{n}_a(\mathbf{r}) = 0 + \tilde{b}_{a,i}^{\ \ \ j} \, ( \partial_j \delta u^i(\mathbf{r}) ) 
  + \sum_b \, \tilde{a}_{ab} \, \frac{ \delta \mu_b(\mathbf{r}) }{ k_B T } \, .
  \label{equation::D_680} 
\end{equation}
On the right-hand side the coarse-graining procedure projects the microscopic factors to the macroscopic ones 
$\tilde{n}_{0,a}(\mathbf{r}) = \tilde{n}_{0,a}$, $\tilde{b}_{a,i}^{\ \ \ j}(\mathbf{r}) = \tilde{b}_{a,i}^{\ \ \ j}$, 
and $\tilde{a}_{ab}(\mathbf{r}) = \tilde{a}_{ab}$. Since the reference state is a perfect crystal which is 
homogeneous in space, these macroscopic functions are constant in space. Hence we find 
$\nabla \tilde{n}_{0,a} = \mathbf{0}$ which implies the first term to be zero. 

Eq.\ \eqref{equation::D_680} is a relation between the macroscopic densities $\delta \tilde{n}_a(\mathbf{r})$ 
and the chemical potentials $\delta \mu_b(\mathbf{r}) = \delta \tilde{\mu}_b(\mathbf{r})$. Both functions are 
defined in the macroscopic function class $\tilde{\mathcal{C}}(\ell)$ which makes the relation invertible. 
Additional free parameters are the vector components of the displacement fields $\delta u^i(\mathbf{r})$. 
We explicitly resolve this relation with respect to the chemical potentials and obtain
\begin{equation}
  \frac{ \delta \mu_a(\mathbf{r}) }{ k_B T } = \sum_b \, \tilde{a}^{-1}_{ab} [ \delta \tilde{n}_b(\mathbf{r}) 
  - \tilde{b}_{b,i}^{\ \ \ j} \, ( \partial_j \delta u^i(\mathbf{r}) ) ] \, .
  \label{equation::D_690} 
\end{equation}
We insert this result into formula \eqref{equation::D_670}. Thus, we rewrite the microscopic densities in 
the form
\begin{equation}
  \begin{split}
    \delta n_a(\mathbf{r}) = &- [ \nabla n_{0,a}(\mathbf{r}) ] \cdot \delta \mathbf{u}(\mathbf{r}) \\
    &+ \Bigl[ b_{a,i}^{\ \ \ j}(\mathbf{r}) - \sum_{bc} \, a_{ab}(\mathbf{r}) \, \tilde{a}^{-1}_{bc} 
    \, \tilde{b}_{c,i}^{\ \ \ j} \Bigr] \, ( \partial_j \delta u^i(\mathbf{r}) ) \\
    &+ \sum_{bc} \, a_{ab}(\mathbf{r}) \, \tilde{a}^{-1}_{bc} \, \delta \tilde{n}_c(\mathbf{r}) 
  \end{split}
  \label{equation::D_700} 
\end{equation}
which are now parameterized by the displacement field $\delta \mathbf{u}(\mathbf{r}) = \delta \tilde{\mathbf{u}}(\mathbf{r})$ 
and the macroscopic densities $\delta \tilde{n}_a(\mathbf{r})$. These microscopic densities already have some 
elements of the form \eqref{equation::D_420} which we want to derive. However, further work must be done.

The basis functions $b_{a,i}^{\ \ \ j}(\mathbf{r})$ defined in \eqref{equation::D_630} together with \eqref{equation::D_640} 
appear to be quite complicated because of the last two factors in Eq.\ \eqref{equation::D_640} which possess a derivative 
$\partial_i$ and a linear factor $( r_1^j - r_2^j )$. We may ask the question if we can extract the main and essential 
contribution. For this purpose in Eq.\ \eqref{equation::D_640} we perform a partial integration and omit some terms. In 
this way we find the essential contribution
\begin{equation}
  d_{0,a,i}^{\ \ \ \ \ j}(\mathbf{r}_1) = \sum_b \int d^dr_2 \, c^{(2)}_{ab}( \mathbf{r}_1, \mathbf{r}_2 )
  \, n_{0,b}(\mathbf{r}_2) \, \delta_i^{\ j}
  \label{equation::D_710} 
\end{equation}
which has a much simpler structure than the original function \eqref{equation::D_640}. Inserting this result into 
Eq.\ \eqref{equation::D_630} and using the Ornstein-Zernike equation \eqref{equation::D_490} we obtain the essential 
contribution of the basis function
\begin{equation}
  b_{0,a,i}^{\ \ \ \ \ j}(\mathbf{r}_1) = n_{0,a}(\mathbf{r}_1) \, \sum_b \int d^dr_2 \, h^{(2)}_{ab}( \mathbf{r}_1, \mathbf{r}_2 )
  \, n_{0,b}(\mathbf{r}_2) \, \delta_i^{\ j}
  \label{equation::D_720} 
\end{equation}
which again has a much simpler structure than the original basis function. We compare this result with the basis function 
$a_{ab}(\mathbf{r})$ defined earlier in Eq.\ \eqref{equation::D_520}. The two basis functions \eqref{equation::D_720} and 
\eqref{equation::D_520} are not independent. Rather we find the relation
\begin{equation}
  b_{0,a,i}^{\ \ \ \ \ j}(\mathbf{r}) = \sum_b \, [ a_{ab}(\mathbf{r}) - n_{0,a}(\mathbf{r}) \, \delta_{ab} ] \, \delta_i^{\ j} \, .
  \label{equation::D_730} 
\end{equation}
In order to proceed we define the difference functions $\Delta d_{a,i}^{\ \ \ j}(\mathbf{r})$ 
and $\Delta b_{a,i}^{\ \ \ j}(\mathbf{r})$ by
\begin{eqnarray}
  d_{a,i}^{\ \ \ j}(\mathbf{r}) &=& d_{0,a,i}^{\ \ \ \ \ j}(\mathbf{r}) + \Delta d_{a,i}^{\ \ \ j}(\mathbf{r}) \, ,
  \label{equation::D_740} \\
  b_{a,i}^{\ \ \ j}(\mathbf{r}) &=& b_{0,a,i}^{\ \ \ \ \ j}(\mathbf{r}) + \Delta b_{a,i}^{\ \ \ j}(\mathbf{r}) \, .
  \label{equation::D_750} 
\end{eqnarray}
Again, the difference functions are related to each other by Eq.\ \eqref{equation::D_630} as it is for the essential 
contributions and for the original functions.

Now, we are ready to calculate the density variations. We obtain
\begin{equation}
  \delta n_a(\mathbf{r}) = - [ \nabla n_{0,a}(\mathbf{r}) ] \cdot \delta \mathbf{u}(\mathbf{r})
  + \frac{ n_{0,a}( \mathbf{r} ) }{ \tilde{n}_{0,a} } 
  \, \delta \tilde{n}_a(\mathbf{r}) + \Delta n_a(\mathbf{r}) 
  \label{equation::D_760} 
\end{equation}
with a correction function $\Delta n_a(\mathbf{r})$. After some manipulations we calculate
\begin{equation}
  \begin{split}
    \Delta n_a(\mathbf{r}) = \,& \Bigl[ \Delta b_{a,i}^{\ \ \ j}(\mathbf{r}) 
    - \sum_{bc} \, a_{ab}(\mathbf{r}) \, \tilde{a}^{-1}_{bc} \, \Delta \tilde{b}_{c,i}^{\ \ \ j} \Bigr] 
    \, ( \partial_j \delta u^i(\mathbf{r}) ) \\
    &- \sum_c \, \Bigl[ \sum_b a_{ab}(\mathbf{r}) \, \tilde{a}^{-1}_{bc} - \frac{ n_{0,a}(\mathbf{r}) }{ \tilde{n}_{0,a} } 
    \, \delta_{ac} \Bigr] \, \delta c_c(\mathbf{r}) 
  \end{split}
  \label{equation::D_770} 
\end{equation}
where in the last term we have defined the \emph{vacancy density}
\begin{equation}
  \delta c_a(\mathbf{r}) = - [ \delta \tilde{n}_a(\mathbf{r}) + \tilde{n}_{0,a} ( \nabla \cdot \delta \mathbf{u}(\mathbf{r}) ) ] 
  \label{equation::D_780} 
\end{equation}
following the authors of Refs.\ \onlinecite{SE93,Sz97,WF10,HF15}. The main result of this subsection is the formula 
\eqref{equation::D_770} for the correction functions of the particle densities. It has two contributions which are written in 
two separate lines. 

The first contribution in the first line is related to a \emph{pure deformation} of the crystal where the number of particles 
on each lattice site is constant. The last factor $\partial_j \delta u^i(\mathbf{r})$ is obtained by linearizing the Jacobi 
matrix \eqref{equation::D_330}. It may be interpreted as the strain tensor which describes the deformations. The related 
macroscopic density variations can be written in the form 
$\delta \tilde{n}_a(\mathbf{r}) = - \tilde{n}_{0,a} ( \nabla \cdot \delta \mathbf{u}(\mathbf{r}) )$. They are obtained from the 
macroscopic densities $\eqref{equation::D_360}$ if we insert the constant densities of the reference state $\tilde{n}_{0,a}$ and 
the Jacobi determinant \eqref{equation::D_320} together with the Jacobi matrix \eqref{equation::D_330} and then linearize with 
respect to the displacement field $\delta \mathbf{u}(\mathbf{r})$.

The second contribution in the second line is related to a \emph{pure change of the number of particles} at each lattice site 
where the crystal structure is not changed and not deformed. This fact arises from the last factor $\delta c_a(\mathbf{r})$ 
which is defined in Eq.\ \eqref{equation::D_780}. If we consider the square bracket terms $[\ldots]$ we clearly see that from 
the particle densities $\delta \tilde{n}_a(\mathbf{r})$ the contribution of the deformation 
$- \tilde{n}_{0,a} ( \nabla \cdot \delta \mathbf{u}(\mathbf{r}) )$ is subtracted. Consequently, $[\ldots] = - \delta c_a(\mathbf{r})$ 
is the particle density of \emph{point defects}. If it is positive there are more than one particle per lattice site which may 
result in particles on \emph{interstitial sites}. On the other hand, if it is negative there are less than one particle per 
lattice site which may result in \emph{vacancies}. Thus, because of the negative sign in Eq.\ \eqref{equation::D_780} 
$\delta c_a(\mathbf{r})$ may be interpreted as \emph{vacancy densities}. We note that the negative sign in the definition is 
a convention of the authors of Refs.\ \onlinecite{SE93,Sz97,WF10,HF15} which we will keep in our paper.

When calculating the indirect correlation function $h^{(2)}_{ab}( \mathbf{r}_1, \mathbf{r}_2 )$ from the direct correlation 
$c^{(2)}_{ab}( \mathbf{r}_1, \mathbf{r}_2 )$ by solving the Ornstein-Zernike equation \eqref{equation::D_490} we must 
correctly fix the boundary conditions. Otherwise the separation into the two contributions is not consistent. For the 
general $h^{(2)}_{ab}( \mathbf{r}_1, \mathbf{r}_2 )$ there appears a particular and a homogeneous solution which 
allows ambiguities. We fix these ambiguities correctly if we restrict all functions to a function space where any 
deformations of the lattice structure are excluded. As a consequence, the indirect correlation function 
$h^{(2)}_{ab}( \mathbf{r}_1, \mathbf{r}_2 )$ describes changes of the number of particles at each lattice site only and 
discards deformations. We note that this choice is the natural and simplest choice for the boundary conditions.

Our results for the density variations \eqref{equation::D_760} are quite close to the formula \eqref{equation::D_420}. 
The difference is the prime on the correction densities $\Delta n^\prime_a(\mathbf{r})$ which is present in 
Eq.\ \eqref{equation::D_420} but not in Eq.\ \eqref{equation::D_760}. For the two equations to be identical we must prove
\begin{equation}
  \Delta n_a(\mathbf{r}) = \Delta \tilde{n}_a(\mathbf{r}) + \Delta n^\prime_a(\mathbf{r}) = \Delta n^\prime_a(\mathbf{r}) \, .
  \label{equation::D_790} 
\end{equation}
For this purpose we apply the coarse-graining procedure \eqref{equation::D_040} to our result for the density corrections 
\eqref{equation::D_770}. If in the first line we replace $\Delta b_{a,i}^{\ \ \ j}(\mathbf{r}) \to \Delta \tilde{b}_{c,i}^{\ \ \ j}$ 
and $a_{ab}(\mathbf{r}) \to \tilde{a}_{ab}$ we clearly see that the two terms cancel each other. Similarly, if in the second 
line we replace $a_{ab}(\mathbf{r}) \to \tilde{a}_{ab}$ and $n_a(\mathbf{r}) \to \tilde{n}_{0,a}$ again the two terms cancel 
each other. Thus, we obtain
\begin{equation}
  \Delta \tilde{n}_a(\mathbf{r}) = 0 + 0 = 0
  \label{equation::D_800} 
\end{equation}
which proves Eq.\ \eqref{equation::D_790}. 

Hence the correction terms for the densities $\Delta n_a(\mathbf{r}) = \Delta n^\prime_a(\mathbf{r})$ are restricted to the 
complement function space $\mathcal{C}^\prime(\ell)$ as it is required. In order to indicate this fact more clearly, in 
the formula \eqref{equation::D_770} on the right-hand side we may project each term separately to the complement function 
space. We may put a prime onto each microscopic function. Thus, we obtain the final result
\begin{equation}
  \begin{split}
    \Delta n^\prime_a(\mathbf{r}) = \,& \Bigl[ \Delta b_{a,i}^{\prime \, \ \ j}(\mathbf{r}) 
    - \sum_{bc} \, a^\prime_{ab}(\mathbf{r}) \, \tilde{a}^{-1}_{bc} \, \Delta \tilde{b}_{c,i}^{\ \ \ j} \Bigr] 
    \, ( \partial_j \delta u^i(\mathbf{r}) ) \\
    &- \sum_c \, \Bigl[ \sum_b a^\prime_{ab}(\mathbf{r}) \, \tilde{a}^{-1}_{bc} - \frac{ n^\prime_{0,a}(\mathbf{r}) }{ \tilde{n}_{0,a} } 
    \, \delta_{ac} \Bigr] \, \delta c_c(\mathbf{r})
  \end{split}
  \label{equation::D_810} 
\end{equation}
together with the vacancy densities \eqref{equation::D_780} which represents the correction terms in our expected, stated, 
and proved formula for the density variations \eqref{equation::D_420}.

\subsection{Extended minimization of the free energy for deformed crystals}
\label{section::04f}
The minimization procedure for the free energy \eqref{equation::D_100} is incomplete. It does not control the deformations 
of the crystal because it does not include the displacement field $\mathbf{u}(\mathbf{r})$. For this reason, the minimization 
procedure must be extended. In the previous two subsections we have done this by solving the necessary conditions of the 
minimization \eqref{equation::D_010} where the special ansatz for the microscopic particle densities \eqref{equation::D_380} 
or \eqref{equation::D_400} is used. In short-hand notation this ansatz may be written as 
$n = n_1(\mathbf{u}, \tilde{n}) + \Delta n^\prime$ where the first contribution indicates the dependence on the displacement 
field $\mathbf{u}(\mathbf{r})$ and the coarse grained densities $\tilde{n}_a(\mathbf{r})$ and where the second contribution 
$\Delta n^\prime_a(\mathbf{r})$ is adjusted to minimize the free energy. Consequently, we may write the extended 
minimization as
\begin{equation}
  \tilde{F}[T,\mathbf{u},\tilde{\mathbf{j}},\tilde{n}] = \min_{\mathbf{j}^\prime, \Delta n^\prime} 
  F[T,\mathbf{j} = \tilde{\mathbf{j}} + \mathbf{j}^\prime,n = n_1(\mathbf{u}, \tilde{n}) + \Delta n^\prime] \, .
  \label{equation::D_820} 
\end{equation}
In the minimization procedure the functions $\mathbf{j}^\prime(\mathbf{r})$ and $\Delta n^\prime_a(\mathbf{r})$ are 
adjusted. While the vector components of the current density $\mathbf{j}^\prime(\mathbf{r})$ may be any functions in 
the complement function space $\mathcal{C}^\prime(\ell)$ the particle densities $\Delta n^\prime_a(\mathbf{r})$ must 
be restricted to an even smaller function space $\mathcal{C}_-^\prime(\ell) \subset \mathcal{C}^\prime(\ell)$ so that 
they do not override the deformation contribution which is the first term in the ansatz \eqref{equation::D_380} or 
\eqref{equation::D_400}.

\section{Taylor expansion of the free energy and the calculation of the elastic constants}
\label{section::05}
In order to investigate and understand the elastic properties of a crystal we expand the free energy $F[T,\ldots]$ 
into a Taylor series up to second order. For an expansion there are two possibilities. First we may consider the 
microscopic free energy functional $F[T,n]$ and perform a \emph{functional expansion} on the \emph{microscopic scale} 
where we start with the reference state with densities $n_{0,a}(\mathbf{r})$ and expand with respect to the microscopic 
deviations $n_a(\mathbf{r}) - n_{0,a}(\mathbf{r})$. Second we may consider the crystal in a state of local thermodynamic 
equilibrium with deformations and macroscopic density changes which are described by the displacement field 
$\mathbf{u}(\mathbf{r})$ and by the macroscopic densities $\tilde{n}_a(\mathbf{r})$. In this case the free energy 
$F[T,\mathbf{u},\tilde{n}] = \int d^dr \, f(T,\partial \mathbf{u}(\mathbf{r}),\tilde{n}(\mathbf{r}))$ is obtained 
from the extended minimization procedure \eqref{equation::D_820} and has a local structure. Again, we start with 
the reference state with zero displacements $\mathbf{u}(\mathbf{r}) = \mathbf{0}$ and constant densities 
$\tilde{n}_{0,a}(\mathbf{r}) = \tilde{n}_{0,a}$. We perform a \emph{local expansion} on the \emph{macroscopic scale} 
with respect to the displacement field $\mathbf{u}(\mathbf{r})$ and the macroscopic density changes 
$\tilde{n}_a(\mathbf{r}) - \tilde{n}_{0,a}$. 

Since we are interested in the elastic properties of the crystals in this section we restrict our considerations 
onto a crystal at rest where the velocity field is $\mathbf{v}(\mathbf{r}) = \mathbf{0}$ and the momentum density is 
$\mathbf{j}(\mathbf{r}) = \mathbf{0}$. Later on we can extend our results to nonzero velocities and nonzero momentum 
densities easily by applying a Galilean transformation.

\subsection{Functional expansion on microscopic scale}
\label{section::05a}
We may use the microscopic densities $n_{0,a}(\mathbf{r})$ of a perfect homogeneous crystal as a \emph{reference} 
state about which we expand the free energy functional $F[T,n]$ into a Taylor series. If $n_a(\mathbf{r})$ are 
arbitrary functions for the particle densities, then up to second order we obtain
\begin{widetext}
\begin{equation}
  \begin{split}
  F[T,n] =& F[T,n_0] + \sum_a \int d^dr \, \frac{ \delta F[T,n] }{ \delta n_a(\mathbf{r}) } \bigg|_{n = n_0} 
  \, [ n_a(\mathbf{r}) - n_{0,a}(\mathbf{r}) ] \\
  &+ \frac{ 1 }{ 2 } \sum_{ab} \int d^dr_1 \int d^dr_2 \, 
  \frac{ \delta^2 F[T,n] }{ \delta n_a(\mathbf{r}_1) \, \delta n_b(\mathbf{r}_2) } \bigg|_{n = n_0} \, 
  [ n_a(\mathbf{r}_1) - n_{0,a}(\mathbf{r}_1) ] \, [ n_b(\mathbf{r}_2) - n_{0,b}(\mathbf{r}_2) ] + \cdots \, .
  \end{split}
  \label{equation::E_010} 
\end{equation}
This formula is quite general from the beginning. However, it is a \emph{functional} Taylor series which implies 
some special problems of functional analysis. It might even be ill defined in some cases as we shall see below. 
The integrals over the space variables $\mathbf{r}$ must converge if we integrate over an infinite volume. For 
this reason the integrands must decay sufficiently fast for $\mathbf{r} \to \infty$. It is most easily achieved 
if we require that the density differences $n_a(\mathbf{r}) - n_{0,a}(\mathbf{r})$ decay sufficiently fast for 
$\mathbf{r} \to \infty$. In the special case where we integrate over a finite volume $V$ we may require that 
$n_a(\mathbf{r}) - n_{0,a}(\mathbf{r}) = 0$ for $\mathbf{r} \in \partial V$ so that boundary terms are zero.

The first functional derivative of the free energy is constant for the reference state. The necessary condition 
\eqref{equation::D_190} together with \eqref{equation::D_080} and \eqref{equation::D_200} implies
\begin{equation}
  \frac{ \delta F[T,n] }{ \delta n_a(\mathbf{r}) } \bigg|_{n = n_0} = \tilde{\mu}_{0,a} = \mu_{0,a} \ .
  \label{equation::E_020} 
\end{equation}
The second functional derivative is calculated by inserting the free energy \eqref{equation::C_090} so that 
we obtain
\begin{equation}
  \begin{split}
    \frac{ \delta^2 F[T,n] }{ \delta n_a(\mathbf{r}_1) \, \delta n_b(\mathbf{r}_2) } \bigg|_{n = n_0} &= 
    \frac{ \delta^2 F_0[T,n] }{ \delta n_a(\mathbf{r}_1) \, \delta n_b(\mathbf{r}_2) } \bigg|_{n = n_0} + 
    \frac{ \delta^2 F_\mathrm{corr}[T,n] }{ \delta n_a(\mathbf{r}_1) \, \delta n_b(\mathbf{r}_2) } \bigg|_{n = n_0} \\
    &= k_B T \, \Bigl[ \frac{ 1 }{ n_{0,a}(\mathbf{r}_1) } \, \delta_{ab} \, \delta( \mathbf{r}_1 - \mathbf{r}_2 ) 
    - c^{(2)}_{ab}(\mathbf{r}_1,\mathbf{r}_2) \Bigr] \, .
  \end{split}
  \label{equation::E_030} 
\end{equation}
Here $c^{(2)}_{ab}(\mathbf{r}_1,\mathbf{r}_2)$ is the two-point direct correlation function which arises from the 
functional derivatives of the correlation energy $F_\mathrm{corr}[T,n]$. In the functional derivatives we always insert 
$n_a(\mathbf{r}) = n_{0,a}(\mathbf{r})$. Thus, in this and the following sections the direct and the indirect correlation 
functions are defined always in the reference state. Now, if we insert these functional derivatives we obtain the 
Taylor series for the free energy
\begin{equation}
  \begin{split}
    F[T,n] = \,& F[T,n_0] + \sum_a \int d^dr \, \mu_{0,a} \, [ n_a(\mathbf{r}) - n_{0,a}(\mathbf{r}) ] \\
    &\hspace{-4mm} + \frac{ 1 }{ 2 } \sum_{ab} \int d^dr_1 \int d^dr_2 \, 
    k_B T \, \Bigl[ \frac{ 1 }{ n_{0,a}(\mathbf{r}_1) } \, \delta_{ab} \, \delta( \mathbf{r}_1 - \mathbf{r}_2 ) 
    - c^{(2)}_{ab}(\mathbf{r}_1,\mathbf{r}_2) \Bigr] \, 
    [ n_a(\mathbf{r}_1) - n_{0,a}(\mathbf{r}_1) ] \, [ n_b(\mathbf{r}_2) - n_{0,b}(\mathbf{r}_2) ] + \cdots \, .
  \end{split}
  \label{equation::E_040} 
\end{equation}
\end{widetext}
Since the chemical potentials $\mu_{0,a}$ are constant in the first order term we may replace the microscopic densities 
by the macroscopic densities, and for this reason we may put a tilde on top of the densities. In the second order term 
we may transform to center of mass and relative coordinates by the formulas
\begin{equation}
  \mathbf{R} = \frac{ 1 }{ 2 } ( \mathbf{r}_1 + \mathbf{r}_2 ) \, , \qquad
  \Delta\mathbf{r} = \mathbf{r}_1 - \mathbf{r}_2 \, .
  \label{equation::E_050} 
\end{equation}
Similarly we transform the integrals
\begin{equation}
  \int d^dr_1 \int d^dr_2 = \int d^dR \int d^d(\Delta r) \, .
  \label{equation::E_060} 
\end{equation}
The direct correlation function $c^{(2)}_{ab}(\mathbf{r}_1,\mathbf{r}_2)$ decays very rapidly to zero for increasing 
distances $\Delta\mathbf{r} = \mathbf{r}_1 - \mathbf{r}_2$ on a scale of a few lattice constants $a$. The same 
is true for the factor $[\cdots]$ in the second-order term of the Taylor expansion \eqref{equation::E_040}. 
Consequently, in the second-order term of the Taylor series the integral over the relative coordinates 
$\int d^d(\Delta r)$ converges very well. The resulting integrand of $\int d^dR$ is a function varying on a 
microscopic scale. This integrand function can be smoothed by the coarse-graining procedure \eqref{equation::D_040}. 
Thus, at the end we expect a free energy of the form
\begin{equation}
  F[T,n] = \int d^dr \, \tilde{f}(\mathbf{r})
  \label{equation::E_070} 
\end{equation}
where $\tilde{f}(\mathbf{r})$ is the \emph{free energy density} which varies with the space coordinate $\mathbf{r}$ 
on a macroscopic scale so that $\tilde{f}(\mathbf{r}) \in \tilde{\mathcal{C}}(\ell)$. In the next subsection this 
free energy density will be determined as a Taylor series up to second order.

\subsection{Local expansion on macroscopic scale}
\label{section::05b}
In Sec.\ \ref{section::04} we have developed the nonequilibrium state of a crystal under \emph{macroscopic} 
deformations with displacements $\mathbf{u}(\mathbf{r})$ and \emph{macroscopic} density changes 
$\tilde{n}_a(\mathbf{r}) - \tilde{n}_{0,a}$. In Subsec.\ \ref{section::04d} we have derived the related 
\emph{microscopic} density change $n_a(\mathbf{r}) - n_{0,a}(\mathbf{r})$ by minimizing the functional free 
energy. We have found the nonlinear formula \eqref{equation::D_400} and the simplified linear formulas 
\eqref{equation::D_410} and \eqref{equation::D_420}. These formulas have two leading terms for the macroscopic 
deformation and density change which have been used already by previous authors \cite{SE93,Sz97,WF10,HF15}. 
Furthermore, they have a third correction term $\Delta n^\prime_a(\mathbf{r})$ which has been calculated 
explicitly in Subsec.\ \ref{section::04e} for the linear case and which is given in the final form by 
Eq.\ \eqref{equation::D_810}. Now, we insert the microscopic density change $n_a(\mathbf{r}) - n_{0,a}(\mathbf{r})$ 
into the functional Taylor expansion \eqref{equation::E_010} or \eqref{equation::E_040} and calculate the 
free energy $F[T,\mathbf{u},\tilde{n}]$ depending on the displacement field $\mathbf{u}(\mathbf{r})$ and on 
the macroscopic density changes $\tilde{n}_a(\mathbf{r}) - \tilde{n}_{0,a}$.

We start with the first-order term in Eq.\ \eqref{equation::E_040} which reads
\begin{equation}
  F^{(1)}[T,n] = \sum_a \int d^dr \, \mu_{0,a} \, [ n_a(\mathbf{r}) - n_{0,a}(\mathbf{r}) ] \, .
  \label{equation::E_080} 
\end{equation}
Since the chemical potential $\mu_{0,a}$ is constant and since in the coarse-graining procedure 
\eqref{equation::D_040} the kernel $w( \mathbf{r} - \mathbf{r}^\prime )$ is normalized to unity according to
\begin{equation}
  \int d^dr \, w( \mathbf{r} - \mathbf{r}^\prime ) = 1
  \label{equation::E_090} 
\end{equation}
we may replace the microscopic densities by the macroscopic ones and obtain 
\begin{equation}
  F^{(1)}[T,\tilde{n}] = \sum_a \int d^dr \, \mu_{0,a} \, [ \tilde{n}_a(\mathbf{r}) - \tilde{n}_{0,a} ]
  \label{equation::E_100} 
\end{equation}
exactly without any approximation. However, in this result we miss an important contribution with a 
stress tensor $\sigma_{0,i}^{\ \ \ k}$ and a strain field $\partial_k u^i$. Rather, we would expect the formula 
\begin{equation}
  \begin{split}
    F^{(1)}[T,\mathbf{u},\tilde{n}] = \int d^dr \, \Bigl[ &\sigma_{0,i}^{\ \ \ k} [ \partial_k u^i(\mathbf{r}) ] \\
    &+ \sum_a \mu_{0,a} \, [ \tilde{n}_a(\mathbf{r}) - \tilde{n}_{0,a} ] \Bigr] \, .
  \end{split}
  \label{equation::E_110} 
\end{equation}
However, since for the reference state the stress tensor $\sigma_{0,i}^{\ \ \ k}$ is constant in space, the 
first term can be rewritten as a divergence term and transformed to a surface integral by the theorem 
of Gauss according to
\begin{equation}
  \begin{split}
    \int_V d^dr \, \sigma_{0,i}^{\ \ \ k} [ \partial_k u^i(\mathbf{r}) ] &= \int_V d^dr 
    \, \partial_k [ \sigma_{0,i}^{\ \ \ k} \, u^i(\mathbf{r}) ] \\
    &= \int_{\partial V} dS_k \, \sigma_{0,i}^{\ \ \ k} \, u^i(\mathbf{r}) = 0 \, .
  \end{split}
  \label{equation::E_120} 
\end{equation}
Thus, the surface integral is zero because we require $u^i(\mathbf{r}) = 0$ for $\mathbf{r} \in \partial V$ where 
the volume $V$ is extended to infinity and hence the surface $\partial V$ is  located at infinity. As a consequence, 
there is no difference between the found formula \eqref{equation::E_100} and the expected formula \eqref{equation::E_110}.

Alternatively, we may insert the linear formula \eqref{equation::D_410} into the microscopic first order term 
\eqref{equation::E_080}. Then, after a Taylor expansion of the displacement field $u^i(\mathbf{r}) = u^i + (\partial_k u^i) \, r^k$,
an averaging over the whole space of volume $V$, and some further manipulations we obtain the formula 
\eqref{equation::E_110} together with the stress tensor
\begin{equation}
  \sigma_{0,i}^{\ \ \ k} = - \frac{ 1 }{ V } \sum_a \int d^dr \, \mu_{0,a} \, [ \partial_i n_{0,a}(\mathbf{r}) ] \, r^k \, .
  \label{equation::E_130} 
\end{equation}
However, this formula relates the stress tensor $\sigma_{0,i}^{\ \ \ k}$ to the chemical potentials $\mu_{0,a}$ 
which is only one degree of freedom for each particle species. All other degrees of freedom which are related to 
nontrivial and nondiagonal terms and which describe deformations are missing. Consequently, this formula is incomplete. 
On the other hand, the integral in Eq.\ \eqref{equation::E_130} is not well defined. Because of the linear factor 
$r^k$ the integrand diverges linearly in the limit $\mathbf{r} \to \infty$ so that strictly speaking the integral 
diverges. Nevertheless, the explicit expression \eqref{equation::E_130} is useful for later formal considerations 
and manipulations to prove some symmetry relations for the elastic constants in Subsec.\ \ref{section::05c}.

The reason for the deficiencies is located in the \emph{functional} Taylor expansion of the free energy. For 
the integrals of the several terms to be well defined we must restrict the function class of the microscopic 
density differences $n_a(\mathbf{r}) - n_{0,a}(\mathbf{r})$. These differences must tend to zero at infinity 
for $\mathbf{r} \to \infty$. Alternatively they must be zero for $\mathbf{r} \in \partial V$ where 
$\partial V$ is the surface of the integration volume. This requirement allows \emph{local} and prohibits 
\emph{global} deformations. Consequently, deformation terms are not present in the first order contribution
\eqref{equation::E_080}. It is not possible to calculate the constant stress tensor $\sigma_{0,ij}$ of the 
homogeneous reference state from the functional Taylor expansion.

In order to proceed with higher-order terms in the Taylor expansion in the first order term of the functional 
expansion \eqref{equation::E_080} we must insert the full nonlinear density differences defined in 
Eq.\ \eqref{equation::D_400} and expand up to higher orders with respect to the displacement field 
$\mathbf{u}(\mathbf{r})$. However, since we have derived the macroscopic formula \eqref{equation::E_100}
\emph{exactly} from the microscopic formula \eqref{equation::E_080} all these terms must drop out. These 
additional contributions are volume integrals of divergence terms which can be transformed into surface 
integrals by the theorem of Gauss which finally are zero. In Eq.\ \eqref{equation::E_120} we have shown 
this explicitly in first order. We have proven the same also in second order so that we expect that 
it is true also in higher orders.

We proceed with the second-order term in the functional Taylor expansion \eqref{equation::E_040}. Here we 
insert the density differences $n_a(\mathbf{r}) - n_{0,a}(\mathbf{r})$ given by the first-order formula 
\eqref{equation::D_410} together with the correction densities $\Delta n^\prime_a(\mathbf{r})$ defined in 
Eq.\ \eqref{equation::D_810}. For convenience of notation we rewrite the correction densities in the form
\begin{equation}
  \begin{split}
    \Delta n^\prime_a(\mathbf{r}) = \,& \Delta B_{a,i}^{\prime \, \ \ j}(\mathbf{r}) 
    \, ( \partial_j u^i(\mathbf{r}) ) \\
    &+ \sum_b \Delta A^\prime_{ab}(\mathbf{r}) \, [ \tilde{n}_b(\mathbf{r}) - \tilde{n}_{0,b} ]
  \end{split}
  \label{equation::E_140} 
\end{equation}
together with the microscopic functions
\begin{equation}
  \begin{split}
    \Delta B_{a,i}^{\prime \, \ \ j}(\mathbf{r}) = \,& \Delta b_{a,i}^{\prime \, \ \ j}(\mathbf{r}) 
    - \sum_{bc} \, a^\prime_{ab}(\mathbf{r}) \, \tilde{a}^{-1}_{bc} \, \Delta \tilde{b}_{c,i}^{\ \ \ j} \\
    &+ \sum_b \Delta A^\prime_{ab}(\mathbf{r}) \, \tilde{n}_{0,b} \, \delta_i^{\ j}
  \end{split}
  \label{equation::E_150} 
\end{equation}
and
\begin{equation}
  \Delta A^\prime_{ab}(\mathbf{r}) = \sum_c a^\prime_{ac}(\mathbf{r}) \, \tilde{a}^{-1}_{cb} 
  - \frac{ n^\prime_a(\mathbf{r}) }{ \tilde{n}_{0,a} } \, \delta_{ab} \, .
  \label{equation::E_160} 
\end{equation}
The last terms in Eqs.\ \eqref{equation::E_140} and \eqref{equation::E_150} with the functions $\Delta A^\prime_{ab}(\mathbf{r})$ 
are obtained from Eq.\ \eqref{equation::D_810} if we insert the vacancy densities \eqref{equation::D_780} together with 
$\delta \tilde{n}_b(\mathbf{r}) = \tilde{n}_b(\mathbf{r}) - \tilde{n}_{0,b}$. We then proceed as described above in the text 
after Eq.\ \eqref{equation::E_040}. We define center of mass and relative coordinates according to \eqref{equation::E_050}. 
We insert the inverse formulas
\begin{equation}
  \mathbf{r}_1 = \mathbf{R} + \Delta\mathbf{r} / 2 \, , \qquad
  \mathbf{r}_2 = \mathbf{R} - \Delta\mathbf{r} / 2
  \label{equation::E_170} 
\end{equation}
and replace the integrals according to \eqref{equation::E_060}. Furthermore, we expand the displacement fields 
$u^i(\mathbf{r}_1)$, $u^i(\mathbf{r}_2)$, the strain tensors $\partial_k u^i(\mathbf{r}_1)$, $\partial_k u^i(\mathbf{r}_2)$, 
and the macroscopic density differences $[ \tilde{n}_a(\mathbf{r}_1) - \tilde{n}_{0,a} ]$, 
$[ \tilde{n}_a(\mathbf{r}_2) - \tilde{n}_{0,a} ]$ depending on $\mathbf{r}_1$ and $\mathbf{r}_2$ into Taylor series with 
respect to the relative coordinates $\Delta \mathbf{r}$ up to first or second order. We keep all terms which are needed.
Then, we obtain the free energy in the form
\begin{equation}
  \begin{split}
    F^{(2)}[T,\mathbf{u},\tilde{n}] = \,& \frac{ 1 }{ 2 } \int d^dR \, \Bigl[ \lambda_{ij}^{\ \ kl} \, [ \partial_k u^i(\mathbf{R}) ] 
    \, [ \partial_l u^j(\mathbf{R}) ] \\
    &- 2 \sum_a \mu_{a,i}^{\ \ \ k} \, \frac{ [ \tilde{n}_a(\mathbf{R}) - \tilde{n}_{0,a} ] }{ \tilde{n}_{0,a} } 
    \, [ \partial_k u^i(\mathbf{R}) ] \\
    &+ \sum_{ab} \nu_{ab} \, \frac{ [ \tilde{n}_a(\mathbf{R}) - \tilde{n}_{0,a} ] }{ \tilde{n}_{0,a} }
    \, \frac{ [ \tilde{n}_b(\mathbf{R}) - \tilde{n}_{0,b} ] }{ \tilde{n}_{0,b} } \Bigr]
  \end{split}
  \label{equation::E_180} 
\end{equation}
where the integral is calculated over the center of mass coordinates $\mathbf{R}$. The elastic parameters 
\begin{equation}
  \Lambda(\mathbf{R}) = ( \lambda_{ij}^{\ \ kl}(\mathbf{R}) , \mu_{a,i}^{\ \ \ k}(\mathbf{R}) , \nu_{ab}(\mathbf{R}) )
  \label{equation::E_190} 
\end{equation}
are given by integrals over the relative coordinates $\Delta \mathbf{r}$ as
\begin{equation}
  \Lambda(\mathbf{R}) = \sum_{ab} \int d^d(\Delta r) 
  \, \frac{ \delta^2 F[T,n] }{ \delta n_a(\mathbf{r}_1) \, \delta n_b(\mathbf{r}_2) } \bigg|_{n = n_0}
  \, [ \ldots ] \, [ \ldots ] \, .
  \label{equation::E_200} 
\end{equation}
In this definition, the elastic parameters are not constant even though in the reference state the crystal is 
perfectly homogeneous with a perfect periodic lattice. Rather, the elastic parameters are microscopic functions 
in the center of mass coordinates $\mathbf{R}$. They are periodic functions which satisfy
\begin{equation}
  \Lambda(\mathbf{R}) = \Lambda( \mathbf{R} + \mathbf{a} ) \quad \mbox{for all $\mathbf{a} \in \mathcal{A}$}
  \label{equation::E_210} 
\end{equation}
where $\mathcal{A}$ is the set of all lattice vectors defined in Eq.\ \eqref{equation::D_220}. However, since 
in the terms of the integral \eqref{equation::E_180} the other factors are macroscopic functions, we may apply 
the coarse-graining procedure \eqref{equation::D_040} and define the macroscopic elastic parameters
\begin{equation}
  \tilde{\Lambda}(\mathbf{R}) = \int d^dR^\prime \, w( \mathbf{R} - \mathbf{R}^\prime ) \, \Lambda(\mathbf{R}^\prime) \, .
  \label{equation::E_220} 
\end{equation}
For our reference state these macroscopic elastic parameters are constant so that $\tilde{\Lambda}(\mathbf{R}) = \tilde{\Lambda}$. 
Hence, we may define elastic constants by averaging over the whole space according to
\begin{equation}
  \Lambda = \frac{ 1 }{ V } \int d^dR^\prime \, \Lambda(\mathbf{R}^\prime)
  \label{equation::E_230} 
\end{equation}
where $V$ is the volume of the whole space. Inserting Eq.\ \eqref{equation::E_200} and applying 
Eq.\ \eqref{equation::E_060} we then obtain the final formula for the elastic constants
\begin{equation}
  \Lambda = \frac{ 1 }{ V } \sum_{ab} \int d^dr_1 \int d^dr_2 
  \, \frac{ \delta^2 F[T,n] }{ \delta n_a(\mathbf{r}_1) \, \delta n_b(\mathbf{r}_2) } \bigg|_{n = n_0}
  [ \ldots ] \, [ \ldots ] \, .
  \label{equation::E_240} 
\end{equation}
In the free energy all terms which depend on a displacement field $u^i(\mathbf{R})$ with no space derivative 
must drop out. The reason is the \emph{translation invariance}. Nevertheless some contributions must be kept 
which can be transformed by a partial integration. These terms are
\begin{eqnarray}
  u^i(\mathbf{R}) \, [ \partial_k \partial_l u^j(\mathbf{R}) ] &\to& - [ \partial_k u^i(\mathbf{R}) ] 
  \, [ \partial_l u^j(\mathbf{R}) ] \, , \qquad
  \label{equation::E_250} \\
  \ [ \partial_k \partial_l u^i(\mathbf{R}) ] \, u^j(\mathbf{R}) &\to& - [ \partial_k u^i(\mathbf{R}) ] 
  \, [ \partial_l u^j(\mathbf{R}) ] \qquad
  \label{equation::E_260}
\end{eqnarray}
which contribute to the first line of the free energy \eqref{equation::E_180} and
\begin{equation}
  [ \partial_k ( \tilde{n}_b(\mathbf{R}) - \tilde{n}_{0,b} ) ] \, u^i(\mathbf{R}) \to 
  - [ \tilde{n}_b(\mathbf{R}) - \tilde{n}_{0,b} ] \, [ \partial_k u^i(\mathbf{R}) ] 
  \label{equation::E_270}
\end{equation}
which contributes to the second line of the free energy \eqref{equation::E_180}. Eventually, only those terms 
are retained which are shown in Eq.\ \eqref{equation::E_180}.

Now, we present our explicit results for the elastic constants including all corrections terms due to the correction 
densities \eqref{equation::E_140}. After a lengthy calculation we obtain the final results
\begin{widetext}
\begin{equation}
  \begin{split}
    \lambda_{ij}^{\ \ kl} = \,& - \frac{ 1 }{ 2 V } \sum_{ab} \int d^dr_1 \int d^dr_2 
    \, \frac{ \delta^2 F[T,n] }{ \delta n_a(\mathbf{r}_1) \, \delta n_b(\mathbf{r}_2) } \bigg|_{n = n_0} 
    \, \Bigl[ [ \partial_i n_{0,a}(\mathbf{r}_1) ] \, [ \partial_j n_{0,b}(\mathbf{r}_2) ] \, ( r_1^k - r_2^k ) \, ( r_1^l - r_2^l ) \\
    &\hspace{73mm} + [ \partial_i n_{0,a}(\mathbf{r}_1) ] \, \Bigr\{ ( r_1^k - r_2^k ) \, \Delta B_{b,j}^{\prime \, \ \ l}(\mathbf{r}_2) 
    + ( r_1^l - r_2^l ) \, \Delta B_{b,j}^{\prime \, \ \ k}(\mathbf{r}_2) \Bigr\} \\
    &\hspace{73mm} - \Bigr\{ \Delta B_{a,i}^{\prime \, \ \ k}(\mathbf{r}_1) \, ( r_1^l - r_2^l ) 
    + \Delta B_{a,i}^{\prime \, \ \ l}(\mathbf{r}_1) \, ( r_1^k - r_2^k ) \Bigr\} \, [ \partial_j n_{0,b}(\mathbf{r}_2) ] \\
    &\hspace{73mm} - \Delta B_{a,i}^{\prime \, \ \ k}(\mathbf{r}_1) \, \Delta B_{b,j}^{\prime \, \ \ l}(\mathbf{r}_2) 
    - \Delta B_{a,i}^{\prime \, \ \ l}(\mathbf{r}_1) \, \Delta B_{b,j}^{\prime \, \ \ k}(\mathbf{r}_2) \Bigr] \, ,
  \end{split}
  \label{equation::E_280} 
\end{equation}
\begin{equation}
  \begin{split}
    \mu_{c,i}^{\ \ \ k} = \,& - \frac{ 1 }{ V } \sum_{ab} \int d^dr_1 \int d^dr_2 
    \, \frac{ \delta^2 F[T,n] }{ \delta n_a(\mathbf{r}_1) \, \delta n_b(\mathbf{r}_2) } \bigg|_{n = n_0} 
    \, \Bigl[ n_{0,a}(\mathbf{r}_1) \, \delta_{ac} + \Delta A^\prime_{ac}(\mathbf{r}_1) \, \tilde{n}_{0,c} \Bigr] \\
    &\hspace{70mm} \times \Bigl[ [ \partial_i n_{0,b}(\mathbf{r}_2) ] \, ( r_1^k - r_2^k ) 
    + \Delta B_{b,i}^{\prime \, \ \ k}(\mathbf{r}_2) \Bigr] \, ,
  \end{split}
  \label{equation::E_290} 
\end{equation}
\begin{equation}
  \begin{split}
    \nu_{cd} = \,& + \frac{ 1 }{ V } \sum_{ab} \int d^dr_1 \int d^dr_2 
    \, \frac{ \delta^2 F[T,n] }{ \delta n_a(\mathbf{r}_1) \, \delta n_b(\mathbf{r}_2) } \bigg|_{n = n_0} 
    \, \Bigl[ n_{0,a}(\mathbf{r}_1) \, \delta_{ac} + \Delta A^\prime_{ac}(\mathbf{r}_1) \, \tilde{n}_{0,c} \Bigr] 
    \, \Bigl[ n_{0,b}(\mathbf{r}_2) \, \delta_{bd} + \Delta A^\prime_{bd}(\mathbf{r}_2) \, \tilde{n}_{0,d} \Bigr] \, .
  \end{split}
  \label{equation::E_300} 
\end{equation}
\end{widetext}

We have calculated the first-order and the second-order terms of the Taylor expansion of the free energy which are 
given by Eqs.\ \eqref{equation::E_110} and \eqref{equation::E_180}. We may put all terms together. Thus, for the 
free energy we obtain
\begin{equation}
  \begin{split}
    F[T,\mathbf{u},\tilde{n}] = \,& F_0[T,\tilde{n}_0] + \int d^dr \, \Bigl[ \sigma_{0,i}^{\ \ \ k} [ \partial_k u^i(\mathbf{r}) ] \\
    &+ \sum_a \mu_{0,a} \, [ \tilde{n}_a(\mathbf{r}) - \tilde{n}_{0,a} ] \\
    &+ \frac{ 1 }{ 2 } \, \lambda_{ij}^{\ \ kl} \, [ \partial_k u^i(\mathbf{r}) ] \, [ \partial_l u^j(\mathbf{r}) ] \\
    &- \sum_a \mu_{a,i}^{\ \ \ k} \, \frac{ [ \tilde{n}_a(\mathbf{r}) - \tilde{n}_{0,a} ] }{ \tilde{n}_{0,a} } 
    \, [ \partial_k u^i(\mathbf{r}) ] \\
    &+ \frac{ 1 }{ 2 } \sum_{ab} \nu_{ab} \, \frac{ [ \tilde{n}_a(\mathbf{r}) - \tilde{n}_{0,a} ] }{ \tilde{n}_{0,a} }
    \, \frac{ [ \tilde{n}_b(\mathbf{r}) - \tilde{n}_{0,b} ] }{ \tilde{n}_{0,b} } \Bigr]
  \end{split}
  \label{equation::E_310} 
\end{equation}
which is a functional on the macroscopic scale and which has a local structure. We may define the local free energy density
\begin{equation}
  \begin{split}
    f(T,\partial \mathbf{u},\tilde{n}) = \,& f_0(T,\tilde{n}_0) + \sigma_{0,i}^{\ \ \ k} \, [ \partial_k u^i ] \\
    &+ \sum_a \mu_{0,a} \, [ \tilde{n}_a - \tilde{n}_{0,a} ] \\
    &+ \frac{ 1 }{ 2 } \, \lambda_{ij}^{\ \ kl} \, [ \partial_k u^i ] \, [ \partial_l u^j ] \\
    &- \sum_a \mu_{a,i}^{\ \ \ k} \, \frac{ [ \tilde{n}_a - \tilde{n}_{0,a} ] }{ \tilde{n}_{0,a} } 
    \, [ \partial_k u^i ] \\
    &+ \frac{ 1 }{ 2 } \sum_{ab} \nu_{ab} \, \frac{ [ \tilde{n}_a - \tilde{n}_{0,a} ] }{ \tilde{n}_{0,a} }
    \, \frac{ [ \tilde{n}_b - \tilde{n}_{0,b} ] }{ \tilde{n}_{0,b} } \, .
  \end{split}
  \label{equation::E_320} 
\end{equation}
This free energy density represents the local thermodynamic equilibrium of a crystal in a deformed state.
Finally, the total free energy may be written as
\begin{equation}
  F[T,\mathbf{u},\tilde{n}] = \int d^dr \, f(T,\partial \mathbf{u}(\mathbf{r}),\tilde{n}(\mathbf{r})) \, .
  \label{equation::E_330} 
\end{equation}
This free energy has the form which we have proposed previously in Eq.\ \eqref{equation::E_070}. 

We note that strictly speaking and according to Eq.\ \eqref{equation::E_070} we must always put a tilde on top 
of the free energy density $\tilde{f} = \tilde{f}(\mathbf{r})$ because it is a macroscopic function which is 
defined within the macroscopic function class $\tilde{\mathcal{C}}(\ell)$. However, in order to simplify the 
notation we omit the tilde and from now on simply write $f = f(\mathbf{r})$. From the context and from the 
concept of local thermodynamic equilibrium it should be clear that the free energy density always is a 
macroscopic function.

\subsection{Implications of translation and rotation invariance on the elastic constants}
\label{section::05c}
For infinitesimal translations and rotations defined by a formula like \eqref{equation::D_280} the 
infinitesimal displacement field is given by
\begin{equation}
  \delta \mathbf{u}(\mathbf{r}) = \delta \mathbf{a} + \delta \boldsymbol{\varphi} \times \mathbf{r} \, .
  \label{equation::E_340} 
\end{equation}
Here $\delta \mathbf{a}$ is an infinitesimal translation vector and $\delta \boldsymbol{\varphi}$ is an 
infinitesimal rotation angle. The related variations of the particle densities are
\begin{equation}
  \delta n_a(\mathbf{r}) = - [ \nabla n_a(\mathbf{r}) ] \cdot \delta \mathbf{u}(\mathbf{r})
  \label{equation::E_350} 
\end{equation}
where Eq.\ \eqref{equation::E_340} is inserted for $\delta \mathbf{u}(\mathbf{r})$. In the next step we 
use the particle densities \eqref{equation::E_350} in order to calculate the variation of the free energy
$\delta F[T,n]$. Sorting for the independent contributions with respect to the 
transformation parameters $\delta \mathbf{a}$ and $\delta \boldsymbol{\varphi}$ we obtain two equations, 
a first equation for the translation invariance
\begin{equation}
  \sum_a \int d^dr \, \frac{ \delta F[T,n] }{ \delta n_a(\mathbf{r}) } \, [ \nabla n_a(\mathbf{r}) ] = \mathbf{0}
  \label{equation::E_360} 
\end{equation}
and a second for the rotation invariance
\begin{equation}
  \sum_a \int d^dr \, \frac{ \delta F[T,n] }{ \delta n_a(\mathbf{r}) } \, [ ( \mathbf{r} \times \nabla ) n_a(\mathbf{r}) ] 
  = \mathbf{0} \, .
  \label{equation::E_370} 
\end{equation}
Next, onto these two equations we apply a second functional derivative with respect to the particle densities. 
In this way we obtain a further first equation for the translation invariance
\begin{equation}
  \begin{split}
    &\sum_b \int d^dr_2 \, \frac{ \delta^2 F[T,n] }{ \delta n_a(\mathbf{r}_1) \delta n_b(\mathbf{r}_2) } 
    \, [ \nabla_2 n_b(\mathbf{r}_2) ] \\
    &+ \sum_b \int d^dr_2 \, \frac{ \delta F[T,n] }{ \delta n_b(\mathbf{r}_2) } 
    \, [ \nabla_2 \, \delta_{ab} \, \delta( \mathbf{r}_1 - \mathbf{r}_2 ) ] = \mathbf{0}
  \end{split}
  \label{equation::E_380} 
\end{equation}
and a further second equation for the rotation invariance
\begin{equation}
  \begin{split}
    &\sum_b \int d^dr_2 \, \frac{ \delta F[T,n] }{ \delta n_a(\mathbf{r}_1) \delta n_b(\mathbf{r}_2) } 
    \, [ ( \mathbf{r}_2 \times \nabla_2 ) n_b(\mathbf{r}_2) ] \\
    &+ \sum_b \int d^dr_2 \, \frac{ \delta F[T,n] }{ \delta n_b(\mathbf{r}_2) } 
    \, [ ( \mathbf{r}_2 \times \nabla_2 ) \, \delta_{ab} \, \delta( \mathbf{r}_1 - \mathbf{r}_2 ) ] = \mathbf{0} \, .
  \end{split}
  \label{equation::E_390} 
\end{equation}
We note that both of these equations have a second term which arises from the product rule when applying the 
second functional derivative onto Eqs.\ \eqref{equation::E_360} and \eqref{equation::E_370}.

We may apply the invariance equations \eqref{equation::E_360}-\eqref{equation::E_390} to our reference state 
by inserting the reference density $n_a(\mathbf{r}) = n_{0,a}(\mathbf{r})$. Then, following Eq.\ \eqref{equation::E_020} 
the first functional derivative of the free energy can be replaced by the constant chemical potential $\mu_{0,a}$ 
of the reference state. Similarly, for the second functional derivative of the free energy we may insert 
\eqref{equation::E_030}. While the invariance equations \eqref{equation::E_360}-\eqref{equation::E_390} are 
written in vector notation, for our following application we must rewrite them in index notation.

In our considerations and investigations we use two different coordinate system. First, we use the laboratory 
frame with coordinates $r^i$ which is a \emph{Cartesian} coordinate system with the metric $g_{ij} = \delta_{ij}$ 
and the inverse metric $g^{ij} = \delta^{ij}$. Second, we use the material fixed frame with coordinates $r_0^i$ 
which is a curvilinear coordinate system following the deformations of the crystal. Both coordinate systems are 
connected with each other by the displacement field $\mathbf{u}(\mathbf{r})$ formula \eqref{equation::D_270}, 
\eqref{equation::D_280}, or \eqref{equation::D_290}. Here, in general the metric tensors $g_{0,ij}$ and $g_0^{ij}$ 
are nontrivial. Thus, in tensors with several indices like the stress tensor and some of the elastic constants 
we must distinguish two types of indices, a Cartesian type and a curvilinear type. However, in the Taylor expansion 
of the free energy \eqref{equation::E_310} the coefficients are defined for infinitely small displacement fields 
in the limit $\mathbf{u}(\mathbf{r}) \to \mathbf{0}$. Consequently, here we can assume that both coordinate 
systems are Cartesian so that the metric tensors $g_{ij} = g_{0,ij} = \delta_{ij}$ and the inverse metric tensors 
$g^{ij} = g_0^{ij} = \delta^{ij}$ are equal and given by Kronecker symbols. Hence, all indices can be lowered 
and risen by Kronecker symbols $\delta_{ij}$ and $\delta^{ij}$, respectively. In this way we rewrite the formulas 
for the free energy \eqref{equation::E_310} and for the free energy density \eqref{equation::E_320}. We use upper 
indices for the strain tensor $\partial^k u^i$ and lower indices for the stress tensor $\sigma_{0,ik}$ and for the 
elastic constants $\lambda_{ij,kl}$ and $\mu_{a,ik}$.

Now, we are ready to prove some symmetry relations for these tensors. First, we consider the stress tensor 
$\sigma_{0,ik}$ defined formally in Eq.\ \eqref{equation::E_130}. If we apply the relation for translation 
invariance \eqref{equation::E_360} we may shift the linear coordinate factor by a constant according to 
$r^k \to r^k - r_0^k$ without changing the result of the integral \eqref{equation::E_130}. On the other hand, 
if we apply the relation for the rotation invariance \eqref{equation::E_370} we obtain
\begin{equation}
  \sigma_{0,ik} - \sigma_{0,ki} = 0 \, .
  \label{equation::E_400} 
\end{equation}
Hence, the stress tensor $\sigma_{0,ik}$ is symmetric in its indices.

Since in Eq.\ \eqref{equation::E_130} the integral diverges and is defined only formally, we may look for 
an alternative route to prove the symmetry of the stress tensor by using the macroscopic free energy 
\eqref{equation::E_310} or the free energy density \eqref{equation::E_320}. In the case of an infinitesimal translation 
and rotation the displacement field is given by the formula \eqref{equation::E_340}. Thus, we calculate the 
strain tensor by differentiation and obtain $\partial^k u^i = \varphi_j \, \epsilon^{ijk}$ which 
is antisymmetric in the indices so that $\partial^k u^i = - \partial^i u^k$. The related first-order term in the 
free energy density \eqref{equation::E_320} is given by $\sigma_{0,ik} \, [ \partial^k u^i ]$. For this term to 
be independent of translations and rotations the stress tensor must by symmetric in the indices. In this way 
we have proven the symmetry relation \eqref{equation::E_400} once again alternatively and independently.

Next we consider the elastic constants $\lambda_{ij,kl}$ and $\mu_{a,ik}$ which are given by the formulas 
\eqref{equation::E_280} and \eqref{equation::E_290}. We apply the translation and the rotation invariance by using 
the symmetry relations \eqref{equation::E_380} and \eqref{equation::E_390}. The calculation is lengthy and 
cumbersome. However, at the end all terms fit together, all calculations are exact, and no approximations are needed. 
Thus, we obtain 
\begin{eqnarray}
  \lambda_{ij,kl} - \lambda_{kl,ij} &=& \delta_{ij} \, \sigma_{0,kl} - \delta_{kl} \, \sigma_{0,ij} \, ,
  \label{equation::E_410} \\
  \mu_{a,ik} - \mu_{a,ki} &=& 0 \, .
  \label{equation::E_420} 
\end{eqnarray}
In a first step the calculation is done with the leading terms of Eqs.\ \eqref{equation::E_280} and 
\eqref{equation::E_290} where the correction terms are omitted. Thus, the symmetry relations \eqref{equation::E_410} 
and \eqref{equation::E_420} are proven for zero corrections $\Delta B^\prime_{a,ik}(\mathbf{r}) = 0$ and 
$\Delta A^\prime_{ab}(\mathbf{r}) = 0$. We note that the formal definition of the stress tensor \eqref{equation::E_130} 
is essential for the formal manipulations in this calculation. 

In a second step we consider the functions $d_{a,ik}(\mathbf{r})$ defined in Eq.\ \eqref{equation::D_640}. Using 
Eq.\ \eqref{equation::E_030} together with $\delta( \mathbf{r}_1 - \mathbf{r}_2 ) \, ( r_1^j - r_2^j ) = 0$ we may 
replace the direct correlation function and rewrite the functions \eqref{equation::D_640} in terms of the second 
functional derivative of the free energy. Now, we can apply the second equation for rotation invariance 
\eqref{equation::E_390} once again. Thus, we obtain
\begin{equation}
  d_{a,ik}(\mathbf{r}) - d_{a,ki}(\mathbf{r}) = 0 \, .
  \label{equation::E_430} 
\end{equation}
Hence, the $d_{a,ik}(\mathbf{r})$ and the related basis functions $b_{a,ik}(\mathbf{r})$ are symmetric in the 
indices. Next, from the definitions \eqref{equation::D_710} and \eqref{equation::D_720} we find explicitly that also 
the essential contributions $d_{0,a,ik}(\mathbf{r})$ and $b_{0,a,ik}(\mathbf{r})$ are symmetric in the indices. 
We proceed with the formulas \eqref{equation::D_740} and \eqref{equation::D_750} and see that the difference functions 
$\Delta d_{a,ik}(\mathbf{r})$ and $\Delta b_{a,ik}(\mathbf{r})$ are symmetric in the indices either. Finally, from 
Eq.\ \eqref{equation::E_150} we find that the correction functions $\Delta B^\prime_{a,ik}(\mathbf{r})$ are symmetric 
in the indices so that 
\begin{equation}
  \Delta B^\prime_{a,ik}(\mathbf{r}) - \Delta B^\prime_{a,ki}(\mathbf{r}) = 0 \, .
  \label{equation::E_440} 
\end{equation}
Now, we apply this symmetry relation in the definitions of the elastic constants \eqref{equation::E_280} and 
\eqref{equation::E_290}. Then, in the symmetry relations \eqref{equation::E_410} and \eqref{equation::E_420} 
all terms drop out which have one or two factors $\Delta B^\prime_{a,ik}(\mathbf{r})$. In Eq.\ \eqref{equation::E_290} 
there remains a term with one correction function $\Delta A^\prime_{ab}(\mathbf{r})$. This term drops out by 
applying Eqs.\ \eqref{equation::E_380} and \eqref{equation::E_390} once again. Eventually, we find that 
for the elastic constants $\lambda_{ij,kl}$ and $\mu_{a,ik}$ the symmetry relations \eqref{equation::E_410} 
and \eqref{equation::E_420} hold also for nonzero correction functions $\Delta B^\prime_{a,ik}(\mathbf{r})$ and 
$\Delta A^\prime_{ab}(\mathbf{r})$.

\subsection{Adding the kinetic energy}
\label{section::05d}
In the previous section we have investigated the free energy of a crystal at rest with deformations and changes 
of the macroscopic density. Now we extend the results by including motions described by the velocity field 
$\mathbf{v}(\mathbf{r})$ and the momentum density $\mathbf{j}(\mathbf{r})$. Because of Galilean invariance 
we simply add the term for the kinetic energy which is given by Eq.\ \eqref{equation::C_060}. Thus, from 
Eqs.\ \eqref{equation::C_030} and \eqref{equation::C_090} we obtain
\begin{equation}
  F[T,\mathbf{j},n] = F[T,n] + F_\mathrm{kin}[\mathbf{j},n] 
  \label{equation::E_450} 
\end{equation}
which is the free energy functional on \emph{microscopic scales}. The Galilean invariance provides the simple 
relation \eqref{equation::D_120} between momentum density $\mathbf{j}(\mathbf{r})$ and the velocity field 
$\mathbf{v}(\mathbf{r})$ where the mass density $\rho(\mathbf{r})$ is defined in Eq.\ \eqref{equation::D_130}. 
Thus, we may rewrite the \emph{microscopic} functional of the kinetic energy 
\eqref{equation::C_060} as
\begin{equation}
  F_\mathrm{kin}[\mathbf{j},n] = \int d^dr \, \frac{ 1 }{ 2 } \, \rho(\mathbf{r}) \, [ \mathbf{v}(\mathbf{r}) ]^2 \, .
  \label{equation::E_460} 
\end{equation}
The integrand of this functional is a microscopic function defined in the full function space 
$\mathcal{C} = \tilde{\mathcal{C}}(\ell) \oplus \mathcal{C}^\prime(\ell)$. Now, we apply the coarse-graining 
procedure \eqref{equation::D_040} to the integrand by substituting 
\begin{equation}
  \rho(\mathbf{r}) \, [ \mathbf{v}(\mathbf{r}) ]^2 \to \tilde{\rho}(\mathbf{r}) \, [ \mathbf{v}(\mathbf{r}) ]^2 \, .
  \label{equation::E_470} 
\end{equation}
We just put a tilde on top of each microscopic function which here is the mass density. Since the velocity 
field $\mathbf{v}(\mathbf{r})$ is a macroscopic function defined in $\tilde{\mathcal{C}}(\ell)$ this must be 
done only for the particle densities $n_a(\mathbf{r}) \to \tilde{n}_a(\mathbf{r})$ and for the mass density 
$\rho(\mathbf{r}) \to \tilde{\rho}(\mathbf{r})$. Because of the normalization of the kernel \eqref{equation::E_090} 
the kinetic energy functional remains unchanged. Thus, we obtain the \emph{macroscopic} functional of the 
kinetic energy
\begin{equation}
  F_\mathrm{kin}[\tilde{\mathbf{j}},\tilde{n}] = \int d^dr \, \frac{ 1 }{ 2 } \, \tilde{\rho}(\mathbf{r}) 
  \, [ \mathbf{v}(\mathbf{r}) ]^2 \, .
  \label{equation::E_480} 
\end{equation}
Next, we change the variables from the velocity field $\mathbf{v}(\mathbf{r})$ to the coarse-grained momentum 
density $\tilde{\mathbf{j}}(\mathbf{r})$ by using the macroscopic relation \eqref{equation::D_170}. Thus, we 
obtain the macroscopic kinetic energy
\begin{equation}
  F_\mathrm{kin}[\tilde{\mathbf{j}},\tilde{n}] = \int d^dr \, f_\mathrm{kin}( \tilde{\mathbf{j}}(\mathbf{r}), \tilde{n}(\mathbf{r}) ) 
  \label{equation::E_490} 
\end{equation}
together with the kinetic energy density
\begin{equation}
  f_\mathrm{kin}(\tilde{\mathbf{j}},\tilde{n}) = 
  \frac{ [ \tilde{\mathbf{j}}(\mathbf{r}) ]^2 }{ 2 \sum_a m_a \, \tilde{n}_a(\mathbf{r}) } \, .
  \label{equation::E_500} 
\end{equation}
In a final step we calculate the total free energy density by adding the results \eqref{equation::E_320} and 
\eqref{equation::E_500}. Thus, we obtain
\begin{equation}
  f(T,\partial \mathbf{u},\tilde{\mathbf{j}},\tilde{n}) = f(T,\partial \mathbf{u},\tilde{n}) 
  + f_\mathrm{kin}(\tilde{\mathbf{j}},\tilde{n}) \, .
  \label{equation::E_510} 
\end{equation}
At the end, the total free energy functional is the integral
\begin{equation}
  \begin{split}
    F[T,\mathbf{u},\tilde{\mathbf{j}},\tilde{n}] 
    &= F[T,\mathbf{u},\tilde{n}] + F_\mathrm{kin}[\tilde{\mathbf{j}},\tilde{n}] \\
    &= \int d^dr \, f(T,\partial \mathbf{u}(\mathbf{r}),\tilde{\mathbf{j}}(\mathbf{r}),\tilde{n}(\mathbf{r})) \, .
  \end{split}
  \label{equation::E_520} 
\end{equation}

\subsection{Thermodynamic relations and conditions for global thermodynamic equilibrium}
\label{section::05e}
In the local thermodynamic equilibrium the free energy density $f$, the temperature $T$, the strain fields 
$\partial_k u^i$, the momentum density $\tilde{\mathbf{j}}$, and the particle densities $\tilde{n}_a$ are local 
thermodynamic variables which are not independent from each other. There exists an equation which relates these 
quantities to each other, a so called \emph{thermodynamic relation} which reads
\begin{equation}
  f = f(T,\partial \mathbf{u},\tilde{\mathbf{j}},\tilde{n}) \, .
  \label{equation::E_530} 
\end{equation}
This equation results from the formula of the free energy density which we have derived in the previous 
subsections and which is given by Eq.\ \eqref{equation::E_510} together with Eqs.\ \eqref{equation::E_320} and 
\eqref{equation::E_500}.

The related \emph{differential} thermodynamic relation is given by 
\begin{equation}
  df = - s \, dT + \sigma_i^{\ k} \, ( \partial_k du^i ) + \mathbf{v} \cdot d\tilde{\mathbf{j}} 
  + \sum_a \mu_a \, d\tilde{n}_a \, .
  \label{equation::E_540} 
\end{equation}
Since the temperature is assumed to be constant the first term on the right-hand side is just zero so that 
the entropy density $s$ is not known, not needed, and not considered in this section. Nevertheless, we keep 
the first term formally because later in Sec.\ \ref{section::08} we extend the theory to the general case 
where the temperature is not constant. From the partial derivatives of Eqs.\ \eqref{equation::E_320} and 
\eqref{equation::E_500} we obtain the explicit formulas
\begin{eqnarray}
  \sigma_i^{\ k} &=& \sigma_{0,i}^{\ \ \ k} + \lambda_{ij}^{\ \ kl} \, ( \partial_l u^j ) 
  - \sum_a \mu_{a,i}^{\ \ \ k} \, \frac{ [ \tilde{n}_a - \tilde{n}_{0,a} ] }{ \tilde{n}_{0,a} } \, , \hspace{9mm}
  \label{equation::E_550} \\
  \mathbf{v} &=& \tilde{\mathbf{j}} / \tilde{\rho} \, ,
  \label{equation::E_560} \\
  \mu_a &=& \mu_{0,a} - \frac{ \mu_{a,i}^{\ \ \ k} }{ \tilde{n}_{0,a} } \, ( \partial_k u^i ) 
  + \sum_b \frac{ \nu_{ab} }{ \tilde{n}_{0,a} } \, \frac{ [ \tilde{n}_b - \tilde{n}_{0,b} ] }{ \tilde{n}_{0,b} } \, .
  \label{equation::E_570} 
\end{eqnarray}
While the free energy density \eqref{equation::E_320} is a Taylor series up to second order, 
Eqs.\ \eqref{equation::E_550} and \eqref{equation::E_570} are first order and hence linear. On the other hand, 
the kinetic energy density \eqref{equation::E_500} and the related formula for the velocity \eqref{equation::E_560} 
are nonlinear.

In order to find the global thermodynamic equilibrium of the crystalline system we must minimize the total 
free energy \emph{functional} $F[T,\partial \mathbf{u},\tilde{\mathbf{j}},\tilde{n}]$ under certain constraints. 
This functional is given by the integral over the free energy \emph{density} defined in Eq.\ \eqref{equation::E_520}. 
The constraints are the integrals 
\begin{equation}
  \int d^dr \, \tilde{\mathbf{j}}(\mathbf{r}) = \mathbf{P} \, , \qquad
  \int d^dr \, \tilde{n}_a(\mathbf{r}) = N_a \, ,
  \label{equation::E_580} 
\end{equation}
respectively, which are related to a conserved total momentum $\mathbf{P}$ and to conserved total particle numbers 
$N_a$. We must derive the necessary conditions for the minimization procedure. For this purpose we consider the 
variation of the free-energy functional \eqref{equation::E_520}. Using Eq.\ \eqref{equation::E_540} we obtain
\begin{equation}
  \begin{split}
    \delta F[T,\mathbf{u},\tilde{\mathbf{j}},\tilde{n}] &= \int d^dr 
    \, \delta f(T,\partial \mathbf{u}(\mathbf{r}),\tilde{\mathbf{j}}(\mathbf{r}),\tilde{n}(\mathbf{r})) \\
    &= \int d^dr \, \bigl[ - s \, \delta T + \sigma_i^{\ k}(\mathbf{r}) \, [ \partial_k \delta u^i(\mathbf{r}) ] \\
    &\hspace{15mm} + \mathbf{v}(\mathbf{r}) \cdot \delta \tilde{\mathbf{j}}(\mathbf{r}) 
    + \sum_a \mu_a(\mathbf{r}) \, \delta \tilde{n}_a(\mathbf{r}) \bigr] \, .
  \end{split}
  \label{equation::E_590} 
\end{equation}
Since the strain fields $\partial_k \delta u^i(\mathbf{r})$ are not but rather the displacement fields 
$\delta u^i(\mathbf{r})$ are the natural variables we must perform a partial integration for the second term. 
The requirement $\delta u^i(\mathbf{r}) = 0$ for $\mathbf{r} \in \partial V$ implies
\begin{equation}
  \int_V d^dr \, \partial_k [ \sigma_i^{\ k}(\mathbf{r}) \, \delta u^i(\mathbf{r}) ] 
  = \int_{\partial V} dS_k \, [ \sigma_i^{\ k}(\mathbf{r}) \, \delta u^i(\mathbf{r}) ] = 0 \, .
  \label{equation::E_600} 
\end{equation}
Thus, we obtain the global thermodynamic relation
\begin{equation}
  \begin{split}
    \delta F[T,\mathbf{u},\tilde{\mathbf{j}},\tilde{n}] &= \int d^dr 
    \, \bigl[ - s \, \delta T - [ \partial_k \, \sigma_i^{\ k}(\mathbf{r}) ] \, \delta u^i(\mathbf{r}) \\
    &\hspace{15mm} + \mathbf{v}(\mathbf{r}) \cdot \delta \tilde{\mathbf{j}}(\mathbf{r}) 
    + \sum_a \mu_a(\mathbf{r}) \, \delta \tilde{n}_a(\mathbf{r}) \bigr] \, .
  \end{split}
  \label{equation::E_610} 
\end{equation}
We note that the temperature $T$ is constant by assumption so that $\delta T = 0$ and the first term of the free 
energy variation \eqref{equation::E_610} can be discarded. Thus, we obtain the nontrivial functional derivatives 
of the free energy and the related necessary conditions for the global thermal equilibrium which read 
\begin{equation}
  \partial_k \, \sigma_i^{\ k}(\mathbf{r}) = 0 \, , \qquad 
  \mathbf{v}(\mathbf{r}) = \mathbf{v} \, , \qquad
  \mu_a(\mathbf{r}) = \mu_a \, .
  \label{equation::E_620} 
\end{equation}
On the left-hand sides of these three equations we must insert our results given by 
Eqs.\ \eqref{equation::E_550}-\eqref{equation::E_570}. On the right-hand sides of the latter two conditions 
there are a constant velocity $\mathbf{v}$ and constant chemical potentials $\mu_a$ which act as Lagrange 
parameters for the constraints of a fixed total momentum $\mathbf{P}$ and fixed total particle numbers $N_a$ 
given in Eq.\ \eqref{equation::E_580}.

The first condition might be interpreted as the divergence of the stress tensor to be zero which might imply 
zero local forces in thermal equilibrium. However, this interpretation is too simple. $\sigma_i^{\ k}$ given 
explicitly in Eq.\ \eqref{equation::E_550} up to linear order is not the total stress tensor of the crystals. 
Rather, it is a stress tensor under the condition that the other variables in the thermodynamic relation 
\eqref{equation::E_540} are constant.

Our approach describes crystals where diffusion of the particles is possible and the number of particles on 
a lattice site may deviate from unity. This physical situation is true in cluster crystals. The deformation 
of the lattice structure of the crystal is \emph{decoupled} from the mean motion of the particles. For this 
reason, the deformation described by the displacement field $\mathbf{u}(\mathbf{r})$ and the particle densities 
$\tilde{n}_a(\mathbf{r})$ relax \emph{independently} to their global equilibrium values. Consequently, for 
these relaxations two necessary conditions are needed which are given by the first and the last condition 
in Eq.\ \eqref{equation::E_620}. Finally, the second condition implies the relaxation of the motions to a 
constant velocity which in most cases is just zero.

The situation is completely different in a normal crystal where strong repulsive interactions at short distances 
imply that the number of particles on each lattice site is forced to unity. Here the motion of the particles is 
strictly coupled to the deformation of the lattice structure. Consequently, the densities $\tilde{n}_a(\mathbf{r})$
are no independent variables. Rather they are determined by the deformation and hence coupled to the displacement 
field $\mathbf{u}(\mathbf{r})$ according to Eq.\ \eqref{equation::D_360}. For infinitesimal deformations this 
equation reads 
\begin{equation}
  \delta \tilde{n}_a(\mathbf{r}) = - \tilde{n}_{0,a} \, [ \nabla \cdot \delta \mathbf{u}(\mathbf{r}) ] 
  = - \tilde{n}_{0,a} \, [ \partial_i \, \delta u^i(\mathbf{r}) ] \, .
  \label{equation::E_630} 
\end{equation}
We insert this result into Eq.\ \eqref{equation::E_610} and then obtain
\begin{equation}
  \begin{split}
    \delta F[T,\mathbf{u},\tilde{\mathbf{j}}] =& \int_V d^dr 
    \, \Bigl[ - s \, \delta T + \mathbf{v}(\mathbf{r}) \cdot \delta \tilde{\mathbf{j}}(\mathbf{r}) \\
    &+ \Bigl( \sigma_i^{\ k}(\mathbf{r}) - \sum_a \mu_a(\mathbf{r}) \, \tilde{n}_{0,a} \, \delta_i^{\ k} \Bigr) 
    \, [ \partial_k \delta u^i(\mathbf{r}) ] \Bigr] \, .
  \end{split}
  \label{equation::E_640} 
\end{equation}
Once again we perform a partial integration for the term with the displacement field. We find a surface 
integral similar like Eq.\ \eqref{equation::E_600} which is zero for the same reason. Thus, we obtain the 
variation of the free energy functional
\begin{equation}
  \begin{split}
    \delta F[T,\mathbf{u},\tilde{\mathbf{j}}] =& \int_V d^dr 
    \, \Bigl[ - s \, \delta T + \mathbf{v}(\mathbf{r}) \cdot \delta \tilde{\mathbf{j}}(\mathbf{r}) \\
    &\hspace{-2mm} - \Bigl\{ \partial_k \Bigl( \sigma_i^{\ k}(\mathbf{r}) 
    - \sum_a \mu_a(\mathbf{r}) \, \tilde{n}_{0,a} \, \delta_i^{\ k} \Bigr) \Bigr\}
    \, \delta u^i(\mathbf{r}) \Bigr] \, .
  \end{split}
  \label{equation::E_650} 
\end{equation}
Eventually, from this result we find the necessary condition for the global thermodynamic equilibrium of a 
\emph{normal crystal}
\begin{equation}
  \partial_k \Bigl( \sigma_i^{\ k}(\mathbf{r}) - \sum_a \mu_a(\mathbf{r}) \, \tilde{n}_{0,a} \, \delta_i^{\ k} \Bigr) 
  = 0 \, , \qquad
  \mathbf{v}(\mathbf{r}) = \mathbf{v} \, .
  \label{equation::E_660} 
\end{equation}
Here, once again the formulas \eqref{equation::E_550}-\eqref{equation::E_570} must be inserted.
Clearly, there are no separate conditions for the chemical potentials $\mu_a(\mathbf{r})$ because the motion of the 
particles is coupled to the deformation of the crystal structure where the particle densities $\tilde{n}_a(\mathbf{r})$ 
are variables which are not independent.

The necessary conditions \eqref{equation::E_660} are derived assuming that both the displacement field 
$\mathbf{u}(\mathbf{r})$ and the momentum density $\tilde{\mathbf{j}}(\mathbf{r})$ are adjusted when minimizing the 
free energy. However, in the later Subsec.\ \ref{section::09d} it turns out that for a normal crystal the time-evolution 
equation for the displacement field $\mathbf{u}(\mathbf{r})$ does not have a dissipative contribution so that the 
displacement field can not relax to equilibrium. This fact arises because in normal crystals the diffusion of particles 
is suppressed. For this reason, in thermal equilibrium the free energy is not minimized with respect to the deformation 
and the displacement field $\mathbf{u}(\mathbf{r})$. Consequently, in Eq.\ \eqref{equation::E_660} the first necessary 
condition is not satisfied and hence must be discarded. On the other hand, the effects of friction and viscosity allow 
the momentum density $\tilde{\mathbf{j}}(\mathbf{r})$ to relax to equilibrium. Thus, in Eq.\ \eqref{equation::E_660} 
the second necessary condition remains valid and hence must be taken into account.

\section{Strain and stress for nonlinear deformations}
\label{section::06}
In order to derive a macroscopic theory within the framework of \emph{continuum mechanics} we must establish a macroscopic 
nonequilibrium state in a so called \emph{local thermodynamic equilibrium}. This concept has been achieved to a certain 
stage with the free energy functional \eqref{equation::E_520} which is written as an integral over a local free energy 
density $f = f(T,\partial \mathbf{u},\tilde{\mathbf{j}},\tilde{n})$ which is a local function of the temperature, the 
strain tensor $\partial_k u^i$, the momentum density $\tilde{\mathbf{j}}$, and the particle densities $\tilde{n}_a$. 
All these local thermodynamic variables should be \emph{slowly varying} functions in the space so that the related 
gradients are small. For the deformation of the crystals this means, that the gradient of the strain tensor 
$\partial_k \partial_l u^i$ which is the \emph{second derivative} of the displacement field $u^i$ is small. In this 
sense the displacement field $\mathbf{u}(\mathbf{r})$ plays an exceptional role within the thermodynamic variables. 

We have derived the Taylor expansion of the free energy up to second order where we have started with a reference 
state which describes a global homogeneous crystal in thermal equilibrium. This means, the nonequilibrium state 
may deviate only slightly from the global equilibrium. In this section we extend the nonequilibrium state into the 
nonlinear regime. More precisely, we want to reach the nonlinear regime on the macroscopic scale where locally 
on a mesoscopic scale $\ell$ we can still use our Taylor expansion and the related linear response formulas 
\eqref{equation::E_550}-\eqref{equation::E_570}.

For this purpose we consider a certain point $P$ in the space which has the coordinate vector $\mathbf{r}_P$ in 
the Cartesian laboratory frame and the related coordinate vector $\mathbf{r}_{0,P}$ in the curvilinear frame of the 
deformed matter. For our nonequilibrium state we choose the latter coordinate frame in such a way so that precisely at 
the point $P$ the displacement field $\mathbf{u}(\mathbf{r}_P)$ and the strain tensor $\partial_k u^i(\mathbf{r}_P)$ 
are zero. Because of the assumption of small gradients the second derivative $\partial_k \partial_l u^i(\mathbf{r}_P)$ 
is small. Consequently, in an epsilon surroundings 
$U_\varepsilon(P) = \{ \mathbf{r} \mid | \mathbf{r} - \mathbf{r}_P | < \varepsilon \}$ we may expand the displacement 
field $\mathbf{u}(\mathbf{r})$ in a Taylor series around $\mathbf{r}_P$ up to second order in the difference 
$\mathbf{r} - \mathbf{r}_P$. For the epsilon we use the mesoscopic scale so that $\varepsilon = \ell$. Here the local 
displacements $\mathbf{u}(\mathbf{r})$ and the local strains $\partial_k u^i(\mathbf{r})$ are small. Similarly, the 
deviations of the densities $\tilde{\mathbf{j}}$ and $\tilde{n}_a$ from the reference values are small. Consequently, 
we may apply our Taylor expansion of the free energy density $f = f(\mathbf{r})$ for $\mathbf{r} \in U_\varepsilon(P)$ 
in the epsilon surroundings.

We repeat this procedure for many points $P_1$, $P_2$, $P_3$, \ldots, $P_n$ in the space. We consider a crystal 
in a volume $V$. We choose a sufficiently large number of points $n$ so that the complete volume $V$ is covered by the 
epsilon surroundings $U_\varepsilon(P_i)$ around these points. In general, this number $n$ may be infinite. However, 
in the case of a finite volume $V$ and in the case of continuous functions the theorem of Heine and Borel tells us 
that a finite number of points $n < \infty$ and hence a finite number of epsilon surroundings is sufficient to cover 
the whole volume. The curvilinear coordinates $\mathbf{r}_{0,P_i}$ of the central points $P_i$ of different epsilon 
surroundings $U_\varepsilon(P_i)$ are related to each other by \emph{linear} coordinate transformations. In this way 
we obtain a nonlinear local thermodynamic theory for the whole volume $V$ or even for the whole space. In the following 
subsections we implement this concept step by step.

\subsection{Nonlinear strain}
\label{section::06a}
Strictly speaking, the deformation of the crystal is described by the displacement field $\mathbf{u}(\mathbf{r})$ 
defined as a function in the whole space. However, this function has too many degrees of freedom. Because of 
translation invariance we may shift this function by a constant value $\mathbf{u}_0$. Hence, a more appropriate 
variable to describe the deformation is the strain tensor $\partial_k u^i(\mathbf{r})$. Alternatively we may 
consider the transformation matrix
\begin{equation}
  e^i_k(\mathbf{r}) = \frac{ \partial r_0^i }{ \partial r^k } = \delta^i_k - \partial_k u^i(\mathbf{r})
  \label{equation::F_010} 
\end{equation}
which differs only by a constant tensor $\delta^i_k$ and a minus sign. We will later see that $e^i_k(\mathbf{r})$ 
is the proper variable for the strain when we derive the time-evolution equations for the dynamics of the system. 
Nevertheless, the transformation matrix still has too many degrees of freedom. It is a $d \times d$ matrix where 
$d$ is the dimension of the space. It includes rotations where, however, a crystal is invariant under rotations. 
Consequently, we need a more appropriate tensor for the strain which only includes deformations but no rotations.

A more appropriate variable for the description of deformations is the metric tensor $g_{0,ij}(\mathbf{r})$ 
of the curvilinear coordinates which is related to the metric tensor $g_{ij} = \delta_{ij}$ of the Cartesian 
coordinates by the transformation formula
\begin{equation}
  g_{0,ij}(\mathbf{r}) = \frac{ \partial r^k }{ \partial r_0^i } \, \frac{ \partial r^l }{ \partial r_0^j } \, g_{kl} 
  = \frac{ \partial r^k }{ \partial r_0^i } \, \frac{ \partial r^l }{ \partial r_0^j } \, \delta_{kl} \, .
  \label{equation::F_020} 
\end{equation}
Here we need the inverse transformation matrices
\begin{equation}
  \frac{ \partial r^k }{ \partial r_0^i } = \delta^k_i + \partial_{0,i} u^k(\mathbf{r})
  \label{equation::F_030} 
\end{equation}
which are calculated from the transformation formula \eqref{equation::D_270}. We insert this result into 
Eq.\ \eqref{equation::F_020} and then obtain the metric tensor
\begin{equation}
  g_{0,ij}(\mathbf{r}) = \delta_{ij} + 2 \, u_{ij}(\mathbf{r})
  \label{equation::F_040} 
\end{equation}
where
\begin{equation}
  u_{ij}(\mathbf{r}) = \frac{ 1 }{ 2 } \, \bigl[ \partial_{0,i} u_j(\mathbf{r}) + \partial_{0,j} u_i(\mathbf{r})
  + ( \partial_{0,i} u^k(\mathbf{r}) ) \, ( \partial_{0,j} u_k(\mathbf{r}) ) \bigr] 
  \label{equation::F_050} 
\end{equation}
is the well known \emph{Lagrange strain tensor}. Because of the last contribution it is a \emph{nonlinear} 
strain tensor. For a simpler notation on the right-hand side of Eq.\ \eqref{equation::F_050} we have lowered 
some indices by using the Cartesian metric $\delta_{ij}$. The metric tensor $g_{0,ij}(\mathbf{r}) = g_{0,ji}(\mathbf{r})$ 
and the Lagrange strain tensor $u_{ij}(\mathbf{r}) = u_{ji}(\mathbf{r})$ are symmetric in the indices and hence 
do not depend on rotations. Consequently, they describe local deformations only.

In Eqs.\ \eqref{equation::F_030} and \eqref{equation::F_050} the derivatives $\partial_{0,i} = \partial / \partial r_0^i$ 
are defined with respect to the curvilinear coordinates $r_0^i$. Unfortunately, these derivatives are not suited 
for our calculations. For this reason, we will not use the inverse transformation matrix, the metric tensor, and 
the Lagrange strain tensor for describing deformations.

A further alternative is the inverse metric tensor $g_0^{ij}(\mathbf{r})$ with upper indices. It is defined by 
the transformation formula
\begin{equation}
  g_0^{ij}(\mathbf{r}) = \frac{ \partial r_0^i }{ \partial r^k } \, \frac{ \partial r_0^j }{ \partial r^l } \, g^{kl} 
  = \frac{ \partial r_0^i }{ \partial r^k } \, \frac{ \partial r_0^j }{ \partial r^l } \, \delta^{kl} \, .
  \label{equation::F_060} 
\end{equation}
Here we may insert the transformation matrix \eqref{equation::F_010}. Then, we obtain
\begin{equation}
  g_0^{ij}(\mathbf{r}) = \delta^{ij} - 2 \, u^{ij}(\mathbf{r})
  \label{equation::F_070} 
\end{equation}
where 
\begin{equation}
  u^{ij}(\mathbf{r}) = \frac{ 1 }{ 2 } \, \bigl[ \partial^i u^j(\mathbf{r}) + \partial^j u^i(\mathbf{r})
  - \partial_k u^i(\mathbf{r}) \, \partial^k u^j(\mathbf{r}) \bigr] 
  \label{equation::F_080} 
\end{equation}
is a nonlinear strain tensor with upper indices which has been defined and used by Grabert and Michel 
\cite{GM83}. (See Eqs.\ (7) and (8) in this reference). On the right-hand side for convenience we have 
lowered and risen some indices by the Cartesian metric tensors $\delta_{ij}$ and $\delta^{ij}$. The 
metric tensor $g_0^{ij}(\mathbf{r}) = g_0^{ji}(\mathbf{r})$ and the nonlinear strain tensor 
$u^{ij}(\mathbf{r}) = u^{ji}(\mathbf{r})$ are symmetric in the indices and hence do not depend on rotations. 
Consequently, they describe local deformations only but exclude rotations. Here, a further advantage is that 
the derivatives $\partial_i = \partial / \partial r^i$ are defined in a natural way with respect to the 
coordinates $r^i$ of the Cartesian laboratory frame. We will use the nonlinear strain tensor with upper 
indices in the next subsection in order to rewrite the Taylor expansion of the free energy density in a 
rotation invariant way.

\subsection{Rotation invariant free energy density}
\label{section::06b}
By applying the Taylor expansion we have derived an explicit formula for the local free energy density of the  
elastic properties of a crystal which is given by Eq.\ \eqref{equation::E_320}. This formula is explicitly 
invariant under translations because it does not depend explicitly on the displacement field $\mathbf{u}$ 
but rather on its derivatives $\partial_k u^i$. However, it is not explicitly invariant under rotations 
since $\partial_k u^i$ is connected to the transformation matrix \eqref{equation::F_010} which includes 
deformations and rotations.

On macroscopic scales Wallace \cite{Wa70} has developed the \emph{thermoelastic theory of stressed crystals}. 
In his equation (1.7) he states that the free energy must depend explicitly on the Lagrange strain tensor 
$u_{ij}$ so that it is explicitly rotation invariant. The reason is that the Lagrange strain tensor is 
connected to the metric tensor $g_{0,ij}$ via Eq.\ \eqref{equation::F_040} which includes only deformations 
but does not depend on rotations. Alternatively, the free energy may depend explicitly on the nonlinear 
strain tensor $u^{ij}$ which is connected to the inverse metric tensor $g_0^{ij}$ via 
Eq.\ \eqref{equation::F_070}. For this reason, now we replace the linear strain $\partial_k u^i$ by the 
nonlinear strain tensor $u^{ij}$ in our formula for the free energy density \eqref{equation::E_320}. 

First, we consider the fourth term of Eq.\ \eqref{equation::E_320} with the elastic constants $\lambda_{ij,kl}$. 
We define symmetric and antisymmetric parts of these constants by
\begin{eqnarray}
  \lambda^s_{ij,kl} &=& ( \lambda_{ij,kl} + \lambda_{kl,ij} ) / 2 \, ,
  \label{equation::F_090} \\
  \lambda^a_{ij,kl} &=& ( \lambda_{ij,kl} - \lambda_{kl,ij} ) / 2 \, ,
  \label{equation::F_100} 
\end{eqnarray}
respectively. Thus, we may split the elastic constants according to
\begin{equation}
  \lambda_{ij,kl} = \lambda^s_{ij,kl} + \lambda^a_{ij,kl} \, .
  \label{equation::F_110} 
\end{equation}
The antisymmetric elastic constants will be nonzero because of the relation \eqref{equation::E_410} which we 
have derived from rotation invariance. On the other hand, the symmetric elastic constant satisfy the Voigt 
symmetry in the indices which is defined by the equations 
\begin{equation}
  \lambda^s_{ij,kl} = \lambda^s_{ji,kl} = \lambda^s_{ij,lk} = \lambda^s_{kl,ij} \, .
  \label{equation::F_120} 
\end{equation}
The first and the second equality sign follows directly from the explicit formula of the elastic constants 
\eqref{equation::E_280}. The last equality sign is a consequence of the definition \eqref{equation::F_090}. 
Now, following Walz and Fuchs \cite{WF10} we may define new elastic constants with four indices by the 
formula
\begin{equation}
  C_{ij,kl} = \lambda^s_{ik,jl} + \lambda^s_{jk,il} - \lambda^s_{ij,kl} \, .
  \label{equation::F_130} 
\end{equation}
The related inverse formula is
\begin{equation}
  \lambda^s_{ij,kl} = [ C_{ik,jl} + C_{jk,il} ] / 2 \, .
  \label{equation::F_140} 
\end{equation}
These new elastic constant also satisfy the Voigt symmetry with respect to their indices which reads 
\begin{equation}
  C_{ij,kl} = C_{ji,kl} = C_{ij,lk} = C_{kl,ij} \, .
  \label{equation::F_150} 
\end{equation}
Consequently, Eqs.\ \eqref{equation::F_130} and \eqref{equation::F_140} represent an invertible mapping 
between the elastic constants $\lambda^s_{ij,kl}$ and $C_{ij,kl}$.

We use the new elastic constants $C_{ij,kl}$ and multiply by two nonlinear strain tensors $u^{ij}$ and $u^{kl}$ 
with upper indices. Then, we calculate
\begin{equation}
  \begin{split}
    C_{ij,kl} \, u^{ij} \, u^{kl} &= C_{ij,kl} \, ( \partial^i u^j ) \, ( \partial^k u^l ) \\
    &= \lambda^s_{ij,kl} \, ( \partial^i u^k ) \, ( \partial^j u^l ) + \partial^i \, w_i \, .
  \end{split}
  \label{equation::F_160} 
\end{equation}
For the first equality sign we insert the nonlinear strain \eqref{equation::F_080}. Since we calculate a 
second-order term of the Taylor series expansion we may omit the nonlinear term of the strain tensor. Because of 
the Voigt symmetry \eqref{equation::F_150} we may turn to the linear unsymmetric strains $u^{ij} \to \partial^i u^j$ 
and $u^{kl} \to \partial^k u^l$. In this way we obtain the term on the right-hand side of the first line of 
Eq.\ \eqref{equation::F_160}. Next we insert the formula for the new elastic constants \eqref{equation::F_130}. 
We rename some indices and perform some calculations. Thus, from the first contribution of 
Eq.\ \eqref{equation::F_130} we obtain the first term in the second line of Eq.\ \eqref{equation::F_160}. 
The latter two terms of Eq.\ \eqref{equation::F_130} result in the divergence term where
\begin{equation}
  w_i = ( \lambda^s_{jk,il} - \lambda^s_{ik,jl} ) \, u^k \, ( \partial^j u^l ) \, .
  \label{equation::F_170} 
\end{equation}
Eventually, the free energy density is integrated over the whole volume of the space. The volume integral 
of the divergence term can be transformed into a surface integral where the surface is in the infinity. 
Since by assumption the displacements and the strains are zero at infinity, this surface integral is zero. 
Hence, in the free energy density we may omit the divergence term. Thus, we may replace in both directions
\begin{equation}
  C_{ij,kl} \, u^{ij} \, u^{kl} \longleftrightarrow \lambda^s_{ij,kl} \, ( \partial^k u^i ) \, ( \partial^l u^j ) 
  \label{equation::F_180} 
\end{equation}
where on the right-hand side we may interchange the index pairs $(i,j) \leftrightarrow (k,l)$ by using the 
Voigt symmetry \eqref{equation::F_120}. We apply this result together with Eq.\ \eqref{equation::F_110} to 
the fourth term of the free energy density \eqref{equation::E_320}. Then, we obtain
\begin{equation}
  \begin{split}
    \frac{ 1 }{ 2 } \, \lambda_{ij,kl} \, ( \partial^k u^i ) \, ( \partial^l u^j ) 
    \to \,& \frac{ 1 }{ 2 } \, \lambda^a_{ij,kl} \, ( \partial^k u^i ) \, ( \partial^l u^j ) \\ 
    &+ \frac{ 1 }{ 2 } \, C_{ij,kl} \, u^{ij} \, u^{kl} \, .
  \end{split}
  \label{equation::F_190} 
\end{equation}
We succeeded to rewrite the symmetric contribution in terms of the nonlinear strain so that this latter 
contribution is explicitly rotation invariant.

In the next step we consider the second term of the free energy density \eqref{equation::E_320}. Thus, 
we take the constant stress tensor $\sigma_{0,ij}$ and multiply by a nonlinear strain $u^{ij}$. We insert 
the nonlinear strain \eqref{equation::F_080} and then calculate
\begin{equation}
  \sigma_{0,ij} \, u^{ij} = \sigma_{0,ik} \, ( \partial^k u^i ) 
  - \frac{ 1 }{ 2 } \sigma_{0,ij} \, \delta_{kl} \, ( \partial^k u^i ) \, ( \partial^l u^j ) \, .
  \label{equation::F_200} 
\end{equation}
We split the nonlinear term into a symmetric and an antisymmetric term. Resolving with respect to the first 
term on the right-hand side we obtain the second term of the free energy density \eqref{equation::E_320}
\begin{equation}
  \begin{split}
    \sigma_{0,ik} \, ( \partial^k u^i ) = &\, \sigma_{0,ij} \, u^{ij} \\
    &+ \frac{ 1 }{ 4 } \, [ \sigma_{0,ij} \, \delta_{kl} + \sigma_{0,kl} \, \delta_{ij} ] \, ( \partial^k u^i ) \, ( \partial^l u^j ) \\
    &+ \frac{ 1 }{ 4 } \, [ \sigma_{0,ij} \, \delta_{kl} - \sigma_{0,kl} \, \delta_{ij} ] \, ( \partial^k u^i ) \, ( \partial^l u^j ) \, .
  \end{split}
  \label{equation::F_210} 
\end{equation}
We may define a correction to the symmetric elastic constants by
\begin{equation}
  \Delta \lambda^s_{ij,kl} = \frac{ 1 }{ 2 } \, [ \sigma_{0,ij} \, \delta_{kl} + \sigma_{0,kl} \, \delta_{ij} ] \, .
  \label{equation::F_220} 
\end{equation}
One easily proves that this correction satisfies the Voigt symmetry \eqref{equation::F_120} for all its four 
indices. Consequently, we may define new corrections $\Delta C_{0,ij,kl}$ by a formula like \eqref{equation::F_130} 
and explicitly obtain
\begin{equation}
  \begin{split}
    \Delta C_{0,ij,kl} &= \frac{ 1 }{ 2 } \, [ \sigma_{0,ik} \, \delta_{jl} + \sigma_{0,jl} \, \delta_{ik} 
    + \sigma_{0,jk} \, \delta_{il} + \sigma_{0,il} \, \delta_{jk} \\ 
    &\hspace{8mm} - \sigma_{0,ij} \, \delta_{kl} - \sigma_{0,kl} \, \delta_{ij} ] \, .
  \end{split}
  \label{equation::F_230} 
\end{equation}
In this way we may rewrite the second term of Eq.\ \eqref{equation::F_210} in terms of the new corrections 
and the nonlinear strain. Thus, we obtain the second term of the free energy density
\begin{equation}
  \begin{split}
    \sigma_{0,ik} \, ( \partial^k u^i ) = &\, \sigma_{0,ij} \, u^{ij} + \frac{ 1 }{ 2 } \, \Delta C_{0,ij,kl} \, u^{ij} \, u^{kl} \\
    &+ \frac{ 1 }{ 4 } \, [ \sigma_{0,ij} \, \delta_{kl} - \sigma_{0,kl} \, \delta_{ij} ] \, ( \partial^k u^i ) \, ( \partial^l u^j ) \, .
  \end{split}
  \label{equation::F_240} 
\end{equation}

Finally, we consider the fifth term of the free energy density \eqref{equation::E_320}. Because of rotation 
invariance and Eq.\ \eqref{equation::E_420} the elastic constant $\mu_{a,ij} = \mu_{a,ji}$ is symmetric in its 
two indices. Thus, we take $\mu_{a,ij}$ and multiply by $u^{ij}$. Then, inserting the nonlinear strain 
\eqref{equation::F_080} we calculate
\begin{equation}
  \mu_{a,ij} \, u^{ij} = \mu_{a,ik} \, ( \partial^k u^{i} ) \, .
  \label{equation::F_250} 
\end{equation}
The nonlinear contribution of the strain tensor can be dropped because the fifth term of the free energy 
density is of second order in the Taylor expansion.

In summary, we have rewritten all terms of the free energy density \eqref{equation::E_320} using the nonlinear 
strain $u^{ij}$ as far as possible. For the second and the fourth term we insert Eqs.\ \eqref{equation::F_240} 
and \eqref{equation::F_190}, respectively. Within the fifth term we use \eqref{equation::F_250}. Thus, as a result 
we obtain
\begin{equation}
  \begin{split}
    f(T,\mathsf{u},\tilde{n}) = \,& f_0(T,\tilde{n}_0) + \sigma_{0,ij} \, u^{ij} 
    + \sum_a \mu_{0,a} \, [ \tilde{n}_a - \tilde{n}_{0,a} ] \\
    &+ \frac{ 1 }{ 2 } \, [ C_{ij,kl} + \Delta C_{0,ij,kl} ] \, u^{ij} \, u^{kl} \\
    &- \sum_a \mu_{a,ij} \, \frac{ [ \tilde{n}_a - \tilde{n}_{0,a} ] }{ \tilde{n}_{0,a} } \, u^{ij} \\
    &+ \frac{ 1 }{ 2 } \sum_{ab} \nu_{ab} \, \frac{ [ \tilde{n}_a - \tilde{n}_{0,a} ] }{ \tilde{n}_{0,a} }
    \, \frac{ [ \tilde{n}_b - \tilde{n}_{0,b} ] }{ \tilde{n}_{0,b} } \, .
  \end{split}
  \label{equation::F_260} 
\end{equation}
This free energy density is written completely in terms of the nonlinear strain tensor $u^{ij}$. Hence, it is 
explicitly translation and rotation invariant. This formula agrees with the Taylor expansion (1.10) of Wallace 
\cite{Wa70}. Beyond that our formula is more general because additionally it depends on the macroscopic particle 
densities $\tilde{n}_a$.

However, there are two contributions more which are not rotation invariant. These are the first term of
Eq.\ \eqref{equation::F_190} and the third term of Eq.\ \eqref{equation::F_240}. They sum up into the excess 
free energy density
\begin{equation}
  \Delta f = \frac{ 1 }{ 4 } \, [ 2 \, \lambda^a_{ij,kl} + \sigma_{0,ij} \, \delta_{kl} 
  - \sigma_{0,kl} \, \delta_{ij} ] \, ( \partial^k u^i ) \, ( \partial^l u^j ) \, .
  \label{equation::F_270} 
\end{equation}
Since the linear strain tensor $\partial^k u^i$ depends on rotations we must require $\Delta f = 0$ in 
order to achieve rotation invariance. As a consequence, the coefficients in square brackets must be zero. 
In this way, rotation invariance implies the condition
\begin{equation}
  2 \, \lambda^a_{ij,kl} + \sigma_{0,ij} \, \delta_{kl} - \sigma_{0,kl} \, \delta_{ij} = 0 
  \label{equation::F_280} 
\end{equation}
which is a relation for the antisymmetric part of the elastic constants $\lambda^a_{ij,kl}$ and the 
stress tensor $\sigma_{0,ij}$. Using the definition \eqref{equation::F_100} we finally obtain and recover 
Eq.\ \eqref{equation::E_410}. Thus, we have proven the relation \eqref{equation::E_410} independently a 
second time as a consequence of rotation invariance.

The kinetic energy density \eqref{equation::E_500} does not depend on the strain tensor at all and hence 
is translation and rotation invariant from the beginning. Thus, from Eq.\ \eqref{equation::E_510} together 
with \eqref{equation::F_260} and \eqref{equation::E_500} we obtain the total free energy density in the 
form $f = f(T,\mathsf{u},\tilde{\mathbf{j}},\tilde{n})$ which is invariant under translations and rotations. 
Calculating the differential we obtain the differential thermodynamic relation
\begin{equation}
  df = - s \, dT + \sigma_{ij} \, du^{ij} + \mathbf{v} \cdot d\tilde{\mathbf{j}} 
  + \sum_a \mu_a \, d\tilde{n}_a
  \label{equation::F_290} 
\end{equation}
which may be compared with Eq.\ \eqref{equation::E_540}. Here, from the explicit derivative of 
Eq.\ \eqref{equation::F_260} we obtain the strain tensor with two lower indices
\begin{equation}
  \sigma_{ij} = \sigma_{0,ij} + [ C_{ij,kl} + \Delta C_{0,ij,kl} ] \, u^{kl} 
  - \sum_a \mu_{a,ij} \, \frac{ [ \tilde{n}_a - \tilde{n}_{0,a} ] }{ \tilde{n}_{0,a} } \, .
  \label{equation::F_300} 
\end{equation}
The velocity $\mathbf{v}$ and the chemical potentials $\mu_a$ are given by Eqs.\ \eqref{equation::E_560} 
and \eqref{equation::E_570}, respectively, as before where now Eq.\ \eqref{equation::F_250} must be inserted.

Alternatively, we may express the free energy density \eqref{equation::F_260} in terms of the Lagrange strain 
$u_{ij}$ with two lower indices. From the chain rule of differential calculus and the inverse of the Jacobi 
matrix \eqref{equation::D_330} we find
\begin{equation}
  \partial_{0,k} u^i = \partial_k u^i + ( \partial_k u^l ) \, ( \partial_l u^i ) + \cdots \, .
  \label{equation::F_310} 
\end{equation}
We insert this result into Eq.\ \eqref{equation::F_050} and then obtain up to second order
\begin{equation}
  \begin{split}
    u_{ij}(\mathbf{r}) = &\, \frac{ 1 }{ 2 } \, \bigl[ \partial_i u_j(\mathbf{r}) + \partial_j u_i(\mathbf{r})
    + ( \partial_i u^k(\mathbf{r}) ) \, ( \partial_j u_k(\mathbf{r}) ) \\
    &+ ( \partial_i u^k(\mathbf{r}) ) \, ( \partial_k u_j(\mathbf{r}) )
    + ( \partial_j u^k(\mathbf{r}) ) \, ( \partial_k u_i(\mathbf{r}) ) \bigr] \, .
  \end{split}
  \label{equation::F_320} 
\end{equation}
This strain tensor can not be obtained from the related tensor with upper indices \eqref{equation::F_080} 
by lowering the two indices because here we have two quadratic terms more. As a consequence, also the 
correction to the elastic constants with upper indices can not be obtained from the related tensor with lower 
indices \eqref{equation::F_230} by rising all for indices indices. Rather, we derive the formula
\begin{equation}
  \begin{split}
    \Delta C_0^{ij,kl} &= - \frac{ 1 }{ 2 } \, [ \sigma_0^{ik} \, \delta^{jl} + \sigma_0^{jl} \, \delta^{ik} 
    + \sigma_0^{jk} \, \delta^{il} + \sigma_0^{il} \, \delta^{jk} \\ 
    &\hspace{10.5mm} + \sigma_0^{ij} \, \delta^{kl} + \sigma_0^{kl} \, \delta^{ij} ] \, .
  \end{split}
  \label{equation::F_330} 
\end{equation}
Comparing the formulas \eqref{equation::F_230} and \eqref{equation::F_330} we find that the first four terms 
change the sign where the last two terms keep the same sign. On the other hand, the remaining elastic constants 
with upper indices $\sigma_0^{ij}$, $C^{ij,kl}$, and $\mu_a^{ij}$ do not have new contributions. 
They are just obtained from the related constants with lower indices by rising all indices with the Cartesian 
inverse metric $\delta^{ij}$. Finally, as a result we obtain an analogous Taylor expansion formula for the free 
energy density like Eq.\ \eqref{equation::F_260} where now the stress tensor and the elastic constants have upper 
indices and the Lagrange strain tensor has lower indices. Similarly, we obtain analogous formulas like 
Eq.\ \eqref{equation::F_290} for the differential thermodynamic relation and Eq.\ \eqref{equation::F_300} for 
the stress tensor with upper indices $\sigma^{ij}$. Like the strain tensors $u^{ij}$ and $u_{ij}$, the stress 
tensors $\sigma^{ij}$ and $\sigma_{ij}$ are not related to each other by simply lowering and rising the 
indices by a Cartesian metric. There are some more terms which differ.

\subsection{Large displacements and deformations on macroscopic scales}
\label{section::06c}
Until now we have assumed that the displacements and deformations are small. For, we need these assumptions for 
the Taylor expansion of the free energy density to be valid. In this case, the metric tensors of the curvilinear 
coordinates \eqref{equation::F_040} and \eqref{equation::F_070} are non trivial. However, they are not far away 
from being Cartesian so that we have approximately $g_{0,ij} \approx \delta_{ij}$ and $g_0^{ij} \approx \delta^{ij}$. 
In this subsection we extend our considerations to large displacements and deformations where we focus on the 
differential thermodynamic relations \eqref{equation::E_540} and \eqref{equation::F_290}.

For this purpose we split the transformation formula from Cartesian to curvilinear coordinates 
\eqref{equation::D_280} into two formulas for two transformations one after the other. Thus, we write 
\begin{eqnarray}
  \mathbf{r}_0 &=& \mathbf{r}_1 - \mathbf{u}_0(\mathbf{r}_1) \, ,
  \label{equation::F_340} \\ 
  \mathbf{r}_1 &=& \mathbf{r} - \mathbf{u}_1(\mathbf{r}) \, .
  \label{equation::F_350} 
\end{eqnarray}
Here, we introduce an intermediate curvilinear coordinate system with coordinates $\mathbf{r}_1$. The main 
idea is the separation of a \emph{large linear} transformation by Eq.\ \eqref{equation::F_340} from a 
\emph{small nonlinear} transformation by Eq.\ \eqref{equation::F_350}. We focus on a small epsilon surroundings 
\begin{equation}
  U_\varepsilon(P) = \{ \mathbf{r} \mid | \mathbf{r} - \mathbf{r}_P | < \varepsilon \}
  \label{equation::F_360} 
\end{equation}
around a point $P$ with coordinates $\mathbf{r}_P$ and apply Taylor expansions to the two displacement fields. 
We expand the first displacement field $\mathbf{u}_0(\mathbf{r}_1)$ up to linear order so that
\begin{equation}
  u_0^p(\mathbf{r}_1) = u_0^p(\mathbf{r}_{1,P}) + [ \partial_k u_0^p(\mathbf{r}_{1,P}) ] \, ( r_1^k - r_{1,P}^k ) \, .
  \label{equation::F_370} 
\end{equation}
On the other hand, we expand the second displacement field $\mathbf{u}_1(\mathbf{r})$ starting with the 
quadratic order
\begin{equation}
  u_1^i(\mathbf{r}) = \frac{ 1 }{ 2 } \, [ \partial_k \partial_l u_1^i(\mathbf{r}_P) ] 
  \, ( r^k - r_P^k ) \, ( r^l - r_P^l ) + \cdots \, .
  \label{equation::F_380} 
\end{equation}
If in Eq.\ \eqref{equation::F_360} $\varepsilon$ is sufficiently small, then the nonlinear second displacement field 
$\mathbf{u}_1(\mathbf{r})$ is much smaller than the linear first displacement field $\mathbf{u}_0(\mathbf{r}_1)$. 
In Eq.\ \eqref{equation::F_370} we may allow that the constant displacement $u_0^p(\mathbf{r}_{1,P})$ and the 
linear strain $\partial_k u_0^p(\mathbf{r}_{1,P})$ are large. On the other hand, for the validity of the concepts 
of continuum mechanics and local thermodynamic equilibrium we must require that in Eq.\ \eqref{equation::F_380} the 
second derivative $\partial_k \partial_l u_1^i(\mathbf{r}_P)$ is sufficiently small or at least not too large. 
In practice the value of $\varepsilon$ will be identified by the mesoscopic length scale $\ell$ which according to 
\eqref{equation::D_030} divides the function space $\mathcal{C} = \tilde{\mathcal{C}}(\ell) \oplus \mathcal{C}^\prime(\ell)$ 
into a macroscopic subspace $\tilde{\mathcal{C}}(\ell)$ and a microscopic complement space $\mathcal{C}^\prime(\ell)$. 
Thus, we may use $\varepsilon = \ell$.

In the laboratory frame the coordinates $\mathbf{r}$ are Cartesian. The metric tensors are given by the 
Kronecker symbols so that for all $\mathbf{r}$ in the whole space we have 
\begin{equation}
  g_{ij}(\mathbf{r}) = \delta_{ij} \, , \qquad g^{ij}(\mathbf{r}) = \delta^{ij} \, .
  \label{equation::F_390} 
\end{equation}
In the intermediate frame the coordinates $r_1^i$ are curvilinear. If we restrict the coordinates to the epsilon 
surroundings $U_\varepsilon(P)$ the nonlinear displacement field $\mathbf{u}_1(\mathbf{r})$ defined in 
\eqref{equation::F_380} is small. Consequently, the intermediate coordinates deviate not that much so that they 
are \emph{approximately} Cartesian. The related metric tensors are calculated by a transformation formula like 
\eqref{equation::F_020}. Consequently, for $\mathbf{r} \in U_\varepsilon(P)$ we obtain approximately
\begin{equation}
  g_{1,ij}(\mathbf{r}) \approx \delta_{ij} \, , \qquad g_1^{ij}(\mathbf{r}) \approx \delta^{ij} \, .
  \label{equation::F_400} 
\end{equation}
Since the linear displacement field $\mathbf{u}_0(\mathbf{r}_1)$ defined in \eqref{equation::F_370} may be large,
for the curvilinear coordinates $r_0^p$ the metric tensors may strongly deviate from Kronecker symbols. Thus, 
for $\mathbf{r}$ in the most areas of the whole space we have
\begin{equation}
  g_{0,pq}(\mathbf{r}) \neq \delta_{pq} \, , \qquad g_1^{pq}(\mathbf{r}) \neq \delta^{pq} \, .
  \label{equation::F_410} 
\end{equation}
Hence, the curvilinear coordinates $r_0^p$ are far from being Cartesian.

We have three different coordinate systems. Consequently, we have three different types of tensor indices 
which must be distinguished. In the previous sections beginning with Subsec.\ \ref{section::04d} we have not 
done this. Rather we have used the indices
\begin{equation}
  i, j, k, l, \ldots
  \label{equation::F_420} 
\end{equation}
in all these sections. The reason was that the displacements and the strains were assumed to be small so 
that the coordinate systems were either exactly Cartesian (laboratory frame) or approximately Cartesian 
(curvilinear coordinates). In this section we may continue with this practice. We use these indices for 
the coordinates $r^i$ of the laboratory frame and for the intermediate coordinates $r_1^i$. This choice 
is seen clearly in the equations for the metric tensors \eqref{equation::F_390} and \eqref{equation::F_400}.

On the other hand, the curvilinear coordinates $r_0^p$ are far from being Cartesian. For this reason in 
this case we use different indices which we denote by
\begin{equation}
  p, q, r, s, \ldots \, .
  \label{equation::F_430} 
\end{equation}
We have already used these indices with all consequences in Eqs.\ \eqref{equation::F_370} and 
\eqref{equation::F_410}. 

Now, we combine the two steps of the transformation into a single step. This means we combine 
Eqs.\ \eqref{equation::F_340} and \eqref{equation::F_350} and compare with \eqref{equation::D_280}. In this 
way we obtain the total displacement field
\begin{equation}
  \mathbf{u}(\mathbf{r}) = \mathbf{u}_1(\mathbf{r}) + \mathbf{u}_0( \mathbf{r} - \mathbf{u}_1(\mathbf{r}) ) \, .
  \label{equation::F_440} 
\end{equation}
We insert the explicit formula of the linear displacement \eqref{equation::F_370} and turn to the index 
notation. In order to simplify the notation we write the constant coefficients $u_0^p = u_0^p(\mathbf{r}_{1,P})$ 
and $\partial_k u_0^p = \partial_k u_0^p(\mathbf{r}_{1,P})$ without arguments. Thus, we find 
\begin{equation}
  u^p(\mathbf{r}) = u_0^p + ( \partial_k u_0^p ) \, ( r^k - r_P^k ) 
  + u_1^p(\mathbf{r}) - ( \partial_k u_0^p ) \, u_1^k(\mathbf{r}) \, .
  \label{equation::F_450} 
\end{equation}
For convenience we define the transformation matrix
\begin{equation}
  e_i^p = \frac{ \partial r_0^p }{ \partial r_1^i } = \delta_i^p - \partial_i u_0^p \, .
  \label{equation::F_460} 
\end{equation}
Then, we may combine the last two terms of Eq.\ \eqref{equation::F_450} and rewrite
\begin{equation}
  u^p(\mathbf{r}) = u_0^p + ( \partial_k u_0^p ) \, ( r^k - r_P^k ) + e_i^p \, u_1^i(\mathbf{r}) \, .
  \label{equation::F_470} 
\end{equation}
This formula may be interpreted as the transformation of the nonlinear displacement field 
$\mathbf{u}_1(\mathbf{r})$ into the total displacement field $\mathbf{u}(\mathbf{r})$. Here, we note that 
$\mathbf{u}_1(\mathbf{r})$ rules the transformation \eqref{equation::F_350} from the laboratory frame 
into the intermediate coordinate system while $\mathbf{u}(\mathbf{r})$ rules the transformation 
\eqref{equation::D_280} from the laboratory frame into the curvilinear coordinate system.

Now, we may derive with respect to the Cartesian coordinates applying the operator 
$\partial_k = \partial / \partial r^k$. Thus, we obtain the transformation formula for the strain tensor 
\begin{equation}
  \partial_k u^p(\mathbf{r}) = \partial_k u_0^p + e_i^p \, \partial_k u_1^i(\mathbf{r}) \, .
  \label{equation::F_480} 
\end{equation}
Next we take the differentials by applying an operator $d$. Thus, from Eqs.\ \eqref{equation::F_470} 
and \eqref{equation::F_480} we obtain
\begin{eqnarray}
  du^p(\mathbf{r}) &=& e_i^p \, du_1^i(\mathbf{r}) \, ,
  \label{equation::F_490} \\
  \partial_k du^p(\mathbf{r}) &=& e_i^p \, \partial_k du_1^i(\mathbf{r}) \, ,
  \label{equation::F_500} 
\end{eqnarray}
respectively. We may compare our results with the tensor formalism which is used commonly in Riemannian 
Geometry and in General Relativity to describe curved spaces with curvilinear coordinates. We find that 
Eqs.\ \eqref{equation::F_490} and \eqref{equation::F_500} are transformation formulas for tensor fields 
of first rank. On the other hand Eqs.\ \eqref{equation::F_470} and \eqref{equation::F_480} do not fit 
into this scheme. Thus, we conclude that the \emph{infinitesimal} displacements $du^p(\mathbf{r})$ 
and strains $\partial_k du^p(\mathbf{r})$ are \emph{tensor fields} with respect to the \emph{contravariant} 
upper index $p$. On the other hand, the \emph{finite} displacements $u^p(\mathbf{r})$ and strains 
$\partial_k u^p(\mathbf{r})$ are no tensor fields because their transformation formulas are more complicated.

We may define the inverse transformation matrix
\begin{equation}
  e_p^i = \frac{ \partial r_1^i }{ \partial r_0^p } = \delta_p^i + e_p^k \, \partial_k u_0^i 
  \label{equation::F_510} 
\end{equation}
so that
\begin{equation}
  e_p^i \, e_j^p = \delta_j^i \, , \qquad e_i^p \, e_q^i = \delta_q^p \, .
  \label{equation::F_520} 
\end{equation}
In order to distinguish the inverse matrix $e_p^i$ from the matrix $e_i^p$ we must look for which index 
type is upper and which is lower. Now, we apply the inverse matrix to the transformation formula 
\eqref{equation::F_470} and resolve with respect to the last term. Thus, we obtain the inverse 
transformation formula for the displacement field
\begin{equation}
  u_1^i(\mathbf{r}) = - e_p^i \, u_0^p - e_p^i \, ( \partial_k u_0^p ) \, ( r^k - r_P^k ) + e_p^i \, u^p(\mathbf{r}) \, .
  \label{equation::F_530} 
\end{equation}
Next, we derive with respect to the Cartesian coordinates by applying the operator 
$\partial_k = \partial / \partial r^k$. Thus, we obtain the inverse transformation formula for 
the strain tensor 
\begin{equation}
  \partial_k u_1^i(\mathbf{r}) = - e_p^i \, ( \partial_k u_0^p ) + e_p^i \, \partial_k u^p(\mathbf{r}) \, .
  \label{equation::F_540} 
\end{equation}
Finally, we take the differentials by applying an operator $d$. Thus, we obtain the inverse 
transformation formulas for the infinitesimal displacement and strain which read
\begin{eqnarray}
  du_1^i(\mathbf{r}) &=& e_p^i \, du^p(\mathbf{r}) \, ,
  \label{equation::F_550} \\
  \partial_k du_1^i(\mathbf{r}) &=& e_p^i \, \partial_k du^p(\mathbf{r}) \, ,
  \label{equation::F_560} 
\end{eqnarray}
respectively. Once again we see that Eqs.\ \eqref{equation::F_550} and \eqref{equation::F_560} are 
inverse transformation formulas for tensor fields where \eqref{equation::F_530} and \eqref{equation::F_540} 
are not. Thus, once again the infinitesimal quantities are tensor fields where on the other hand the 
finite quantities are not.

\subsection{Free energy density for large deformations beyond the Taylor expansion}
\label{section::06d}
In Sec.\ \ref{section::05} we have derived the Taylor expansion series up to second order for the free energy 
density. The main formula is \eqref{equation::E_320} which yields $f = f(T,\partial \mathbf{u},\tilde{n})$.
This formula is valid under the assumption that the displacement field $\mathbf{u}(\mathbf{r})$, the 
strain tensor $\partial_k u^i(\mathbf{r})$, and the deviations of the particle densities $\tilde{n}_a(\mathbf{r})$ 
from the reference densities $\tilde{n}_{0,a}$ are small. The assumptions are valid if we calculate the free 
energy density for strains and deformations in the intermediate coordinate system with coordinates $\mathbf{r_1}$. 
More precisely, we insert just the nonlinear displacement field $\mathbf{u}_1(\mathbf{r})$ defined in 
Eq.\ \eqref{equation::F_380} and the related nonlinear strain tensor $\partial_k u_1^i(\mathbf{r})$. Then, we obtain 
the free energy density $f = f(T,\partial \mathbf{u}_1,\tilde{n})$ in the epsilon surroundings $U_\varepsilon(P)$ 
of the point $P$ with coordinates $\mathbf{r}_P$.

This procedure can be repeated for many points $P_i$ and many epsilon surroundings $U_\varepsilon(P_i)$ with 
indices $i=1,2,\ldots,n$ where $n$ might be infinite. The many epsilon surroundings should cover the whole 
volume $V$ of the crystal or even the whole space. In this way, we obtain the free energy density 
$f = f(T,\partial \mathbf{u}_1,\tilde{n})$ for all points in the whole space. While the laboratory frame with 
coordinates $\mathbf{r}$ is global and hence the same for all epsilon surroundings, the intermediate coordinate 
system with coordinates $\mathbf{r}_1$ is different for each epsilon surroundings. More precisely, the nonlinear 
displacement field $\mathbf{u}_1(\mathbf{r})$ and hence the nonlinear strain tensor $\partial_k u_1^i(\mathbf{r})$ 
are separately defined for each epsilon surroundings $U_\varepsilon(P_i)$.

On the other hand, the curvilinear coordinates $\mathbf{r}_0$ are global and hence the same for all epsilon 
surroundings $U_\varepsilon(P_i)$. Consequently, we should apply the linear coordinate transformation 
\eqref{equation::F_340} defined by the linear displacement field \eqref{equation::F_370}. We transform from 
the weakly deformed intermediate coordinate system to the strongly deformed curvilinear coordinate system. For the 
displacement fields and the strain tensors the transformation formulas are given by Eqs.\ \eqref{equation::F_470} 
and \eqref{equation::F_480} while the inverse transformation formulas are given by Eqs.\ \eqref{equation::F_530} and 
\eqref{equation::F_540}. If we take the free energy density $f = f(T,\partial \mathbf{u}_1,\tilde{n})$ and insert the 
inverse transformation formula \eqref{equation::F_540} for the nonlinear strain tensor $\partial_k u_1^i(\mathbf{r})$ 
then we obtain obtain the free energy density $f = f(T,\partial \mathbf{u},\tilde{n})$ depending on the full 
strain tensor $\partial_k u^i(\mathbf{r})$ which is \emph{globally} defined in the whole space. Nevertheless, we 
obtain different formulas with different Taylor series for each epsilon surroundings. However, by construction 
the results will overlap \emph{continuously} between neighboring epsilon surroundings. In this way we obtain 
a free energy density $f = f(\mathbf{r})$ which is a \emph{continuous} function in $\mathbf{r}$ over the whole 
space.

On the other hand, since we are developing a theory of continuum mechanics we know that the free energy density 
$f = f(\mathbf{r})$ must be a continuous function. Hence, the Taylor expansion in the epsilon surroundings of the 
points $P_i$ is not necessary. Rather the function values $f_i = f(\mathbf{r}_{P_i})$ at these points with 
coordinates $\mathbf{r}_{P_i}$ are sufficient. In between we can extrapolate the free energy density by spline 
functions. We conclude that we must calculate the free energy density at these points.

We take a particular point $P$ with coordinates $\mathbf{r}_P$. Here the inverse transformation formulas 
\eqref{equation::F_530} and \eqref{equation::F_540} imply zero values for the nonlinear displacement field 
$\mathbf{u}_1(\mathbf{r}_P) = \mathbf{0}$ and the nonlinear strain $\partial_k u_1^i(\mathbf{r}_P) = 0$. 
Furthermore, the densities have the values of the reference state so that $\tilde{n}_a(\mathbf{r}_P) = \tilde{n}_{0,a}$. 
Consequently, all higher order terms of the Taylor series expansion are zero so that only the zeroth order term 
survives which is the free energy of the reference state. Thus, from Eq.\ \eqref{equation::E_320} we obtain 
$f(\mathbf{r}_P) = f_0$. The reference state is homogeneous in space. Consequently, we find the total free energy 
\begin{equation}
  F_0 = \int_V d^dr \, f_0 = V \, f_0 = V \, f(\mathbf{r}_P) \, .
  \label{equation::F_570} 
\end{equation}
Vice versa we calculate the free energy density
\begin{equation}
  f(\mathbf{r}_P) = f_0 = F_0 / V \, .
  \label{equation::F_580} 
\end{equation}
The reference state can be considered for any constant densities $\tilde{n}_a$ and any regular, homogeneous, 
and perfect lattice structure. Linear deformations defined in Eqs.\ \eqref{equation::F_340} together with 
a linear displacement field \eqref{equation::F_370} are allowed because they keep the homogeneity of the
lattice structure, so that the system remains in a global thermodynamic equilibrium, where, nevertheless, 
it is a different global equilibrium. Thus, we may calculate the total free energy $F[T,\mathbf{u},\tilde{n}]$ 
and then obtain the free energy density $f = f(T,\partial\mathbf{u},\tilde{n})$ by the formula
\begin{equation}
  f(T,\partial\mathbf{u},\tilde{n}) = F[T,\mathbf{u},\tilde{n}] / V \, .
  \label{equation::F_590} 
\end{equation}
Eventually, we may add the kinetic energy density \eqref{equation::E_500} in order to include the effects 
of motion and the dependence on the momentum density $\tilde{\mathbf{j}}$. Thus, finally from 
Eq.\ \eqref{equation::E_510} we obtain $f = f(T,\partial\mathbf{u},\tilde{\mathbf{j}},\tilde{n})$.

The \emph{local} thermodynamic equilibrium means that we first calculate the free energy density for the 
\emph{global} thermodynamic equilibrium and then use this result in an epsilon surroundings $U_\varepsilon(P)$ 
of a point $P$. We repeat this procedure for any point in the space.

\subsection{Thermodynamic relation and necessary equilibrium conditions for strong deformations}
\label{section::06e}
The differential thermodynamic relation \eqref{equation::E_540} has been derived for small displacement 
fields and small strain tensors. In the term $\sigma_i^{\ k} ( \partial_k u^i )$ the indices are assumed 
to be Cartesian. Hence, we may use this formula in an epsilon surroundings $U_\varepsilon(P)$ for the 
intermediate coordinate system. We insert the infinitesimal nonlinear displacements $d\mathbf{u}_1(\mathbf{r})$ 
and strains $\partial_k du_1^i(\mathbf{r})$ so that 
\begin{equation}
  df = - s \, dT + \sigma_i^{\ k} \, ( \partial_k du_1^i ) + \mathbf{v} \cdot d\tilde{\mathbf{j}} 
  + \sum_a \mu_a \, d\tilde{n}_a \, .
  \label{equation::F_600} 
\end{equation}
Now, we may use the inverse transformation formulas \eqref{equation::F_550} and \eqref{equation::F_560} 
in order to rewrite the stress-strain term. Thus, we calculate
\begin{equation}
  \begin{split}
    \sigma_i^{\ k} \, ( \partial_k du_1^i ) &= \sigma_i^{\ k} \, ( \partial_k ( e_p^i \, du^p ) )
    = \sigma_i^{\ k} \, e_p^i \, ( \partial_k du^p ) \\
    &= \sigma_p^{\ k} \, ( \partial_k du^p ) \, .
  \end{split}
  \label{equation::F_610} 
\end{equation}
For the second equality sign we assume that the transformation matrix $e_i^p$ defined in 
Eq.\ \eqref{equation::F_460} and the inverse transformation matrix $e_p^i$ defined in 
Eq.\ \eqref{equation::F_510} are constant in space so that they commute with all the operators 
of spatial derivatives $\partial_k$, $\partial_{1,k}$, and $\partial_{0,k}$. This assumption is 
essential here and is guaranteed by the linear transformation \eqref{equation::F_340} together with 
the linear displacement field \eqref{equation::F_370}.

For the last equality sign in Eq.\ \eqref{equation::F_610} we have defined the stress tensor 
$\sigma_p^{\ k}(\mathbf{r})$ with one lower index in the curvilinear coordinate system by the 
transformation formula
\begin{equation}
  \sigma_p^{\ k}(\mathbf{r}) = e_p^i \, \sigma_i^{\ k} (\mathbf{r}) \, .
  \label{equation::F_620} 
\end{equation}
The related inverse transformation formula is 
\begin{equation}
  \sigma_i^{\ k}(\mathbf{r}) = e_i^p \, \sigma_p^{\ k} (\mathbf{r}) \, .
  \label{equation::F_630} 
\end{equation}
These transformation formulas for the stress tensor may be compared with the transformation formulas 
\eqref{equation::F_500} and \eqref{equation::F_560} for the infinitesimal strain tensor.
Then, as a consequence from Eq.\ \eqref{equation::F_600} we obtain the differential thermodynamic 
relation in the curvilinear coordinate system
\begin{equation}
  df = - s \, dT + \sigma_p^{\ k} \, ( \partial_k du^p ) + \mathbf{v} \cdot d\tilde{\mathbf{j}} 
  + \sum_a \mu_a \, d\tilde{n}_a \, .
  \label{equation::F_640} 
\end{equation}
We have derived this equation for a particular epsilon surroundings $U_\varepsilon(P)$ around a 
particular point $P$. We repeat this calculation for many epsilon surroundings $U_\varepsilon(P_i)$ 
around points $P_i$ which cover the whole space. In all cases we obtain the same thermodynamic relation 
\eqref{equation::F_640} in the same form. Thus, we conclude that Eq.\ \eqref{equation::F_640} is valid 
in the whole space where the deformations may be strong on the macroscopic scale.

Similarly, we consider the necessary conditions for the global thermodynamic equilibrium. For small 
displacement fields and small stress tensors we have derived the conditions \eqref{equation::E_620}. 
Once again, we may use these conditions in an epsilon surroundings $U_\varepsilon(P)$ for the 
intermediate coordinate system. Then, we transform to the curvilinear coordinate system by inserting 
the inverse transformation formula \eqref{equation::F_630} for the stress tensor. Thus, we calculate 
\begin{equation}
  0 = \partial_k \, \sigma_i^{\ k}(\mathbf{r}) = \partial_k \, [ e_i^p \, \sigma_p^{\ k} (\mathbf{r}) ]
  = e_i^p \, [ \partial_k \, \sigma_p^{\ k} (\mathbf{r}) ] \, .
  \label{equation::F_650} 
\end{equation}
For the last equality sign we have used once again that the transformation matrix $e_i^p$ is constant 
in space so that we may commute the spatial derivative and the transformation matrix. As a consequence 
we obtain the necessary conditions
\begin{equation}
  \partial_k \, \sigma_p^{\ k}(\mathbf{r}) = 0 \, , \qquad 
  \mathbf{v}(\mathbf{r}) = \mathbf{v} \, , \qquad
  \mu_a(\mathbf{r}) = \mu_a \, .
  \label{equation::F_660} 
\end{equation}
We may repeat the same arguments as below Eq.\ \eqref{equation::F_640}. As a result these necessary 
conditions for the global thermodynamic equilibrium are valid not only in a small epsilon surroundings 
but rather in the whole space. They are valid for strong deformations on macroscopic scales.

The necessary conditions can be derived alternatively by starting with the variation of the free 
energy inserting the infinitesimal free energy density in the form \eqref{equation::F_640} which is valid 
for strongly deformed crystals. Again we perform a partial integration by using the integration theorem 
of Gauss where the surface terms are zero. Thus, we proceed in analogy to Eqs.\ \eqref{equation::E_590} 
and \eqref{equation::E_610} where now the index $i$ of the intermediate coordinate system is replaced by 
the index $p$ of the curvilinear coordinate system. Thus, we calculate 
\begin{equation}
  \begin{split}
    \delta F[T,\mathbf{u},\tilde{\mathbf{j}},\tilde{n}] &= \int d^dr 
    \, \delta f(T,\partial \mathbf{u}(\mathbf{r}),\tilde{\mathbf{j}}(\mathbf{r}),\tilde{n}(\mathbf{r})) \\
    &= \int d^dr \, \bigl[ - s \, \delta T + \sigma_p^{\ k}(\mathbf{r}) \, [ \partial_k \delta u^p(\mathbf{r}) ] \\
    &\hspace{15mm} + \mathbf{v}(\mathbf{r}) \cdot \delta \tilde{\mathbf{j}}(\mathbf{r}) 
    + \sum_a \mu_a(\mathbf{r}) \, \delta \tilde{n}_a(\mathbf{r}) \bigr] \\
    &= \int d^dr \, \bigl[ - s \, \delta T - [ \partial_k \, \sigma_p^{\ k}(\mathbf{r}) ] \, \delta u^p(\mathbf{r}) \\
    &\hspace{15mm} + \mathbf{v}(\mathbf{r}) \cdot \delta \tilde{\mathbf{j}}(\mathbf{r}) 
    + \sum_a \mu_a(\mathbf{r}) \, \delta \tilde{n}_a(\mathbf{r}) \bigr] 
  \end{split}
  \label{equation::F_670} 
\end{equation}
and recover the necessary conditions in the form \eqref{equation::F_660}. Eqs.\ \eqref{equation::F_660} and 
\eqref{equation::F_670} are now valid for strong macroscopic displacements and deformations.

\subsection{Strain and stress tensors in different representations}
\label{section::06f}
We define the space-dependent transformation matrix
\begin{equation}
  e_i^p(\mathbf{r}) = \frac{ \partial r_0^p }{ \partial r^i } = \delta_i^p - \partial_i u^p(\mathbf{r})
  \label{equation::F_680} 
\end{equation}
which describes the transformation from the Cartesian coordinates $\mathbf{r}$ to the curvilinear coordinates 
$\mathbf{r}_0$. We may use this to calculate the inverse metric tensor. Thus, from Eq.\ \eqref{equation::F_060} 
we obtain
\begin{equation}
  g_0^{pq}(\mathbf{r}) = e_k^p(\mathbf{r}) \, e_l^q(\mathbf{r}) \, \delta^{kl} \, .
  \label{equation::F_690} 
\end{equation}
Next we calculate the differential
\begin{equation}
  dg_0^{pq}(\mathbf{r}) = de_k^p(\mathbf{r}) \, e_l^q(\mathbf{r}) \, \delta^{kl} 
  + e_k^p(\mathbf{r}) \, de_l^q(\mathbf{r}) \, \delta^{kl} \, .
  \label{equation::F_700} 
\end{equation}
Now, applying the formula \eqref{equation::F_680} for the differential 
transformation matrix and using Eq.\ \eqref{equation::F_070} we obtain an equation for the infinitesimal 
nonlinear strain tensor with two upper indices which is given by
\begin{equation}
  du^{pq}(\mathbf{r}) = \frac{ 1 }{ 2 } \, \bigl( [ \partial^k du^p(\mathbf{r}) ] \, e_k^q(\mathbf{r}) 
  + e_k^p(\mathbf{r}) \, [ \partial^k du^q(\mathbf{r}) ] \bigl) \, .
  \label{equation::F_710} 
\end{equation}
This formula transforms the infinitesimal linear strain tensor $\partial^k du^p(\mathbf{r})$ into the related 
nonlinear tensor $du^{pq}(\mathbf{r})$ where the latter tensor is symmetric in its indices because of rotation 
invariance.

We rewrite the differential thermodynamic relation \eqref{equation::F_290} in terms of the infinitesimal strain 
tensor $du^{pq}$ and define the stress tensor $\sigma_{pq}$ with two lower indices in the curvilinear coordinate 
system so that
\begin{equation}
  df = - s \, dT + \sigma_{pq} \, du^{pq} + \mathbf{v} \cdot d\tilde{\mathbf{j}} 
  + \sum_a \mu_a \, d\tilde{n}_a \, .
  \label{equation::F_720} 
\end{equation}
We insert Eq.\ \eqref{equation::F_710} for the infinitesimal strain tensor, use the symmetry relation 
$\sigma_{pq} = \sigma_{qp}$, and calculate
\begin{equation}
  \begin{split}
    \sigma_{pq}(\mathbf{r}) \, du^{pq}(\mathbf{r}) 
    &= \sigma_{pq}(\mathbf{r}) \, e_k^q(\mathbf{r}) \, [ \partial^k du^p(\mathbf{r}) ] \\
    &= \sigma_{pk}(\mathbf{r}) \, [ \partial^k du^p(\mathbf{r}) ] \, .
    \label{equation::F_730} 
  \end{split}
\end{equation}
The last equality sign results from the requirement that Eq.\ \eqref{equation::F_720} must be equivalent to 
Eq.\ \eqref{equation::F_640}. Thus, we obtain a transformation formula between the stress tensor 
$\sigma_{pq}(\mathbf{r})$ with two lower curvilinear indices and the stress tensor $\sigma_{pk}(\mathbf{r})$ 
with one lower curvilinear index and one lower Cartesian index. This formula is given by 
\begin{equation}
  \sigma_{pk}(\mathbf{r}) = \sigma_{pq}(\mathbf{r}) \, e_k^q(\mathbf{r}) \, .
  \label{equation::F_740} 
\end{equation}
Next, in analogy we may define a stress tensor with two lower Cartesian indices by the formula
\begin{equation}
  \sigma_{kl}(\mathbf{r}) = e_k^p(\mathbf{r}) \, \sigma_{pl}(\mathbf{r}) = e_k^p(\mathbf{r}) \, \sigma_{pq}(\mathbf{r}) \, e_l^q(\mathbf{r}) \, .
  \label{equation::F_750} 
\end{equation}
Unfortunately, for the related infinitesimal strain tensor with two upper Cartesian indices there does not 
exist a total differential so that we cannot write a differential thermodynamic relation for $df$ in this case.

We note that Cartesian indices might be risen and lowered by Kronecker symbols $\delta^{kl}$ and $\delta_{kl}$. 
Thus we need not care about if Cartesian indices are upper or lower. However, for the curvilinear indices it 
is important that they are lower for the stress tensors in the above formulas.

The metric tensors $g_{0,pq}(\mathbf{r})$ and $g_0^{pq}(\mathbf{r})$ are matrices which are inverse to each other. 
Thus, for the differentials of these tensors we have the well known relation 
\begin{equation}
  dg_{0,pq}(\mathbf{r}) = - g_{0,pr}(\mathbf{r}) \, dg_0^{rs}(\mathbf{r}) \, g_{0,sq}(\mathbf{r}) \, .
  \label{equation::F_760} 
\end{equation}
The Lagrange strain tensor $u_{pq}(\mathbf{r})$ is related to the metric tensor $g_{0,pq}(\mathbf{r})$ by 
Eq.\ \eqref{equation::F_040}. Similarly, the Grabert-Michel strain tensor $u^{pq}(\mathbf{r})$ is related to the 
inverse metric tensor $g_0^{pq}(\mathbf{r})$ by Eq.\ \eqref{equation::F_070}. Thus, we obtain the related equation 
for the infinitesimal strain tensors 
\begin{equation}
  du_{pq}(\mathbf{r}) = g_{0,pr}(\mathbf{r}) \, du^{rs}(\mathbf{r}) \, g_{0,sq}(\mathbf{r})
  \label{equation::F_770} 
\end{equation}
where on the right-hand side the sign has become positive. This equation means that for the infinitesimal 
strain tensors $du_{pq}(\mathbf{r})$ and $du^{pq}(\mathbf{r})$ we can lower and rise the indices as usual for 
tensor fields in curvilinear coordinates or curved spaces by using the metric tensors $g_{0,pq}(\mathbf{r})$ 
and $g_0^{pq}(\mathbf{r})$, respectively. We claim the same also for the stress tensor so that we define a stress 
tensor with two upper curvilinear indices by an analogous formula which reads 
\begin{equation}
  \sigma^{pq}(\mathbf{r}) = g^{0,pr}(\mathbf{r}) \, \sigma_{rs}(\mathbf{r}) \, g^{0,sq}(\mathbf{r}) \, .
  \label{equation::F_780} 
\end{equation}
As a consequence we find
\begin{equation}
  \sigma^{pq}(\mathbf{r}) \, du_{pq}(\mathbf{r}) = \sigma_{pq}(\mathbf{r}) \, du^{pq}(\mathbf{r})
  \label{equation::F_790} 
\end{equation}
so that we can rewrite the differential thermodynamic relation \eqref{equation::F_720} accordingly 
in the form
\begin{equation}
  df = - s \, dT + \sigma^{pq} \, du_{pq} + \mathbf{v} \cdot d\tilde{\mathbf{j}} 
  + \sum_a \mu_a \, d\tilde{n}_a \, .
  \label{equation::F_800} 
\end{equation}
Here, in the second term the stress tensor has two upper indices where the strain tensor has two lower 
indices and represents the Lagrange strain.

We may derive transformation formulas for stress tensors with two upper indices similar like 
Eqs.\ \eqref{equation::F_740} and \eqref{equation::F_750}. First, we note that for the stress tensor 
with two Cartesian indices it does not matter if the indices are upper or lower. We just have 
$\sigma^{kl}(\mathbf{r}) = \sigma_{kl}(\mathbf{r})$ because we can rise and lower the indices with Kronecker 
symbols. Thus, we find a stress tensor with one upper curvilinear index by the transformation
\begin{equation}
  \sigma^{kp}(\mathbf{r}) = \sigma^{kl}(\mathbf{r}) \, e_l^p(\mathbf{r}) \, .
  \label{equation::F_810} 
\end{equation}
Moreover, we find a stress tensor with two upper curvilinear indices by
\begin{equation}
  \sigma^{pq}(\mathbf{r}) = e_k^p(\mathbf{r}) \, \sigma^{kq}(\mathbf{r}) 
  = e_k^p(\mathbf{r}) \, \sigma^{kl}(\mathbf{r}) \, e_l^q(\mathbf{r}) \, .
  \label{equation::F_820} 
\end{equation}
This latter stress tensor is the same as that one which is defined in Eq.\ \eqref{equation::F_780}. One 
can prove this fact by using the formula \eqref{equation::F_690} which expresses the inverse metric 
tensor in terms of two transformation matrices.

We have defined five different stress tensors $\sigma^{pq}(\mathbf{r})$, $\sigma^{kp}(\mathbf{r})$, 
$\sigma^{kl}(\mathbf{r}) = \sigma_{kl}(\mathbf{r})$, $\sigma_{pk}(\mathbf{r})$, and $\sigma_{pq}(\mathbf{r})$ 
which are distinguished by the property which asks how many upper or lower curvilinear indices the 
stress tensor has. We have derived several transformation formulas which relate these stress tensors 
with each other. In the literature there exists a variety of different stress tensors which are named 
as Cauchy, Kirchhoff, first Piola-Kirchhoff, second Piola-Kirchhoff stress tensor, and some more. We 
may ask the question to which extent the names can be attached to our five stress tensors.

First, the Cauchy stress tensor is defined in the laboratory frame with Cartesian coordinates. 
Hence, it is identified by our stress tensor $\sigma^{kl}(\mathbf{r}) = \sigma_{kl}(\mathbf{r})$. For 
the other stress tensors the so called \emph{deformation gradient} $\mathsf{F}$ is needed, which in our 
notation is the inverse transformation matrix 
\begin{equation}
  e_p^i(\mathbf{r}) = \frac{ \partial r^i }{ \partial r_0^p } = \delta_p^i + \partial_{0,p} u^i(\mathbf{r})
   = \delta_p^i + e_p^k(\mathbf{r}) \, \partial_k u^i(\mathbf{r}) \, . 
  \label{equation::F_830} 
\end{equation}
The Kirchhoff stress tensor is obtained, if we multiply the Cauchy stress tensor by the determinant 
of the deformation gradient $\det \mathsf{F}$ which counts for a volume change by the deformation. In our 
notation it is the determinant of the inverse transformation matrix \eqref{equation::F_830} or the inverse 
of the Jacobi determinant \eqref{equation::D_320}. Next, the \emph{first} Piola-Kirchhoff stress tensor 
is obtained by multiplying additionally by \emph{one} inverse deformation gradient matrix $\mathsf{F}^{-1}$, 
while the \emph{second} Piola-Kirchhoff stress tensor is obtained by multiplying additionally by \emph{two} 
inverse deformation gradient matrices. In our notation this means that we multiply additionally by one or two 
transformation matrices \eqref{equation::F_680}, respectively, which are inverse to \eqref{equation::F_830}. 
As a result we find that the name Piola stands for one or two upper curvilinear indices, respectively. 

In our transformation formulas the Jacobi determinant is missing so that we must omit the name 
\emph{Kirchhoff}. In this way we arrive at the following identification of our stress tensors which is 
given by
\begin{eqnarray}
  \sigma^{pq}(\mathbf{r}) &=& \mbox{second Piola stress tensor} \, ,
  \label{equation::F_840} \\
  \sigma^{kp}(\mathbf{r}) &=& \mbox{first Piola stress tensor} \, ,
  \label{equation::F_850} \\
  \sigma^{kl}(\mathbf{r}) &=& \sigma_{kl}(\mathbf{r}) = \mbox{Cauchy stress tensor} \, ,
  \label{equation::F_860} \\
  \sigma_{pk}(\mathbf{r}) &=& \mbox{minus first Piola stress tensor} \, ,
  \label{equation::F_870} \\
  \sigma_{pq}(\mathbf{r}) &=& \mbox{minus second Piola stress tensor} \, . \hspace{8mm}
  \label{equation::F_880}
\end{eqnarray}
For the last two stress tensors with one or two lower curvilinear indices we have extended the 
naming in a logical way. Here we multiply one or two times by the inverse transformation matrix 
\eqref{equation::F_830}, respectively. This means that we multiply one or two times by the inverse 
of the inverse deformation gradient matrix $(\mathsf{F}^{-1})^{-1} = \mathsf{F}$.

There is an important general difference of our five stress tensors compared with the usual 
stress tensors in the literature. In our case we consider crystals where diffusion of particles is 
possible. Examples are the so called \emph{cluster crystals} where the interactions do not have a 
hard core so that several particles may be on one lattice position. This means, that in our 
definition the macroscopic particle densities $\tilde{n}_a$ are kept constant. One clearly sees 
this in the several differential thermodynamic relations \eqref{equation::F_640}, \eqref{equation::F_720}, 
and \eqref{equation::F_800}. On the other hand, in normal crystals where strong repulsive interactions 
with hard cores prevent particle diffusion the macroscopic particle densities $\tilde{n}_a$ are not 
constant but rather change with the deformation according to the formula \eqref{equation::D_360}. In this 
latter case there will be additional terms in all stress tensors similar as we have found in our discussions 
at the end of Subsec.\ \ref{section::05e} in Eqs.\ \eqref{equation::E_640}-\eqref{equation::E_660}.

In their work Grabert and Michel \cite{GM83} use three stress tensors which agree with our three 
stress tensors \eqref{equation::F_860}-\eqref{equation::F_880} with two lower indices. Our transformation 
formula \eqref{equation::F_750} together with the transformation matrix \eqref{equation::F_680} which 
relates the three different stress tensors with each other is equivalent to Eq.\ (19) of Grabert and Michel.

Finally, we may ask the question which of the five stress tensors is the natural variable for our theory. 
In our calculations we first encountered $\sigma_p^{\ k}(\mathbf{r})$ which has one lower curvilinear 
index. Hence, the natural variable is the \emph{minus first Piola} stress tensor. The necessary conditions 
for the global thermodynamic equilibrium \eqref{equation::F_660} represent one of the most important 
results of our calculations up to now. The first of these conditions is formulated for the stress tensor 
$\sigma_p^{\ k}(\mathbf{r})$ with one lower curvilinear index. In order to transform to the other 
four strain tensors we must multiply either by the transformation matrix \eqref{equation::F_680} or 
by the inverse matrix \eqref{equation::F_830}. Both transformation matrices are locally defined and 
depend on the position coordinates. For this reason, the transformation matrices $e_i^p(\mathbf{r})$ and 
$e_p^i(\mathbf{r})$ do not commute with the partial derivative $\partial_k = \partial / \partial r^k$ in 
Eq.\ \eqref{equation::F_660}. Hence a calculation like Eq.\ \eqref{equation::F_650} is not possible here. 

The natural variable for the conjugate strain tensor is the transformation matrix $e_k^p(\mathbf{r})$ 
defined in Eq.\ \eqref{equation::F_680} which has one upper curvilinear index. The related differential 
is the infinitesimal linear strain tensor $de_k^p(\mathbf{r}) = - \partial_k du^p(\mathbf{r})$. In the 
next section we shall derive the time-evolution equations for the dynamics of the crystal system on 
macroscopic scales. There again we shall find that $\partial_k du^p(\mathbf{r})$ is the natural variable 
for the strain and $\sigma_p^{\ k}(\mathbf{r})$ is the natural variable for the stress.

\section{Projection operators and the time-evolution equations}
\label{section::07}
The time-evolution equations of continuum mechanics are derived from the microscopic theory in several 
steps. First one selects and specifies the relevant variables which describe the relevant properties at 
large space scales and large time scales. Here we choose the coarse grained densities of the conserved 
quantities. Furthermore for describing the deformation of a crystal we add the displacement field. This 
has been done carefully in the preceding sections. In a second step we define the projection operators 
$\mathsf{P}$ and $\mathsf{Q}$ which project the statistical distribution function when acting to the left 
or any physical variable when acting to the right either to the subspace of the relevant variables or to 
its orthogonal complement. In a third step, the projection operators are used to derive a master equation 
for the relevant distribution function from the microscopic time-evolution equation. Finally, the master 
equation is multiplied by the relevant variables and the integral is calculated over all degrees of freedom 
of the whole phase space. As a result, time-evolution equations for the average relevant variables are 
obtained which are the equations of continuum mechanics.

The procedure is described in several text books and several publications. We follow our own way presented 
in our previous paper \cite{Ha16} where we have derived the hydrodynamic equations for simple fluids. 
Here we extend the derivations and considerations to include the coarse graining and the displacement 
field in order to describe the elasticity of crystals. However, we restrict our calculations to the 
isothermal case where the temperature is constant and where the effects of heat transport and warming 
by friction are neglected. The general case is considered in the next section.

\subsection{Projection operators}
\label{section::07a}
The statistical properties of a classical many-particle system are described by the distribution function 
$\hat{\varrho} = \varrho(\Gamma)$ which depends on all coordinates and momenta of the particles 
$\Gamma = ( \mathbf{r}_{ai},\mathbf{p}_{ai} )$ in the whole phase space. For an exact microscopic description
with \emph{maximum information} the statistical weight is located at a single point in the phase space which 
might be realized by delta functions of coordinates and momenta. We denote this exact distribution function 
by the plain $\hat{\varrho} = \varrho(\Gamma)$. On the other hand, we may maximize the entropy under certain 
constraints in order to obtain the relevant distribution function $\hat{\tilde{\varrho}} = \tilde{\varrho}(\Gamma)$ 
with \emph{minimum information}. We find that this latter distribution function is given by the formula 
\eqref{equation::B_130} which describes the physical system in a nonequilibrium state with a local thermodynamic 
equilibrium at each space point. We use the tilde to distinguish the relevant 
$\hat{\tilde{\varrho}} = \tilde{\varrho}(\Gamma)$ from the microscopic $\hat{\varrho} = \varrho(\Gamma)$.

We denote the relevant variables by $\hat{\tilde{x}}_i(\mathbf{r}) = \tilde{x}_i(\mathbf{r},\Gamma)$ where 
the hat indicates that these are \emph{microscopic} quantities defined in the phase space. The index $i$ 
counts the several variables. The argument $\mathbf{r}$ makes the variable a local function in the coordinate 
space. Here, the tilde is used to indicate that the coarse graining procedure \eqref{equation::D_040} has 
been applied so that the relevant variables are slowly varying functions in space. The averages of the relevant 
variables $\tilde{x}_i(\mathbf{r})$ are calculated with respect to the distribution functions according to 
\begin{equation}
  \begin{split}
    \tilde{x}_i(\mathbf{r}) = \langle \hat{\tilde{x}}_i(\mathbf{r}) \rangle 
    &= \mathrm{Tr} \{ \hat{\varrho} \, \hat{\tilde{x}}_i(\mathbf{r}) \} 
    = \int d\Gamma \, \varrho(\Gamma) \, \tilde{x}_i(\mathbf{r},\Gamma) \\
    &= \mathrm{Tr} \{ \hat{\tilde{\varrho}} \, \hat{\tilde{x}}_i(\mathbf{r}) \} 
    = \int d\Gamma \, \tilde{\varrho}(\Gamma) \, \tilde{x}_i(\mathbf{r},\Gamma) \, .
  \end{split}
  \label{equation::G_010}
\end{equation}
We use the trace $\mathrm{Tr} \{ \ldots \}$ as a short-hand notation for the phase-space integral 
$\int d\Gamma \, \{ \ldots \}$. Since we always consider the grand canonical ensemble the integration 
over the phase space includes a summation over all possible values for the particle numbers. There are 
two possibilities to calculate the averages. We may use the microscopic distribution function 
$\hat{\varrho} = \varrho(\Gamma)$ which is done in the first line. Alternatively, we may use the relevant 
distribution function $\hat{\tilde{\varrho}} = \tilde{\varrho}(\Gamma)$ which is done in the second line. 
Here, for the \emph{relevant} variables $\hat{\tilde{x}}_i(\mathbf{r}) = \tilde{x}_i(\mathbf{r},\Gamma)$ 
we obtain the same result so that the first line equals the second line. These equalities are the 
constraints when maximizing the entropy to obtain the relevant distribution function.

In general, for any other physical variable $\hat{Y}(\mathbf{r}) = Y(\mathbf{r},\Gamma)$ this equality 
usually does not hold so that in most cases we have
\begin{equation}
  Y(\mathbf{r}) = \langle \hat{Y}(\mathbf{r}) \rangle 
  = \mathrm{Tr} \{ \hat{\tilde{\varrho}} \, \hat{Y}(\mathbf{r}) \} 
  \neq \mathrm{Tr} \{ \hat{\varrho} \, \hat{Y}(\mathbf{r}) \} \, .
  \label{equation::G_020}
\end{equation}
However, we may define a projection operator $\mathsf{P}[\tilde{x}]$ which projects onto the subspace of 
the relevant variables. Thus, we may write
\begin{equation}
  Y(\mathbf{r}) = \langle \hat{Y}(\mathbf{r}) \rangle 
  = \mathrm{Tr} \{ \hat{\tilde{\varrho}} \, \hat{Y}(\mathbf{r}) \} 
  = \mathrm{Tr} \{ \hat{\varrho} \, \mathsf{P}[\tilde{x}] \, \hat{Y}(\mathbf{r}) \} \, .
  \label{equation::G_030}
\end{equation}
Here and in the following we use the convention that average values 
$\langle \ldots \rangle = \mathrm{Tr} \{ \hat{\tilde{\varrho}} \ldots \}$ are calculated always with the 
\emph{relevant} distribution function $\hat{\tilde{\varrho}} = \tilde{\varrho}(\Gamma)$.

There are many definitions around for the projection operator $\mathsf{P}[\tilde{x}]$. We use either 
the Kawasaki-Gunton projection operator \cite{KG73} or the Grabert projection operator \cite{Gr82}. 
If it acts to the \emph{left} then it is the \emph{Kawasaki-Gunton} projection operator. On the other hand, 
if it acts to the \emph{right} onto the physical variable then it is the \emph{Grabert} projection operator. 
Both operators are equivalent to each other and are doing the same thing. The basic condition which a 
projection operator must satisfy is
\begin{equation}
  \mathsf{P}[\tilde{x}] \, \mathsf{P}[\tilde{x}] = \mathsf{P}[\tilde{x}] \, .
  \label{equation::G_040}
\end{equation}
On the other hand, there are two special cases for how the projection operator acts. First, for the 
distribution function we find
\begin{equation}
  \hat{\varrho} \, \mathsf{P}[\tilde{x}] = \hat{\tilde{\varrho}} \, .
  \label{equation::G_050}
\end{equation}
Second, for the relevant variables we have
\begin{equation}
  \mathsf{P}[\tilde{x}] \, \hat{\tilde{x}}_i(\mathbf{r}) = \hat{\tilde{x}}_i(\mathbf{r}) \, .
  \label{equation::G_060}
\end{equation}
These two equations are constraints which the projection operator must satisfy. 

Here and in the following we define the Grabert Projection operator. Acting to the right on any physical 
variable $\hat{Y}(\mathbf{r})$ we define
\begin{equation}
  \begin{split}
    \mathsf{P}[\tilde{x}] \, \hat{Y}(\mathbf{r}) = \,& \biggl( 1 + \sum_i \int d^dr_1 
    \, [ \hat{\tilde{x}}_i(\mathbf{r}_1) - \tilde{x}_i(\mathbf{r}_1) ] 
    \, \frac{ \delta }{ \delta \tilde{x}_i(\mathbf{r}_1) } \biggr) \\
    &\times \mathrm{Tr} \{ \hat{\tilde{\varrho}} \, \hat{Y}(\mathbf{r}) \} \, .
  \end{split}
  \label{equation::G_070}
\end{equation}
The relevant distribution $\hat{\tilde{\varrho}} = \tilde{\varrho}(\Gamma)$ defined in 
Eq.\ \eqref{equation::B_130} for a local thermodynamic equilibrium is a functional of the relevant variables. 
For this reason we denote the projection operator $\mathsf{P}[\tilde{x}]$ with the argument $[\tilde{x}]$. 
Beyond that it is important to note that here and below in all definitions of the Grabert projection operator 
the functional derivatives $\delta / \delta \tilde{x}_i(\mathbf{r}_1)$ act only onto the relevant distribution 
$\hat{\tilde{\varrho}}$ but never onto the physical variable $\hat{Y}(\mathbf{r})$. The physical variable is 
considered to be constant.

In our previous paper \cite{Ha16} we have defined the projection operator $\mathsf{P}[x]$ by Eq.\ \eqref{equation::G_070} 
where we have inserted the microscopic variables $\hat{x}_i(\mathbf{r}) = x_i(\mathbf{r},\Gamma)$ with no tilde 
and where we have not cared about coarse graining. Here, we first extend the projection operator for the liquid 
state by including coarse graining. This is done in Eq.\ \eqref{equation::G_070} by using the relevant variables 
with tilde. We may rewrite the projection operator in terms of the microscopic variables with no tilde. For 
this purpose we apply the coarse graining procedure \eqref{equation::D_040} to the relevant variables. Thus, we 
obtain
\begin{widetext}
\begin{equation}
  \mathsf{P}[\tilde{x}] \, \hat{Y}(\mathbf{r}) = \biggl( 1 + \sum_i \int d^dr_1 \int d^dr_2 
  \, [ \hat{x}_i(\mathbf{r}_1) - x_i(\mathbf{r}_1) ] 
  \, w( \mathbf{r}_1 - \mathbf{r}_2 ) \, \frac{ \delta }{ \delta x_i(\mathbf{r}_2) } \biggr) 
  \, \mathrm{Tr} \{ \hat{\tilde{\varrho}} \, \hat{Y}(\mathbf{r}) \} \, .
  \label{equation::G_080}
\end{equation}
This operator is a generalization of the standard Grabert projection operator \cite{Gr82} by inserting 
an integration kernel $w( \mathbf{r}_1 - \mathbf{r}_2 )$ in between and using a double integral. For the 
projection property \eqref{equation::G_040} to hold the integral kernel $w( \mathbf{r}_1 - \mathbf{r}_2 )$ 
must satisfy Eq.\ \eqref{equation::D_050} which is the projection property for integral kernels. We may define 
an even more general projection operator 
\begin{equation}
  \mathsf{P}[\tilde{x}] \, \hat{Y}(\mathbf{r}) = \biggl( 1 + \sum_{ij} \int d^dr_1 \int d^dr_2 
  \, [ \hat{x}_i(\mathbf{r}_1) - x_i(\mathbf{r}_1) ] 
  \, F_{ij}( \mathbf{r}_1, \mathbf{r}_2 ) \, \frac{ \delta }{ \delta x_j(\mathbf{r}_2) } \biggr) 
  \, \mathrm{Tr} \{ \hat{\tilde{\varrho}} \, \hat{Y}(\mathbf{r}) \} 
  \label{equation::G_090}
\end{equation}
where $F_{ij}( \mathbf{r}_1, \mathbf{r}_2 )$ are any functions with two indices and two independent arguments 
$\mathbf{r}_1$ and $\mathbf{r}_2$ which satisfy the projection condition
\begin{equation}
  F_{ik}( \mathbf{r}_1, \mathbf{r}_3 ) = \sum_j \int d^dr_2 \, F_{ij}( \mathbf{r}_1, \mathbf{r}_2 ) 
  \, F_{jk}( \mathbf{r}_2, \mathbf{r}_3 ) \, .
  \label{equation::G_100} 
\end{equation}
Now, for the relevant variables $x_i(\mathbf{r})$ we may insert the particle densities $n_a(\mathbf{r})$ and 
the momentum densities $\mathbf{j}(\mathbf{r})$. Then, we obtain the projection operator for a many-particles 
system in the fluid state
\begin{equation}
  \begin{split}
    \mathsf{P}[\tilde{n},\tilde{\mathbf{j}}] \, \hat{Y}(\mathbf{r}) = \,& \biggl( 1 
    + \sum_a \int d^dr_1 \int d^dr_2 \, [ \hat{n}_a(\mathbf{r}_1) - n_a(\mathbf{r}_1) ] 
    \, w( \mathbf{r}_1 - \mathbf{r}_2 ) \, \frac{ \delta }{ \delta n_a(\mathbf{r}_2) } \\
    &\hspace{4.5mm} + \sum_k \int d^dr_1 \int d^dr_2 \, [ \hat{j}^k(\mathbf{r}_1) - j^k(\mathbf{r}_1) ] 
    \, w( \mathbf{r}_1 - \mathbf{r}_2 ) \, \frac{ \delta }{ \delta j^k(\mathbf{r}_2) } \biggr) 
    \, \mathrm{Tr} \{ \hat{\tilde{\varrho}} \, \hat{Y}(\mathbf{r}) \} \, .
  \end{split}
  \label{equation::G_110}
\end{equation}
\end{widetext}

We may apply the projection operator onto the microscopic variables. In the general case we apply 
$\mathsf{P}[\tilde{x}]$ defined in Eq.\ \eqref{equation::G_080} onto the general microscopic variables 
$\hat{x}_i(\mathbf{r}) = x_i(\mathbf{r},\Gamma)$. After some short calculations we obtain
\begin{equation}
  \mathsf{P}[\tilde{x}] \, \hat{x}_i(\mathbf{r}) = x_i(\mathbf{r}) 
  + [ \hat{\tilde{x}}_i(\mathbf{r}) - \tilde{x}_i(\mathbf{r}) ] \, .
  \label{equation::G_120}
\end{equation}
The first term is the average of the microscopic densities $x_i(\mathbf{r})$. It originates from the 
first term in the projection operator \eqref{equation::G_080} which is a factor unity times the average. 
The second term in square brackets represents the coarse-grained \emph{fluctuations} of the densities 
on macroscopic scales. For comparison, we apply the projection operator onto the average microscopic 
densities $x_i(\mathbf{r})$. In this case only the first term of the Eq.\ \eqref{equation::G_080} 
is active so that we simply obtain
\begin{equation}
  \mathsf{P}[\tilde{x}] \, x_i(\mathbf{r}) = x_i(\mathbf{r}) \, .
  \label{equation::G_130}
\end{equation}
As a consequence, from the difference of Eqs.\ \eqref{equation::G_120} and \eqref{equation::G_130} 
we find
\begin{equation}
  \mathsf{P}[\tilde{x}] \, [ \hat{x}_i(\mathbf{r}) - x_i(\mathbf{r}) ] = 
  \hat{\tilde{x}}_i(\mathbf{r}) - \tilde{x}_i(\mathbf{r}) \, .
  \label{equation::G_140}
\end{equation}
Thus, the microscopic density fluctuations $\hat{x}_i(\mathbf{r}) - x_i(\mathbf{r})$ are projected onto 
the coarse-grained macroscopic density fluctuations $\hat{\tilde{x}}_i(\mathbf{r}) - \tilde{x}_i(\mathbf{r})$.

Similarly, in our special case we apply $\mathsf{P}[\tilde{n},\tilde{\mathbf{j}}]$ defined in 
Eq.\ \eqref{equation::G_110} onto the microscopic particle densities 
$\hat{n}_a(\mathbf{r}) = n_a(\mathbf{r},\Gamma)$ and the microscopic momentum density 
$\hat{\mathbf{j}}(\mathbf{r}) = \mathbf{j}(\mathbf{r},\Gamma)$. Here we obtain the related equations
\begin{eqnarray}
  \mathsf{P}[\tilde{n},\tilde{\mathbf{j}}] \,\, \hat{n}_a(\mathbf{r}) &=& n_a(\mathbf{r}) 
  + [ \hat{\tilde{n}}_a(\mathbf{r}) - \tilde{n}_a(\mathbf{r}) ] \, ,
  \label{equation::G_150} \\
  \mathsf{P}[\tilde{n},\tilde{\mathbf{j}}] \,\, \hat{\mathbf{j}}(\mathbf{r}) &=& \mathbf{j}(\mathbf{r}) 
  + [ \hat{\tilde{\mathbf{j}}}(\mathbf{r}) - \tilde{\mathbf{j}}(\mathbf{r}) ]
  \label{equation::G_160}
\end{eqnarray}
where on the right-hand sides there are the related averages of the microscopic densities plus 
the coarse-grained density \emph{fluctuations} on macroscopic scales. Further projection equations 
analogous to Eqs.\ \eqref{equation::G_130} and \eqref{equation::G_140} can be derived also in our 
special case. As a result Eq.\ \eqref{equation::G_140} and its analogs of our special case provide a 
representation of the coarse-graining procedure \eqref{equation::D_040} within the framework of 
projection operators applied onto the \emph{fluctuations} of the microscopic variables.

Until now the projection operator performs the projection to the relevant conserved densities and provides 
the coarse-graining procedure. Hence, this projection operator may be applied to derive the hydrodynamic 
equations of a simple liquid. This has been done in our previous publication \cite{Ha16} where now the 
procedure is extended by including the coarse graining. However, the effects of deformation and the 
displacement field $\mathbf{u}(\mathbf{r})$ are not included up to now. For this reason, a further extension 
of the projection operator is necessary which we shall do next.

In the linearized theory the density variations $\delta n_a(\mathbf{r})$ related to deformations were 
obtained by solving the homogeneous integral equation \eqref{equation::D_530}. Unfortunately, the solution 
of this equation provides deformations only with infinitely large wave lengths. For this reason, in order to 
handle deformations with finite wavelengths we rather consider the related eigenvalue equation
\begin{equation}
  \begin{split}
    \sum_b \int d^dr_2 \, \frac{ \delta^2 F[T,n] }{ \delta n_a(\mathbf{r}_1) \delta n_b(\mathbf{r}_2) } \bigg|_{n = n_0} 
    b_{ib}(\mathbf{k},\mathbf{r}_2) = \\
    = \lambda_i(\mathbf{k}) \, b_{ia}(\mathbf{k},\mathbf{r}_1) \, .
  \end{split}
  \label{equation::G_170} 
\end{equation}
Here the eigenvalues $\lambda_i(\mathbf{k})$ are real. They depend explicitly on the wave vector $\mathbf{k}$ 
and on the polarization $i$ of the deformation. On the other hand, the eigenfunctions are complex. They can 
be orthonormalized according to
\begin{equation}
  \sum_a \int d^dr \, [ b_{ia}(\mathbf{k}_1,\mathbf{r}) ]^* \, b_{ja}(\mathbf{k}_2,\mathbf{r}) = 
  \delta_{ij} \, ( 2 \pi )^d \, \delta( \mathbf{k}_1 - \mathbf{k}_2 ) \, .
  \label{equation::G_180} 
\end{equation}
Here and in the following we always assume that the eigenfunctions are orthonormalized and satisfy the 
condition \eqref{equation::G_180}. 

For nonzero wave vectors $\mathbf{k}$ the eigenfunctions may be written in terms of a Bloch ansatz as 
\begin{equation}
  b_{ia}(\mathbf{k},\mathbf{r}) = u_{ia}(\mathbf{k},\mathbf{r}) \, \exp( i \mathbf{k} \cdot \mathbf{r} )
  \label{equation::G_190} 
\end{equation}
where $u_{ia}(\mathbf{k},\mathbf{r})$ are periodic functions on the crystal lattice satisfying
\begin{equation}
  u_{ia}(\mathbf{k},\mathbf{r}) = u_{ia}( \mathbf{k}, \mathbf{r} + \mathbf{a} ) 
  \quad \mbox{for all $\mathbf{a} \in \mathcal{A}$} \, .
  \label{equation::G_200} 
\end{equation}
In the special case $\mathbf{k} = \mathbf{0}$ we already know the solution. Translation invariance provides 
the eigenvalues $\lambda_i(\mathbf{k}) = 0$ and the eigenfunctions 
\begin{equation}
  b_{ia}(\mathbf{0},\mathbf{r}) = u_{ia}(\mathbf{0},\mathbf{r}) = c_i^k \, [ \partial_k n_{0,a}(\mathbf{r}) ]
  \label{equation::G_210} 
\end{equation}
which are just linear combinations of the spatial derivatives of the microscopic particle densities 
of the reference state. The coefficients $c_i^k$ are $d \times d$ matrices where $d$ is the number 
of dimension in the space. They guarantee the correct polarizations of the eigenfunctions in the limit 
$\mathbf{k} \to \mathbf{0}$. The number of the polarizations is $d$. There are as many polarizations 
as dimensions in the space.

For later considerations we define the inverse coefficient matrix $d_k^i$ by the conditions
\begin{equation}
  c_i^k \, d_k^j = \delta_i^j \, , \qquad d_k^i \, c_i^l = \delta_k^l \, .
  \label{equation::G_220} 
\end{equation}
Furthermore we define the normalization matrix $\mathcal{N}_{kl}$ and the inverse normalization matrix 
$\mathcal{N}^{kl}$ by
\begin{equation}
  \mathcal{N}_{kl} = d_k^i \, \delta_{ij} \, d_l^j \, , \qquad
  \mathcal{N}^{kl} = c_i^k \, \delta^{ij} \, c_j^l \, ,
  \label{equation::G_230} 
\end{equation}
respectively. Then, from the orthonormalization condition of the eigenfunctions \eqref{equation::G_180} 
we infer the normalization matrix
\begin{equation}
  \mathcal{N}_{kl} = \frac{ 1 }{ V } \sum_a \int_V d^dr \, [ \partial_k n_{0,a}(\mathbf{r}) ] 
  \, [ \partial_l n_{0,a}(\mathbf{r}) ] \, .
  \label{equation::G_240} 
\end{equation}
On the other hand, from Eqs.\ \eqref{equation::G_220} and \eqref{equation::G_230} we prove the relations
\begin{equation}
  \mathcal{N}_{ij} \, \mathcal{N}^{jk} = \delta_i^k \, , \qquad
  \mathcal{N}^{ij} \, \mathcal{N}_{jk} = \delta^i_k \, .
  \label{equation::G_250} 
\end{equation}

For small nonzero wave vectors in the limit $\mathbf{k} \to \mathbf{0}$ we expect \emph{acoustic phonons} 
as the essential deformation modes. Hence, up to linear order the eigenvalues may be written as
\begin{equation}
  \lambda_i(\mathbf{k}) = a_{il} \, k^l
  \label{equation::G_260} 
\end{equation}
where the linear coefficients $a_{il}$ are related to the sound velocities. On the other hand 
up to linear order we write the Bloch functions as
\begin{equation}
  u_{ia}(\mathbf{k},\mathbf{r}) = c_i^k \, \Bigl( [ \partial_k n_{0,a}(\mathbf{r}) ] 
  - \Delta B_{a,k}^{\prime \, \ \ l}(\mathbf{r}) \, i k_l \Bigr)
  \label{equation::G_270} 
\end{equation}
which must be inserted into the formula for the eigenfunctions \eqref{equation::G_190}.
This formula is an extension of Eq.\ \eqref{equation::G_210} for small nonzero wave vectors.
Here, $\Delta B_{a,k}^{\prime \, \ \ l}(\mathbf{r})$ are real functions which later will be identified 
with the corrections functions defined in Eq.\ \eqref{equation::E_150}.

Now, we may use the eigenfunctions $b_{ia}(\mathbf{k},\mathbf{r})$ as basis functions of the function 
subspace related to the deformations. We define the integration kernel 
\begin{widetext}
\begin{equation}
  \begin{split}
    F_{ab}( \mathbf{r}_1, \mathbf{r}_2 ) &= \sum_{ij} \int \frac{ d^dk }{ ( 2 \pi )^d } \, b_{ia}(\mathbf{k},\mathbf{r}_1) 
    \, \delta^{ij} \, \theta( \Lambda^2 - \mathbf{k}^2 ) \, [ b_{jb}(\mathbf{k},\mathbf{r}_2) ]^* \\
    &= \sum_{ij} \int \frac{ d^dk }{ ( 2 \pi )^d } \, u_{ia}(\mathbf{k},\mathbf{r}_1) 
    \, \delta^{ij} \, \theta( \Lambda^2 - \mathbf{k}^2 ) \, \exp( i \mathbf{k} \cdot [ \mathbf{r}_1 - \mathbf{r}_2 ] ) 
    \, [ u_{jb}(\mathbf{k},\mathbf{r}_2) ]^* 
  \end{split}
  \label{equation::G_280} 
\end{equation}
where the second line is obtained by inserting the Bloch ansatz \eqref{equation::G_190}.
Similar like in the integration kernel for coarse graining \eqref{equation::D_060} here the Heaviside theta 
function again restricts the wave vectors $\mathbf{k}$ to the maximum absolute value $\Lambda = 2\pi / \ell$ 
which is related to the mesoscopic length $\ell$. Using the orthonormalization \eqref{equation::G_180} one 
can easily prove that the integration kernel \eqref{equation::G_280} satisfies a projection condition similar 
like \eqref{equation::G_100}. Hence, we may use this kernel to extend our projection operator in order to 
include the deformations of a crystal. Thus, we define the extended projection operator
\begin{equation}
  \begin{split}
    \mathsf{P}[\tilde{n},\tilde{\mathbf{j}},\mathbf{u}] \, \hat{Y}(\mathbf{r}) = \,& \biggl( 1 
    + \sum_a \int d^dr_1 \int d^dr_2 \, [ \hat{n}_a(\mathbf{r}_1) - n_a(\mathbf{r}_1) ] 
    \, w( \mathbf{r}_1 - \mathbf{r}_2 ) \, \frac{ \delta }{ \delta n_a(\mathbf{r}_2) } \\
    &\hspace{4.5mm} + \sum_k \int d^dr_1 \int d^dr_2 \, [ \hat{j}^k(\mathbf{r}_1) - j^k(\mathbf{r}_1) ] 
    \, w( \mathbf{r}_1 - \mathbf{r}_2 ) \, \frac{ \delta }{ \delta j^k(\mathbf{r}_2) } \\
    &\hspace{4.5mm} + \sum_{ab} \int d^dr_1 \int d^dr_2 \, [ \hat{n}_a(\mathbf{r}_1) - n_a(\mathbf{r}_1) ] 
    \, F_{ab}( \mathbf{r}_1, \mathbf{r}_2 ) \, \frac{ \delta }{ \delta n_b(\mathbf{r}_2) } \biggr) 
    \, \mathrm{Tr} \{ \hat{\tilde{\varrho}} \, \hat{Y}(\mathbf{r}) \} \, .
  \end{split}
  \label{equation::G_290}
\end{equation}
\end{widetext}
Beyond the unity at the beginning we have three nontrivial contributions to this projection operator which 
project onto the function subspaces of the coarse-grained particle densities $\tilde{n}_a(\mathbf{r})$, 
the coarse-grained momentum densities $\tilde{\mathbf{j}}(\mathbf{r})$, and the displacement field 
$\mathbf{u}(\mathbf{r})$, respectively. For the projection operator to be consistent it is important that 
all three function subspaces are orthogonal to each other. For the second term it is easy to prove that 
this term is orthogonal to the other two terms. The reason is that the second term involves the momentum 
densities while the other terms involve the particle densities. These are just different variables. However, 
the first and the third term both involve the particle densities. Here we must consider the basis functions 
and prove the orthogonality by calculating the scalar product \eqref{equation::D_020}.

For the third term the basis functions are defined in Eq.\ \eqref{equation::G_190}. For the first term the 
basis functions are
\begin{equation}
  a(\mathbf{k},\mathbf{r}) = \exp( i \mathbf{k} \cdot \mathbf{r} ) 
  \label{equation::G_300} 
\end{equation}
because the Fourier representation of the integration kernel \eqref{equation::D_060} may be rewritten as 
\begin{equation}
  w( \mathbf{r}_1 - \mathbf{r}_2 ) = \int \frac{ d^dk }{ ( 2 \pi )^d } \, a(\mathbf{k},\mathbf{r}_1) 
  \, \theta( \Lambda^2 - \mathbf{k}^2 ) \, [ a(\mathbf{k},\mathbf{r}_2) ]^* \ .
  \label{equation::G_310} 
\end{equation}
Consequently, for the proof of orthogonality the relevant scalar product is 
\begin{equation}
  \begin{split}
    \langle a(\mathbf{k}_1) \mid b_{ia}(\mathbf{k}_2) \rangle &=
    \int d^dr \, [ a(\mathbf{k}_1,\mathbf{r}) ]^* \, b_{ia}(\mathbf{k}_2,\mathbf{r}) \\
    &= \int d^dr \, u_{ia}(\mathbf{k}_2,\mathbf{r}) \, \exp( - i [ \mathbf{k}_1 - \mathbf{k}_2 ] \cdot \mathbf{r} ) \\
    &= 0 \, .
  \end{split}
  \label{equation::G_320} 
\end{equation}
We must prove the last equality sign so that the scalar product is zero. For different wave vectors 
$\mathbf{k}_1 \neq \mathbf{k}_2$ this is done by the oscillations of the exponential function. Hence, there 
remains the scalar product with equal wave vectors
\begin{equation}
  \langle a(\mathbf{k}) \mid b_{ia}(\mathbf{k}) \rangle = \int d^dr \, u_{ia}(\mathbf{k},\mathbf{r}) = 0 
  \label{equation::G_330} 
\end{equation}
which must be proven to be zero. In the special case $\mathbf{k} = \mathbf{0}$ this is easily done. 
Inserting Eq.\ \eqref{equation::G_210} and using the integral theorem of Gauss we calculate 
\begin{equation}
  \begin{split}
    \langle a(\mathbf{0}) \mid b_{ia}(\mathbf{0}) \rangle &= \int_V d^dr \, c_i^k \, [ \partial_k n_{0,a}(\mathbf{r}) ] \\
    &= c_i^k \, \int_{\partial V} dS_k \, n_{0,a}(\mathbf{r}) = 0 \, .
  \end{split}
  \label{equation::G_340} 
\end{equation}
The last equality sign follows exactly if we choose a volume $V$ which fits with the lattice structure 
so that periodic boundary conditions can be applied. In the other cases the surface integral is less than 
proportional to the volume $V$ which is sufficient for orthogonality. 

Next we extend the proof to small nonzero wave vectors $\mathbf{k}$ up to linear order. We insert the 
extended Bloch function \eqref{equation::G_270} into the orthogonality condition \eqref{equation::G_330}. 
The first term is zero according to Eq.\ \eqref{equation::G_340}. The second term is zero because the 
correction function $\Delta B_{a,k}^{\prime \, \ \ l}(\mathbf{r})$ is defined within the complement 
function space $\mathcal{C}^\prime(\ell)$ so that the integral of this function over the whole space is 
zero. Hence, both terms are zero so that the last equality sign of Eq.\ \eqref{equation::G_330} is proven.

In the general case for larger nonzero wave vectors $\mathbf{k}$ we do not have an explicit proof. However, 
as long as the wave vector $\mathbf{k}$ is sufficiently small the continuity implies that in leading order 
the scalar product \eqref{equation::G_330} may be proportional to the square of $\mathbf{k}$. Consequently, 
if it is nonzero then it is at least small. In any case, if necessary we may orthogonalize the basis functions 
$b_{ia}(\mathbf{k},\mathbf{r})$ by using the method of Gram and Schmidt \cite{CK12}. For this purpose we add 
spatially constant values to the Bloch functions $u_{ia}(\mathbf{k},\mathbf{r})$ which are adjusted in such 
a way so that the orthogonality condition \eqref{equation::G_330} is satisfied. In this way we make the 
projection operator \eqref{equation::G_290} consistent so that all three nontrivial contributions are 
orthogonal to each other.

The maximum absolute value $\Lambda = 2\pi / \ell$ confines the wave vector $\mathbf{k}$ to small values 
so that we may perform an approximation in the projection operator. Starting with the integral kernel 
$\eqref{equation::G_280}$ in the second line in the Bloch functions we replace the argument by 
$\mathbf{k} = \mathbf{0}$. Then we may use the leading-order Bloch functions defined explicitly in 
Eq.\ \eqref{equation::G_210}. Thus, in leading-order approximation we obtain the integral kernel
\begin{widetext}
\begin{equation}
  \begin{split}
    F_{ab}( \mathbf{r}_1, \mathbf{r}_2 ) &\approx 
    \sum_{ij} \int \frac{ d^dk }{ ( 2 \pi )^d } \, u_{ia}(\mathbf{0},\mathbf{r}_1) 
    \, \delta^{ij} \, \theta( \Lambda^2 - \mathbf{k}^2 ) 
    \, \exp( i \mathbf{k} \cdot [ \mathbf{r}_1 - \mathbf{r}_2 ] ) \, [ u_{jb}(\mathbf{0},\mathbf{r}_2) ]^* \\
    &= \sum_{ijkl} \, c_i^k \, [ \partial_k n_{0,a}(\mathbf{r}_1) ] \, \delta^{ij} 
    \, \biggl( \int \frac{ d^dk }{ ( 2 \pi )^d } \, \theta( \Lambda^2 - \mathbf{k}^2 ) 
    \, \exp( i \mathbf{k} \cdot [ \mathbf{r}_1 - \mathbf{r}_2 ] ) \biggr) \, c_j^l 
    \, [ \partial_l n_{0,b}(\mathbf{r}_2) ] \, .
  \end{split}
  \label{equation::G_350} 
\end{equation}
\end{widetext}
The integral over the wave vector is confined to the Heaviside theta function and the exponential function. 
Hence, it can be evaluated explicitly by the formula \eqref{equation::D_060} so that it results in the 
integration kernel $w( \mathbf{r}_1 - \mathbf{r}_2 )$ of the coarse-graining procedure \eqref{equation::D_040}. 
The coefficients $c_i^k$ and $c_j^l$ arising from the linear combinations in Eq.\ \eqref{equation::G_210} can 
be combined into the inverse normalization matrix $\mathcal{N}^{kl}$ by using the formula \eqref{equation::G_230}. 
As a result we obtain a simple formula for the integral kernel in leading-order approximation which reads
\begin{equation}
  F_{ab}( \mathbf{r}_1, \mathbf{r}_2 ) = \sum_{kl} \, [ \partial_k n_{0,a}(\mathbf{r}_1) ] \, \mathcal{N}^{kl} 
  \, w( \mathbf{r}_1 - \mathbf{r}_2 ) \, [ \partial_l n_{0,b}(\mathbf{r}_2) ] \, .
  \label{equation::G_360}
\end{equation}
Inserting this integral kernel into the third line of Eq.\ \eqref{equation::G_290} we finally obtain the projection 
operator $\mathsf{P}[\tilde{n},\tilde{\mathbf{j}},\mathbf{u}]$ in leading-order approximation. 

For the following calculations we need an explicit expression for the displacement field 
$\hat{\mathbf{u}}(\mathbf{r}) = \mathbf{u}(\mathbf{r},\Gamma)$ in terms of the microscopic particle densities 
$\hat{n}_a(\mathbf{r}) = n_a(\mathbf{r},\Gamma)$. A formula which is compatible with the projection operator 
\eqref{equation::G_290} and the integral kernel \eqref{equation::G_360} in leading-order approximation reads
\begin{equation}
  \begin{split}
    \hat{u}^k(\mathbf{r}_1) = \,& u^k(\mathbf{r}_1) - \sum_{lb} \int d^dr_2 \, \mathcal{N}^{kl} \, w( \mathbf{r}_1 - \mathbf{r}_2 ) \\
    &\times [ \partial_l n_{0,b}(\mathbf{r}_2) ] \, [ \hat{n}_b(\mathbf{r}_2) - n_b(\mathbf{r}_2) ] \, .
  \end{split}
  \label{equation::G_370}
\end{equation}
The definition guarantees that $\mathbf{u}(\mathbf{r}_1) = \langle \hat{\mathbf{u}}(\mathbf{r}_1) \rangle$ is indeed 
the average displacement field if the average is calculated with the relevant distribution function 
$\hat{\tilde{\varrho}} = \tilde{\varrho}(\Gamma)$ according to Eqs.\ \eqref{equation::G_010}-\eqref{equation::G_030}.
Now, we are ready to investigate the action of the projection operator $\mathsf{P}[\tilde{n},\tilde{\mathbf{j}},\mathbf{u}]$ 
onto the microscopic particle densities $\hat{n}_a(\mathbf{r}) = n_a(\mathbf{r},\Gamma)$. Using 
Eqs.\ \eqref{equation::G_290}, \eqref{equation::G_360}, and \eqref{equation::G_370} after some calculations 
we obtain
\begin{equation}
  \begin{split}
    \mathsf{P}[\tilde{n},\tilde{\mathbf{j}},\mathbf{u}] \,\, \hat{n}_a(\mathbf{r}) = \,& 
    n_a(\mathbf{r}) + [ \hat{\tilde{n}}_a(\mathbf{r}) - \tilde{n}_a(\mathbf{r}) ] \\
    &- [ \nabla n_{0,a}(\mathbf{r}) ] \cdot [ \hat{\mathbf{u}}(\mathbf{r}) - \mathbf{u}(\mathbf{r}) ] \, .
  \end{split}
  \label{equation::G_380}
\end{equation}
We note that nearly all calculations are exact. The only exception is the approximation in the first line of 
the integral kernel \eqref{equation::G_350}. Eq.\ \eqref{equation::G_380} is an extension of our previous result 
\eqref{equation::G_150} including the effects of deformation and the displacement field. On the right-hand side, 
the first term is the average of the microscopic particle densities $n_a(\mathbf{r})$ as expected. While the 
second term recovers the \emph{fluctuations} of the coarse-grained macroscopic particle densities 
$\hat{\tilde{n}}_a(\mathbf{r}) - \tilde{n}_a(\mathbf{r})$ the third term is the well known leading contribution 
caused by the displacement field $\hat{\mathbf{u}}(\mathbf{r}) = \mathbf{u}(\mathbf{r},\Gamma)$. Contributions 
to the particle densities in the form 
$\delta n_a(\mathbf{r}) = - [ \nabla n_{0,a}(\mathbf{r}) ] \cdot \delta \mathbf{u}(\mathbf{r})$ arise from the 
first order Taylor expansion of the deformed density. These expressions have been used by earlier authors like 
Szamel and Ernst \cite{SE93,Sz97} and like Walz and Fuchs \cite{WF10} in their derivations of the elastic equations.

Next, we apply the projection operator $\mathsf{P}[\tilde{n},\tilde{\mathbf{j}},\mathbf{u}]$ onto the microscopic 
momentum density $\hat{\mathbf{j}}(\mathbf{r}) = \hat{\mathbf{j}}(\mathbf{r},\Gamma)$. Because of orthogonality we 
just recover Eq.\ \eqref{equation::G_160}. On the right-hand side, the first term is the average of the microscopic 
momentum density $\mathbf{j}(\mathbf{r})$ as expected. The second term recovers the \emph{fluctuations} of the 
coarse-grained macroscopic momentum density $\hat{\tilde{\mathbf{j}}}(\mathbf{r}) - \tilde{\mathbf{j}}(\mathbf{r})$ 
which stays within the macroscopic function space $\tilde{\mathcal{C}}(\ell)$. A third term involving the displacement 
field does not occur. Thus, nothing new happens.

Finally, we apply the projection operator onto the coarse-grained macroscopic densities 
$\hat{\tilde{n}}_a(\mathbf{r}) = \tilde{n}_a(\mathbf{r},\Gamma)$ and 
$\hat{\tilde{\mathbf{j}}}(\mathbf{r}) = \tilde{\mathbf{j}}(\mathbf{r},\Gamma)$. Once again, nothing new 
happens. We just recover equations like Eq.\ \eqref{equation::G_060}. A new macroscopic function is 
the displacement field $\hat{\mathbf{u}}(\mathbf{r}) = \mathbf{u}(\mathbf{r},\Gamma)$ which is defined in 
Eq.\ \eqref{equation::G_370}. We apply the projection operator $\mathsf{P}[\tilde{n},\tilde{\mathbf{j}},\mathbf{u}]$ 
onto Eq.\ \eqref{equation::G_370} and use Eq.\ \eqref{equation::G_380}. Then, after some calculations and by 
using the orthonormalization conditions \eqref{equation::G_180} we obtain
\begin{equation}
  \mathsf{P}[\tilde{n},\tilde{\mathbf{j}},\mathbf{u}] \,\, \hat{\mathbf{u}}(\mathbf{r}) = \hat{\mathbf{u}}(\mathbf{r})
  \label{equation::G_390}
\end{equation}
as we expect. Thus, we have checked that $\mathsf{P}[\tilde{n},\tilde{\mathbf{j}},\mathbf{u}]$ indeed is a 
projection operator for the relevant variables $\hat{\tilde{n}}_a(\mathbf{r}) = \tilde{n}_a(\mathbf{r},\Gamma)$, 
$\hat{\tilde{\mathbf{j}}}(\mathbf{r}) = \tilde{\mathbf{j}}(\mathbf{r},\Gamma)$, and
$\hat{\mathbf{u}}(\mathbf{r}) = \mathbf{u}(\mathbf{r},\Gamma)$.

If we compare the right-hand side of Eq.\ \eqref{equation::G_380} with the linearized formula of the 
microscopic density change \eqref{equation::D_410} then we miss a correction term $\Delta n^\prime_a(\mathbf{r})$ 
which is defined by Eq.\ \eqref{equation::D_810} or which is defined by Eq.\ \eqref{equation::E_140} together 
with Eqs.\ \eqref{equation::E_150} and \eqref{equation::E_160}. More precisely, we are missing a correction 
term which involves the correction function $\Delta B_{a,k}^{\prime \, \ \ m}(\mathbf{r})$. For this fact 
the reason is the approximation which we have used in Eq.\ \eqref{equation::G_350} to obtain the projection 
operator \eqref{equation::G_360} in leading approximation. As a consequence, on the right-hand side of 
Eq.\ \eqref{equation::G_380} there is only the leading term.

In order to include the correction terms we must extend the approximation up to the next order. 
For this purpose we use the Bloch function up to linear order in $\mathbf{k}$ which is defined in 
Eq.\ \eqref{equation::G_270}. Inserting this Bloch function into the second line of 
Eq.\ \eqref{equation::G_280} we obtain the extended integral kernel
\begin{equation}
  \begin{split}
    F_{ab}( \mathbf{r}_1, \mathbf{r}_2 ) = \,& \sum_{kl} \, \Bigl( [ \partial_k n_{0,a}(\mathbf{r}_1) ] 
    - \Delta B_{a,k}^{\prime \, \ \ m}(\mathbf{r}_1) \, \overrightarrow{\partial}_{1,m} \Bigr) \\ 
    &\times \, \mathcal{N}^{kl} \, w( \mathbf{r}_1 - \mathbf{r}_2 ) \\
    &\times \, \Bigl( [ \partial_l n_{0,b}(\mathbf{r}_2) ] - \overleftarrow{\partial}_{2,n} 
    \, \Delta B_{b,l}^{\prime \, \ \ n}(\mathbf{r}_2) \Bigr) \, .
  \end{split}
  \label{equation::G_400}
\end{equation}
Here, the operator $\overrightarrow{\partial}_{1,m}$ is a derivative with respect to the coordinates 
$\mathbf{r}_1$ which is applied to the \emph{right} onto the integration kernel $w( \mathbf{r}_1 - \mathbf{r}_2 )$.
Similarly, the operator $\overleftarrow{\partial}_{2,n}$ is a derivative with respect to the coordinates 
$\mathbf{r}_2$ which is applied to the \emph{left} onto the integration kernel $w( \mathbf{r}_1 - \mathbf{r}_2 )$.
Finally, we insert this integral kernel into the definition \eqref{equation::G_290} in order to obtain 
the extended version of the projection operator $\mathsf{P}[\tilde{n},\tilde{\mathbf{j}},\mathbf{u}]$.

In the next step we extend the definition of the displacement field \eqref{equation::G_370} and write
\begin{equation}
  \begin{split}
    \hat{u}^k(\mathbf{r}_1) = \,& u^k(\mathbf{r}_1) - \sum_{lb} \int d^dr_2 \, \mathcal{N}^{kl} 
    \, w( \mathbf{r}_1 - \mathbf{r}_2 ) \\
    &\times \Bigl( [ \partial_l n_{0,b}(\mathbf{r}_2) ] - \overleftarrow{\partial}_{2,n} 
    \, \Delta B_{b,l}^{\prime \, \ \ n}(\mathbf{r}_2) \Bigr) \\
    &\times [ \hat{n}_b(\mathbf{r}_2) - n_b(\mathbf{r}_2) ] \, .
  \end{split}
  \label{equation::G_410}
\end{equation}
Once again, we may apply the projection operator onto the microscopic particle densities 
$\hat{n}_a(\mathbf{r}) = n_a(\mathbf{r},\Gamma)$. After some calculations we obtain
\begin{equation}
  \begin{split}
    \mathsf{P}[\tilde{n},\tilde{\mathbf{j}},\mathbf{u}] \,\, \hat{n}_a(\mathbf{r}) = \,& 
    n_a(\mathbf{r}) + [ \hat{\tilde{n}}_a(\mathbf{r}) - \tilde{n}_a(\mathbf{r}) ] \\
    &- [ \nabla n_{0,a}(\mathbf{r}) ] \cdot [ \hat{\mathbf{u}}(\mathbf{r}) - \mathbf{u}(\mathbf{r}) ] \\
    &+ \Delta B_{b,k}^{\prime \, \ \ l}(\mathbf{r}) \, [ \partial_l \hat{u}^k(\mathbf{r}) - \partial_l u^k(\mathbf{r}) ] \, .
  \end{split}
  \label{equation::G_420}
\end{equation}
Clearly, the last term is the expected correction term. In the definition of the extended Bloch 
function \eqref{equation::G_270} the correction function $\Delta B_{a,k}^{\prime \, \ \ l}(\mathbf{r})$ 
was introduced as an unknown function. Now, if we compare the right-hand side of 
Eq.\ \eqref{equation::G_420} with the linearized formula of the microscopic density change 
\eqref{equation::D_410} together with \eqref{equation::E_140} we clearly see that 
$\Delta B_{a,k}^{\prime \, \ \ l}(\mathbf{r})$ is the same function. Thus, we conclude that 
the correction function is given by the formula \eqref{equation::E_150} together with 
Eq.\ \eqref{equation::E_160} also in the present case.

We may split the extended displacement field defined in Eq.\ \eqref{equation::G_410} into two 
contributions according to
\begin{equation}
  \hat{u}^k(\mathbf{r}) = \hat{u}_0^k(\mathbf{r}) + \partial_n \Delta \hat{u}_0^{kn}(\mathbf{r})  \, .
  \label{equation::G_430}
\end{equation}
The first term $\hat{u}_0^k(\mathbf{r})$ is the displacement field in leading order approximation 
defined in Eq.\ \eqref{equation::G_370}. The second term may be written as the spatial derivative 
of the correction function
\begin{equation}
  \begin{split}
    \Delta \hat{u}^{kn}(\mathbf{r}_1) = \,& - \sum_{lb} \int d^dr_2 \, \mathcal{N}^{kl} \, w( \mathbf{r}_1 - \mathbf{r}_2 ) \\
    &\times \, \Delta B_{a,l}^{\prime \, \ \ n}(\mathbf{r}_2) \, [ \hat{n}_b(\mathbf{r}_2) - n_b(\mathbf{r}_2) ] \, .
  \end{split}
  \label{equation::G_440}
\end{equation}
All three functions $\hat{u}^k(\mathbf{r})$, $\hat{u}_0^k(\mathbf{r})$, and $\Delta \hat{u}^{kn}(\mathbf{r})$ 
are spatially slowly varying functions which are defined within the macroscopic function subspace 
$\tilde{\mathcal{C}}(\ell)$. The spatial derivative of a slowly varying function is small. Thus, 
we conclude that in Eq.\ \eqref{equation::G_430} the second term is small compared to the first term. 
Hence, it is a small correction.

In the next subsection we derive the equation of motions for the macroscopic continuum mechanics of 
crystals by using the projection operator $\mathsf{P}[\tilde{n},\tilde{\mathbf{j}},\mathbf{u}]$ to 
separate the slow relevant variables from the remaining fast variables. In continuum mechanics the 
concept of the local thermodynamic equilibrium is used where gradient terms are neglected. As a 
consequence, the second term of the extended displacement field \eqref{equation::G_430} is a 
gradient term which is beyond the scope of continuum mechanics. Thus, we conclude that for our 
derivations in the next subsection the leading approximation is sufficient. We shall use the 
projection operator with the leading-approximation integration kernel \eqref{equation::G_360} and 
the leading-approximation displacement field \eqref{equation::G_370}. We may neglect the correction 
terms in this context.

We may ask the question whether the extended integration kernel \eqref{equation::G_400} really satisfies 
the projection condition \eqref{equation::G_100}. There is indeed an inaccuracy. In the definition of 
the extended Bloch functions \eqref{equation::G_270} the polarization coefficients $c_i^k$ are not 
constant. Rather they depend on the wave vector $\mathbf{k}$ so that we must expand up to linear 
order. Similarly, the inverse coefficient matrix $d_k^i$ and the normalization matrices $\mathcal{N}_{kl}$ 
and $\mathcal{N}^{kl}$ depend on the wave vector, too. Hence, for our formulas above we need the 
inverse normalization matrix up to first order which we write as
\begin{equation}
  \mathcal{N}^{kl}(\mathbf{k}) = \mathcal{N}^{kl} + \Delta \mathcal{N}^{klm} \, i k_m 
  \label{equation::G_450}
\end{equation}
with some correction coefficients $\Delta \mathcal{N}^{klm}$. Since the Fourier integral is explicitly 
performed, we must replace $\mathbf{k} \to -i \nabla$ and 
$\mathcal{N}^{kl}(\mathbf{k}) \to \mathcal{N}^{kl}(-i \nabla)$. Consequently, in our definitions 
of the extended integration kernel \eqref{equation::G_400} and of the extended displacement field 
\eqref{equation::G_410} we must apply the substitution 
\begin{equation}
  \begin{split}
    \mathcal{N}^{kl} \, w( \mathbf{r}_1 - \mathbf{r}_2 ) &\to 
    [ \mathcal{N}^{kl}( - i \nabla_1 ) \, w( \mathbf{r}_1 - \mathbf{r}_2 ) ] \\ 
    &= [ ( \mathcal{N}^{kl} + \Delta \mathcal{N}^{klm} \, \partial_{1,m} ) \, w( \mathbf{r}_1 - \mathbf{r}_2 ) ] \, .
  \end{split}
  \label{equation::G_460}
\end{equation}
The square brackets indicate that the spatial derivative $\partial_{1,m}$ is confined and applies only 
onto the integral kernel $w( \mathbf{r}_1 - \mathbf{r}_2 )$ but not beyond.

We may recalculate the projection condition \eqref{equation::G_100} for the extended integral kernel 
\eqref{equation::G_400} together with the substitution \eqref{equation::G_460}. We expect that now the 
condition is satisfied up to linear order in the gradient terms. Furthermore, in the decomposition of 
the displacement field \eqref{equation::G_430} there will appear additional gradient terms. However, for 
the same reason as argued above these additional terms will be small so that they can be neglected in 
the next subsection, too.

\subsection{Derivation of the time-evolution equations}
\label{section::07b}
For the derivation of the time-evolution equations of continuum mechanics we need some mathematical tools. 
The most important tool is the projection operator which projects any physical variables onto the space 
of the relevant variables. This projection operator is defined in Eq.\ \eqref{equation::G_290} where for the 
coarse-graining of the densities we use the integral kernel \eqref{equation::D_060} and for the displacement 
field we use the integral kernel in leading-order approximation \eqref{equation::G_360}. Once we consider 
time-dependent phenomena the average relevant variables in general $\tilde{x}_i(\mathbf{r},t)$ and hence the 
average relevant variables in detail $\tilde{n}_a(\mathbf{r},t)$, $\tilde{\mathbf{j}}(\mathbf{r},t)$, and 
$\mathbf{u}(\mathbf{r},t)$ will depend on the space coordinates $\mathbf{r}$ and additionally on the time 
$t$. As a consequence, the projection operator becomes implicitly time dependent according to
\begin{equation}
  \mathsf{P}(t) = \mathsf{P}[\tilde{x}(t)] = \mathsf{P}[\tilde{n}(t),\tilde{\mathbf{j}}(t),\mathbf{u}(t)] \, .
  \label{equation::G_470}
\end{equation}
On the other hand, we need the projection operator for the orthogonal complement space which projects
any physical variables onto the space of the remaining irrelevant variables. We denote this projection 
operator as
\begin{equation}
  \mathsf{Q}(t) = \mathsf{Q}[\tilde{x}(t)] = \mathsf{Q}[\tilde{n}(t),\tilde{\mathbf{j}}(t),\mathbf{u}(t)] \, .
  \label{equation::G_480}
\end{equation}
We require the two spaces to be orthogonal and the two operators to sum up to the unity operator so that 
\begin{equation}
  \mathsf{P}(t) + \mathsf{Q}(t) = \mathsf{1} \, .
  \label{equation::G_490}
\end{equation}
While $\mathsf{P}(t)$ is defined in Eq.\ \eqref{equation::G_290} we may interpret Eq.\ \eqref{equation::G_490} 
as the defining equation for the orthogonal projection operator $\mathsf{Q}(t)$.

Next, we define the Liouville operator $\mathsf{L}$. Acting to the right onto any physical variable 
$\hat{Y}(\mathbf{r}) = Y(\mathbf{r},\Gamma)$ it is defined by a Poisson bracket according to
\begin{equation}
  \mathsf{L} \, \hat{Y}(\mathbf{r}) = i \, \{ \hat{H} , \hat{Y}(\mathbf{r}) \}
  \label{equation::G_500}
\end{equation}
where $\hat{H}$ is the Hamilton function defined in Eq.\ \eqref{equation::B_020}. The time-evolution of the 
microscopic distribution function $\hat{\varrho}(t) = \varrho(t,\Gamma)$ is ruled by the Liouville equation 
\begin{equation}
  \partial_t \hat{\varrho}(t) = \hat{\varrho}(t) \, i \, \mathsf{L} = \{ \hat{H} , \hat{\varrho}(t) \}
  \label{equation::G_510}
\end{equation}
where here the Liouville operator acts to the left which causes an opposite sign. The Liouville equation is 
a consequence of the Hamilton equations of motion defined in Eq.\ \eqref{equation::B_010} which rule the 
time evolution of the particle coordinates and momenta $\Gamma = ( \mathbf{r}_{ai},\mathbf{p}_{ai} )$.

Now, the projection operators imply a decomposition of the microscopic distribution function into 
a relevant and a remaining part according to
\begin{equation}
  \begin{split}
    \hat{\varrho}(t) &= \hat{\varrho}(t) \, [ \mathsf{P}(t) + \mathsf{Q}(t) ] \\
    &= \hat{\varrho}(t) \, \mathsf{P}(t) + \hat{\varrho}(t) \, \mathsf{Q}(t) 
    = \hat{\tilde{\varrho}}(t) + \hat{\varrho}^\prime(t) \, .
  \end{split}
  \label{equation::G_520}
\end{equation}
As a consequence, we find two separate time-evolution equations for the relevant $\hat{\tilde{\varrho}}(t)$ 
and the remaining $\hat{\varrho}^\prime(t)$, respectively, which are coupled to each other. We solve the 
second equation explicitly for $\hat{\varrho}^\prime(t)$ and insert the solution into the first equation. 
In this way, the remaining distribution function $\hat{\varrho}^\prime(t)$ is eliminated. As a result,
we obtain the \emph{master equation} for the relevant distribution function 
$\hat{\tilde{\varrho}}(t) = \tilde{\varrho}(t,\Gamma)$ which reads
\begin{equation}
  \begin{split}
    \partial_t \hat{\tilde{\varrho}}(t) = \,& \hat{\tilde{\varrho}}(t) \, i \, \mathsf{L} \, \mathsf{P}(t) \\
    &+ \int_{t_0}^t dt^\prime \, \hat{\tilde{\varrho}}(t^\prime) \, i \, \mathsf{L} \, \mathsf{Q}(t^\prime)
    \, \mathsf{U}(t^\prime,t) \, i \, \mathsf{L} \, \mathsf{P}(t) \\
    &+ \hat{\varrho}(t_0) \, \mathsf{Q}(t_0) \, \mathsf{U}(t_0,t) \, i \, \mathsf{L} \, \mathsf{P}(t)
  \end{split}
  \label{equation::G_530}
\end{equation}
together with a time-evolution operator
\begin{equation}
  \mathsf{U}(t_0,t) = \mathsf{T} \, \exp\Bigl\{ i \int_{t_0}^t dt^\prime \, \mathsf{L} \, \mathsf{Q}(t^\prime) \Bigr\}
  \label{equation::G_540}
\end{equation}
which by the projection operator $\mathsf{Q}(t)$ is restricted to the subspace of the remaining irrelevant 
variables. On the right-hand side of the master equation there are three terms. The first term is restricted 
to the subspace of the relevant variables. It provides the slowly varying \emph{reversible} contributions. 
The second term describes an interaction from the relevant variables to the remaining variables and back to 
the relevant variables. It provides the \emph{irreversible} contribution including dissipation and memory 
effects. Finally, the third term provides the fast varying \emph{fluctuating} contribution which arises 
solely from the remaining irrelevant variables. More details for the derivation of Eqs.\ \eqref{equation::G_530} 
and \eqref{equation::G_540} are found in our previous paper \cite{Ha16}.

In the framework of linear-response theory the projection operators $\mathsf{P}$ and $\mathsf{Q}$ are 
constant in time. In this case the master equation has a simpler form and was first derived by Nakajima 
\cite{Na58} and by Zwanzig \cite{Zw60}. Beyond that the Zwanzig-Mori formalism was derived 
\cite{Zw60,Zw61,Zw01,Mo65a,Mo65b} which provides an elaborate theory for correlation functions and 
susceptibilities in global thermodynamic equilibrium. In the full nonlinear case with time-dependent 
projection operators $\mathsf{P}(t)$ and $\mathsf{Q}(t)$ and for physical systems far from equilibrium 
the master equation \eqref{equation::G_530} together with the time-evolution operator 
\eqref{equation::G_540} was first derived by Robertson \cite{Ro66}.

In order to derive the time-evolution equations for continuum mechanics we consider the time-dependent 
averages of the relevant variables which are defined by
\begin{equation}
  \begin{split}
    \tilde{x}_i(\mathbf{r},t) &= \langle \hat{\tilde{x}}_i(\mathbf{r}) \rangle_t 
    = \mathrm{Tr} \{ \hat{\tilde{\varrho}}(t) \, \hat{\tilde{x}}_i(\mathbf{r}) \} \\
    &= \int d\Gamma \, \tilde{\varrho}(t,\Gamma) \, \tilde{x}_i(\mathbf{r},\Gamma) \, .
  \end{split}
  \label{equation::G_550}
\end{equation}
We take the time derivative
\begin{equation}
  \partial_t \tilde{x}_i(\mathbf{r},t) = \mathrm{Tr} \{ [ \partial_t \hat{\tilde{\varrho}}(t) ] 
  \, \hat{\tilde{x}}_i(\mathbf{r}) \} \, .
  \label{equation::G_560}
\end{equation}
On the right-hand side we insert the time derivative of the relevant distribution function which is 
given by the right-hand side of the master equation \eqref{equation::G_530}. In this way, we obtain 
the time-evolution equations for the average relevant variables $\tilde{x}_i(\mathbf{r},t)$. These 
are the equations of continuum mechanics which we want to derive.

The time-evolution equation following from Eqs.\ \eqref{equation::G_560} and \eqref{equation::G_530} 
was derived by several authors in several different forms. A presentation of the linear-response 
theory and of the full nonlinear theory can be found in the book by Fick and Sauermann \cite{FS90}. 
We prefer an alternative form of the time-evolution equation, the so called GENERIC form of 
\"Ottinger and Grmela \cite{GO97A,GO97B,Ot05}. We have derived this time-evolution equation in our 
previous paper \cite{Ha16}. It is given by Eq.\ (88) therein together with the Poisson bracket (140) 
for the reversible term, the memory function (85) for the dissipative term, and the fluctuating 
forces (73). The time-evolution equation was derived in most generality. In this section we consider 
a simplified physical situation where the temperature $T$ is assumed to be constant. This means we 
consider an approximation where the effects of heat and entropy and their production and transport 
are omitted. In a crystal the elastic properties are usually dominated by the potential energies due 
to the interactions between the particles where on the other hand thermal effects and heat play a 
minor role. For this reason, the assumption of constant temperature and the related approximations 
are justified. Another reason is that we want to stay within the framework of conventional 
density-functional theory which anticipates a constant temperature.

As a consequence, we write the time-evolution equations for the relevant variables in the form
\begin{equation}
  \begin{split}
    \partial_t \tilde{x}_i(\mathbf{r},t) = &\, \{ \tilde{x}_i(\mathbf{r},t) , F[\tilde{x}(t)] \} \\ 
    &- \sum_j \int d^dr^\prime \int_{t_0}^t dt^\prime \, M_{ij}( \mathbf{r},t; \mathbf{r}^\prime,t^\prime )
    \, \frac{ \delta F[\tilde{x}(t^\prime)] }{ \delta \tilde{x}_j(\mathbf{r}^\prime,t^\prime) } \\
    &+ f_i(\mathbf{r},t) \, .
  \end{split}
  \label{equation::G_570}
\end{equation}
In the original GENERIC form on the right-hand side the energy $E[\tilde{x}(t)]$ and the entropy 
$S[\tilde{x}(t)]$ appear as functionals in the first and second term, respectively. Since we consider a 
reduced form with constant temperature $T$ we have replaced $E[\tilde{x}(t)] \to F[\tilde{x}(t)]$ in the 
first term and $S[\tilde{x}(t)] \to - F[\tilde{x}(t)] / T$ in the second term, respectively. Hence, in 
this section the free energy $F[\tilde{x}(t)]$ is used as functional in both terms. The substitutions can 
be justified by considering the thermodynamic relations and Legendre transformations.

On the right-hand side the first term is a macroscopic Poisson bracket which, alternatively, can be 
written in the form
\begin{equation}
  \{ \tilde{x}_i(\mathbf{r},t) , F[\tilde{x}(t)] \} = \sum_j \int d^dr^\prime 
  \, L_{ij}( \mathbf{r}, \mathbf{r}^\prime;t )
  \, \frac{ \delta F[\tilde{x}(t)] }{ \delta \tilde{x}_j(\mathbf{r}^\prime,t) }
  \label{equation::G_580}
\end{equation}
where
\begin{equation}
  \begin{split}
    L_{ij}( \mathbf{r}, \mathbf{r}^\prime; t ) &= \{ \tilde{x}_i(\mathbf{r},t) , \tilde{x}_j(\mathbf{r}^\prime,t) \} \\
    &= \mathrm{Tr} \{ \hat{\tilde{\varrho}}(t) 
    \, \{ \hat{\tilde{x}}_i(\mathbf{r}) , \hat{\tilde{x}}_j(\mathbf{r}^\prime) \} \}
  \end{split}
  \label{equation::G_590}
\end{equation}
is the \emph{Poisson matrix}. In the first line it is the macroscopic Poisson bracket of the average densities 
while in the second line it is expressed as the average of the microscopic Poisson brackets of the densities.
The latter two equations may be compared with Eqs.\ (140) and (149) in our previous paper \cite{Ha16}. 
Below the Poisson brackets will be explained in more detail and calculated explicitly.

In Eq.\ \eqref{equation::G_570} the second term includes nonlocal effects in space and time which are 
represented by the integral kernel
\begin{equation}
  \begin{split}
    M_{ij}( \mathbf{r},t; \mathbf{r}^\prime,t^\prime ) = \,& (k_B T)^{-1} \\ 
    &\times \, ( \, \mathsf{Q}(t^\prime) \mathsf{L} \hat{\tilde{x}}_j(\mathbf{r}^\prime) \mid \mathsf{U}(t^\prime,t)
    \, \mathsf{Q}(t) \mathsf{L} \hat{\tilde{x}}_i(\mathbf{r}) \, )_{t^\prime} \, .
  \end{split}
  \label{equation::G_600}
\end{equation}
This formula may be compared with Eq.\ (85) in our previous paper \cite{Ha16}. The integral kernel is a 
complicated correlation function which is defined in terms of a Mori scalar product \cite{Mo65a,Mo65b} 
extended to nonequilibrium and defined by
\begin{equation}
  ( \, \hat{Y}_1(\mathbf{r}_1) \mid \hat{Y}_2(\mathbf{r}_2) \, )_t = \mathrm{Tr} \{ \hat{\varrho}(t) 
  \, [ \hat{Y}_1(\mathbf{r}_1) ]^* \, \hat{Y}_2(\mathbf{r}_2) \} \, .
  \label{equation::G_610}
\end{equation}
In classical physics the Mori scalar product is just a correlation function. The more complicated version 
for quantum physics has been defined by Eq.\ (76) in our previous publication \cite{Ha16}. Finally, 
in Eq.\ \eqref{equation::G_570} the third and last term is the fluctuating force which is defined by
\begin{equation}
  f_i(\mathbf{r},t) = \mathrm{Tr} \{ \hat{\varrho}(t_0) \, \mathsf{Q}(t_0) \, \mathsf{U}(t_0,t) 
  \, i \, \mathsf{L} \, \hat{\tilde{x}}_i(\mathbf{r}) \} \, .
  \label{equation::G_620}
\end{equation}
This formula may be compared with Eq.\ (73) in our previous paper \cite{Ha16}. We note that the first 
factor $\hat{\varrho}(t_0)$ is a \emph{microscopic} distribution function at the initial time $t_0$. 
For a full and correct treatment of the fluctuations it is important that this distribution function 
is a pure-state distribution function which is represented by multidimensional delta functions of 
coordinates and momenta located at one point in the phase space.

We may now consider an ensemble of many pure initial states. We perform an average over these initial 
states which we denote by a long bar over the quantities to be averaged. We assume that the average 
of the microscopic distribution functions of these pure states results into a relevant distribution 
function according to 
\begin{equation}
  \overline{ \hat{\varrho}(t_0) } = \hat{\tilde{\varrho}}(t_0) \, .
  \label{equation::G_630}
\end{equation}
Then, for the fluctuating forces defined in Eq.\ \eqref{equation::G_620} we may prove that the averages 
and the correlations are
\begin{eqnarray}
  \overline{ f_i(\mathbf{r},t) } &=& 0 \, ,
  \label{equation::G_640} \\
  \overline{ f_i(\mathbf{r},t) \, f_j(\mathbf{r}^\prime,t^\prime) } &=& ( k_B T ) 
  \, M_{ij}( \mathbf{r},t; \mathbf{r}^\prime,t^\prime ) \, ,
  \label{equation::G_650}
\end{eqnarray}
respectively. The proof has been performed in our previous paper \cite{Ha16} in subsection 4.5 
around the Eqs.\ (236)-(242) therein. For a general nonequilibrium state Eqs.\ \eqref{equation::G_640} 
and \eqref{equation::G_650} are valid only approximately. However, they become exact in the limiting 
case of a local thermodynamic equilibrium. We note that the averages \eqref{equation::G_640} and the 
correlations \eqref{equation::G_650} are the foundations for the introduction of Gaussian stochastic 
forces which we will define later.

\subsection{Reversible terms}
\label{section::07c}
In order to evaluate the reversible terms explicitly we must first calculate the macroscopic Poisson 
brackets for all relevant variables. We start with the particle densities and momentum densities defined in 
Eqs.\ \eqref{equation::B_120} and \eqref{equation::B_110}, respectively. Using the Poisson brackets for the 
classical mechanics of particles with coordinates and momenta $\Gamma = ( \mathbf{r}_{ai},\mathbf{p}_{ai} )$ 
we calculate directly on a microscopic scale
\begin{eqnarray}
  \{ \hat{n}_a(\mathbf{r}) , \hat{n}_b(\mathbf{r}^\prime) \} &=& 0 \, ,
  \label{equation::G_660} \\
  \{ \hat{\mathbf{j}}(\mathbf{r}) , \hat{n}_b(\mathbf{r}^\prime) \} &=& - \hat{n}_b(\mathbf{r}) 
  \, \nabla \, \delta( \mathbf{r} - \mathbf{r}^\prime ) \, ,
  \label{equation::G_670} \\
  \{ \hat{j}_i(\mathbf{r}) , \hat{j}_k(\mathbf{r}^\prime) \} &=& - [ \hat{j}_k(\mathbf{r}) \, \partial_i 
  + \partial_k \, \hat{j}_i(\mathbf{r}) ] \, \delta( \mathbf{r} - \mathbf{r}^\prime ) \, . \hspace{10mm} 
  \label{equation::G_680}
\end{eqnarray}
Now, we apply the coarse-graining procedure \eqref{equation::D_040} twice on both spatial arguments, first 
on $\mathbf{r}$ and second on $\mathbf{r}^\prime$. Using the projection property \eqref{equation::D_050} 
we find related Poisson brackets on macroscopic scales where the delta functions are replaced by the 
coarse-graining integral kernels. Thus, we obtain
\begin{eqnarray}
  \{ \hat{\tilde{n}}_a(\mathbf{r}) , \hat{\tilde{n}}_b(\mathbf{r}^\prime) \} &=& 0 \, ,
  \label{equation::G_690} \\
  \{ \hat{\tilde{\mathbf{j}}}(\mathbf{r}) , \hat{\tilde{n}}_b(\mathbf{r}^\prime) \} &=& - \hat{\tilde{n}}_b(\mathbf{r}) 
  \, \nabla \, w( \mathbf{r} - \mathbf{r}^\prime ) \, ,
  \label{equation::G_700} \\
  \{ \hat{\tilde{j}}_i(\mathbf{r}) , \hat{\tilde{j}}_k(\mathbf{r}^\prime) \} &=& - [ \hat{\tilde{j}}_k(\mathbf{r}) \, \partial_i 
  + \partial_k \, \hat{\tilde{j}}_i(\mathbf{r}) ] \, w( \mathbf{r} - \mathbf{r}^\prime ) \, . \hspace{10mm} 
  \label{equation::G_710}
\end{eqnarray}
In the next step we take the displacement field in leading-order approximation \eqref{equation::G_370} and 
calculate all Poisson brackets for all possible combinations. First, we calculate  
\begin{eqnarray}
  \{ \hat{u}^k(\mathbf{r}) , \hat{\tilde{n}}_b(\mathbf{r}^\prime) \} &=& 0 \, ,
  \label{equation::G_720} \\
  \{ \hat{u}^k(\mathbf{r}) , \hat{u}^l(\mathbf{r}^\prime) \} &=& 0
  \label{equation::G_730}
\end{eqnarray}
where Eq.\ \eqref{equation::G_660} provides the zeros on the right-hand sides. Second, for the remaining 
combination we obtain a more complicated formula. In this case we calculate
\begin{widetext}
\begin{equation}
  \begin{split}
    \{ \hat{u}^k(\mathbf{r}) , \hat{\tilde{j}}_i(\mathbf{r}^\prime) \} &= 
    - \sum_{lb} \int d^dr_2 \, \mathcal{N}^{kl} \, w( \mathbf{r} - \mathbf{r}_2 ) 
    \, [ \partial_l n_{0,b}(\mathbf{r}_2) ] \, \{ \hat{n}_b(\mathbf{r}_2) , \hat{\tilde{j}}_i(\mathbf{r}^\prime) \} \\
    &= - \sum_{lb} \int d^dr_2 \, \int d^dr_3 \, \mathcal{N}^{kl} \, w( \mathbf{r} - \mathbf{r}_2 ) 
    \, [ \partial_l n_{0,b}(\mathbf{r}_2) ] \, \{ \hat{n}_b(\mathbf{r}_2) , \hat{j}_i(\mathbf{r}_3) \}
    \, w( \mathbf{r}_3 - \mathbf{r}^\prime ) \\
    &= + \sum_{lb} \int d^dr_2 \, \int d^dr_3 \, \mathcal{N}^{kl} \, w( \mathbf{r} - \mathbf{r}_2 ) 
    \, [ \partial_l n_{0,b}(\mathbf{r}_2) ] \, \partial_{2,i} \hat{n}_b(\mathbf{r}_2)
    \, \delta( \mathbf{r}_2 - \mathbf{r}_3 ) \, w( \mathbf{r}_3 - \mathbf{r}^\prime ) \\
    &\approx + \sum_{lb} \int d^dr_2 \, \mathcal{N}^{kl} \, w( \mathbf{r} - \mathbf{r}_2 ) 
    \, [ \partial_l n_{0,b}(\mathbf{r}_2) ] \, [ \partial_i \hat{n}_b(\mathbf{r}_2) ]
    \, w( \mathbf{r}_2 - \mathbf{r}^\prime ) \, .
  \end{split}
  \label{equation::G_740}
\end{equation}
For the equality sign from the second to the third line we use Eq.\ \eqref{equation::G_670}. The spatial 
derivative $\partial_{2,i}$ acts on all functions to the right which have an argument $\mathbf{r}_2$. In 
the last line we confine this spatial derivative to the microscopic density $\hat{n}_b(\mathbf{r}_2)$. 
The remaining term has a spatial derivative of a slowly varying function 
$\partial_{2,i} \, w( \mathbf{r}_2 - \mathbf{r}^\prime )$ which can be neglected in good approximation.

We proceed to calculate the macroscopic Poisson brackets by using the formula \eqref{equation::G_590}. 
We calculate the average with respect to the time-dependent relevant distribution function 
$\hat{\tilde{\varrho}}(t) = \tilde{\varrho}(t,\Gamma)$. Thus, from Eq.\ \eqref{equation::G_740} we 
obtain and calculate
\begin{equation}
  \begin{split}
    \{ u^k(\mathbf{r},t) , \tilde{j}_i(\mathbf{r}^\prime,t) \} &\approx 
    \sum_{lb} \int d^dr_2 \, \mathcal{N}^{kl} \, w( \mathbf{r} - \mathbf{r}_2 ) 
    \, [ \partial_l n_{0,b}(\mathbf{r}_2) ] \, [ \partial_i n_b(\mathbf{r}_2) ]
    \, w( \mathbf{r}_2 - \mathbf{r}^\prime ) \\ 
    &\approx \sum_l \int d^dr_2 \, \mathcal{N}^{kl} \, w( \mathbf{r} - \mathbf{r}_2 ) 
    \, \Bigl( \frac{ 1 }{ V } \sum_b \int d^dr_3 \, [ \partial_l n_{0,b}(\mathbf{r}_3) ] 
    \, [ \partial_i \hat{n}_{0,b}(\mathbf{r}_3) ] \Bigr) \, w( \mathbf{r}_2 - \mathbf{r}^\prime ) \\
    &= \sum_l \int d^dr_2 \, \mathcal{N}^{kl} \, w( \mathbf{r} - \mathbf{r}_2 ) \, \mathcal{N}_{li}
    \, w( \mathbf{r}_2 - \mathbf{r}^\prime ) 
    = \delta^k_{\ i} \, w( \mathbf{r} - \mathbf{r}^\prime )
    \, .
  \end{split}
  \label{equation::G_750}
\end{equation}
\end{widetext}
For the approximation from the first to the second line we have used the fact that 
$w( \mathbf{r} - \mathbf{r}_2 )$ and $w( \mathbf{r}_2 - \mathbf{r}^\prime )$ are slowly varying functions 
in space with respect to $\mathbf{r}_2$ where on the other hand the particle densities $n_{0,b}(\mathbf{r}_2)$ 
are fast varying functions on the microscopic scale. For the equality sign from the second to the third 
line we have used the definition of the normalization matrix \eqref{equation::G_240}. Finally, for the 
last equality sign we have used Eqs.\ \eqref{equation::D_050} and \eqref{equation::G_250}.

The calculation of the Poisson brackets may be redone by starting with the extended displacement field 
\eqref{equation::G_410} which includes the correction term with the function 
$\Delta B_{a,k}^{\prime \, \ \ l}(\mathbf{r})$. While the Poisson brackets \eqref{equation::G_720} and 
\eqref{equation::G_730} remain unchanged in Eq.\ \eqref{equation::G_740} the correction term is a next order 
term in the gradient expansion implied by Eq.\ \eqref{equation::G_430}. While in the leading term all gradients 
act on fast varying functions in the correction term one gradient acts on the slowly varying function 
$w( \mathbf{r} - \mathbf{r}_2 )$. Consequently, the correction term will be smaller by a factor $a / \ell$ 
where $a$ is the microscopic length scale of the mean particle distances and $\ell$ is the mesoscopic length 
scale of the coarse graining. Hence, the correction term can be neglected in good approximation so that the 
final result \eqref{equation::G_750} remains unchanged. Thus, the Poisson brackets 
\eqref{equation::G_720}-\eqref{equation::G_750} are not affected by the correction term.

The Poisson bracket \eqref{equation::G_750} has been derived only for infinitesimally small deformations in 
linear response. However, an extension to strong deformations and large displacement fields on macroscopic 
scales is possible. For this purpose we again use the concept of an epsilon surroundings which is defined 
in Eq.\ \eqref{equation::F_360}. According to Eq.\ \eqref{equation::F_380} we define the small nonlinear 
displacement field $\mathbf{u}_1(\mathbf{r},t)$. Then, we assume that Eq.\ \eqref{equation::G_750} has 
been calculated for this displacement field so that it is valid for coordinates 
$\mathbf{r}, \mathbf{r}^\prime \in U_\varepsilon(P)$. In the next step we use the linear transformation 
\eqref{equation::F_340} together with Eq.\ \eqref{equation::F_370} in order to transform to the large 
displacement field $\mathbf{u}(\mathbf{r},t)$. For finite displacements the transformation formula is 
Eq.\ \eqref{equation::F_470}. However for displacement fields within a Poisson bracket we may use the 
transformation formula for the infinitesimal differentials \eqref{equation::F_490}. Consequently, we 
may just multiply by a constant matrix $e_k^p$ which is defined in Eq.\ \eqref{equation::F_460}. In this 
way, from \eqref{equation::G_750} we obtain 
\begin{equation}
  \begin{split}
    \{ u^p(\mathbf{r},t) , \tilde{j}_i(\mathbf{r}^\prime,t) \} &= e_k^p 
    \, \{ u^k(\mathbf{r},t) , \tilde{j}_i(\mathbf{r}^\prime,t) \} \\
    &= e_k^p \, \delta^k_{\ i} \, w( \mathbf{r} - \mathbf{r}^\prime )
    = e_i^p \, w( \mathbf{r} - \mathbf{r}^\prime )
  \end{split}
  \label{equation::G_760}
\end{equation}
which is valid for $\mathbf{r}, \mathbf{r}^\prime \in U_\varepsilon(P)$ where $p$ and $i$ are any indices. 
We may perform this transformation for any epsilon surroundings for any point $P$ at coordinate $\mathbf{r}_P$ 
and time $t_P$. As a result we may replace the constant transformation matrix $e_i^p$ by the space and time 
dependent matrix 
\begin{equation}
  e_i^p(\mathbf{r},t) = \frac{ \partial r_0^p }{ \partial r^i } = \delta_i^p - \partial_i u^p(\mathbf{r},t) \, .
  \label{equation::G_770} 
\end{equation}

The other macroscopic Poisson brackets are obtained straight forwardly. At the end, all macroscopic 
Poisson brackets are given by
\begin{eqnarray}
  \{ \tilde{n}_a(\mathbf{r},t) , \tilde{n}_b(\mathbf{r}^\prime,t) \} &=& 0 \, ,
  \label{equation::G_780} \\
  \{ \tilde{\mathbf{j}}(\mathbf{r},t) , \tilde{n}_b(\mathbf{r}^\prime,t) \} &=& - \tilde{n}_b(\mathbf{r},t) 
  \, \nabla \, w( \mathbf{r} - \mathbf{r}^\prime ) \, ,
  \label{equation::G_790} \\
  \{ \tilde{j}_i(\mathbf{r},t) , \tilde{j}_k(\mathbf{r}^\prime,t) \} &=& - [ \tilde{j}_k(\mathbf{r},t) \, \partial_i  \nonumber \\
  &&\hspace{4mm} + \partial_k \tilde{j}_i(\mathbf{r},t) ] \, w( \mathbf{r} - \mathbf{r}^\prime ) \, , \hspace{10mm}
  \label{equation::G_800} \\
  \{ u^p(\mathbf{r},t) , \tilde{n}_b(\mathbf{r}^\prime,t) \} &=& 0 \, ,
  \label{equation::G_810} \\
  \{ u^p(\mathbf{r},t) , u^q(\mathbf{r}^\prime,t) \} &=& 0 \, ,
  \label{equation::G_820} \\
  \{ u^p(\mathbf{r},t) , \tilde{j}_i(\mathbf{r}^\prime,t) \} &=& e_i^p(\mathbf{r},t) \, w( \mathbf{r} - \mathbf{r}^\prime ) \, .
  \label{equation::G_830}
\end{eqnarray}
The first three Poisson brackets are well known since a long time. The latter three involving the 
displacement field must be evaluated carefully. Especially, the last one is difficult to evaluate 
including the transformation to strong deformations and large displacements. In the intermediate 
stages some approximations are needed. However, at the end in the limit of large scales for continuum 
mechanics and the local thermodynamic equilibrium all Poisson brackets are exact.

As further ingredients we need the functional derivatives of the free energy functional 
$F[T,\mathbf{u},\tilde{\mathbf{j}},\tilde{n}]$. From the global differential thermodynamic relation 
\eqref{equation::F_670} which was derived for strong deformations and large displacement fields we obtain
\begin{eqnarray}
  \frac{ \delta F[T,\mathbf{u}(t),\tilde{\mathbf{j}}(t),\tilde{n}(t)] }{ \delta \tilde{n}_a(\mathbf{r},t) } &=&
  \mu_a(\mathbf{r},t) \, ,
  \label{equation::G_840} \\
  \frac{ \delta F[T,\mathbf{u}(t),\tilde{\mathbf{j}}(t),\tilde{n}(t)] }{ \delta \tilde{\mathbf{j}}(\mathbf{r},t) } &=&
  \mathbf{v}(\mathbf{r},t) \, ,
  \label{equation::G_850} \\
  \frac{ \delta F[T,\mathbf{u}(t),\tilde{\mathbf{j}}(t),\tilde{n}(t)] }{ \delta u^p(\mathbf{r},t) } &=&
  - \partial_k \, \sigma_p^{\ k}(\mathbf{r},t) \, .
  \label{equation::G_860}
\end{eqnarray}
Now, we are ready to evaluate the Poisson brackets with the free energy functional \eqref{equation::G_580} 
to obtain the reversible terms of Eq.\ \eqref{equation::G_570}. Thus, we calculate
\begin{eqnarray}
  \{ \tilde{n}_a(\mathbf{r},t) , F(t) \} &=& - \nabla \cdot [ \tilde{n}_a(\mathbf{r},t) \, \mathbf{v}(\mathbf{r},t) ] \, ,
  \label{equation::G_870} \\
  \{ \tilde{j}_i(\mathbf{r},t) , F(t) \} &=& - \partial_k \, [ \tilde{j}_i(\mathbf{r},t) \, v^k(\mathbf{r},t) ] \nonumber \\
  &&- \tilde{j}_k(\mathbf{r},t) \, [ \partial_i \, v^k(\mathbf{r},t) ] \nonumber \\
  &&- \sum_a \tilde{n}_a(\mathbf{r},t) \, [ \partial_i \, \mu_a(\mathbf{r},t) ] \hspace{10mm} \nonumber \\
  &&+ e_i^p(\mathbf{r},t) \, [ \partial_k \, \sigma_p^{\ k}(\mathbf{r},t) ] \, ,
  \label{equation::G_880} \\
  \{ u^p(\mathbf{r},t) , F(t) \} &=& e_i^p(\mathbf{r},t) \, v^i(\mathbf{r},t)
  \label{equation::G_890}
\end{eqnarray}
where for a shorter notation we write the free energy functional as 
$F(t) = F[T,\mathbf{u}(t),\tilde{\mathbf{j}}(t),\tilde{n}(t)]$. The right-hand side of Eq.\ \eqref{equation::G_880} 
can be simplified. First, from the relation of Galilean invariance \eqref{equation::D_170} we infer
\begin{equation}
  \tilde{j}_i(\mathbf{r},t) \, v^k(\mathbf{r},t) = \tilde{\rho}(\mathbf{r},t) \, v_i(\mathbf{r},t) \, v^k(\mathbf{r},t) \ .
  \label{equation::G_900}
\end{equation}
Second, we define the grand canonical potential density $\omega$ by the Legendre transformation
\begin{equation}
  \omega = f - \mathbf{v} \cdot \tilde{\mathbf{j}} - \sum_a \mu_a \, \tilde{n}_a = - p \ .
  \label{equation::G_910}
\end{equation}
In standard thermodynamics \cite{LL05} for a homogeneous system this quantity is usually identified as 
minus the pressure $p$. As a consequence, from the differential thermodynamic relation for strong 
deformations \eqref{equation::F_640} we obtain
\begin{equation}
  d\omega = - s \, dT + \sigma_p^{\ k} \, ( \partial_k du^p ) - \tilde{\mathbf{j}} \cdot d\mathbf{v} 
  - \sum_a \tilde{n}_a \, d\mu_a = - dp \, .
  \label{equation::G_920} 
\end{equation}
From this formula for constant temperature we find 
\begin{equation}
  - \partial_i p = - \tilde{j}_k \, ( \partial_i v^k ) - \sum_a \tilde{n}_a \, ( \partial_i \mu_a ) 
  + \sigma_p^{\ k} \, ( \partial_k \partial_i u^p ) \, .
  \label{equation::G_930} 
\end{equation}
The spatial derivative of the transformation matrix \eqref{equation::G_770} is 
$\partial_k e_i^p = - \partial_k \partial_i u^p$. Thus, we rewrite 
\begin{equation}
  - \partial_i p = - \tilde{j}_k \, ( \partial_i v^k ) - \sum_a \tilde{n}_a \, ( \partial_i \mu_a ) 
  - \sigma_p^{\ k} \, ( \partial_k e_i^p ) \, .
  \label{equation::G_940} 
\end{equation}
Now, Eqs.\ \eqref{equation::G_900} and \eqref{equation::G_940} can be used to rewrite the right-hand 
side of the Poisson bracket \eqref{equation::G_880}. After a short calculation we finally obtain
\begin{equation}
  \begin{split}
    \{ \tilde{j}_i(\mathbf{r},t) , F(t) \} = - \partial_k \bigl[ &\tilde{\rho}(\mathbf{r},t) 
    \, v_i(\mathbf{r},t) \, v^k(\mathbf{r},t) \\
    &+ p(\mathbf{r},t) \, \delta_i^{\ k} - \sigma_i^{\ k}(\mathbf{r},t) \bigr] 
  \end{split}
  \label{equation::G_950} 
\end{equation}
where 
\begin{equation}
  \sigma_i^{\ k}(\mathbf{r},t) = e_i^p(\mathbf{r},t) \, \sigma_p^{\ k}(\mathbf{r},t)
  \label{equation::G_960} 
\end{equation}
is the \emph{Cauchy} stress tensor with two Cartesian indices and no curvilinear index.
Next, we define the \emph{reversible} contributions of the particle current densities and the 
momentum current density by
\begin{eqnarray}
  \tilde{\mathbf{j}}_{\mathrm{rev},a}(\mathbf{r},t) &=& \tilde{n}_a(\mathbf{r},t) \, \mathbf{v}(\mathbf{r},t) \, ,
  \label{equation::G_970} \\
  \tilde{\Pi}_\mathrm{rev}^{ik}(\mathbf{r},t) &=& \tilde{\rho}(\mathbf{r},t) \, v^i(\mathbf{r},t) 
  \, v^k(\mathbf{r},t) \nonumber \\
  &&+ p(\mathbf{r},t) \, \delta^{ik} - \sigma^{ik}(\mathbf{r},t) \, ,
  \label{equation::G_980}
\end{eqnarray}
respectively. Then, the first two Poisson brackets \eqref{equation::G_870} and \eqref{equation::G_880} 
can be rewritten in the more compact forms
\begin{eqnarray}
  \{ \tilde{n}_a(\mathbf{r},t) , F(t) \} &=& - \nabla \cdot \tilde{\mathbf{j}}_{\mathrm{rev},a}(\mathbf{r},t) \, ,
  \label{equation::G_990} \\
  \{ \tilde{j}_i(\mathbf{r},t) , F(t) \} &=& - \partial_k \, \tilde{\Pi}_\mathrm{rev}^{ik}(\mathbf{r},t) \, ,
  \label{equation::G_A00}
\end{eqnarray}
respectively, where the third one \eqref{equation::G_890} is already in an optimum form.

Finally, we may write down the time-evolution equations of the continuum mechanics of an elastic crystal. 
Inserting the above Poisson brackets into the GENERIC equations \eqref{equation::G_570} we obtain
\begin{eqnarray}
  \partial_t \, \tilde{n}_a(\mathbf{r},t) &=& - \nabla \cdot \tilde{\mathbf{j}}_a(\mathbf{r},t) \, ,
  \label{equation::G_A10} \\
  \partial_t \, \tilde{j}^i(\mathbf{r},t) &=& - \partial_k \, \tilde{\Pi}^{ik}(\mathbf{r},t) \, ,
  \label{equation::G_A20} \\
  \partial_t \, u^p(\mathbf{r},t) &=& e_i^p(\mathbf{r},t) \, v^i(\mathbf{r},t) \, .
  \label{equation::G_A30}
\end{eqnarray}
Until now only the \emph{reversible} contributions are taken into account. Thus, in the present form 
Eqs.\ \eqref{equation::G_A10}-\eqref{equation::G_A30} represent the time-evolution equations for 
an \emph{ideal crystal} with no dissipation and no fluctuations. In the next subsections we shall 
derive further contributions to the current densities so that further terms will occur on the 
right-hand sides of the equations.

We note that the Poisson brackets and the time-evolution equations are \emph{exact} for continuum 
mechanics in the limit of large length scales where the concept of the local thermodynamic equilibrium 
may be applied. In our derivations resulting in the explicit expressions for the current densities 
no approximations are made. On the other hand, it is important to note that the current densities 
\eqref{equation::G_970} and \eqref{equation::G_980} are defined with Cartesian indices in the 
laboratory frame. For the consistency of the theory this fact is an essential result because here 
we find the Cauchy stress tensor with two Cartesian indices which is defined in Eq.\ \eqref{equation::G_960}. 
A curvilinear index occurs only in the time-evolution equation of the displacement field \eqref{equation::G_A30}. 

It is worth while to rewrite this latter equation in a more explicit form. We insert the transformation 
matrix \eqref{equation::G_770} and then obtain
\begin{equation}
  \partial_t \, u^p(\mathbf{r},t) = v^p(\mathbf{r},t) - [ v^k(\mathbf{r},t) \, \partial_k ] \, u^p(\mathbf{r},t) \, .
  \label{equation::G_A40}
\end{equation}
In vector notation we may rewrite the displacement field and the velocity field as 
\begin{equation}
  \mathbf{u}(\mathbf{r},t) = u^p(\mathbf{r},t) \, \mathbf{e}_p \, , \qquad
  \mathbf{v}(\mathbf{r},t) = v^k(\mathbf{r},t) \, \mathbf{e}_k \, .
  \label{equation::G_A50}
\end{equation}
In all cases $\mathbf{e}_p = \delta_p^{\ k} \, \mathbf{e}_k$ are the orthogonal basis vectors of the 
Cartesian laboratory frame. Now, we put the nonlinear term to the left-hand side. Then, we rewrite 
Eq.\ \eqref{equation::G_A40} as
\begin{equation}
  \frac{ d \mathbf{u}(\mathbf{r},t) }{ d t } = \frac{ \partial \mathbf{u}(\mathbf{r},t) }{ \partial t } 
  + [ \mathbf{v}(\mathbf{r},t) \cdot \nabla ] \, \mathbf{u}(\mathbf{r},t) = \mathbf{v}(\mathbf{r},t) \, .
  \label{equation::G_A60}
\end{equation}
The two terms on the left-hand side are interpreted as the \emph{total} or \emph{substantial} time 
derivative which moves with the velocity field. This equation provides an important physical result. 
The displacement field $\mathbf{u}(\mathbf{r},t)$ describes the motion and deformation of the lattice 
structure of the crystal system. Eq.\ \eqref{equation::G_A60} tells us that the velocity field of 
this motion is exactly $\mathbf{v}(\mathbf{r},t)$. On the other hand, from Galilean invariance and 
from the relations \eqref{equation::D_170} and \eqref{equation::G_970} we know that the material of 
the crystal moves on average with the velocity field $\mathbf{v}(\mathbf{r},t)$. Both velocity fields 
are the same, the velocity for the lattice structure and the velocity for the material. Consequently, 
the material is fixed to the crystal structure. Both move in parallel with each other.

In the next subsection we will extend the theory by including the possible diffusion of the particles. 
Then, we will observe that in principle and in most generality the motion of the lattice structure and 
the motion of the material are decoupled and independent from each other.

The pressure variable $p$ is formally defined as a thermodynamic quantity where we identify $\omega = -p$ 
in Eq.\ \eqref{equation::G_910}. It is not the real physical pressure which can be measured in an 
experiment. Similarly, $\sigma^{ij}$ is not the real physical stress tensor. Rather it is defined 
formally in our differential thermodynamic relations \eqref{equation::F_640} or \eqref{equation::G_920} 
together with the transformation \eqref{equation::G_960}. Since Eq.\ \eqref{equation::G_A20} is the 
continuity equation for the local conservation of momentum we may obtain the true physical values 
from the momentum current density defined in Eq.\ \eqref{equation::G_980}. We rewrite the momentum 
current density in two ways according to 
\begin{equation}
  \begin{split}
    \Pi_\mathrm{rev}^{ik} &= \tilde{\rho} \, v^i \, v^k - \sigma_\mathrm{tot}^{ik} \\
    &= \tilde{\rho} \, v^i \, v^k + p_\mathrm{tot} \, \delta^{ik} - \sigma^{\prime ik}
  \end{split}
  \label{equation::G_A70} 
\end{equation}
where $\sigma^{\prime ik}$ is the irreducible contribution of the stress tensor which has a zero 
trace so that $\sigma_i^{\prime\, i} = \delta_{ik} \, \sigma^{\prime ik} = 0$. If we compare the first 
line of Eq.\ \eqref{equation::G_A70} with Eq.\ \eqref{equation::G_980} then we obtain the \emph{total} 
stress tensor in Cartesian coordinates
\begin{equation}
  \sigma_\mathrm{tot}^{ik} = - p \, \delta^{ik} + \sigma^{ik} \, .
  \label{equation::G_A80}
\end{equation}
On the other hand we decompose the Cauchy stress tensor into a scalar and an irreducible contribution
according to
\begin{equation}
  \sigma^{ik} = \sigma_0 \, \delta^{ik} + \sigma^{\prime ik} \, .
  \label{equation::G_A90}
\end{equation}
Then, from the second line of Eq.\ \eqref{equation::G_A70} we infer the \emph{total} pressure
\begin{equation}
  p_\mathrm{tot} = p - \sigma_0 \, .
  \label{equation::G_B00}
\end{equation}
The total stress tensor and the total pressure defined in Eqs.\ \eqref{equation::G_A80} and 
\eqref{equation::G_B00}, respectively, are real physical quantities which can be measured in an 
experiment.

\subsection{Irreversible terms: effects of diffusion and dissipation}
\label{section::07d}
The second line of the general time-evolution equation \eqref{equation::G_570} provides the 
irreversible contributions with diffusion and dissipation effects. The functional derivatives of 
the free energy have been calculated already in Eqs.\ \eqref{equation::G_840}-\eqref{equation::G_860}. 
Here we must consider the integral kernels in detail which are defined in Eq.\ \eqref{equation::G_600}. 
These integral kernels include effects of memory and of spatial nonlocalities.

The relevant variables $\tilde{x}_i(\mathbf{r},t)$ must be chosen in a way so that we achieve a 
separation of the scales of space and time. As a consequence, the operator $\mathsf{P}(t)$ projects 
onto the subspace of \emph{slowly} varying variables in space and time while the operator $\mathsf{Q}(t)$ 
projects onto the subspace of \emph{fast} varying variables. In the definition \eqref{equation::G_600} 
the projection operator $\mathsf{Q}(t)$ implies that the integral kernel 
$M_{ij}( \mathbf{r},t; \mathbf{r}^\prime,t^\prime )$ is a correlation function of fast varying variables. 
Hence this correlation function decays quickly to zero with increasing distances 
$\vert \mathbf{r} - \mathbf{r}^\prime \vert$ or $\vert t - t^\prime \vert$ in space and time. In the 
limit of continuum mechanics and perfect space and time scale separation we assume
\begin{equation}
  M_{ij}( \mathbf{r},t; \mathbf{r}^\prime,t^\prime ) = 2 \, M_{ij}(\mathbf{r},t) 
  \, w( \mathbf{r} - \mathbf{r}^\prime ) \, \delta( t - t^\prime )
  \label{equation::G_B10}
\end{equation}
where 
\begin{equation}
  M_{ij}(\mathbf{r},t) = M_{ij}(T,\partial \mathbf{u}(\mathbf{r},t),\tilde{\mathbf{j}}(\mathbf{r},t),\tilde{n}(\mathbf{r},t)) 
  \label{equation::G_B20}
\end{equation}
is the \emph{Onsager matrix} which is a function of the temperature, the displacement field, and the coarse 
grained densities similar like the free energy density \eqref{equation::E_510} is. The dependence on the 
space and time variables $\mathbf{r}$ and $t$ is implicit via the arguments. If some or all 
$\hat{\tilde{x}}_i(\mathbf{r})$ are conserved densities then we may rewrite the expressions 
\begin{equation}
  \mathsf{Q}(t) \mathsf{L} \hat{\tilde{x}}_i(\mathbf{r}) = i \, \mathsf{Q}(t) \, [ \partial_k \hat{\tilde{j}}_i^k(\mathbf{r}) ] 
  = i \, \partial_k \, [ \mathsf{Q}(t) \, \hat{\tilde{j}}_i^k(\mathbf{r}) ] 
  \label{equation::G_B30}
\end{equation}
as divergences of related current densities. As a consequence, in the integral kernel there appear up to 
two spatial derivatives $\partial_k$. Thus, in the diagonal case for the correlation of two conserved 
quantities we find
\begin{equation}
  \begin{split}
    M_{ij}( \mathbf{r},t; \mathbf{r}^\prime,t^\prime ) &= \partial_k \partial^\prime_l 
    \, N_{ij}^{kl}( \mathbf{r},t; \mathbf{r}^\prime,t^\prime ) \\
    &= - 2 \, \partial_k \, N_{ij}^{kl}(\mathbf{r},t) \, \partial_l 
    \, w( \mathbf{r} - \mathbf{r}^\prime ) \, \delta( t - t^\prime ) \, .
  \end{split}
  \label{equation::G_B40}
\end{equation}
Hence, the Onsager matrix may be written as
\begin{equation}
  M_{ij}(\mathbf{r},t) = - \, \partial_k \, N_{ij}^{kl}(\mathbf{r},t) \, \partial_l \, .
  \label{equation::G_B50}
\end{equation}
In the offdiagonal case where a conserved quantity is correlated with a nonconserved quantity or vice versa there 
may be only one partial derivative, either $\partial_k$ on the left-hand side or $\partial_l$ on 
the right-hand side according to
\begin{eqnarray}
  M_{ij}(\mathbf{r},t) &=& \partial_k \, K_{ij}^k(\mathbf{r},t) \, ,
  \label{equation::G_B60} \\
  M_{ij}(\mathbf{r},t) &=& - \, K_{ji}^l(\mathbf{r},t) \, \partial_l \, ,
  \label{equation::G_B70}
\end{eqnarray}
respectively. In Eqs.\ \eqref{equation::G_B10} and \eqref{equation::G_B40} an additional factor $2$ 
occurs. This factor is needed in the second line of the time-evolution equation \eqref{equation::G_570} 
because the time integral extends only over half of the peak of the delta function. The reason is the 
upper boundary $t$ of the time integral. If we insert one of these integral kernels then we obtain the 
time-evolution equation
\begin{equation}
  \begin{split}
    \partial_t \tilde{x}_i(\mathbf{r},t) = &\, \{ \tilde{x}_i(\mathbf{r},t) , F[\tilde{x}(t)] \} \\ 
    &- \sum_j M_{ij}(\mathbf{r},t)
    \, \frac{ \delta F[\tilde{x}(t)] }{ \delta \tilde{x}_j(\mathbf{r},t) } \\
    &+ f_i(\mathbf{r},t) 
  \end{split}
  \label{equation::G_B80}
\end{equation}
where the Onsager matrix may have diagonal blocks of the forms \eqref{equation::G_B20} and \eqref{equation::G_B50} 
and offdiagonal blocks of the forms \eqref{equation::G_B60} and \eqref{equation::G_B70}.

Now, we evaluate the blocks of the Onsager matrix in detail. First, we consider the correlation between 
two densities $\tilde{n}_a(\mathbf{r},t)$ and $\tilde{n}_b(\mathbf{r}^\prime,t^\prime)$. Then, we identify 
\begin{equation}
  M_{n_a,n_b}(\mathbf{r},t) = - \, \partial_k \, D_{ab}^{kl}(\mathbf{r},t) \, \partial_l 
  \label{equation::G_B90}
\end{equation}
where $D_{ab}^{kl}(\mathbf{r},t)$ is the transport-coefficient matrix for particle diffusion. Second, an 
offdiagonal correlation between a density $\tilde{n}_a(\mathbf{r},t)$ and the momentum current density 
$\tilde{\mathbf{j}}(\mathbf{r}^\prime,t^\prime)$ does not occur because of time inversion invariance so that 
\begin{equation}
  M_{n_a,j^k}(\mathbf{r},t) = 0 \, .
  \label{equation::G_C00}
\end{equation}
Third, we consider the offdiagonal correlation between a particle density $\tilde{n}_a(\mathbf{r},t)$ and 
the displacement field $\mathbf{u}(\mathbf{r}^\prime,t^\prime)$. Here we find 
\begin{equation}
  M_{n_a,u^p}(\mathbf{r},t) = \partial_k \, K_a^{kp}(\mathbf{r},t) 
  \label{equation::G_C10}
\end{equation}
where $K_a^{kp}(\mathbf{r},t)$ are offdiagonal transport coefficients. Now, we insert the resulting 
Onsager matrices \eqref{equation::G_B90}-\eqref{equation::G_C10} and the functional derivatives of 
the free energy \eqref{equation::G_840}-\eqref{equation::G_860} into the second line of 
Eq.\ \eqref{equation::G_B80}. Thus, for the particle densities we obtain the irreversible contribution 
to the time-evolution equation
\begin{equation}
  \partial_t \, \tilde{n}_a = \ldots + \sum_b \partial_k \, D_{ab}^{kl} \, \partial_l \, \mu_b 
  + \partial_k \, K_a^{kp} \, \partial_l \, \sigma_p^{\ l} \, .
  \label{equation::G_C20}
\end{equation}
For a shorter notation we have omitted the space-time arguments $\mathbf{r}$ and $t$. If we define 
the irreversible or dissipative parts of the particle current densities
\begin{equation}
  \tilde{j}_{\mathrm{diss},a}^k = - \sum_b D_{ab}^{kl} \, ( \partial_l \, \mu_b ) 
  - K_a^{kp} \, ( \partial_l \, \sigma_p^{\ l} ) 
  \label{equation::G_C30}
\end{equation}
then we may rewrite the time-evolution equation \eqref{equation::G_C20} in the form \eqref{equation::G_A10} 
which is the continuity equation for particle number conservation. 
In order to obtain the results with indices $p$ in the curvilinear coordinates we first consider an 
epsilon surroundings $U_\varepsilon(P)$ around a given point $P$ with radius $\varepsilon = \ell$ and 
evaluate all terms in the intermediate coordinate system. In this way we obtain the Onsager matrices at 
the point $P$ in local thermodynamic equilibrium. Then we perform the \emph{linear} transformation to the 
curvilinear coordinate system by using the \emph{constant} transformation matrix $e_i^p$ defined in 
Eq.\ \eqref{equation::F_460}. In this way for the second term in Eq.\ \eqref{equation::G_C30} we calculate
\begin{equation}
  \begin{split}
    K_a^{ki} \, \partial_l \, \sigma_i^{\ l} &= K_a^{ki} \, \partial_l \, e_i^p \, \sigma_p^{\ l} 
    = K_a^{ki} \, e_i^p \, \partial_l \, \sigma_p^{\ l} \\
    &= K_a^{kp} \, \partial_l \, \sigma_p^{\ l} \, .
  \end{split}
  \label{equation::G_C40}
\end{equation}
Here it is important that the transformation matrix $e_i^p$ is constant so that it can be interchanged 
with the spatial derivative $\partial_l$. Finally, we cover the whole space with many epsilon surroundings 
so that the resulting formula \eqref{equation::G_C30} can be extended to the whole space.

Following Eq.\ \eqref{equation::D_150} we may add up the particle densities $\tilde{n}_a$ into the mass 
density $\tilde{\rho} = \sum_a m_a \, \tilde{n}_a$. Similarly, we may add up the particle current densities 
$\tilde{\mathbf{j}}_a$ into the mass current density $\tilde{\mathbf{j}} = \sum_a m_a \, \tilde{\mathbf{j}}_a$. 
However, Galilean invariance implies that $\tilde{\mathbf{j}}$ is identified as the momentum density. As a 
consequence, the conservation of momentum implies that here the dissipative terms must be zero. Hence, 
we find the constraint condition
\begin{equation}
  \tilde{\mathbf{j}}_\mathrm{diss} = \sum_a m_a \, \tilde{\mathbf{j}}_{\mathrm{diss},a} = \mathbf{0} \, .
  \label{equation::G_C50}
\end{equation}
Then, inserting Eq.\ \eqref{equation::G_C30} we obtain constraint conditions for the diagonal and the 
offdiagonal transport coefficients of particle diffusion which read
\begin{equation}
  \sum_a m_a \, D_{ab}^{kl} = 0 \, , \qquad \sum_a m_a \, K_a^{kp} = 0 \, ,
  \label{equation::G_C60}
\end{equation}
respectively. Physically these constraints mean that particles may diffuse relative to each other to 
change their concentrations. However, the diffusion of the total material and the total mass is zero.
In the special case $n=1$ where only one species of particles is present the constraint 
conditions imply that there are no nonzero transport coefficients of particle diffusion at all so that 
$D_{ab}^{kl} = 0$ and $K_a^{kp} = 0$ in this case.

In the next step we consider the correlations between two momentum densities $\tilde{j}^i(\mathbf{r},t)$ 
and $\tilde{j}^j(\mathbf{r}^\prime,t^\prime)$. Then, we obtain the diagonal block of the Onsager matrix 
\begin{equation}
  M_{j^i,j^j}(\mathbf{r},t) = - \, \partial_k \, \Lambda^{ik,jl}(\mathbf{r},t) \, \partial_l \, .
  \label{equation::G_C70}
\end{equation}
The offdiagonal blocks of the Onsager matrix involving one momentum density are zero because of time 
inversion invariance. Thus, from the second line of the time-evolution equation \eqref{equation::G_B80} 
we obtain 
\begin{equation}
  \partial_t \, \tilde{j}^i = \ldots + \partial_k \, \Lambda^{ik,jl} \, \partial_l \, v_j \, .
  \label{equation::G_C80}
\end{equation}
If we define the irreversible or dissipative contribution of the momentum current density
\begin{equation}
  \tilde{\Pi}_\mathrm{diss}^{ik} = - \Lambda^{ik,jl} \, ( \partial_l \, v_j )
  \label{equation::G_C90}
\end{equation}
then we may rewrite the time-evolution equation \eqref{equation::G_C80} in the form \eqref{equation::G_A20} 
which is the continuity equation for momentum conservation. The irreversible contribution 
\eqref{equation::G_C90} represents the viscous effects within a crystal which leads to dissipation. Here 
the tensor $\Lambda^{ik,jl}$ includes all viscosity parameters. In the special case of polycrystalline 
or amorphous materials on macroscopic scales we have an isotropic symmetry. In this case the viscous tensor 
simplifies according to 
\begin{equation}
  \Lambda^{ik,jl} = \eta \Bigl( \delta^{ij} \delta^{kl} + \delta^{il} \delta^{kj} 
  - \frac{ 2 }{ d } \, \delta^{ik} \delta^{jl} \Bigr) + \zeta \, \delta^{ik} \delta^{jl}
  \label{equation::G_D00}
\end{equation}
so that the dissipative momentum current density becomes
\begin{equation}
  \tilde{\Pi}_\mathrm{diss}^{ik} = - \eta \Bigl( \partial^k v^i + \partial^i v^k 
  - \frac{ 2 }{ d } \, \delta^{ik} \, ( \partial_n v^n ) \Bigr) 
  - \zeta \, \delta^{ik} \, ( \partial_n v^n )
  \label{equation::G_D10}
\end{equation}
which is equivalent to the well known result for a liquid. Here $\eta$ is the shear viscosity while 
$\zeta$ is the volume viscosity.

In the final step we first consider the offdiagonal correlation between the displacement field 
$\mathbf{u}(\mathbf{r},t)$ and the particle density $\tilde{n}_a(\mathbf{r}^\prime,t^\prime)$. 
In analogy to Eq.\ \eqref{equation::G_C10} we obtain
\begin{equation}
  M_{u^p,n_a}(\mathbf{r},t) = - K_a^{kp}(\mathbf{r},t) \, \partial_k \, .
  \label{equation::G_D20}
\end{equation}
Second, the offdiagonal correlation between the displacement field $\mathbf{u}(\mathbf{r},t)$ and 
the momentum current density $\tilde{\mathbf{j}}(\mathbf{r}^\prime,t^\prime)$ is zero because of 
time inversion invariance. Thus, we have
\begin{equation}
  M_{u^p,j^k}(\mathbf{r},t) = 0 \, .
  \label{equation::G_D30}
\end{equation}
Third, the correlation between two displacement fields $u^p(\mathbf{r},t)$ and 
$u^q(\mathbf{r}^\prime,t^\prime)$ implies directly
\begin{equation}
  M_{u^p,u^q}(\mathbf{r},t) = \zeta^{pq}(\mathbf{r},t) \, .
  \label{equation::G_D40}
\end{equation}
Now, we insert the resulting Onsager matrices \eqref{equation::G_D20}-\eqref{equation::G_D40} and the 
functional derivatives of the free energy \eqref{equation::G_840}-\eqref{equation::G_860} into the 
second line of Eq.\ \eqref{equation::G_B80}. Thus, for the displacement field we obtain the 
irreversible contribution to the time-evolution equation
\begin{equation}
  \partial_t \, u^p = \ldots + \sum_a K_a^{kp} \, \partial_k \, \mu_a 
  + \zeta^{pq} \, \partial_k \, \sigma_q^{\ k} \, .
  \label{equation::G_D50}
\end{equation}
Here again we may apply the concept of local thermodynamic equilibrium and the epsilon surroundings. 
In analogy to Eq.\ \eqref{equation::G_C40} for the second term we may calculate
\begin{equation}
  \begin{split}
    e_i^p \, \zeta^{ij} \, \partial_k \, \sigma_j^{\ k} &= e_i^p \, \zeta^{ij} \, \partial_k \, e_j^q \, \sigma_q^{\ k} 
    = e_i^p \, \zeta^{ij} \, e_j^q \, \partial_k \, \sigma_q^{\ k} \\
    &= \zeta^{pq} \, \partial_k \, \sigma_p^{\ k} 
  \end{split}
  \label{equation::G_D60}
\end{equation}
where $e_i^p$ and $e_j^q$ are the \emph{constant} transformation matrices for the \emph{linear} 
transformation from the intermediate coordinate system to the curvilinear coordinate system. Thus, 
$e_j^q$ commutes with $\partial_k$. Eventually, the whole space is covered by many epsilon surroundings 
so that Eq.\ \eqref{equation::G_D50} is written for the curvilinear coordinate system and extended to 
the whole space.

We may define the velocity field for the diffusive motion 
$\mathbf{v}_\mathrm{diff}(\mathbf{r},t) = v_\mathrm{diff}^i(\mathbf{r},t) \, \mathbf{e}_i$ with the 
Cartesian components 
\begin{equation}
  v_\mathrm{diff}^i(\mathbf{r},t) = e_p^i(\mathbf{r},t) \, v_\mathrm{diff}^p(\mathbf{r},t) 
  \label{equation::G_D70}
\end{equation}
where 
\begin{equation}
  v_\mathrm{diff}^p(\mathbf{r},t) = - \sum_a K_a^{kp} \, \partial_k \, \mu_a 
  - \zeta^{pq} \, \partial_k \, \sigma_q^{\ k} 
  \label{equation::G_D80}
\end{equation}
are the velocity components in the curvilinear coordinate system and $e_p^i(\mathbf{r},t)$ is the local 
inverse transformation matrix which is inverse to the matrix defined in Eq.\ \eqref{equation::G_770}. 
Then we may add the reversible term of Eq.\ \eqref{equation::G_A30} and the irreversible term of 
Eq.\ \eqref{equation::G_D50}. Thus, as a result we obtain the total time-evolution equation for the 
displacement field 
\begin{equation}
  \partial_t \, u^p(\mathbf{r},t) = e_i^p(\mathbf{r},t) \, [ v^i(\mathbf{r},t) - v_\mathrm{diff}^i(\mathbf{r},t) ] \, .
  \label{equation::G_D90}
\end{equation}
For convenience we additionally define a further velocity field 
$\mathbf{v}_\mathrm{struc}(\mathbf{r},t) = v_\mathrm{struc}^i(\mathbf{r},t) \, \mathbf{e}_i$ with 
the Cartesian components
\begin{equation}
  v_\mathrm{struc}^i(\mathbf{r},t) = v^i(\mathbf{r},t) - v_\mathrm{diff}^i(\mathbf{r},t) \, .
  \label{equation::G_E00}
\end{equation}
Then, Eq.\ \eqref{equation::G_D90} may be rewritten as
\begin{equation}
  \partial_t \, u^p(\mathbf{r},t) = e_i^p(\mathbf{r},t) \, v_\mathrm{struc}^i(\mathbf{r},t) \, .
  \label{equation::G_E10}
\end{equation}
Now, we may proceed in a similar way as we have done for the reversible equation 
\eqref{equation::G_A30}. In analogy to Eq.\ \eqref{equation::G_A60} we obtain 
\begin{equation}
  \frac{ d \mathbf{u}(\mathbf{r},t) }{ d t } = \frac{ \partial \mathbf{u}(\mathbf{r},t) }{ \partial t } 
  + [ \mathbf{v}_\mathrm{struc}(\mathbf{r},t) \cdot \nabla ] \, \mathbf{u}(\mathbf{r},t) = 
  \mathbf{v}_\mathrm{struc}(\mathbf{r},t) \, .
  \label{equation::G_E20}
\end{equation}
This equation tells us the meaning of the velocity field $\mathbf{v}_\mathrm{struc}(\mathbf{r},t)$: 
It describes the local motion of the lattice structure of the crystal. On the other hand, from 
Eq.\ \eqref{equation::G_E00} we obtain an equation for the vector fields which reads
\begin{equation}
  \mathbf{v}_\mathrm{struc}(\mathbf{r},t) + \mathbf{v}_\mathrm{diff}(\mathbf{r},t) = \mathbf{v}(\mathbf{r},t) \, .
  \label{equation::G_E30}
\end{equation}
Since $\mathbf{v}(\mathbf{r},t)$ describes the average motion of the material in the laboratory 
frame, this latter equation tells us that $\mathbf{v}_\mathrm{diff}(\mathbf{r},t)$ describes the 
diffusive motion of the material relative to the lattice structure. The main consequence of 
Eq.\ \eqref{equation::G_E30} is the fact that the motion of the lattice structure is \emph{decoupled} 
from the motion of the material whenever diffusive motion of the particles is possible.

\subsection{Additional reversible terms: reactive contributions}
\label{section::07e}
In our previous publication \cite{Ha16} in Subsec.\ 4.2 we have considered \emph{reactive contributions} 
for a simple fluid. These are additional reversible terms which are obtained from the memory function. 
Analogously, we may consider reactive contributions also for the elastic properties of a solid crystal. 
Thus, in the limit of continuum mechanics and perfect space and time scale separation we may extend 
Eq.\ \eqref{equation::G_B10} and rewrite the memory function as
\begin{equation}
  \begin{split}
    M_{ij}( \mathbf{r},t; \mathbf{r}^\prime,t^\prime ) = &\, 2 \, M^\prime_{ij}(\mathbf{r},t) 
    \, w( \mathbf{r} - \mathbf{r}^\prime ) \, \varepsilon( t - t^\prime ) \, \delta( t - t^\prime ) \\ 
    &+2 \, M^{\prime\prime}_{ij}(\mathbf{r},t) \, w( \mathbf{r} - \mathbf{r}^\prime ) \, \delta( t - t^\prime ) \, .
  \end{split}
  \label{equation::G_E40}
\end{equation}
Here the first term is new and describes an additional reversible contribution which is parameterized 
by the antisymmetric \emph{reactive} matrix 
\begin{equation}
  M^\prime_{ij}(\mathbf{r},t) = - M^\prime_{ji}(\mathbf{r},t) \, .
  \label{equation::G_E50}
\end{equation}
The sign function $\varepsilon( t - t^\prime )$ defined by 
$\varepsilon( t - t^\prime ) = \pm 1$ for $t - t^\prime \,^>_<\, 0$ causes that the first term has 
different signs for $t > t^\prime$ and $t < t^\prime$. The second term is the dissipative contribution 
which is parameterized by the symmetric \emph{dissipative} matrix 
\begin{equation}
  M^{\prime\prime}_{ij}(\mathbf{r},t) = + M^{\prime\prime}_{ji}(\mathbf{r},t) \, .
  \label{equation::G_E60}
\end{equation}
This latter term has been considered in the previous subsection where the dissipative matrix is the 
Onsager matrix so that $M^{\prime\prime}_{ij}(\mathbf{r},t) = M_{ij}(\mathbf{r},t)$. The sign function 
$\varepsilon( t - t^\prime )$ together with the symmetry conditions \eqref{equation::G_E50} and 
\eqref{equation::G_E60} guarantees the symmetry property of the memory function 
\begin{equation}
  M_{ij}( \mathbf{r},t; \mathbf{r}^\prime,t^\prime ) = M_{ji}( \mathbf{r}^\prime,t^\prime; \mathbf{r},t )
  \label{equation::G_E70}
\end{equation}
which follows from the defining equations \eqref{equation::G_600} and \eqref{equation::G_650}. The 
symmetry conditions \eqref{equation::G_E50} and \eqref{equation::G_E60} are straight forward for 
nonconserved quantities. In the case of conserved quantities the matrices have structures like 
Eqs.\ \eqref{equation::G_B50}-\eqref{equation::G_B70} with additional spatial differential operators 
$\partial_k$ and $\partial_l$. In the symmetry conditions \eqref{equation::G_E50} and \eqref{equation::G_E60} 
for each spatial differential operator an additional sign change must be considered because for each 
differential operator a partial integration must be performed. In all cases the particular symmetry 
conditions \eqref{equation::G_E50} and \eqref{equation::G_E60} must be written in a form so that 
eventually Eq.\ \eqref{equation::G_E70} is satisfied.

The time inversion invariance implies that some of the matrix elements are zero. However, the symmetry 
conditions  \eqref{equation::G_E50} and \eqref{equation::G_E60} imply that this fact is disjunct for the 
\emph{reactive} matrix $M^\prime_{ij}(\mathbf{r},t)$ versus the \emph{dissipative} matrix 
$M^{\prime\prime}_{ij}(\mathbf{r},t)$. This means that those matrix elements which are zero for 
$M^{\prime\prime}_{ij}(\mathbf{r},t) = M_{ij}(\mathbf{r},t)$ are nonzero for $M^\prime_{ij}(\mathbf{r},t)$ 
and vice versa. Thus, in contrast to Eqs.\ \eqref{equation::G_C00} and \eqref{equation::G_D30} the 
offdiagonal \emph{reactive} block matrices 
\begin{eqnarray}
  M^\prime_{n_a,j^j}(\mathbf{r},t) &=& - \, \partial_k \, \phi_a^{k,jl}(\mathbf{r},t) \, \partial_l \, , 
  \label{equation::G_E80} \\
  M^\prime_{u^p,j^j}(\mathbf{r},t) &=& - \, \theta^{p,jl}(\mathbf{r},t) \, \partial_l  
  \label{equation::G_E90}
\end{eqnarray}
may be nonzero so that there are additional nonzero reactive terms in the time-evolution equations 
\eqref{equation::G_C20} and \eqref{equation::G_D50}. Similarly, because of Eq.\ \eqref{equation::G_E50} 
the transposed matrices 
\begin{eqnarray}
  M^\prime_{j^j,n_a}(\mathbf{r},t) &=& + \, \partial_l \, \phi_a^{k,jl}(\mathbf{r},t) \, \partial_k \, ,
  \label{equation::G_F00} \\
  M^\prime_{j^j,u^p}(\mathbf{r},t) &=& - \, \partial_l \, \theta^{p,jl}(\mathbf{r},t) 
  \label{equation::G_F10}
\end{eqnarray}
may be nonzero so that there are additional nonzero reactive terms in the time-evolution equation 
\eqref{equation::G_C80}. In the latter two formulas one must be aware that for each differential operator 
$\partial_k$ and $\partial_l$ a partial integration must be performed so that for each differential operator 
there appears an additional minus sign. In analogy to Eq.\ \eqref{equation::G_C60} the parameters of one of 
the block matrices must satisfy the constraint condition 
\begin{equation}
  \sum_a m_a \, \phi_a^{k,jl} = 0 \, . 
  \label{equation::G_F20}
\end{equation}
Beyond that, all other reactive matrix elements are zero. As a result we obtain \emph{reactive} contributions 
to the particle current densities
\begin{equation}
  \tilde{j}_{\mathrm{reac},a}^k = - \, \phi_a^{k,jl} \, ( \partial_l \, v_j ) \, ,
  \label{equation::G_F30}
\end{equation}
to the momentum current densities
\begin{equation}
  \tilde{\Pi}_\mathrm{reac}^{ik} = + \sum_a \phi_a^{l,ik} \, ( \partial_l \, \mu_a )
  + \theta^{p,ik} \, ( \partial_l \, \sigma_p^{\ l} ) \, ,
  \label{equation::G_F40}
\end{equation}
and to the veloctiy of the reactive motion of the particles relative to the lattice structure
\begin{equation}
  v_\mathrm{reac}^p(\mathbf{r},t) =  - \, \theta^{p,jl} \, ( \partial_l \, v_j ) \, .
  \label{equation::G_F50}
\end{equation}
In a system with only one species of particles the reactive terms parameterized by $\phi_a^{k,jl}$ do 
no occur because the constraint condition \eqref{equation::G_F20} forces the parameter to zero.
The reactive terms parameterized by $\theta^{p,jl}$ have been found previously by Miserez \cite{Mi21} 
and by Mabillard and Gaspard \cite{MG21}.

We must compare the leading reversible terms of Subsec.\ \ref{section::07c} with the additional reversible 
terms or reactive terms of the present subsection. For this purpose we compare the Poisson matrix 
\eqref{equation::G_590} defined explicitly by the Poisson brackets \eqref{equation::G_780}-\eqref{equation::G_830} 
with the memory function \eqref{equation::G_E40}. If we consider the conserved quantities $\tilde{n}_a(\mathbf{r},t)$ 
and $\tilde{\mathbf{j}}(\mathbf{r},t)$ only we see that in the Poisson brackets there is \emph{one} gradient operator 
$\partial_k$ where in the memory function the Onsager matrices \eqref{equation::G_B90} and \eqref{equation::G_C70} 
and similarly the offdiagonal reactive matrices \eqref{equation::G_E80} and \eqref{equation::G_F00} provide 
\emph{two} gradient operators $\partial_k$ and $\partial_l$. This fact extends analogously also to the 
displacement field $\mathbf{u}(\mathbf{r},t)$. Eventually, in the time-evolution equations the additional 
reversible terms or reactive terms have one gradient operator more compared to the leading reversible terms. 
Since continuum mechanics is a gradient expansion, the additional reversible terms may be interpreted as 
higher-order terms which may be neglected. For this reason, in our time-evolution equations for the nonlinear 
elasticity of solid crystals we do not consider these additional reversible or reactive terms.

\subsection{Fluctuating terms}
\label{section::07f}
The third line of the general time-evolution equation \eqref{equation::G_570} or \eqref{equation::G_B80} 
provides the contributions of the fluctuating forces. These are defined in Eq.\ \eqref{equation::G_620}. 
In the case of conserved quantities we may again apply Eq.\ \eqref{equation::G_B30} so that the fluctuating 
forces may be rewritten as divergences 
\begin{equation}
  f_i(\mathbf{r},t) = - \partial_k \, g_i^k(\mathbf{r},t)
  \label{equation::G_F60}
\end{equation}
where
\begin{equation}
  g_i^k(\mathbf{r},t) = \mathrm{Tr} \{ \hat{\varrho}(t_0) \, \mathsf{Q}(t_0) \, \mathsf{U}(t_0,t) 
  \, \hat{\tilde{j}}_i^k(\mathbf{r}) \}
  \label{equation::G_F70}
\end{equation}
are the related fluctuating currents. On average the fluctuating forces and the fluctuating currents 
are zero so that in analogy to Eq.\ \eqref{equation::G_640} we have
\begin{equation}
  \overline{ f_i(\mathbf{r},t) } = 0 \quad \mbox{and} \quad \overline{ g_i^k(\mathbf{r},t) } = 0 \, .
  \label{equation::G_F80}
\end{equation}
In Eqs.\ \eqref{equation::G_B10} and \eqref{equation::G_B40} we have expressed the integral kernels 
in terms of delta functions in space and time in order to handle the fast decays of the correlations 
in space and time correctly in the limit of continuum mechanics. We may insert these integral kernels 
into Eq.\ \eqref{equation::G_650} in order to calculate the correlations of the fluctuating forces 
and fluctuating currents. Thus, we obtain
\begin{equation}
  \overline{ f_i(\mathbf{r},t) \, f_j(\mathbf{r}^\prime,t^\prime) } = 2 \, ( k_B T ) 
  \, M_{ij}(\mathbf{r},t) \, w( \mathbf{r} - \mathbf{r}^\prime ) \, \delta( t - t^\prime )
  \label{equation::G_F90}
\end{equation}
in general and
\begin{equation}
  \overline{ g_i^k(\mathbf{r},t) \, g_j^l(\mathbf{r}^\prime,t^\prime) } = 2 \, ( k_B T ) 
  \, N_{ij}^{kl}(\mathbf{r},t) \, w( \mathbf{r} - \mathbf{r}^\prime ) \, \delta( t - t^\prime )
  \label{equation::G_G00}
\end{equation}
for conserved quantities. Similarly, using the offdiagonal Onsager matrices \eqref{equation::G_B60}
and \eqref{equation::G_B70} we obtain the offdiagonal correlations
\begin{equation}
  \overline{ g_i^k(\mathbf{r},t) \, f_j(\mathbf{r}^\prime,t^\prime) } = - 2 \, ( k_B T ) 
  \, K_{ij}^k(\mathbf{r},t) \, w( \mathbf{r} - \mathbf{r}^\prime ) \, \delta( t - t^\prime )
  \label{equation::G_G10}
\end{equation}
where on the right-hand side the minus sign arises from the minus sign in Eq.\ \eqref{equation::G_F60}.

Now, we may find the averages and correlations of the fluctuating forces and currents related to 
the relevant variables $\tilde{n}_a(\mathbf{r},t)$, $\tilde{\mathbf{j}}(\mathbf{r},t)$, and 
$\mathbf{u}(\mathbf{r},t)$ of our elastic crystal. First, since the particle densities are conserved 
quantities, we find the fluctuating forces and currents
\begin{equation}
  f_{n_a}(\mathbf{r},t) = - \partial_k \, g_{n_a}^k(\mathbf{r},t) \, , \qquad
  g_{n_a}^k(\mathbf{r},t) = \tilde{j}_{\mathrm{fluc},a}^k(\mathbf{r},t) \, .
  \label{equation::G_G20}
\end{equation}
Thus, we obtain the averages and correlations
\begin{eqnarray}
  \overline{ \tilde{j}_{\mathrm{fluc},a}^k(\mathbf{r},t) } &=& 0 \, ,
  \label{equation::G_G30} \\
  \overline{ \tilde{j}_{\mathrm{fluc},a}^k(\mathbf{r},t) \, \tilde{j}_{\mathrm{fluc},b}^l(\mathbf{r}^\prime,t^\prime) } &=& 
  2 \, ( k_B T ) \, D_{ab}^{kl}(\mathbf{r},t) \nonumber \\ 
  &&\times \, w( \mathbf{r} - \mathbf{r}^\prime ) \, \delta( t - t^\prime ) \, , \hspace{10mm}
  \label{equation::G_G40}
\end{eqnarray}
respectively, where $D_{ab}^{kl}(\mathbf{r},t)$ is the transport-coefficient matrix for particle diffusion
which occurred in Eqs.\ \eqref{equation::G_B90}, \eqref{equation::G_C20}, and \eqref{equation::G_C30}.

Second, since the momentum density is a conserved quantity, we find the fluctuating forces and currents
\begin{equation}
  f_{j^i}(\mathbf{r},t) = - \partial_k \, g_{j^i}^k(\mathbf{r},t) \, , \qquad
  g_{j^i}^k(\mathbf{r},t) = \tilde{\Pi}_\mathrm{fluc}^{ik}(\mathbf{r},t) \, .
  \label{equation::G_G50}
\end{equation}
Thus, we obtain the averages and correlations
\begin{eqnarray}
  \overline{ \tilde{\Pi}_\mathrm{fluc}^{ik}(\mathbf{r},t) } &=& 0 \, ,
  \label{equation::G_G60} \\
  \overline{ \tilde{\Pi}_\mathrm{fluc}^{ik}(\mathbf{r},t) \, \tilde{\Pi}_\mathrm{fluc}^{jl}(\mathbf{r}^\prime,t^\prime) } &=& 
  2 \, ( k_B T ) \, \Lambda^{ik,jl}(\mathbf{r},t) \nonumber \\ 
  &&\times \, w( \mathbf{r} - \mathbf{r}^\prime ) \, \delta( t - t^\prime ) \, , \hspace{10mm}
  \label{equation::G_G70}
\end{eqnarray}
respectively, where $\Lambda^{ik,jl}(\mathbf{r},t)$ is the viscosity tensor which occurred in 
Eqs.\ \eqref{equation::G_C70}-\eqref{equation::G_C90}. For polycrystalline or amorphous materials 
which on macroscopic scales support an isotropic symmetry we may use the special viscosity tensor 
\eqref{equation::G_D00} which is parameterized by the shear viscosity $\eta$ and the volume 
viscosity $\zeta$.

Third, the displacement field is a normal quantity which is not conserved. The related fluctuating 
force may be written as
\begin{equation}
  f_{u^p}(\mathbf{r},t) = - v_\mathrm{fluc}^p(\mathbf{r},t) \, .
  \label{equation::G_G80}
\end{equation}
where $v_\mathrm{fluc}^p(\mathbf{r},t)$ are the components of the fluctuating velocity with an 
upper index in the curvilinear coordinate system. Thus, we find the averages and correlations
\begin{eqnarray}
  \overline{ v_\mathrm{fluc}^p(\mathbf{r},t) } &=& 0 \, ,
  \label{equation::G_G90} \\
  \overline{ v_\mathrm{fluc}^p(\mathbf{r},t) \, v_\mathrm{fluc}^q(\mathbf{r}^\prime,t^\prime) } &=& 
  2 \, ( k_B T ) \, \zeta^{pq}(\mathbf{r},t) \nonumber \\ 
  &&\times \, w( \mathbf{r} - \mathbf{r}^\prime ) \, \delta( t - t^\prime ) \, , \hspace{10mm}
  \label{equation::G_H00}
\end{eqnarray}
respectively, where $\zeta^{pq}(\mathbf{r},t)$ is the transport-coefficient matrix which already 
occurred in Eqs.\ \eqref{equation::G_D40}, \eqref{equation::G_D50}, and \eqref{equation::G_D80}.

Furthermore, we may consider the offdiagonal correlations for our three different fluctuating 
forces and currents. Here, we find
\begin{eqnarray}
  \overline{ \tilde{j}_{\mathrm{fluc},a}^k(\mathbf{r},t) \, \tilde{\Pi}^{jl}(\mathbf{r}^\prime,t^\prime) } &=& 0 \, ,
  \label{equation::G_H10} \\
  \overline{ \tilde{j}_{\mathrm{fluc},a}^k(\mathbf{r},t) \, v_\mathrm{fluc}^p(\mathbf{r}^\prime,t^\prime) } &=& 
  2 \, ( k_B T ) \, K_a^{kp}(\mathbf{r},t) \nonumber \\ 
  &&\times \, w( \mathbf{r} - \mathbf{r}^\prime ) \, \delta( t - t^\prime ) \, , \hspace{10mm}
  \label{equation::G_H20} \\
  \overline{ \tilde{\Pi}_\mathrm{fluc}^{ik}(\mathbf{r},t) \, v_\mathrm{fluc}^p(\mathbf{r}^\prime,t^\prime) } &=& 0 
  \label{equation::G_H30}
\end{eqnarray}
where $K_a^{kp}(\mathbf{r},t)$ is the offdiagonal transport-coeffient matrix which already occurred 
in Eqs.\ \eqref{equation::G_C10}-\eqref{equation::G_C30} and in Eqs.\ \eqref{equation::G_D20}, 
\eqref{equation::G_D50}, and \eqref{equation::G_D80}. Two of these offdiagonal correlations 
which involve the momentum current density are zero because of time inversion invariance.

Eventually, in the limit of continuum mechanics where the large scales in space and time are 
separated perfectly from the short scales we may interpret all fluctuating forces and currents as 
\emph{Gaussian stochastic forces} and \emph{currents} where the averages and the correlations 
are the defining equations.

In analogy to Eq.\ \eqref{equation::G_D70} we may define the velocity field for the fluctuative motion 
$\mathbf{v}_\mathrm{fluc}(\mathbf{r},t) = v_\mathrm{fluc}^i(\mathbf{r},t) \, \mathbf{e}_i$ with the 
Cartesian components 
\begin{equation}
  v_\mathrm{fluc}^i(\mathbf{r},t) = e_p^i(\mathbf{r},t) \, v_\mathrm{fluc}^p(\mathbf{r},t) 
  \label{equation::G_H40}
\end{equation}
where the components with an upper curvilinear index are defined in Eq.\ \eqref{equation::G_G80}.
We may subtract the Cartesian components from Eq.\ \eqref{equation::G_E00} so that the velocity components 
for the motion of the lattice structure become
\begin{equation}
  v_\mathrm{struc}^i(\mathbf{r},t) = v^i(\mathbf{r},t) - v_\mathrm{diff}^i(\mathbf{r},t)
  - v_\mathrm{fluc}^i(\mathbf{r},t) \, .
  \label{equation::G_H50}
\end{equation}
As a consequence, the time-evolution equations for the displacement field in the forms \eqref{equation::G_E10} 
and \eqref{equation::G_E20} remain valid when fluctuations are added. Moreover, we may extend 
Eq.\ \eqref{equation::G_E30} by adding the fluctuating velocity on the left-hand side so that
\begin{equation}
  \mathbf{v}_\mathrm{struc}(\mathbf{r},t) + \mathbf{v}_\mathrm{diff}(\mathbf{r},t) 
  + \mathbf{v}_\mathrm{fluc}(\mathbf{r},t) = \mathbf{v}(\mathbf{r},t) \, .
  \label{equation::G_H60}
\end{equation}
Thus, there are two sources which decouple the motion of the material from the motion of the lattice 
structure. First, there is the \emph{diffusive motion} of the material relative to the lattice structure
with veloctiy $\mathbf{v}_\mathrm{diff}(\mathbf{r},t)$. Second, there is the \emph{fluctuative motion} 
of the material relative to the lattice structure with velocity $\mathbf{v}_\mathrm{fluc}(\mathbf{r},t)$.

\subsection{Final equations of continuum mechanics}
\label{section::07g}
Now, we have derived and found all terms which contribute. Thus, we may combine all previous results 
and write down the complete time-evolution equations for the continuum mechanics of elastic crystals. 
There are three equations which read
\begin{eqnarray}
  \partial_t \, \tilde{n}_a &=& - \nabla \cdot \tilde{\mathbf{j}}_a \, ,
  \label{equation::G_H70} \\
  \partial_t \, \tilde{j}^i &=& - \partial_k \, \tilde{\Pi}^{ik} \, ,
  \label{equation::G_H80} \\
  \partial_t \, u^p &=& v_\mathrm{struc}^p \, .
  \label{equation::G_H90}
\end{eqnarray}
For the convenience of a shorter notation we have omitted the space and time arguments $\mathbf{r}$ 
and $t$. The first two equations have been stated already in Eqs.\ \eqref{equation::G_A10} and 
\eqref{equation::G_A20}. They are continuity equations for the conservation of the particle numbers of the 
several particle species and for the conservation of the momentum density, respectively. On the right-hand 
sides the particle current densities are given by
\begin{equation}
  \tilde{j}_a^k = \tilde{n}_a \, v^k 
  - \sum_b D_{ab}^{kl} \, ( \partial_l \, \mu_b ) - K_a^{kp} \, ( \partial_l \, \sigma_p^{\ l} ) 
  + \tilde{j}_{\mathrm{fluc},a}^k \, ,
  \label{equation::G_I00} \\
\end{equation}
and the momentum densities are given by
\begin{equation}
  \tilde{\Pi}^{ik} = [ \tilde{\rho} \, v^i \, v^k + p \, \delta^{ik} - \sigma^{ik} ]
  - \Lambda^{ik,jl} \, ( \partial_l \, v_j ) 
  + \tilde{\Pi}_\mathrm{fluc}^{ik} \, .
  \label{equation::G_I10}
\end{equation}
One clearly sees the reversible, the dissipative, and the fluctuation terms.

The third equation represents the time-evolution of the components of the displacement field. On the 
right-hand side there are the velocity components for the motion of the lattice structure. Here the 
upper index refers to the curvilinear coordinate system of the deformed crystal lattice. The 
velocity components are given by
\begin{equation}
  \begin{split}
    v_\mathrm{struc}^p = \,& v^p - v_\mathrm{diff}^p - v_\mathrm{fluc}^p \\
    = \,& ( \delta_i^p - \partial_i u^p ) \, v^i
    + \sum_a K_a^{kp} \, ( \partial_k \, \mu_a ) + \zeta^{pq} \, ( \partial_k \, \sigma_q^{\ k} ) \\
    &- v_\mathrm{fluc}^p \, .
  \end{split}
  \label{equation::G_I20}
\end{equation}

The third equation may be written alternatively as a continuity equation. For this purpose we take 
the spatial derivative by applying an operator $\partial_k$. Thus, we obtain the time-evolution 
equation for the \emph{strain tensor} $\partial_k u^p$ which reads
\begin{equation}
  \partial_t \, ( \partial_k u^p ) = - \partial_l \, J_k^{pl}
  \label{equation::G_I30}
\end{equation}
where on the right-hand side the current with three indices is defined by 
\begin{equation}
  J_k^{pl} = - v_\mathrm{struc}^p \, \delta_k^{\ l} \, . 
  \label{equation::G_I40}
\end{equation}
This last definition may appear trivial. However, in this way all three equations for the 
continuum mechanics of an elastic crystal may be written as \emph{continuity equations} for 
\emph{three conserved quantities} which are the particle densities $\tilde{n}_a$, the momentum density 
$\tilde{j}$, and the strain field $\partial_k u^p$. We note that the free energy density defined 
in Eq.\ \eqref{equation::E_510} exactly depends on these three variables. Thus, we conclude that 
in our theory the strain field $\partial_k u^p$ is a more natural variable than the displacement 
field $\mathbf{u} = u^p \, \mathbf{e}_p$.

\subsection{Time-evolution of the free energy density}
\label{section::07h}
The free energy density $f = f(T,\partial \mathbf{u},\tilde{\mathbf{j}},\tilde{n})$ is a function of the 
relevant variables. Thus, its variation with space and time is not independent. Rather it is determined 
by the space and time dependence of its arguments. The time-evolution equation is obtained from the 
differential thermodynamic relation \eqref{equation::F_640} if we replace the total differentials $d$ by 
the partial time derivatives $\partial_t$ so that 
\begin{equation}
  \partial_t f = - s \, \partial_t T + \sigma_p^{\ k} \, ( \partial_k \partial_t u^p ) 
  + v_i \, \partial_t \tilde{j}^i + \sum_a \mu_a \, \partial_t \tilde{n}_a \, .
  \label{equation::G_I50} 
\end{equation}
On the right-hand side we insert the time evolution equations which we have derived for the relevant 
variables in the present section and summarized in the previous subsection. In order to regroup and rewrite 
the terms we additionally need the differential thermodynamic relation written in terms of the space 
derivatives $\nabla$ which is
\begin{equation}
  \nabla f = - s \, \nabla T + \sigma_p^{\ k} \, ( \partial_k \nabla u^p ) + v^i \, \nabla \tilde{j}^i 
  + \sum_a \mu_a \, \nabla \tilde{n}_a \, .
  \label{equation::G_I60} 
\end{equation}
Furthermore, we need the formula of the Legendre transformation \eqref{equation::G_910} which defines the 
pressure $p$. We note that in the isothermal case which is considered in the present section the temperature 
$T$ is constant so that the first terms on the right-hand sides of Eqs.\ \eqref{equation::G_I50} and 
\eqref{equation::G_I60} are zero. Then, after some longer calculations we obtain the time-evolution equation 
in the form
\begin{equation}
  \partial_t f = - \nabla \cdot \mathbf{j}_F - R \, .
  \label{equation::G_I70}
\end{equation}
The resulting equation is an extended continuity equation with a free energy current density $\mathbf{j}_F$ 
and a negative source term $R$. We split the free energy current density into three contributions according to
\begin{equation}
  \mathbf{j}_F = \mathbf{j}_{F,\mathrm{rev}} + \mathbf{j}_{F,\mathrm{diss}} + \mathbf{j}_{F,\mathrm{fluc}} \, .
  \label{equation::G_I80}
\end{equation}
Thus, we obtain the reversible contribution
\begin{equation}
  j_{F,\mathrm{rev}}^k = ( f + p ) \, v^k - v_i \, \sigma^{ik} \, ,
  \label{equation::G_I90}
\end{equation}
the dissipative contribution
\begin{equation}
  \begin{split}
    j_{F,\mathrm{diss}}^k = \,& - \sum_a \mu_a \, \Bigl( \sum_b D_{ab}^{kl} \, ( \partial_l \, \mu_b ) 
    + K_a^{kq} \, ( \partial_l \, \sigma_q^{\ l} ) \Bigr) \\
    &- v_i \, \Lambda^{ik,jl} \, ( \partial_l \, v_j ) \\
    &- \sigma_p^{\ k} \, \Bigl( \sum_b K_b^{lp} \, ( \partial_l \, \mu_b ) 
    + \zeta^{pq} \, ( \partial_l \, \sigma_q^{\ l} ) \Bigr) \, ,
  \end{split}
  \label{equation::G_J00}
\end{equation}
and the fluctuating contribution
\begin{equation}
  j_{F,\mathrm{fluc}}^k = \sum_a \mu_a \, \tilde{j}_{a,\mathrm{fluc}}^k + v_i \, \tilde{\Pi}_\mathrm{fluc}^{ik} 
  + \sigma_p^{\ k} \, v_\mathrm{fluc}^p \, .
  \label{equation::G_J10}
\end{equation}
If no dissipations and no fluctuations are present, then Eq.\ \eqref{equation::G_I70} would be a normal 
continuity equation so that the free energy would be a conserved quantity. Hence, for the source term 
there is no reversible contribution. As a consequence, we split the source term only into two contributions 
according to
\begin{equation}
  R = R_\mathrm{diss} + R_\mathrm{fluc} \, .
  \label{equation::G_J20}
\end{equation}
Thus, we obtain the dissipative contribution 
\begin{equation}
  \begin{split}
    R_\mathrm{diss} = \,& \sum_a ( \partial_k \, \mu_a ) \, \Bigl( \sum_b D_{ab}^{kl} \, ( \partial_l \, \mu_b ) 
    + K_a^{kq} \, ( \partial_l \, \sigma_q^{\ l} ) \Bigr) \\
    &+ ( \partial_k \, v_i ) \, \Lambda^{ik,jl} \, ( \partial_l \, v_j ) \\
    &+ ( \partial_k \, \sigma_p^{\ k} ) \, \Bigl( \sum_b K_b^{lp} \, ( \partial_l \, \mu_b ) 
    + \zeta^{pq} \, ( \partial_l \, \sigma_q^{\ l} ) \Bigr) 
  \end{split}
  \label{equation::G_J30}
\end{equation}
and the fluctuating contribution
\begin{equation}
  R_\mathrm{fluc} = - \sum_a ( \partial_k \, \mu_a ) \, \tilde{j}_{a,\mathrm{fluc}}^k 
  - ( \partial_k \, v_i ) \, \tilde{\Pi}_\mathrm{fluc}^{ik} 
  - ( \partial_k \sigma_p^{\ k} ) \, v_\mathrm{fluc}^p \, .
  \label{equation::G_J40}
\end{equation}

The current density defined in Eqs.\ \eqref{equation::G_I80}-\eqref{equation::G_J10} has the typical 
form of an energy current. Here it is the free-energy current. On the other hand, there is a source 
term defined by Eqs.\ \eqref{equation::G_J20}-\eqref{equation::G_J40} so that the free energy is not 
a conserved quantity. Rather we see the second law of thermodynamics. The dissipative contribution 
\eqref{equation::G_J30} is positive definite and may be interpreted as the heat produced by friction 
and dissipation. In the time-evolution equation \eqref{equation::G_I70} it appears with negative sign 
which implies that the free energy density decreases continuously. The decrease is stopped in thermal 
equilibrium by the fluctuating contribution \eqref{equation::G_J40}.

We note that in the restricted theory of the present section where the temperature is constant the 
free energy plays the role of the entropy where, however, the sign is opposite. Thus, in thermal 
equilibrium the free energy is \emph{minimized} while in nonequilibrium it \emph{decreases} continuously 
with time bothered by the fluctuations. In the extended theory of the next section which includes the 
coarse-grained energy density as a relevant variable we find a normal continuity equation for the 
energy density and an extended continuity equation with positive source for the entropy density.
\vspace{30mm}

\section{Extension to the general case}
\label{section::08}
In the previous sections we have developed the continuum mechanics of nonlinear elastic crystals in the 
isothermal case for constant temperatures $T$ where the effects of heat transport and warming by friction 
are excluded. This restriction may be a good approximation because the elastic effects of crystals are 
dominated by the potential energy resulting from the interaction energies between the particles. Nevertheless,
for a complete nonlinear elasticity theory we must extend our approach. Thus, in this section we allow 
the temperature $T = T(\mathbf{r},t)$ to be space and time dependent and include the effects of heat 
transport and warming by friction. In order to do this as a further variable we must include the energy 
density $\varepsilon = \varepsilon(\mathbf{r},t)$. Since the energy is a conserved quantity the related 
energy density is a relevant variable for the macroscopic properties of a crystal.

\subsection{Local thermodynamic equilibrium}
\label{section::08a}
For the local thermodynamic equilibrium the general statistical approach was presented in our previous paper 
\cite{Ha16}. The phase-space distribution function \eqref{equation::B_130} must be generalized to 
\begin{equation}
  \begin{split}
    \hat{\varrho} = Z^{-1} \ \exp\Bigl( - \int d^dr \ \beta(\mathbf{r}) \, \Bigl[ \hat{\varepsilon}(\mathbf{r}) 
    - \mathbf{v}(\mathbf{r}) \cdot \hat{\mathbf{j}}(\mathbf{r}) \\ 
    - \sum_a \mu_a(\mathbf{r}) \, \hat{n}_a(\mathbf{r}) \Bigr] \Bigr) 
  \end{split}
  \label{equation::H_010} 
\end{equation}
where now $\beta(\mathbf{r}) = 1 / k_B \, T(\mathbf{r})$ is a space dependent function. In the next step the 
partition function $Z$ and the grand canonical thermodynamic potential $-k_B \ln Z$ are calculated. Finally, 
after a Legendre transformation with respect to all Lagrange parameters the entropy 
$S = S[x] = S[\varepsilon,\mathbf{j},n]$ is obtained. It is a functional of the relevant variables $x_i(\mathbf{r})$ 
which are the energy density $\varepsilon(\mathbf{r})$, the momentum density $\mathbf{j}(\mathbf{r})$, and the 
particle densities $n_a(\mathbf{r})$. Taking the differential in analogy to Eq.\ \eqref{equation::C_020} we 
obtain the thermodynamic relation 
\begin{equation}
  \begin{split}
  d S = \,& \int d^dr \  \frac{ 1 }{ T(\mathbf{r}) } \, d \varepsilon(\mathbf{r}) 
  - \int d^dr \ \frac{ \mathbf{v}(\mathbf{r}) }{ T(\mathbf{r}) } \cdot d \mathbf{j}(\mathbf{r}) \\
  &- \sum_a \int d^dr \ \frac{ \mu_a(\mathbf{r}) }{ T(\mathbf{r}) } \, d n_a(\mathbf{r}) \, .
  \end{split}
  \label{equation::H_020} 
\end{equation}

In the extended theory the entropy $S = S[\varepsilon,\mathbf{j},n]$ is the density functional which replaces 
the free energy of the conventional theory. We note that $S$ is a microscopic functional where the relevant 
variables $\varepsilon(\mathbf{r})$, $\mathbf{j}(\mathbf{r})$, and $n_a(\mathbf{r})$ are microscopic functions 
defined in the full function class $\mathcal{C} = \tilde{\mathcal{C}}(\ell) \oplus \mathcal{C}^\prime(\ell)$. 
In order to obtain the local thermodynamic equilibrium the entropy is maximized under certain constraints which 
keep the macroscopic variables constant. In this way the minimization of the free energy \eqref{equation::D_100} 
is replaced by the maximization of the entropy 
\begin{equation}
  \tilde{S}[\tilde{\varepsilon},\tilde{\mathbf{j}},\tilde{n}] = \max_{\varepsilon^\prime, \mathbf{j}^\prime, n^\prime} 
  S[\tilde{\varepsilon} + \varepsilon^\prime, \mathbf{j} = \tilde{\mathbf{j}} + \mathbf{j}^\prime, n = \tilde{n} + n^\prime] \, .
  \label{equation::H_030} 
\end{equation}
From the differential thermodynamic relation \eqref{equation::H_020} we obtain the related necessary conditions
\begin{equation}
  \frac{ \delta S }{ \delta \varepsilon(\mathbf{r}) } = \frac{ 1 }{ T(\mathbf{r}) } \, , \quad
  \frac{ \delta S }{ \delta \mathbf{j}(\mathbf{r}) } = - \frac{ \mathbf{v}(\mathbf{r}) }{ T(\mathbf{r}) } \, , \quad
  \frac{ \delta S }{ \delta n_a(\mathbf{r}) } = - \frac{ \mu_a(\mathbf{r}) }{ T(\mathbf{r}) } 
  \label{equation::H_040} 
\end{equation}
which may be compared with Eq.\ \eqref{equation::D_010}.

The maximization procedure \eqref{equation::H_030} may be useful for a liquid. For a crystal we must additionally 
include the displacement field $\mathbf{u}(\mathbf{r})$ to describe the deformations. Thus, in the latter case we 
must replace the extended minimization procedure \eqref{equation::D_820} by the extended maximization procedure
\begin{widetext}
\begin{equation}
  \tilde{S}[\tilde{\varepsilon},\mathbf{u},\tilde{\mathbf{j}},\tilde{n}] = \max_{\varepsilon^\prime, \mathbf{j}^\prime, \Delta n^\prime} 
  S[\tilde{\varepsilon} + \varepsilon^\prime, \mathbf{j} = \tilde{\mathbf{j}} + \mathbf{j}^\prime, 
  n = n_1(\mathbf{u}, \tilde{n}) + \Delta n^\prime] \, .
  \label{equation::H_050} 
\end{equation}
\end{widetext}
Again, for the microscopic particle densities $n_a(\mathbf{r})$ we must insert the ansatz \eqref{equation::D_380} 
or \eqref{equation::D_400} which is parameterized by the displacement field $\mathbf{u}(\mathbf{r})$ and the 
macroscopic densities $\tilde{n}_a(\mathbf{r})$. Subsequently, we may repeat the calculations of our previous 
Secs.\ \ref{section::04}, \ref{section::05}, and \ref{section::06} where we replace the free energy $F$ and the 
minimization procedure by the entropy $S$ and the maximization procedure. Formally all these calculations can be 
done. However, in practice there is a severe problem. There does not exist an explicit formula for the entropy 
functional $S[\varepsilon,\mathbf{j},n]$.

For the \emph{classical} interacting many-particle system we have found an explicit formula for the grand canonical 
thermodynamic potential \eqref{equation::B_230} in terms of the connected Mayer diagrams. Applying the Legendre 
transformation \eqref{equation::C_010} we obtain the free energy functional $F[T,\mathbf{j},n]$ in terms of 
the \emph{irreducible} Mayer diagrams given by the formulas \eqref{equation::C_030}-\eqref{equation::C_060}. 
In this way the countable infinite number of diagrams is considerably reduced which is advantageous for any 
practical calculation using a perturbation series. On the other hand, the density-functional theory provides 
many free energy functionals written down as ansatzes which have been applied and approved in many practical 
calculations.

However, in order to obtain the entropy functional $S[\varepsilon,\mathbf{j},n]$ a further Legendre 
transformation is performed where the local temperature $T(\mathbf{r})$ is replaced by the local energy density 
$\varepsilon(\mathbf{r})$. Unfortunately, in this case no simple perturbation series expansion can be 
obtained in terms of some specialized irreducible Mayer diagrams. On the other hand, in a recent extension 
of the dynamic density-functional theory Anero \emph{et al.}\ \cite{AET13} provide three explicit formulas 
for the entropy functional. The most promising candidate is an entropy functional for a hard-sphere system. 
However, these functionals have not been applied and tested in practice in numerical calculations. Thus, the 
density-functional theory does not provide an elaborated and established ansatz for $S[\varepsilon,\mathbf{j},n]$.

The situation is completely different for \emph{quantum-field theories} of interacting many-particle systems 
\cite{AGD63,FW71,LL09}. In this case we start with a \emph{density matrix} $\hat{\varrho}$ which is given by 
Eq.\ \eqref{equation::H_010} where now all microscopic quantities indicated by a hat are quantum operators 
defined in a Hilbert space. Again, in a next step we calculate the partition function $Z$ and the grand canonical 
thermodynamic potential $-k_B \ln Z$. As a result, for both quantities we obtain a perturbation series in terms 
of Feynman diagrams. De Dominicis and Martin \cite{DM64a} have shown by a Legendre transformation that the 
entropy $S = S[G,\Gamma]$ can be rewritten as a functional of the exact Green function $G$ and the exact vertex 
function $\Gamma$. Furthermore, they have derived an explicit perturbation series in terms of irreducible 
Feynman diagrams where the lines are identified by $G$ and the vertices by $\Gamma$ \cite{DM64b}. In this 
way, for the entropy an explicit formula is available.

Unfortunately, the quantum mechanical entropy functional is too general. The relevant variables $x_i(\mathbf{r})$ 
which are the energy density $\varepsilon(\mathbf{r})$, the momentum density $\mathbf{j}(\mathbf{r})$, and the 
particle densities $n_a(\mathbf{r})$ can be expressed explicitly in terms of the exact Green function $G$ and 
the exact vertex function $\Gamma$. Nevertheless, $G$ and $\Gamma$ can be eliminated by a constrained maximization 
procedure. In this way we calculate
\begin{equation}
  S[\varepsilon,\mathbf{j},n] = \max_{G,\Gamma} S[G,\Gamma] \, \Big\vert_{\varepsilon,\mathbf{j},n\ \mathrm{fixed}} \, .
  \label{equation::H_060} 
\end{equation}
Because of quantum effects the extremum need not be a maximum. Rather it will be a saddle point. At the end 
the resulting entropy functional $S[\varepsilon,\mathbf{j},n]$ will be implicitly defined. This means there 
will be no simple perturbation series in terms of irreducible Feynman diagrams. Hence, also the methods of 
the quantum-field theory do not yield an explicit expression for the entropy functional which might be useful 
for explicit calculations.

Nevertheless, for practical calculations there is a possibility to circumvent all these difficulties. We need 
neither the \emph{microscopic} entropy functional $S[\varepsilon,\mathbf{j},n]$ nor the \emph{microscopic} 
energy density $\varepsilon(\mathbf{r})$. Rather, we can start with the macroscopic free energy 
$F[T,\mathbf{u},\tilde{\mathbf{j}},\tilde{n}]$ which in Sec.\ \ref{section::05} we have calculated explicitly in 
terms of a Taylor series expansion. In Eq.\ \eqref{equation::E_520} for the local thermodynamic equilibrium we find 
that the free energy can be written as the integral of a free energy density 
$f(T,\partial \mathbf{u},\tilde{\mathbf{j}},\tilde{n})$. Originally, the temperature $T$ is assumed to be 
constant \emph{globally}. However, it suffices that the temperature $T$ is constant \emph{locally} in an epsilon 
surroundings. Thus, we may generalize our result for the free energy density to slowly varying temperatures 
$T(\mathbf{r})$ which are defined within the macroscopic function class $\tilde{\mathcal{C}}(\ell)$. 

In the next step we perform a Legendre transformation where the temperature $T$ is replaced in favor of the entropy 
density $s$. This Legendre transformation is defined by
\begin{equation}
  \tilde{\varepsilon}(s,\partial \mathbf{u},\tilde{\mathbf{j}},\tilde{n}) = f(T,\partial \mathbf{u},\tilde{\mathbf{j}},\tilde{n}) + T \, s \, .
  \label{equation::H_070} 
\end{equation}
In this way we obtain the macroscopic energy density
\begin{equation}
  \tilde{\varepsilon} = \tilde{\varepsilon}(s,\partial \mathbf{u},\tilde{\mathbf{j}},\tilde{n}) 
  = \frac{ [ \tilde{\mathbf{j}}(\mathbf{r}) ]^2 }{ 2 \sum_a m_a \, \tilde{n}_a(\mathbf{r}) } 
  + \tilde{\varepsilon}_\mathrm{in}(s,\partial \mathbf{u},\tilde{n}) 
  \label{equation::H_080} 
\end{equation}
which may be interpreted as a local thermodynamic relation. Right of the second equality sign we have applied the 
Galilean invariance and split the energy density into a kinetic part and an internal part in agreement with 
Eqs.\ \eqref{equation::E_500} and \eqref{equation::E_510}. From Eqs.\ \eqref{equation::H_070} and 
\eqref{equation::F_640} we obtain the differential thermodynamic relation
\begin{equation}
  d\tilde{\varepsilon} = T \, ds + \sigma_p^{\ k} \, ( \partial_k du^p ) + \mathbf{v} \cdot d\tilde{\mathbf{j}} 
  + \sum_a \mu_a \, d\tilde{n}_a \, .
  \label{equation::H_090} 
\end{equation}
Similarly, from Eqs.\ \eqref{equation::F_720} and \eqref{equation::F_800} we obtain equivalent differential 
thermodynamic relations with alternative representations of the strain and stress tensors. Calculating the integral 
we obtain the energy functional 
\begin{equation}
  E[s,\mathbf{u},\tilde{\mathbf{j}},\tilde{n}] = \int d^dr \, \tilde{\varepsilon}(s,\partial \mathbf{u},\tilde{\mathbf{j}},\tilde{n}) \, .
  \label{equation::H_100} 
\end{equation}

Alternatively, the thermodynamic relation \eqref{equation::H_080} can be resolved with respect to the entropy 
density. Thus, we obtain
\begin{equation}
  s = s(\tilde{\varepsilon},\partial \mathbf{u},\tilde{\mathbf{j}},\tilde{n}) 
  = s(\tilde{\varepsilon}_\mathrm{in},\partial \mathbf{u},\tilde{n}) \, .
  \label{equation::H_110} 
\end{equation}
The Galilean invariance implies that the entropy density $s$ does not depend on the energy density 
$\tilde{\varepsilon}$ and the momentum density $\tilde{\mathbf{j}}$ separately and independently. Rather, it 
depends on the combination
\begin{equation}
  \tilde{\varepsilon}_\mathrm{in} = \tilde{\varepsilon} 
  - \frac{ [ \tilde{\mathbf{j}}(\mathbf{r}) ]^2 }{ 2 \sum_a m_a \, \tilde{n}_a(\mathbf{r}) } 
  \label{equation::H_120} 
\end{equation}
which is interpreted as the \emph{internal} energy density. In Eq.\ \eqref{equation::H_110} this fact is 
indicated by the functional form after the second equality sign. The differential thermodynamic relation 
\eqref{equation::H_090} can be resolved with respect to the differential entropy density. Thus, we obtain
\begin{equation}
  ds = \frac{ 1 }{ T } \, d\tilde{\varepsilon} - \frac{ \sigma_p^{\ k} }{ T } \, ( \partial_k du^p ) 
  - \frac{ \mathbf{v} }{ T } \cdot d\tilde{\mathbf{j}} - \sum_a \frac{ \mu_a }{ T } \, d\tilde{n}_a \, .
  \label{equation::H_130} 
\end{equation}
Calculating the integrals of the densities $\tilde{\varepsilon}$ and $s$ we obtain the energy functional 
\begin{equation}
  E[\tilde{\varepsilon}] = \int d^dr \, \tilde{\varepsilon} 
  \label{equation::H_140} 
\end{equation}
and the entropy functional 
\begin{equation}
  S[\tilde{\varepsilon},\mathbf{u},\tilde{\mathbf{j}},\tilde{n}] 
  = \int d^dr \, s(\tilde{\varepsilon},\partial \mathbf{u},\tilde{\mathbf{j}},\tilde{n}) 
  \label{equation::H_150} 
\end{equation}
which are needed for the GENERIC formalism of Grmela and \"Ottinger \cite{GO97A,GO97B,Ot05} in order to 
derive the time-evolution equations for the relevant variables.

A further quantity which we need is the grand canonical thermodynamic potential density which is defined by 
the Legendre transformation
\begin{equation}
  \omega = \tilde{\varepsilon} - T \, s - \mathbf{v} \cdot \tilde{\mathbf{j}} - \sum_a \mu_a \, \tilde{n}_a = - p \ .
  \label{equation::H_160}
\end{equation}
Because of \eqref{equation::H_070} the above equation is equivalent to our previous definition of the grand 
canonical thermodynamic potential density \eqref{equation::G_910} where as expected and required the differential 
thermodynamic relation \eqref{equation::G_920} remains unchanged. As before in the isothermal case also here 
in the general case the standard thermodynamics of a homogeneous system implies that the grand canonical 
thermodynamic potential density $\omega$ is identified by minus the pressure $p$. We note that the pressure is 
needed in the reversible contributions of the momentum current density \eqref{equation::G_980} and of the energy 
current density \eqref{equation::H_280}. The above definition of the pressure was used in the macroscopic and 
phenomenological approaches of Grabert and Michel \cite{GM83} and of Temmen \emph{et al.}\ \cite{Te00}. Moreover, 
in both approaches there appear similar differential thermodynamic relations as our Eq.\ \eqref{equation::H_090}.

\subsection{Projection operators}
\label{section::08b}
The projection operator $\mathsf{P}[\tilde{x}]$ has been established in Subsec.\ \ref{section::07a} where the 
general formulas are given by Eqs.\ \eqref{equation::G_080} or \eqref{equation::G_090}. In the isothermal case the 
specific and explicit formula for our elastic crystal is represented by Eq.\ \eqref{equation::G_290}. The latter 
formula may be extended to the general case which is considered in this section. For this purpose, we add a 
further contribution for the energy density $\varepsilon(\mathbf{r})$ as an additional relevant variable. 
Thus, we obtain
\begin{widetext}
\begin{equation}
  \begin{split}
    \mathsf{P}[\tilde{n},\tilde{\mathbf{j}},\tilde{\varepsilon},\mathbf{u}] \, \hat{Y}(\mathbf{r}) = \,& \biggl( 1 
    + \sum_a \int d^dr_1 \int d^dr_2 \, [ \hat{n}_a(\mathbf{r}_1) - n_a(\mathbf{r}_1) ] 
    \, w( \mathbf{r}_1 - \mathbf{r}_2 ) \, \frac{ \delta }{ \delta n_a(\mathbf{r}_2) } \\
    &\hspace{4.5mm} + \sum_k \int d^dr_1 \int d^dr_2 \, [ \hat{j}^k(\mathbf{r}_1) - j^k(\mathbf{r}_1) ] 
    \, w( \mathbf{r}_1 - \mathbf{r}_2 ) \, \frac{ \delta }{ \delta j^k(\mathbf{r}_2) } \\
    &\hspace{4.5mm} + \int d^dr_1 \int d^dr_2 \, [ \hat{\varepsilon}(\mathbf{r}_1) - \varepsilon(\mathbf{r}_1) ] 
    \, w( \mathbf{r}_1 - \mathbf{r}_2 ) \, \frac{ \delta }{ \delta \varepsilon(\mathbf{r}_2) } \\
    &\hspace{4.5mm} + \sum_{ab} \int d^dr_1 \int d^dr_2 \, [ \hat{n}_a(\mathbf{r}_1) - n_a(\mathbf{r}_1) ] 
    \, F_{ab}( \mathbf{r}_1, \mathbf{r}_2 ) \, \frac{ \delta }{ \delta n_b(\mathbf{r}_2) } \biggr) 
    \, \mathrm{Tr} \{ \hat{\tilde{\varrho}} \, \hat{Y}(\mathbf{r}) \} \, .
  \end{split}
  \label{equation::H_170}
\end{equation}
\end{widetext}
In the three terms for the three densities $n_a(\mathbf{r})$, $\mathbf{j}(\mathbf{r})$, and $\varepsilon(\mathbf{r})$ 
the integration kernel $w( \mathbf{r}_1 - \mathbf{r}_2 )$ is inserted which is defined in Eq.\ \eqref{equation::D_060} 
and which serves for the coarse graining. Thus, the actual relevant variables are the coarse grained densities so that 
in the argument of the projection operator we may write a tilde on top of the three densities. The last term is the 
contribution for the displacement field $\mathbf{u}(\mathbf{r})$. Here, the integration kernel 
$F_{ab}( \mathbf{r}_1, \mathbf{r}_2 )$ remains unchanged. It is defined either by the formula \eqref{equation::G_360} 
or by \eqref{equation::G_400}. The latter formula includes the correction terms which we have found in 
Sec.\ \ref{section::04} when minimizing the free energy. Nevertheless, the first formula with no correction terms 
should be used for the derivation of the time-evolution equation. For, the correction terms imply gradient terms 
in the time-evolution equations which may be discarded.

The orthogonal projection operator $\mathsf{Q}[\tilde{x}]$ is defined by Eq.\ \eqref{equation::G_490}. This latter 
projection operator occurs in the formulas for the memory functions and for the fluctuating forces. It affects 
the transport coefficients and the related Onsager matrix. As a result, we have extended the projection operators 
$\mathsf{P}[\tilde{x}]$ and $\mathsf{Q}[\tilde{x}]$ to the general case which are needed to derive the time-evolution 
equations for the relevant variables.

\subsection{Time-evolution equations}
\label{section::08c}
In the general case there are two possibilities to extend the time-evolution equation \eqref{equation::G_B80}. 
First, we may add the energy density $\tilde{\varepsilon}(\mathbf{r},t)$ to the relevant variables so that we 
identify $\tilde{x}_i(\mathbf{r},t)$ by $\tilde{n}_a(\mathbf{r},t)$, $\tilde{\mathbf{j}}(\mathbf{r},t)$, 
$\tilde{\varepsilon}(\mathbf{r},t)$, and $\mathbf{u}(\mathbf{r},t)$. In this case we obtain the original 
GENERIC equation \cite{GO97A,GO97B,Ot05}
\begin{equation}
  \begin{split}
    \partial_t \tilde{x}_i(\mathbf{r},t) = &\, \{ \tilde{x}_i(\mathbf{r},t) , E[\tilde{x}(t)] \} \\ 
    &+ \sum_j M_{ij}(\mathbf{r},t)
    \, \frac{ \delta S[\tilde{x}(t)] }{ \delta \tilde{x}_j(\mathbf{r},t) } \\
    &+ f_i(\mathbf{r},t) \, .
  \end{split}
  \label{equation::H_180}
\end{equation}
Here, explicit formulas for the functionals $E[\tilde{x}(t)]$ and $S[\tilde{x}(t)]$ are given by 
Eqs.\ \eqref{equation::H_140} and \eqref{equation::H_150}, respectively. Since the entropy density 
$s(\mathbf{r},t)$ is not independent but rather related to the other relevant variables by the 
thermodynamic relation \eqref{equation::H_110} from the differential thermodynamic relation 
\eqref{equation::H_130} we obtain a further time-evolution equation for the entropy density.

Second, we may add the entropy density $s(\mathbf{r},t)$ to the relevant variables so that we identify 
$\tilde{x}_i(\mathbf{r},t)$ by $\tilde{n}_a(\mathbf{r},t)$, $\tilde{\mathbf{j}}(\mathbf{r},t)$, $s(\mathbf{r},t)$, 
and $\mathbf{u}(\mathbf{r},t)$. In this case we obtain the time-evolution equation 
\begin{equation}
  \begin{split}
    \partial_t \tilde{x}_i(\mathbf{r},t) = &\, \{ \tilde{x}_i(\mathbf{r},t) , E[\tilde{x}(t)] \} \\ 
    &- \sum_j M_{ij}(\mathbf{r},t)
    \, \frac{ \delta E[\tilde{x}(t)] }{ \delta \tilde{x}_j(\mathbf{r},t) } \\
    &+ f_i(\mathbf{r},t) + \frac{ R(\mathbf{r},t) }{ T(\mathbf{r},t) } \, \delta_{i,s} \, .
  \end{split}
  \label{equation::H_190}
\end{equation}
Here, the explicit formula for the energy functional $E[\tilde{x}(t)]$ is given by Eq.\ \eqref{equation::H_100}.
In this case the energy density $\tilde{\varepsilon}(\mathbf{r},t)$ is not independent but rather related to 
the other relevant variables by the thermodynamic relation \eqref{equation::H_080}. Thus, from the differential 
thermodynamic relation \eqref{equation::H_090} we obtain a further time-evolution equation for the energy density. 
The last term in Eq.\ \eqref{equation::H_190} is a source term which is present only in the entropy equation. 
It causes the second law of thermodynamics, includes fluctuations, and must be chosen such that the energy is 
conserved. Thus, we obtain 
\begin{equation}
  \begin{split}
    R(\mathbf{r},t) = \,& \sum_{ij} \frac{ \delta E[\tilde{x}(t)] }{ \delta \tilde{x}_i(\mathbf{r},t) } 
    \, M_{ij}(\mathbf{r},t) \, \frac{ \delta E[\tilde{x}(t)] }{ \delta \tilde{x}_j(\mathbf{r},t) } \\
    &- \sum_i \frac{ \delta E[\tilde{x}(t)] }{ \delta \tilde{x}_i(\mathbf{r},t) } \, f_i(\mathbf{r},t) \, .
  \end{split}
  \label{equation::H_200} 
\end{equation}

We note that both approaches described above are equivalent and eventually lead to the same time-evolution 
equations for all relevant variables. Eq. \eqref{equation::H_180} on the one hand and Eq.\ \eqref{equation::H_190} 
together with \eqref{equation::H_200} on the other hand can be transformed into each other. For the 
hydrodynamics of a simple liquid this transformation is described explicitly in Subsec.\ 4.4 of our 
previous publication \cite{Ha16} and by the formulas (209)-(218) therein. For the continuum mechanics 
of an elastic crystal this transformation is extended straight forwardly where the displacement field 
$\mathbf{u}(\mathbf{r},t)$ must be included as an additional relevant variable. The Onsager matrices 
$M_{ij}(\mathbf{r},t)$ are different in the two different time-evolution equations \eqref{equation::H_180} 
and \eqref{equation::H_190} because the entropy density is replaced by the energy density and vice versa. 
With respect to the physical units they differ by a factor of $T(\mathbf{r},t)$. Nevertheless, one may 
derive a transformation formula for the two versions of the Onsager matrix.

However, in the present work we prefer to use the second approach. In this case the Galilean invariance implies 
that the time-evolution equations have a simpler structure. Most of the time-evolution equations have the form 
of a continuity equation. We recover the final equations \eqref{equation::G_H70}-\eqref{equation::G_H90} of 
Subsec.\ \ref{section::07g} which are the continuity equations for the particle densities $\tilde{n}_a(\mathbf{r},t)$, 
the continuity equation for the momentum density $\tilde{\mathbf{j}}(\mathbf{r},t)$, and a time-evolution equation 
for the displacement field $\mathbf{u}(\mathbf{r},t)$. Here, in the extended version of the theory for the general 
case we must add a continuity equation for the energy density 
\begin{equation}
  \partial_t \tilde{\varepsilon} = - \nabla \cdot \tilde{\mathbf{j}}_E \, .
  \label{equation::H_210} 
\end{equation}
Furthermore, we must add a time-evolution equation for the entropy density
\begin{equation}
  \partial_t s = - \nabla \cdot \mathbf{q} + R / T 
  \label{equation::H_220} 
\end{equation}
where on the right-hand side the last term is a source term which represents the entropy production rate by 
dissipation and by fluctuations. We note that Eqs.\ \eqref{equation::H_210} and \eqref{equation::H_220} are 
not independent from each other. They are related to each other by the differential thermodynamic relations 
\eqref{equation::H_090} and \eqref{equation::H_130}. Furthermore, we note that in the general case the 
time-evolution equation for the entropy \eqref{equation::H_220} plays a similar role like the time-evolution 
equation for the free energy density \eqref{equation::G_I70} does in the isothermal case.

\subsection{Reversible terms}
\label{section::08d}
The reversible contributions to the current densities and to the velocity of the lattice structure on the right-hand 
sides of the time-evolution equations \eqref{equation::G_H70}-\eqref{equation::G_H90}, \eqref{equation::H_210}, and 
\eqref{equation::H_220} are calculated by the Poisson bracket of the first term on the right-hand side of one of 
the general time-evolution equations \eqref{equation::H_180} or \eqref{equation::H_190}. For this purpose we need 
the Poisson brackets of all relevant variables. For the variables of the isothermal case we have derived the 
Poisson brackets \eqref{equation::G_780}-\eqref{equation::G_830}. In the general case we must add some more 
Poisson brackets which involve the entropy density $s(\mathbf{r},t)$ and which read
\begin{eqnarray}
  \{ \tilde{n}_a(\mathbf{r},t) , s(\mathbf{r}^\prime,t) \} &=& 0 \, ,
  \label{equation::H_230} \\
  \{ \tilde{\mathbf{j}}(\mathbf{r},t) , s(\mathbf{r}^\prime,t) \} &=& - s(\mathbf{r},t) 
  \, \nabla \, w( \mathbf{r} - \mathbf{r}^\prime ) \, , \hspace{10mm}
  \label{equation::H_240} \\
  \{ s(\mathbf{r},t) , s(\mathbf{r}^\prime,t) \} &=& 0 \, ,
  \label{equation::H_250} \\
  \{ u^p(\mathbf{r},t) , s(\mathbf{r}^\prime,t) \} &=& 0 \, .
  \label{equation::H_260}
\end{eqnarray}
The first three of these Poisson brackets are the coarse grained versions of the microscopic Poisson brackets 
which have been derived in our previous paper \cite{Ha16} and which are given by Eqs.\ (195)-(197) therein. 
Here, the delta functions $\delta( \mathbf{r} - \mathbf{r}^\prime )$ have been replaced by the coarse-grained 
integration kernels $w( \mathbf{r} - \mathbf{r}^\prime )$. According to Eqs.\ \eqref{equation::G_370} or 
\eqref{equation::G_410} the displacement field $\hat{u}^p(\mathbf{r},t)$ is expressed in terms of the microscopic 
density $\hat{n}_a(\mathbf{r},t)$. Thus, from Eq.\ (195) of Ref.\ \onlinecite{Ha16} we furthermore obtain the 
last Poisson bracket \eqref{equation::H_260}. Further Poisson brackets involving the energy density 
$\tilde{\varepsilon}(\mathbf{r},t)$ may be calculated by applying the differential thermodynamic relation 
\eqref{equation::H_090} to Poisson brackets.

The succeeding calculation is done similar as for the isothermal case in Subsec.\ \ref{section::07c}. We 
recover Eqs.\ \eqref{equation::G_870}-\eqref{equation::G_890} where now the free energy $F(t)$ is replaced 
by the energy $E(t)$. As a result we recover the reversible current densities \eqref{equation::G_970} and 
\eqref{equation::G_980}. Furthermore, we recover the right-hand side of Eq.\ \eqref{equation::G_A30}. This 
means we recover the reversible contribution to the velocity of the lattice structure which is just the 
velocity of the matter. There are no changes to the previous results. Additionally, for the general case we 
calculate the reversible contributions to the entropy current density and the energy density which are
\begin{eqnarray}
  \mathbf{q}_\mathrm{rev}(\mathbf{r},t) &=& s(\mathbf{r},t) \, \mathbf{v}(\mathbf{r},t) \, ,
  \label{equation::H_270} \\ 
  \tilde{j}^k_{E,\mathrm{rev}}(\mathbf{r},t) &=& [ \tilde{\varepsilon}(\mathbf{r},t) 
  + p(\mathbf{r},t) ] \, v^k(\mathbf{r},t) \nonumber \\
  &&- v_i(\mathbf{r},t) \, \sigma^{ik}(\mathbf{r},t) \, ,
  \label{equation::H_280}
\end{eqnarray}
respectively. Since there are no reversible contributions to the source terms for the entropy production 
rate we obtain $R_\mathrm{rev}(\mathbf{r},t) = 0$.

For the reversible terms the results are quite robust. We nearly obtain the same formulas for the current 
densities and for the entropy production rate which are known from the hydrodynamics of a simple liquid. 
For this purpose we may compare our present results with Eqs.\ (185), (187), (188), (206), and (207) of our 
previous publication \cite{Ha16}. For the elasticity theory of a crystal we obtain an extension only in the 
momentum current density \eqref{equation::G_980} and the energy current density \eqref{equation::H_280}. 
In both cases there is an additional term involving the Cauchy stress tensor $\sigma^{ik}(\mathbf{r},t)$ 
with two Cartesian indices which describes the elastic properties of the crystal.

\subsection{Dissipative terms}
\label{section::08e}
The dissipative terms are obtained from the second term on the right-hand side of the time-evolution 
equation \eqref{equation::H_190}. Some offdiagonal terms of the Onsager matrix $M_{ij}(\mathbf{r},t)$ are 
zero because of the invariance under time reversal. These are the terms correlating the momentum density 
with the other variables. The functional derivatives of the energy $E[\tilde{x}(t)]$ are calculated by using 
the formulas \eqref{equation::H_090} and \eqref{equation::H_100}. All further calculations proceed similar as 
in Subsec.\ \ref{section::07d}. Eventually, we obtain the dissipative contributions to the current densities
\begin{eqnarray}
  \tilde{j}_{\mathrm{diss},a}^k &=& - \sum_b D_{ab}^{kl} \, ( \partial_l \, \mu_b ) 
  - G_a^{kl} \, ( \partial_l \, T ) \nonumber \\
  &&- K_a^{kp} \, ( \partial_l \, \sigma_p^{\ l} ) \, ,
  \label{equation::H_290} \\
  \tilde{\Pi}_\mathrm{diss}^{ik} &=& - \Lambda^{ik,jl} \, ( \partial_l \, v_j ) \, ,
  \label{equation::H_300} \\
  q_\mathrm{diss}^k &=& - \sum_b G_b^{lk} \, ( \partial_l \, \mu_b ) \nonumber \\
  &&- \alpha^{kl} \, ( \partial_l \, T ) 
  - \xi^{kp} \, ( \partial_l \, \sigma_p^{\ l} ) \, .
  \label{equation::H_310}
\end{eqnarray}
Furthermore, we obtain the velocity field for the diffusive motion
\begin{eqnarray}
  v_\mathrm{diff}^p(\mathbf{r},t) &=& - \sum_a K_a^{kp} \, ( \partial_k \, \mu_a ) 
  \nonumber \\
  &&- \xi^{kp} \, ( \partial_k \, T ) 
  - \zeta^{pq} \, ( \partial_k \, \sigma_q^{\ k} ) \, .
  \label{equation::H_320}
\end{eqnarray}
We recover all dissipative terms which we have derived previously for the isothermal case in 
Subsec.\ \ref{section::07d}. In the present formulas for the general case we find some more dissipative 
terms which are related to additional elements of the Onsager matrix which involve the entropy density 
$s(\mathbf{r},t)$. These additional elements are
\begin{eqnarray}
  M_{n_a,s}(\mathbf{r},t) &=& - \, \partial_k \, G_a^{kl}(\mathbf{r},t) \, \partial_l \, ,
  \label{equation::H_330} \\
  M_{s,s}(\mathbf{r},t) &=& - \, \partial_k \, \alpha^{kl}(\mathbf{r},t) \, \partial_l \, ,
  \label{equation::H_340} \\
  M_{s,u^p}(\mathbf{r},t) &=& + \, \partial_k \, \xi^{kp}(\mathbf{r},t) \, .
  \label{equation::H_350}
\end{eqnarray}
Thus, we obtain three more transport coefficients which are $G_a^{kl}$, $\alpha^{kl}$, and $\xi^{kp}$. 
The Galilean invariance imposes constraints onto all elements of the Onsager matrix which are related to 
the particle densities $\tilde{n}_a(\mathbf{r},t)$. Hence, beyond Eq.\ \eqref{equation::G_C60} we obtain 
the additional constraints 
\begin{equation}
  \sum_a m_a \, G_a^{kl} = 0 \, .
  \label{equation::H_360}
\end{equation}

The energy density $\tilde{\varepsilon} = \tilde{\varepsilon}(s,\partial \mathbf{u},\tilde{\mathbf{j}},\tilde{n})$ 
is not independent but rather depends on the other relevant variables because of the thermodynamic relation 
\eqref{equation::H_080}. Thus, the related time-evolution equation \eqref{equation::H_210} is not independent. 
Rather, it is related to and can be obtained from the time-evolution equations of the other relevant variables. 
We may proceed similar as in Subsec.\ \ref{section::07h} where we have derived the time-evolution equation for 
the free energy density. From Eq.\ \eqref{equation::H_090} we obtain the thermodynamic relation for the time
derivatives
\begin{equation}
  \partial_t \tilde{\varepsilon} = T \, \partial_t s + \sigma_p^{\ k} \, ( \partial_k \partial_t u^p ) 
  + \mathbf{v} \cdot \partial_t \tilde{\mathbf{j}} + \sum_a \mu_a \, \partial_t \tilde{n}_a \, .
  \label{equation::H_370} 
\end{equation}
On the right-hand side we insert the time-evolution equations \eqref{equation::G_H70}-\eqref{equation::G_H90} 
and \eqref{equation::H_220}. For the space derivatives on the right-hand side we need the thermodynamic relation 
\begin{equation}
  \nabla \tilde{\varepsilon} = T \, \nabla s + \sigma_p^{\ k} \, ( \partial_k \nabla u^p ) 
  + \mathbf{v} \cdot \nabla \tilde{\mathbf{j}} + \sum_a \mu_a \, \nabla \tilde{n}_a \, .
  \label{equation::H_380} 
\end{equation}
The entropy production rate $R$ on the right-hand side of the entropy equation \eqref{equation::H_220} 
must be chosen properly so that finally the right-hand side of Eq.\ \eqref{equation::H_370} can be written 
as minus the divergence of an energy current density $\mathbf{j}_E$ so that the continuity equation 
\eqref{equation::H_210} is obtained. Then, after some calculations we obtain the dissipative contribution 
of the energy current density
\begin{equation}
  \tilde{j}_{E,\mathrm{diss}}^k = \sum_a \mu_a \, \tilde{j}_{\mathrm{diss},a}^k + v_i \, \tilde{\Pi}_\mathrm{diss}^{ik}
  + T \, q_\mathrm{diss}^k + \sigma_p^{\ k} \, v_\mathrm{diff}^p 
  \label{equation::H_390}
\end{equation}
and the dissipative contribution of the entropy production rate
\begin{equation}
  \begin{split}
    R_\mathrm{diss} = &- \sum_a ( \partial_k \mu_a ) \, \tilde{j}_{\mathrm{diss},a}^k 
    - ( \partial_k v_i ) \, \tilde{\Pi}_\mathrm{diss}^{ik} - ( \partial_k T ) \, q_\mathrm{diss}^k \\
    &- ( \partial_k \sigma_p^{\ k} ) \, v_\mathrm{diff}^p \, .
  \end{split}
  \label{equation::H_400}
\end{equation}
We note that all reversible contributions drop out because they have been determined properly in the 
previous subsection.

Explicit formulas are obtained if we insert the dissipative current densities 
\eqref{equation::H_290}-\eqref{equation::H_310} and the diffusive velocity field \eqref{equation::H_320}. 
Thus, as final results we obtain the dissipative energy current density
\begin{widetext}
\begin{equation}
  \begin{split}
    \tilde{j}_{E,\mathrm{diss}}^k = \,& - \sum_a \mu_a \, \Bigl( \sum_b D_{ab}^{kl} \, ( \partial_l \, \mu_b ) 
    + G_a^{kl} \, ( \partial_l \, T ) + K_a^{kq} \, ( \partial_l \, \sigma_q^{\ l} ) \Bigr) \\
    &- v_i \, \Lambda^{ik,jl} \, ( \partial_l \, v_j ) \\
    &- T \, \Bigl( \sum_b G_b^{lk} \, ( \partial_l \, \mu_b ) 
    + \alpha^{kl} \, ( \partial_l \, T ) + \xi^{kq} \, ( \partial_l \, \sigma_q^{\ l} ) \Bigr) \\
    &- \sigma_p^{\ k} \, \Bigl( \sum_b K_b^{lp} \, ( \partial_l \, \mu_b ) 
    + \xi^{lp} \, ( \partial_l \, T ) + \zeta^{pq} \, ( \partial_l \, \sigma_q^{\ l} ) \Bigr) 
  \end{split}
  \label{equation::H_410}
\end{equation}
and the dissipative entropy production rate
\begin{equation}
  \begin{split}
    R_\mathrm{diss} = \,& \sum_a ( \partial_k \mu_a ) \, \Bigl( \sum_b D_{ab}^{kl} \, ( \partial_l \, \mu_b ) 
    + G_a^{kl} \, ( \partial_l \, T ) + K_a^{kq} \, ( \partial_l \, \sigma_q^{\ l} ) \Bigr) \\
    &+ ( \partial_k v_i ) \, \Lambda^{ik,jl} \, ( \partial_l \, v_j ) \\
    &+ ( \partial_k T ) \, \Bigl( \sum_b G_b^{lk} \, ( \partial_l \, \mu_b ) 
    + \alpha^{kl} \, ( \partial_l \, T ) + \xi^{kq} \, ( \partial_l \, \sigma_q^{\ l} ) \Bigr) \\
    &+ ( \partial_k \sigma_p^{\ k} ) \, \Bigl( \sum_b K_b^{lp} \, ( \partial_l \, \mu_b ) 
    + \xi^{lp} \, ( \partial_l \, T ) + \zeta^{pq} \, ( \partial_l \, \sigma_q^{\ l} ) \Bigr) \, .
  \end{split}
  \label{equation::H_420}
\end{equation}
\end{widetext}
We note that the energy current density $\mathbf{j}_{E,\mathrm{diss}}$ is a \emph{linear} form of the 
gradients of the intensive thermodynamic variables. This formal result is expected for any current density. 
On the other hand, the entropy production rate $R_\mathrm{diss}$ is a \emph{quadratic} form of the gradients 
of the intensive thermodynamic variables. This form is also expected from general nonequilibrium statistical 
mechanics. It is positive definite and implies a monotonic increase of the entropy density which represents 
the second law of thermodynamics.

In Subsec.\ \ref{section::06f} we have found five different stress tensors which are classified by 
Eqs.\ \eqref{equation::F_840}-\eqref{equation::F_880}. Here, in this subsection in the dissipative contributions 
of all current densities and the entropy production rate there appears only one stress tensor. It is 
the \emph{minus first Piola} stress tensor $\sigma_p^{\ k}$ and its divergence $\partial_k \, \sigma_p^{\ k}$ 
with one lower curvilinear index $p$ and one Cartesian index $k$. (A Cartesian index may be upper or lower, 
this makes no difference.) This result is very important. It guarantees that our theory of continuum mechanics 
is valid in the \emph{nonlinear} regime for strong deformations.

Grabert and Michel \cite{GM83} present their nonlinear time-evolution equations in their Eq.\ (24). Since the 
paper is rather short and the notation is different, it takes some effort to compare the results of Grabert and 
Michel with our results. One must be very careful to treat the nonlinearities correctly. After some longer 
calculations we find that our current densities agree with those of Grabert and Michel. We prove this for the 
reversible contributions \eqref{equation::G_970}, \eqref{equation::G_980}, \eqref{equation::H_280} and for the 
dissipative contributions \eqref{equation::H_290}, \eqref{equation::H_300}, \eqref{equation::H_410}. Since only 
one particle species is considered several dissipative parameters are just zero because of the constraints 
\eqref{equation::G_C60} and \eqref{equation::H_360}. Furthermore, we find that our time-evolution equation for 
the displacement field \eqref{equation::G_H90} agrees with the related equation within (24) of Grabert and 
Michel if we insert the velocity of the lattice structure \eqref{equation::G_I20} and use the diffusive part 
\eqref{equation::H_320}. At the end, we only find two minor errors. In the last two lines of Eq.\ (24) of 
Ref.\ \onlinecite{GM83} which are part of the time-evolution equation for the energy density one index is wrong and one 
contribution from the offdiagonal element $\theta^{\alpha\beta}$ of the Onsager matrix is missing. However, 
these minor errors can be corrected easily. In the second last line we must replace $\delta^{\nu\mu}$ by 
$\delta^{\nu\beta}$. Furthermore, in the last line we must add the term 
$( \partial / \partial x^\alpha ) \, \theta^{\mu\beta} S^{\mu\alpha} T^{-1} \partial T / \partial x^\beta$. 
This latter term is written as $\partial_k \, \sigma_p^{\ k} \, \xi^{lp} \, ( \partial_l \, T )$ in our 
notation and hence corresponds to our contribution $- \sigma_p^{\ k} \, \xi^{lp} \, ( \partial_l \, T )$ in 
the energy current density \eqref{equation::H_410}.

Grabert and Michel \cite{GM83} define three different stress tensors which are denoted by $S^{\alpha\beta}$, 
$\lambda^{\alpha\beta}$, and $\Lambda^{\alpha\beta}$ and which correspond to our stress tensors 
$\sigma_{kl}$, $\sigma_{pk}$, $\sigma_{pq}$ classified in Eqs.\ \eqref{equation::F_860}-\eqref{equation::F_880}, 
respectively. In their Eq.\ (24) in the dissipative contributions only the stress tensor $\lambda^{\alpha\beta}$ 
and its divergence $\partial_\beta \lambda^{\alpha\beta}$ occur. This result agrees exactly with our result. 
Thus, we confirm that Grabert and Michel have treated the nonlinearities of the elasticity theory correctly.

Nevertheless, there is a difference. Grabert and Michel \cite{GM83} have considered a crystal with only 
\emph{one species} of particles. This is a special case of our theory where the constraints \eqref{equation::G_C60} 
and \eqref{equation::H_360} imply that all transport coefficients related to explicit diffusion of particles are 
zero. Thus, in the theory of Grabert and Michel the transport coefficients $D_{ab}^{kl}$, $G_a^{kl}$ , and $K_a^{kp}$ 
are not present. Our theory may be interpreted as a generalization which considers crystals with several different 
species of particles and hence includes explicit diffusion effects where the particles move relative to each other.

\subsection{Reactive terms}
\label{section::08f}
Also in the general case there are additional reversible terms which come from the reactive contributions 
of the memory function \eqref{equation::G_E40}. Thus, Subsec.\ \ref{section::07e} may be applied and extended 
to the general case. Starting with the time-evolution equation in the form \eqref{equation::H_190} we expect 
nonzero contributions for the additional offdiagonal reactive block matrices 
\begin{eqnarray}
  M^\prime_{s,j^j}(\mathbf{r},t) &=& - \, \partial_k \, \chi^{k,jl}(\mathbf{r},t) \, \partial_l \, , 
  \label{equation::H_430} \\
  M^\prime_{j^j,s}(\mathbf{r},t) &=& + \, \partial_l \, \chi^{k,jl}(\mathbf{r},t) \, \partial_k \ .
  \label{equation::H_440}
\end{eqnarray}
These block matrices imply \emph{reactive} contributions to the entropy density 
\begin{equation}
  q_\mathrm{reac}^k = - \, \chi^{k,jl} \, ( \partial_l \, v_j ) 
  \label{equation::H_450}
\end{equation}
and to the momentum current densities
\begin{equation}
  \tilde{\Pi}_\mathrm{reac}^{ik} = \ldots 
  + \chi^{l,ik} \, ( \partial_l \, T )
  \label{equation::H_460}
\end{equation}
where the three dots indicate the two terms given in Eq.\ \eqref{equation::G_F40}. The reactive terms 
parameterized by $\chi^{k,jl}$ have been found previously by Mabillard and Gaspard \cite{MG21}. Strictly 
speaking there is a difference because we consider the \emph{entropy} density as a relevant variable where 
Mabillard and Gaspard consider the \emph{energy} density. However, the different reactive contributions are 
equivalent and can be transformed into each other.

Nevertheless, also in the general case in the time-evolution equations the reactive terms have one gradient 
operator more compared to the leading reversible terms. For this reason within the view of a gradient expansion 
the reactive terms are of higher order and may be neglected. Thus, also in the general case we do not consider 
the reactive terms in our time-evolution equations for the nonlinear elasticity of solid crystals.

\subsection{Fluctuating terms}
\label{section::08g}
On the right-hand side of the time-evolution equation \eqref{equation::H_190} the third term provides the 
fluctuating forces. In the case of conserved quantities we obtain fluctuating current densities. For the isothermal 
case the fluctuating contributions are presented in Subsec.\ \ref{section::07f}. All these results remain valid 
also in the general case. Here we consider the additional fluctuating contributions which occur only in the general 
case. A further independent relevant variable is the entropy density $s(\mathbf{r},t)$. We find the related 
fluctuating forces and currents 
\begin{equation}
  f_{s}(\mathbf{r},t) = - \partial_k \, g_{s}^k(\mathbf{r},t) \, , \qquad
  g_{s}^k(\mathbf{r},t) = q_\mathrm{fluc}^k(\mathbf{r},t) \, .
  \label{equation::H_470}
\end{equation}
Thus, for the fluctuating contribution of the entropy current density $\mathbf{q}_\mathrm{fluc}(\mathbf{r},t)$ 
we obtain the averages and correlations
\begin{eqnarray}
  \overline{ q_\mathrm{fluc}^k(\mathbf{r},t) } &=& 0 \, ,
  \label{equation::H_480} \\
  \overline{ q_\mathrm{fluc}^k(\mathbf{r},t) \, q_\mathrm{fluc}^l(\mathbf{r}^\prime,t^\prime) } &=& 
  2 \, ( k_B T ) \, \alpha^{kl}(\mathbf{r},t) \nonumber \\ 
  &&\times \, w( \mathbf{r} - \mathbf{r}^\prime ) \, \delta( t - t^\prime ) \, , \hspace{10mm}
  \label{equation::H_490}
\end{eqnarray}
respectively, where $\alpha^{kl}(\mathbf{r},t)$ is the coefficient matrix for the entropy transport which 
already occurred in Eqs.\ \eqref{equation::H_310} and \eqref{equation::H_340}.

Furthermore, we consider the offdiagonal correlations of the entropy density with the other fluctuating 
forces and currents. Here, we find
\begin{eqnarray}
  \overline{ \tilde{j}_{\mathrm{fluc},a}^k(\mathbf{r},t) \, q_\mathrm{fluc}^l(\mathbf{r}^\prime,t^\prime) } &=& 
  2 \, ( k_B T ) \, G_a^{kl}(\mathbf{r},t) \nonumber \\ 
  &&\times \, w( \mathbf{r} - \mathbf{r}^\prime ) \, \delta( t - t^\prime ) \, , \hspace{10mm}
  \label{equation::H_500} \\
  \overline{ \tilde{\Pi}_\mathrm{fluc}^{ik}(\mathbf{r},t) \, q_\mathrm{fluc}^l(\mathbf{r}^\prime,t^\prime) } &=& 0 \, ,
  \label{equation::H_510} \\
  \overline{ q_\mathrm{fluc}^k(\mathbf{r},t) \, v_\mathrm{fluc}^p(\mathbf{r}^\prime,t^\prime) } &=& 
  2 \, ( k_B T ) \, \xi^{kp}(\mathbf{r},t) \nonumber \\ 
  &&\times \, w( \mathbf{r} - \mathbf{r}^\prime ) \, \delta( t - t^\prime ) \hspace{10mm}
  \label{equation::H_520}
\end{eqnarray}
where $G_a^{kl}(\mathbf{r},t)$ and $\xi^{kp}(\mathbf{r},t)$ are offdiagonal block matrices of transport coefficients 
which already occurred in the dissipative contributions \eqref{equation::H_290}-\eqref{equation::H_320}. One of these 
offdiagonal correlations which involves the momentum current density is zero because of time inversion invariance.

Since the energy density $\tilde{\varepsilon} = \tilde{\varepsilon}(s,\partial \mathbf{u},\tilde{\mathbf{j}},\tilde{n})$ 
is not independent but rather depends on the other relevant variables here we must proceed in a similar way as for 
the dissipaive terms in the preceding subsection. For the fluctuating contributions we obtain similar formulas as 
Eqs.\ \eqref{equation::H_390} and \eqref{equation::H_400}. Thus, we obtain the fluctuating contribution of the 
energy current density
\begin{equation}
  \tilde{j}_{E,\mathrm{fluc}}^k = \sum_a \mu_a \, \tilde{j}_{a,\mathrm{fluc}}^k + v_i \, \tilde{\Pi}_\mathrm{fluc}^{ik} 
  + T \, q_\mathrm{fluc}^k + \sigma_p^{\ k} \, v_\mathrm{fluc}^p \, .
  \label{equation::H_530}
\end{equation}
Furthermore, we obtain the fluctuating contribution of the entropy production rate
\begin{equation}
  \begin{split}
    R_\mathrm{fluc} = &- \sum_a ( \partial_k \mu_a ) \, \tilde{j}_{a,\mathrm{fluc}}^k 
    - ( \partial_k v_i ) \, \tilde{\Pi}_\mathrm{fluc}^{ik} - ( \partial_k T ) \, q_\mathrm{fluc}^k \\ 
    &- ( \partial_k \sigma_p^{\ k} ) \, v_\mathrm{fluc}^p \, .
  \end{split}
  \label{equation::H_540}
\end{equation}
These two fluctuating contributions satisfy the continuity equation for the energy density \eqref{equation::H_210}
and the time-evolution equation for the entropy density \eqref{equation::H_220}.

Since the reversible contribution is zero the entropy production rate has two terms 
\begin{equation}
  R = R_\mathrm{diss} + R_\mathrm{fluc} \, .
  \label{equation::H_550}
\end{equation}
The dissipative term $R_\mathrm{diss}$ is a positive definite quadratic form given in Eq.\ \eqref{equation::H_420}. 
As a consequence, it is always positive or at least zero. This term provides the second law of thermodynamics. 
Starting in a nonequilibrium state the physical system moves to a thermodynamic equilibrium where the entropy 
increases monotonically. However, once the equilibrium is reached, the increase of the entropy must stop. 
Nevertheless, even in thermal equilibrium fluctuations are present so that the dissipative term $R_\mathrm{diss}$ 
will stick at a \emph{positive} nonzero value. Here, the fluctuating contribution $R_\mathrm{fluc}$ comes into play. 
On average this latter contribution will have a \emph{negative} nonzero value so that it cancels the dissipative 
contribution exactly. As a result the total $R = R_\mathrm{diss} + R_\mathrm{fluc}$ is zero on average so that 
in the presence of fluctuations the second law of thermodynamics is not in conflict with the thermal equilibrium.

Grabert and Michel \cite{GM83} have not considered fluctuations. Nevertheless, their theory can be extended 
straightforwardly by fluctuating terms using the results of the present subsection. Since they consider only one 
species of particles the constraint conditions \eqref{equation::G_C60} and \eqref{equation::H_360} imply that in 
this case all correlations involving the transport-coefficient matrices $D_{ab}^{kl}$, $G_a^{kl}$ , and $K_a^{kp}$ 
are zero and may be discarded.

\section{Discussions}
\label{section::09}

\subsection{Hybrid dynamic density-functional theory}
\label{section::09a}
Our general time-evolution equations in the form \eqref{equation::G_B80} may be interpreted as 
the dynamic equations of a \emph{dynamic density-functional theory} 
\cite{Ev79,Mu89,DFM90,Fr93,MT99,AE04,Yo05,EL09,St18,VLW20}. It is a differential equation with respect 
to the time $t$ for the general densities $\tilde{x}_i(\mathbf{r},t)$ where on the right-hand side the 
Helmholtz free energy $F[\tilde{x}(t)]$ is the \emph{density functional}. In global thermodynamic 
equilibrium the free energy is minimized under the constraints that the conserved quantities are 
fixed. The time evolution equations \eqref{equation::G_B80} provide the relaxation of the densities 
$\tilde{x}_i(\mathbf{r},t)$ toward equilibrium values and fluctuations around them. 

However, the situation is more complicated because we deal with continuum mechanics where we apply 
the concept of \emph{local thermodynamic equilibrium}. This means we perform coarse graining where 
we separate effects on large scales from effects on small scales. For the microscopic densities 
$x_i(\mathbf{r},t)$ as functions of the space coordinates $\mathbf{r}$ this means that we split 
the function space $\mathcal{C} = \tilde{\mathcal{C}}(\ell) \oplus \mathcal{C}^\prime(\ell)$ into 
a subspace of slowly varying functions $\tilde{\mathcal{C}}(\ell)$ and a subspace of fast varying 
functions $\mathcal{C}^\prime(\ell)$. Here $\ell$ is the characteristic length which separates 
the scales. It must be well above the mean distance between the particles $a$ and well below the 
characteristic length $L$ of the macroscopic phenomena in the continuum mechanics.

The coarse grained densities $\tilde{x}_i(\mathbf{r},t)$ written with a tilde are functions defined 
in the macroscopic function space $\tilde{\mathcal{C}}(\ell)$. Consequently, our time-evolution 
equation \eqref{equation::G_B80} determines the dynamics on macroscopic length scales above $\ell$. 
For small scales below $\ell$ there is the concept of local thermodynamic equilibrium which is 
realized by the locally constrained minimization procedure \eqref{equation::D_100}. Here the 
\emph{microscopic} free energy functional $F[x(t)]$ depending on the microscopic densities 
$x_i(\mathbf{r},t) = \tilde{x}_i(\mathbf{r},t) + \tilde{x}^\prime_i(\mathbf{r},t)$ 
is minimized by adjusting the fast varying parts of the densities $\tilde{x}^\prime_i(\mathbf{r},t)$ 
defined in the complement function space $\mathcal{C}^\prime(\ell)$ where the slowly varying parts 
of the densities $\tilde{x}_i(\mathbf{r},t)$ defined in the macroscopic function space 
$\tilde{\mathcal{C}}(\ell)$ are kept fixed as constraints. This constrained minimization procedure 
is done for each time $t$. As a result we obtain the \emph{macroscopic} free energy functional 
$F[\tilde{x}(t)]$ which implicitly is a function of time. Then, for the dynamics on large scales we 
insert $F[\tilde{x}(t)]$ into our time-evolution equation \eqref{equation::G_B80} in order to calculate 
the time dependence of the coarse grained densities $\tilde{x}_i(\mathbf{r},t)$ on large scales.

As a consequence, our theory for the time-evolution of continuum mechanics is a \emph{hybrid} dynamic 
density-functional theory. We use a \emph{dynamic} density-functional theory for \emph{large} scales 
where we use a \emph{static} density-functional theory locally for \emph{small} scales. This means 
that we apply the dynamic equation \eqref{equation::G_B80} on large scales where we apply the 
constraint minimization procedure \eqref{equation::D_100} on small scales. In this way we obtain the 
hydrodynamic equations for a liquid which we have derived in our previous paper \cite{Ha16}.

Until now, only the coarse-graining procedure for the densities \eqref{equation::D_040} is taken 
into account. In order to obtain the continuum-mechanics equations for an elastic crystal we must 
extend the theory and define and include the displacement field $\mathbf{u}(\mathbf{r},t)$ properly 
which describes the deformations of the crystal structure. On the one hand, for the \emph{statics} 
this is done in Sec.\ \ref{section::04} by solving explicitly the locally constraint minimization 
procedure which can be written in the form \eqref{equation::D_820}. On the other hand, for the 
\emph{dynamics} this is done in Sec.\ \ref{section::07} by extending the projection operator and 
by explicitly deriving a time-evolution equation for the displacement field.

Originally, the dynamic density-functional theory \cite{MT99,AE04} was designed to describe dynamic 
phenomena of interacting many particles on \emph{microscopic} scales. In principle, our theory can 
be extended to microsopic scales either. In this case we forget about the local thermodynamic 
equilibrium and the locally constrained minimization procedure. Rather, we write down the time-evolution 
equations \eqref{equation::G_B80} directly for the microscopic densities $x_i(\mathbf{r},t)$ and 
use the microscopic free energy $F[x(t)]$ for the functional derivatives. However, we can not 
guarantee that the time-evolution equations provide a stable solution. Starting for a given time 
with initial values the densities may diverge as the time evolves and become strongly varying on 
short scales so that eventually the solution does not represent a local thermodynamic equilibrium 
and hence does not make any physical sense. This problem may arise because we have derived our 
time-evolution equations for Newtonian mechanics which is defined in Sec.\ \ref{section::02}. The 
problem may be due to kinetic effects. It may be related to the conservation of momentum and the 
Galilean invariance.

The original dynamic density-functional theory \cite{MT99,AE04} was derived for the Brownian 
motion of colloidal particles in a solvent. In this case the total momentum is not conserved and 
there is no Galilean invariance. As a consequence, the momentum density $\mathbf{j}(\mathbf{r},t)$ 
is no relevant variable and must be discarded. Furthermore, on the microscopic level there is no 
displacement field $\mathbf{u}(\mathbf{r},t)$ so that it must be discarded either. Thus, the only 
relevant variables that remain are the microscopic particle densities $n_a(\mathbf{r},t)$. Hence, 
in the free energy functional \eqref{equation::C_030} the kinetic energy \eqref{equation::C_060} 
must be omitted so that the free energy reduces to Eq.\ \eqref{equation::C_090}. In order to obtain 
the reversible terms we may use the Poisson brackets of the densities defined in 
Eq.\ \eqref{equation::G_660}. Since these Poisson brackets are zero in all cases, the reversible 
terms are all zero. Thus, our general time-evolution equations \eqref{equation::G_B80} reduce to
\begin{equation}
  \begin{split}
    \partial_t n_a(\mathbf{r},t) = - \sum_b M_{n_a,n_b}(\mathbf{r},t)
    \, \frac{ \delta F[n(t)] }{ \delta n_b(\mathbf{r},t) }
    + f_{n_a}(\mathbf{r},t) \, .
  \end{split}
  \label{equation::I_010}
\end{equation}
Since also in Brownian dynamics the particle numbers are conserved quantities, we may use 
the Onsager matrix \eqref{equation::G_B90} which reads
\begin{equation}
  M_{n_a,n_b}(\mathbf{r},t) = - \, \partial_k \, D_{ab}^{kl}(\mathbf{r},t) \, \partial_l \, .
  \label{equation::I_020}
\end{equation}
The fluctuating forces may be expressed in terms of fluctuating currents according to 
\eqref{equation::G_G20} where the averages and the correlations are ruled by 
Eqs.\ \eqref{equation::G_G30} and \eqref{equation::G_G40}, respectively. However, in 
Eq.\ \eqref{equation::G_G40} the macroscopic integral kernel $w( \mathbf{r} - \mathbf{r}^\prime )$ 
must be replaced by a microscopic delta function $\delta( \mathbf{r} - \mathbf{r}^\prime )$.

The strength of the dissipation effects and of the fluctuations is parameterized by the 
transport matrix $D_{ab}^{kl}(\mathbf{r},t)$. Since momentum conservation and Galilean 
invariance are absent, the constraint conditions \eqref{equation::G_C60} do not apply.
The original approaches \cite{Ev79,Mu89,DFM90,Fr93,MT99,AE04,Yo05,EL09} explicitly derive
\begin{equation}
  D_{ab}^{kl}(\mathbf{r},t) = \Gamma_a \, n_a(\mathbf{r},t) \, \delta_{ab} \, \delta^{kl} \, .
  \label{equation::I_030}
\end{equation}
This transport matrix is diagonal in two respects. First, with respect to the species indices it 
is proportional to the unit matrix $\delta_{ab}$ which means that the diffusion of each particle 
species is independent. Second, with respect to the space indices it is proportional to the unit 
matrix $\delta^{kl}$ which means that the system is isotropic in space. The main physical effects 
are described by the constant parameters $\Gamma_a$ which are defined for each particle species $a$. 
Eventually, if we put Eqs.\ \eqref{equation::I_010}-\eqref{equation::I_030} together we obtain the 
well known equations of the dynamic density-functional theory 
\cite{Ev79,Mu89,DFM90,Fr93,MT99,AE04,Yo05,EL09} which for a multi-component system read
\begin{equation}
  \partial_t n_a(\mathbf{r},t) = 
  \nabla \cdot \Gamma_a \, n_a(\mathbf{r},t) \, \nabla \, \frac{ \delta F[n(t)] }{ \delta n_a(\mathbf{r},t) }
    + f_{n_a}(\mathbf{r},t) \, .
  \label{equation::I_040}
\end{equation}
The fluctuating forces defined in Eqs.\ \eqref{equation::G_G20}-\eqref{equation::G_G40} are either 
omitted or taken into account whether the deterministic or whether the stochastic version of the 
theory is considered. A detailed review of the different versions of the dynamical density-functional 
theory and an overview of the historical development is given by te Vrugt \emph{et al.}\ \cite{VLW20}.

The transport matrix \eqref{equation::I_030} is not constant. Rather, it depends linearly on 
the particle densities $n_a(\mathbf{r},t)$. As a consequence, there is a \emph{nonlinear noise} 
which appears in the correlations of the fluctuating currents \eqref{equation::G_G40}. On the other 
hand, the functional form \eqref{equation::I_030} appears to be very special. Nevertheless, it 
becomes clear if one considers a dilute gas of particles with weak interactions in the ideal limit. 
If we take the free energy functional of an ideal gas which is given by Eq.\ \eqref{equation::C_040} 
we calculate the chemical potentials by the functional derivatives 
\begin{equation}
  \mu_a(\mathbf{r},t) = \frac{ \delta F_0[n(t)] }{ \delta n_a(\mathbf{r},t) } = 
  k_B T \, \ln[ ( \lambda_a )^d \, n_a(\mathbf{r},t) ] \, .
  \label{equation::I_050}
\end{equation}
Then, from Eqs.\ \eqref{equation::I_010}-\eqref{equation::I_030} or directly from 
Eq.\ \eqref{equation::I_040} we obtain
\begin{equation}
  \partial_t n_a(\mathbf{r},t) = D_a \, \nabla^2 n_a(\mathbf{r},t) + f_{n_a}(\mathbf{r},t) 
  \label{equation::I_060}
\end{equation}
which precisely are diffusion equations with stochastic forces where $D_a = \Gamma_a \, k_B T$ 
are the diffusion constants. Thus, the dynamic density-functional theory is designed to recover 
the standard diffusion for each particle species. This fact strongly supports the functional
form of the transport matrix \eqref{equation::I_030}.

Alternatively, the transport matrix \eqref{equation::I_030} can be calculated within the 
projection-operator formalism \cite{Yo05,EL09}. From the general integral kernel 
\eqref{equation::G_600} together with the space- and time-scale separation \eqref{equation::G_B10} 
and the identification \eqref{equation::G_B90} we obtain explicit formulas for the transport 
matrix which are given by Eqs.\ \eqref{equation::I_690} and \eqref{equation::I_760} below in 
Subsec.\ \ref{section::09g}. Here, we must evaluate a special correlation function of two particle 
current densities $\hat{j}_a^k(\mathbf{r},t)$ and $\hat{j}_b^l(\mathbf{r}^\prime,t^\prime)$ for 
which we insert microscopic expressions similar like Eq.\ \eqref{equation::B_110}. In order to 
proceed we need three assumptions \cite{Yo05,EL09} which lead to approximations. First, the 
coordinates $\mathbf{r}_{ai}$ must evolve much slower in time than the momenta $\mathbf{p}_{ai}$ 
so that the motion of the particles is strongly overdamped. Second, the coordinates $\mathbf{r}_{ai}$ 
and the momenta $\mathbf{p}_{ai}$ must be statistically independent from each other. Third, there 
may be no correlations between the momenta $\mathbf{p}_{ai}$ of different particles and of different 
space directions. Then, as a result we obtain the transport matrix \eqref{equation::I_030} together 
with the parameters $\Gamma_a = \tau_a / m_a$ where $m_a$ are the masses and $\tau_a$ are the 
relaxation times of the particles of species $a$. Again, the result is obtained for a dilute 
gas of weakly interacting particles in the ideal limit where we find the diffusion constants 
$D_a = k_B T \, \tau_a / m_a$. 

Thus, we conclude that within our framework we may derive the original equations of the dynamic 
density-functional theory. Nevertheless, while the transport matrix $\eqref{equation::I_030}$ is 
derived in the limit of an ideal gas where the many particles form a dilute and weakly interacting 
system, the dynamic density-functional theory is extended and applied to dense and strongly interacting 
many-particle systems as liquids and crystalline solids with success where good results are obtained. 
Within and beyond that the theory is applied to investigate dynamic phenomena on \emph{microscopic} 
scales. 

Marconi and Tarazona \cite{MT99} apply the theory to the motion of hard rods in \emph{one} 
dimension while Stopper \emph{et al.}\ \cite{St18} consider the motion of hard disks in \emph{two} 
dimensions. Both calculate density profiles as functions of the time. While initially these 
profiles show oscillations and nearest neighbor effects on microscopic scales, for longer times 
the microscopic structures decay and smooth out so that the density profiles converge. Thus, 
we conclude that the time-evolution equations of the dynamic density-functional theory defined 
in Eqs.\ \eqref{equation::I_010}-\eqref{equation::I_030} are stable. 

The reason for this stability may be the fact that only particle densities $n_a(\mathbf{r},t)$ 
are involved. The conjugate thermodynamic variables are the chemical potentials 
\begin{equation}
  \mu_a(\mathbf{r},t) = \frac{ \delta F[n(t)] }{ \delta n_a(\mathbf{r},t) } 
  \label{equation::I_070}
\end{equation}
which may compete with \emph{one-particle} potentials $U_a(\mathbf{r},t)$. There is no reason why 
the motion of particles in a one-particle potential may be unstable. Furthermore, there is no 
conflict with a local thermodynamic equilibrium. Consequently, we need not require the chemical 
potentials $\mu_a(\mathbf{r},t)$ to be macroscopic functions defined in the function subspaces 
$\tilde{\mathcal{C}}(\ell)$. Rather, they may vary on microscopic scales so that they are defined 
in the full function space $\mathcal{C}$. Thus, we conclude that on the right-hand side of 
Eq.\ \eqref{equation::I_010} microscopically varying chemical potentials \eqref{equation::I_070} 
do not destabilize the time-evolution of the particle densities $n_a(\mathbf{r},t)$.

Until now we have discussed only the isothermal case. The reason is that the conventional static and 
dynamic density-functional theories are defined for constant temperatures $T$. However, the considerations 
and interpretations can be extended also to the general case. Here, the entropy $S[\varepsilon,\mathbf{j},n]$ 
is interpreted as the density functional where the energy density $\varepsilon = \varepsilon(\mathbf{r},t)$ 
is an additional variable. First, on microscopic scales the variables are adjusted by a constrained 
maximization of the entropy according to Eq.\ \eqref{equation::H_030} for a liquid and according to 
Eq.\ \eqref{equation::H_050} for a crystal. In this way we obtain a local thermodynamic equilibrium 
with a locally varying temperature $T = T(\mathbf{r},t)$. Second, on the macroscopic scales the 
GENERIC equation \eqref{equation::H_180} is used for the time-evolution of the relevant variables, 
including the energy density $\tilde{\varepsilon} = \tilde{\varepsilon}(\mathbf{r},t)$. While for a 
liquid the relevant variables are given by the coarse-grained densities of the conserved quantities 
only for a crystal additionally the displacement field $\mathbf{u} = \mathbf{u}(\mathbf{r},t)$ must be 
taken into account. Eventually, as a result we obtain a hybrid dynamic density-functional theory also 
for the general case which, however, is written and interpreted in an extended form.

The dynamic density-functional theory has been extended by Anero \emph{et al.}\ \cite{AET13} to include 
non-isothermal situations with non-constant temperatures $T = T(\mathbf{r},t)$. The particle density 
$n = n(\mathbf{r},t)$ of one particle species and the energy density $\varepsilon = \varepsilon(\mathbf{r},t)$ 
are considered as relevant variables. As a result, time-evolution equations are obtained similar like our 
Eq.\ \eqref{equation::H_180} where the entropy $S = S[\varepsilon,n]$ represents the density functional. 
However, in their approach Anero \emph{et al.}\ do not consider the momentum density 
$\mathbf{j} = \mathbf{j}(\mathbf{r},t)$ as a relevant variable. Two alternative extensions of the 
dynamic density-functional theory have been presented by Wittkowski \emph{et al.}\ \cite{WLB12,WLB13} 
where either the energy density $\varepsilon = \varepsilon(\mathbf{r},t)$ or the entropy density 
$s = s(\mathbf{r},t)$ is considered as an additional relevant variable. However, these latter two 
approaches are less connected to our theory and to our time-evolution equations which are either 
Eq.\ \eqref{equation::H_180} or Eq.\ \eqref{equation::H_190} together with Eq.\ \eqref{equation::H_200}. 
We note that the dynamic density-functional theories are designed to describe the physical systems on 
\emph{microscopic} scales where, on the other hand, we apply coarse graining and consider continuum 
mechanics to describe the physical systems on \emph{macroscopic} scales.

\subsection{Langevin and Fokker-Planck equations}
\label{section::09b}
The averages \eqref{equation::G_F80} and the correlations \eqref{equation::G_F90}-\eqref{equation::G_G10} 
imply that the fluctuating forces and currents are \emph{Gaussian stochastic} forces and currents. As 
a consequence our time-evolution equations \eqref{equation::G_B80} may be interpreted as \emph{Langevin} 
equations. These equations provide a stochastic description. An alternative is a \emph{Fokker-Planck} 
equation which provides a statistical description in terms of a distribution function $P[\tilde{x},t]$ 
which evolves with time. There exists a mapping between a Langevin equation and a Fokker-Planck equation 
in both directions which is well known since a long time \cite{GH71a,GH71b,Gr73}.

In our previous paper \cite{Ha16} in Sec.\ 6 we have shown and investigated that the time-evolution 
equation of the GENERIC formalism can be written as Langevin equation and mapped to a Fokker-Planck 
equation. An important question is under which conditions the Fokker-Planck equation has a simple 
Boltzmann distribution function $P_\mathrm{eq}[\tilde{x}] = Z^{-1} \exp( - F[\tilde{x}] / k_B T )$ 
as a stationary solution for the global thermodynamic equilibrium. For this purpose the conditions of 
micro reversibility and detailed balance must be satisfied. These conditions require a certain structure 
of the terms on the right-hand sides of the Langevin equations. The reversible terms must be written in 
terms of an antisymmetric Poisson matrix as in Eq.\ \eqref{equation::G_580}. The dissipative terms must 
be written in terms of a symmetric Onsager matrix as the second term in Eq.\ \eqref{equation::G_B80}. 
The fluctuating forces must be Gaussian stochastic forces where the strength of the fluctuations is 
ruled by the Onsager matrix. All these conditions are satisfied by our time-evolution equations which 
in Sec.\ \ref{section::07} we have derived for the continuum mechanics of a deformed crystal.

However, there are additional complications. We have \emph{nonlinear noise} because the Onsager matrix 
is not constant but rather depends on the relevant variables $\tilde{x}_i(\mathbf{r},t)$. Consequently, 
there must be some additional terms which involve the functional divergence of the Onsager matrix. 
Furthermore, the Poisson matrix is not constant but depends also on the relevant variables 
$\tilde{x}_i(\mathbf{r},t)$. This fact can be seen explicitly in our Poisson brackets 
\eqref{equation::G_780}-\eqref{equation::G_830}. Here, the relevant variables appear linearly on 
the right-hand sides. Hence, some further terms are needed which involve the functional divergence 
of the Poisson matrix. All these additional terms are not present in the time-evolution equations 
which are derived within the framework of the GENERIC formalism.

Moreover, the precise form of a Langevin equation depends on the discretization of the time 
derivative. Two version are commonly used, the It\^o form and the Stratonovich form \cite{St63,vK07}. 
As a consequence, the terms involving the functional divergences of the Poisson matrix and the 
Onsager matrix are not unique but differ for the It\^o form and for the Stratonovich form. 

The problem may be solved in the following way. If all three terms on the right-hand side of 
a Langevin equation are written as the divergence of a current then the related functional 
divergences of the Poisson matrix and of the Onsager matrix are zero. This fact has been proven in 
Sec.\ 6 of our previous paper \cite{Ha16}. In this case the above described complications are gone. 
Moreover, there is even no difference between the It\^o form and the Stratonovich form.

Now, we turn to our final equations for the continuum mechanics of a deformed crystal which we 
have presented in Eqs.\ \eqref{equation::G_H70}-\eqref{equation::G_H90} in Subsec.\ \ref{section::07g}. 
The first two equations for the coarse-grained particle densities $\tilde{n}_a(\mathbf{r},t)$ 
and for the coarse-grained momentum density $\tilde{\mathbf{j}}(\mathbf{r},t)$ are written as 
\emph{continuity equations} where on the right-hand sides there are divergences of current densities. 
Hence, these two equations satisfy the requirement. Unfortunately, the third equation for the 
displacement field $\mathbf{u}(\mathbf{r},t)$ does not satisfy the requirement. Thus, on a first 
sight the continuum mechanics of a deformed crystal fails with respect to this aspect. However, 
we have shown that alternatively we may consider the spatial derivative of the displacement 
field which is the strain tensor $\partial_k u^p(\mathbf{r},t)$. For this latter quantity we 
find the continuity equation \eqref{equation::G_I30} where the current is explicitly given by 
Eq.\ \eqref{equation::G_I40}. Thus, we conclude that if we consider $\tilde{n}_a(\mathbf{r},t)$, 
$\tilde{\mathbf{j}}(\mathbf{r},t)$, and $\partial_k u^p(\mathbf{r},t)$ as the relevant variables 
then all requirements are satisfied.

Moreover, following Eq.\ \eqref{equation::E_520} the free energy is written as the integral 
\begin{equation}
  F[T,\partial \mathbf{u},\tilde{\mathbf{j}},\tilde{n}] = \int d^dr 
  \, f(T,\partial \mathbf{u}(\mathbf{r}),\tilde{\mathbf{j}}(\mathbf{r}),\tilde{n}(\mathbf{r})) 
  \label{equation::I_080} 
\end{equation}
where again $\tilde{n}_a(\mathbf{r})$, $\tilde{\mathbf{j}}(\mathbf{r})$, and $\partial_k u^p(\mathbf{r})$ 
are the natural variables. Thus, the related Fokker-Planck equation must be written for these 
variables. Since all requirements are satisfied the Fokker-Planck equation has a Boltzmann 
distribution function as a stationary solution which reads
\begin{equation}
  \begin{split}
    P_\mathrm{eq}[\partial \mathbf{u},\tilde{\mathbf{j}},\tilde{n}] = 
    \,& Z^{-1} \, \exp\Bigl\{ - \Bigl( F[T,\partial \mathbf{u},\tilde{\mathbf{j}},\tilde{n}] 
    - \mathbf{v} \cdot \mathbf{P}[\tilde{\mathbf{j}}] \\
    &\hspace{20mm} - \sum_a \mu_a \, N_a[\tilde{n}] \Bigr) / k_B T \Big\} 
  \end{split}
  \label{equation::I_090} 
\end{equation}
and which describes the global thermodynamic equilibrium. Beyond the free energy $F$ there are 
two further contributions in the exponent. These are the momentum $\mathbf{P}$ and the particle 
numbers $N_a$ which are conserved quantities and which are defined by 
\begin{equation}
  \mathbf{P}[\tilde{\mathbf{j}}] = \int d^dr \, \tilde{\mathbf{j}}(\mathbf{r}) \, , \qquad
  N_a[\tilde{n}] = \int d^dr \, \tilde{n}_a(\mathbf{r}) \, ,
  \label{equation::I_100} 
\end{equation}
respectively. The global equilibrium is described by three constant parameters, the temperature 
$T$, the velocity $\mathbf{v}$, and the chemical potentials $\mu_a$.

For using the distribution functions $P[\tilde{x},t]$ or $P_\mathrm{eq}[\tilde{x}]$ in order to 
calculate averages and correlation functions we must know precisely the integration measure. Here 
it is just the measure of a functional integral which factorizes according to 
\begin{equation}
  \begin{split}
    \mathcal{D}[\partial \mathbf{u},\tilde{\mathbf{j}},\tilde{n}] &= \mathcal{D}(\partial \mathbf{u})
    \, \mathcal{D} \tilde{\mathbf{j}} \, \mathcal{D} \tilde{n} \\
    &= \Bigl( \prod_{\mathbf{r},kp} d[\partial_k u^p(\mathbf{r})] \Bigr)
    \, \Bigl( \prod_{\mathbf{r},i} d[\tilde{j}^i(\mathbf{r})] \Bigr) \\
    &\hspace{4mm} \times \, \Bigl( \prod_{\mathbf{r},a} d[\tilde{n}_a(\mathbf{r})] \Bigr) \, .
  \end{split}
  \label{equation::I_110} 
\end{equation}
Each integration variable for each position $\mathbf{r}$ and for each index $k$, $p$, $i$, or $a$ 
is independent. For each integration variable we integrate from $-\infty$ to $+\infty$. Especially, 
all the $d \times d$ components of the strain tensor $\partial_k u^p(\mathbf{r})$ are considered 
to be independent. Here we do not care about that the strain tensor may be a spatial derivative 
of a displacement field.

We have restricted the discussion to the isothermal case where we started with the Langevin equations 
of Sec.\ \ref{section::07}. Analogously, we can extend the procedure to the general case where we start 
with the Langevin equations of Sec.\ \ref{section::08}. More precisely we start with the Langevin equations 
in the GENERIC form \eqref{equation::H_180} and derive the related Fokker-Planck equation. In this latter 
case the discussion is closer to Sec.\ 6 of our previous publication \cite{Ha16} where now 
$\tilde{n}_a(\mathbf{r},t)$, $\tilde{\mathbf{j}}(\mathbf{r},t)$, $\tilde{\varepsilon}(\mathbf{r},t)$, 
and $\partial_k u^p(\mathbf{r},t)$ are the relevant variables. For the global thermal equilibrium we 
find a Boltzmann distribution function as a stationary solution which in Ref.\ \onlinecite{Ha16} is given 
for a liquid by Eq.\ (276) together with Eq.\ (274). For a crystal this distribution function reads 
\begin{equation}
  \begin{split}
    P_\mathrm{eq}[\tilde{\varepsilon},\partial \mathbf{u},\tilde{\mathbf{j}},\tilde{n}] = 
    \,& Z^{-1} \, \exp\Bigl\{ - \Bigl( E[\tilde{\varepsilon}] 
    - T \, S[\tilde{\varepsilon},\partial \mathbf{u},\tilde{\mathbf{j}},\tilde{n}] \\
    &\hspace{10mm} - \mathbf{v} \cdot \mathbf{P}[\tilde{\mathbf{j}}] 
    - \sum_a \mu_a \, N_a[\tilde{n}] \Bigr) / k_B T \Big\} 
  \end{split}
  \label{equation::I_120} 
\end{equation}
where the coarse-grained entropy functional $S$ is given by Eq.\ \eqref{equation::H_150}. On the other 
hand, the conserved quantities are given by the energy $E$ defined in \eqref{equation::H_140} and by 
the momentum $\mathbf{P}$ and the particle numbers $N_a$ defined in Eq.\ \eqref{equation::I_100}. The 
measure of the functional integral \eqref{equation::I_110} must be extended by the additional factor 
$\mathcal{D} \tilde{\varepsilon} = \prod_\mathbf{r} d[\tilde{\varepsilon}(\mathbf{r})]$ to include 
the energy density.

Eventually, we find that all considerations and all conclusions which we have derived and described 
for the hydrodynamics of a liquid in Sec.\ 6 of our previous paper \cite{Ha16} may be applied also 
and can be transferred directly to the continuum mechanics of an elastic crystal of our present paper. 
The only and essential requirement is that we consider the strain tensor $\partial_k u^p(\mathbf{r},t)$ 
as the natural variable but not the displacement field $\mathbf{u}(\mathbf{r},t)$.

\subsection{Time-evolution equations for the microscopic densities}
\label{section::09c}
We have derived explicit formulas for the displacement field 
$\hat{\mathbf{u}}(\mathbf{r}) = \mathbf{u}(\mathbf{r},\Gamma)$ depending on the phase-space variables 
$\Gamma = ( \mathbf{r}_{ai},\mathbf{p}_{ai} )$. Whether we consider only the leading-order approximation 
or whether we include the correction terms these formulas are given by Eqs.\ \eqref{equation::G_370} 
or \eqref{equation::G_410}, respectively. However, once we consider time-dependent phenomena ruled by 
the time-evolution equations of continuum mechanics all the average values become time dependent, 
including the average particle densities $n_a(\mathbf{r},t)$ and the average displacement field 
$\mathbf{u}(\mathbf{r},t)$. As a consequence the formulas \eqref{equation::G_370} and \eqref{equation::G_410} 
become implicitly time dependent. Thus, including the correction terms we find
\begin{equation}
  \begin{split}
    \hat{u}^k(\mathbf{r}_1) = \,& u^k(\mathbf{r}_1,t) - \sum_{lb} \int d^dr_2 \, \mathcal{N}^{kl} 
    \, w( \mathbf{r}_1 - \mathbf{r}_2 ) \\
    &\times \Bigl( [ \partial_l n_{0,b}(\mathbf{r}_2) ] - \overleftarrow{\partial}_{2,n} 
    \, \Delta B_{a,l}^{\prime \, \ \ n}(\mathbf{r}_2) \Bigr) \\
    &\times [ \hat{n}_b(\mathbf{r}_2) - n_b(\mathbf{r}_2,t) ] 
  \end{split}
  \label{equation::I_130}
\end{equation}
which on the right-hand side depends \emph{implicitly} on the time $t$.

However, we have derived our time-evolution equation with the assumption that all relevant variables 
$\hat{\tilde{x}}_i(\mathbf{r}) = \tilde{x}_i(\mathbf{r},\Gamma)$ which are functions of the phase-space 
variables $\Gamma = ( \mathbf{r}_{ai},\mathbf{p}_{ai} )$ do not depend on the time $t$. This fact follows  
from Eqs.\ \eqref{equation::G_550} and \eqref{equation::G_560}. If we calculate the time derivatives 
of the average relevant variables defined in Eq.\ \eqref{equation::G_550} we find
\begin{equation}
  \begin{split}
    \partial_t \tilde{x}_i(\mathbf{r},t) &= \partial_t \langle \hat{\tilde{x}}_i(\mathbf{r}) \rangle_t 
    = \partial_t \mathrm{Tr} \{ \hat{\tilde{\varrho}}(t) \, \hat{\tilde{x}}_i(\mathbf{r}) \} \\
    &= \mathrm{Tr} \{ [ \partial_t \hat{\tilde{\varrho}}(t) ] \, \hat{\tilde{x}}_i(\mathbf{r}) \} \, .
  \end{split}
  \label{equation::I_140}
\end{equation}
The last equality sign follows from Eq.\ \eqref{equation::G_560} which requires
\begin{equation}
  \partial_t \hat{\tilde{x}}_i(\mathbf{r}) = \partial_t \tilde{x}_i(\mathbf{r},\Gamma) = 0 \ .
  \label{equation::I_150}
\end{equation}
For most relevant variables this last condition is trivial. However, it is not trivial for the 
displacement field defined in Eq.\ \eqref{equation::I_130} because on the right-hand side we 
have an implicit time dependence via the average particle densities $n_a(\mathbf{r},t)$ and the 
average displacement field $\mathbf{u}(\mathbf{r},t)$. Thus, inserting the displacement field 
\eqref{equation::I_130} into the condition \eqref{equation::I_150} we obtain the linear integral 
equation 
\begin{equation}
  \begin{split}
    \partial_t u^k(\mathbf{r}_1,t) = &\, - \sum_{lb} \int d^dr_2 \, \mathcal{N}^{kl} 
    \, w( \mathbf{r}_1 - \mathbf{r}_2 ) \\
    &\times \Bigl( [ \partial_l n_{0,b}(\mathbf{r}_2) ] - \overleftarrow{\partial}_{2,n} 
    \, \Delta B_{a,l}^{\prime \, \ \ n}(\mathbf{r}_2) \Bigr) \\
    &\times [ \partial_t n_b(\mathbf{r}_2,t) ] \, .
  \end{split}
  \label{equation::I_160}
\end{equation}
On the left-hand side the time derivative of the displacement field $\partial_t \mathbf{u}(\mathbf{r},t)$ 
is the inhomogeneity of the linear equation. If we solve this linear equation with respect to 
$\partial_t n_a(\mathbf{r},t)$ then we obtain a time-evolution equation for the \emph{microscopic} 
density field which is a supplement to the macroscopic time-evolution equations for the continuum 
mechanics \eqref{equation::G_H70}-\eqref{equation::G_H90} and \eqref{equation::H_210}.

For more generality we consider the variations of the averages $\delta \mathbf{u}(\mathbf{r},t)$ 
and $\delta n_a(\mathbf{r},t)$. These variations may not be independent. Rather, they are related 
to each other by the requirement that the displacement field \eqref{equation::I_130} does not depend 
on these more general variations. Thus, we obtain a similar but more general linear integral equation 
which reads
\begin{equation}
  \begin{split}
    \delta u^k(\mathbf{r}_1,t) = &\, - \sum_{lb} \int d^dr_2 \, \mathcal{N}^{kl} 
    \, w( \mathbf{r}_1 - \mathbf{r}_2 ) \\
    &\times \Bigl( [ \partial_l n_{0,b}(\mathbf{r}_2) ] - \overleftarrow{\partial}_{2,n} 
    \, \Delta B_{a,l}^{\prime \, \ \ n}(\mathbf{r}_2) \Bigr) \\
    &\times \, \delta n_b(\mathbf{r}_2,t) \, .
  \end{split}
  \label{equation::I_170}
\end{equation}

In Sec.\ \ref{section::04} we have considered the minimization of the free energy for a solid crystal. 
From the necessary conditions of the minimization procedure we have derived linear-response equations 
for the particle densities $\delta n_a(\mathbf{r},t)$ where on the right-hand sides the chemical potentials 
$\delta \mu_a(\mathbf{r},t)$ are the inhomogeneities. These linear-response equations are given by 
Eq.\ \eqref{equation::D_450}. The chemical potentials may be replaced in favor of the coarse-grained 
densities $\delta \tilde{n}_a(\mathbf{r},t)$ and the displacement field $\delta \mathbf{u}(\mathbf{r},t)$ 
by using Eq.\ \eqref{equation::D_690}. We have calculated an explicit solution of the linear-response 
equations which is given by Eq.\ \eqref{equation::D_760} together with Eq.\ \eqref{equation::D_790} and 
the correction term 
\eqref{equation::D_810}. The correction term may be rewritten in the more compact form \eqref{equation::E_140} 
together with the correction functions \eqref{equation::E_150} and \eqref{equation::E_160}. Here, we 
write this solution as
\begin{equation}
  \begin{split}
    \delta n_a(\mathbf{r},t) = \,& - \Bigl( [ \partial_i n_{0,a}(\mathbf{r}) ] 
    - \Delta B_{a,i}^{\prime \, \ \ j}(\mathbf{r}) \, \overrightarrow{\partial}_j \Bigr) 
    \, \delta u^i(\mathbf{r},t) \\
    &+ \sum_b \, \Bigl( \frac{ n_{0,a}( \mathbf{r} ) }{ \tilde{n}_{0,a} } \, \delta_{ab}
    + \Delta A^\prime_{ab}(\mathbf{r}) \Bigr) \, \delta \tilde{n}_b(\mathbf{r},t) \, .
  \end{split}
  \label{equation::I_180}
\end{equation}
We may ask the question to which extent these particle densities also solve our present linear integral 
equation \eqref{equation::I_170}. First, we see that the first line of Eq.\ \eqref{equation::I_180} is 
the \emph{particular} solution of Eq.\ \eqref{equation::I_170}. As long as the correction functions 
$\Delta B_{a,i}^{\prime \, \ \ j}(\mathbf{r})$ are omitted the proof follows from the definition of 
the normalization matrix \eqref{equation::G_240} and its inverse matrix in Eq.\ \eqref{equation::G_250}.
If the correction functions are included then we must use the substitution \eqref{equation::G_460} 
so that the proof remains valid up to leading order in the gradient expansion.

Second, we see that the second line of Eq.\ \eqref{equation::I_180} is the 
\emph{homogeneous} solution of Eq.\ \eqref{equation::I_170}. As long as the correction functions 
$\Delta A^\prime_{ab}(\mathbf{r})$ and $\Delta B_{a,i}^{\prime \, \ \ j}(\mathbf{r})$ are omitted 
the proof follows from the orthogonality of the functions $\partial_l n_{0,b}(\mathbf{r})$ and 
$n_{0,a}( \mathbf{r} )$ in the scalar product \eqref{equation::D_020}. If the correction functions 
are added the orthogonality should not be disturbed. Thus, we conclude that the microscopic density 
variations defined in Eq.\ \eqref{equation::I_180} are indeed the correct and general solution of the 
linear integral equation \eqref{equation::I_170}. Since they are derived as the solutions of the 
necessary conditions they are compatible with the constrained minimization of the free energy 
\eqref{equation::D_820} in the isothermal case or the constrained maximization of the entropy 
\eqref{equation::H_050} in the general case where all the macroscopic fields and the displacement 
field are kept fixed.

Now, we consider the special case where the general variation $\delta$ is replaced by the partial 
time derivative $\partial_t$. Then, in analogy we obtain 
\begin{equation}
  \begin{split}
    \partial_t n_a(\mathbf{r},t) = \,& - \Bigl( [ \partial_i n_{0,a}(\mathbf{r}) ] 
    - \Delta B_{a,i}^{\prime \, \ \ j}(\mathbf{r}) \, \overrightarrow{\partial}_j \Bigr) 
    \, [ \partial_t u^i(\mathbf{r},t) ] \\
    &+ \sum_b \, \Bigl( \frac{ n_{0,a}( \mathbf{r} ) }{ \tilde{n}_{0,a} } \, \delta_{ab}
    + \Delta A^\prime_{ab}(\mathbf{r}) \Bigr) \, [ \partial_t \tilde{n}_b(\mathbf{r},t) ] 
  \end{split}
  \label{equation::I_190}
\end{equation}
which is the general solution of Eq.\ \eqref{equation::I_160}. Clearly, Eq.\ \eqref{equation::I_190} 
is the time-evolution equation for the microscopic particle densities $n_a(\mathbf{r},t)$. Again, by 
construction this equation is compatible with the constrained minimization of the free energy 
\eqref{equation::D_820} or the constrained maximization of the entropy \eqref{equation::H_050} where 
all the macroscopic fields and the displacement field are kept fixed.

We note that we have made extensive use of the concept of local thermodynamic equilibrium. In all the 
equations of this subsection the factors within the \emph{large} brackets $( \cdots )$ are expressed 
in terms of the functions $n_{0,a}( \mathbf{r} )$, $\partial_k n_{0,a}( \mathbf{r} )$, 
$\Delta A^\prime_{ab}(\mathbf{r})$, and $\Delta B_{a,i}^{\prime \, \ \ j}(\mathbf{r})$. These functions 
are defined in the global thermodynamic equilibrium for a homogeneous crystal with a perfect periodic 
lattice structure. At a point $P$ in the space time with coordinates $\mathbf{r}_P$ and time $t_P$ this 
homogeneous crystal in equilibrium is adjusted \emph{tangentially} to the nonlinearly deformed crystal in 
the nonequilibrium. Thus, all the above equations are restricted to an epsilon surroundings 
\begin{equation}
  U_\varepsilon(P) = \{ ( \mathbf{r}, t ) \mid | \mathbf{r} - \mathbf{r}_P | < \varepsilon_r 
  \enskip \mbox{and} \enskip \vert t - t_P \vert < \varepsilon_t \} 
  \label{equation::I_200} 
\end{equation}
in space and time where $\varepsilon_r = \ell$ is identified by the mesoscopic length scale and 
$\varepsilon_r / \varepsilon_t = c_s$ is the mean sound velocity. However, we may repeat the 
considerations for many epsilon surroundings around many space-time points $( \mathbf{r}_P, t_P )$. 
In this way, the final results can be extended to the whole space.

Finally, we ask the question whether there exists a time-evolution equation also for the momentum density 
$\mathbf{j}(\mathbf{r},t)$. Here, we refer to Galilean invariance which implies the constraint condition 
\eqref{equation::D_120} where the mass density $\rho(\mathbf{r},t)$ is defined in 
Eq.\ \eqref{equation::D_130}. For time dependent fields this constraint condition may be written as 
\begin{equation}
  \mathbf{j}(\mathbf{r},t) = \rho(\mathbf{r},t) \, \mathbf{v}(\mathbf{r},t)
  = \sum_a m_a \, n_a(\mathbf{r},t) \, \mathbf{v}(\mathbf{r},t) \ .
  \label{equation::I_210}
\end{equation}
We note that the microscopic momentum density $\mathbf{j}(\mathbf{r},t)$, the microscopic mass density 
$\rho(\mathbf{r},t)$, and the microscopic particle densities $n_a(\mathbf{r},t)$ vary on short scales 
in the space and show the microscopic lattice structure of the crystal. On the other hand, the velocity 
field $\mathbf{v}(\mathbf{r},t)$ varies only slowly on large and macroscopic scales. As a consequence 
the microscopic momentum density $\mathbf{j}(\mathbf{r},t)$ is not an independent function. Rather, the 
time-evolution of $\mathbf{j}(\mathbf{r},t)$ follows directly from the time-evolution of $n_a(\mathbf{r},t)$. 
Hence, the same is true also for the related time-evolution equations.

\subsection{Normal crystals with no defects and no diffusion}
\label{section::09d}
The time-evolution equations \eqref{equation::G_H70}-\eqref{equation::G_H90} include the effects of 
the diffusion of particles in a natural way. This fact is clearly seen in the explicit formulas of the 
current densities and the lattice-structure velocity \eqref{equation::G_I00}-\eqref{equation::G_I20}. 
In Eq.\ \eqref{equation::G_I00} the diagonal coefficients $D_{ab}^{kl}$ describe the explicit 
diffusion of the particles. On the other hand, in Eq.\ \eqref{equation::G_I20} the diagonal 
coefficients $\zeta^{pq}$ describe the implicit effect of the diffusion which results in a decoupling 
of the motion of the lattice structure from the motion of the material. Furthermore, there are the 
off-diagonal coefficients $K_a^{kp}$ which couple the two effects with each other.

However, in a normal crystal there are strong repulsive interaction forces between the particles. These 
interactions strongly suppress any diffusion effects but rather force the particles to sit on lattice 
sites. There must be exactly one particle on each lattice site. The strong repulsive interactions do 
not allow two or more particles sitting on the same site. As a consequence, the material is coupled 
strictly to the lattice structure where no diffusion is possible. Exceptions may be point defects like 
\emph{vacancies} which are empty lattice sites or \emph{interstitial} particles on positions in between 
the lattice sites. In Eq.\ \eqref{equation::D_780} the vacancy densities $c_a(\mathbf{r},t)$ are 
defined which are positive if vacancies are dominant and negative if interstitial particles are dominant.

However, these point defects will have large activation energies $\Delta E_a$ which for usual temperatures 
$T$ suppress their numbers per volume strongly by a factor $\exp( - \Delta E_a / k_B T)$. As a result, 
the vacancy densities $c_a(\mathbf{r},t)$ will be constrained to values close to zero. On the other hand 
there will be large energy barriers $\Delta E_a$ which the point defects must overcome when moving and 
hopping through the lattice structure. Thus, the diffusion of the point defects will be suppressed by a 
similar exponential factor $\exp( - \Delta E_a / k_B T)$. As a consequence all the transport coefficients 
$D_{ab}^{kl}$, $K_a^{kp}$, and $\zeta^{pq}$ are strongly suppressed by the exponential factor. 

In the ideal limit of no point defects and no diffusion we may set all these coefficients to zero so that
\begin{equation}
  D_{ab}^{kl} = 0 \, , \qquad K_a^{kp} = 0 \, , \qquad \zeta^{pq} = 0 \ .
  \label{equation::I_220}
\end{equation}
Then, in the particle current densities \eqref{equation::G_I00} only the first term survives 
which is the reversible contribution. Thus, we find the exact relation
\begin{equation}
  \tilde{\mathbf{j}}_a(\mathbf{r},t) = \tilde{n}_a(\mathbf{r},t) \, \mathbf{v}(\mathbf{r},t) \ .
  \label{equation::I_230}
\end{equation}
Now, we try to solve the continuity equations \eqref{equation::G_H70} \emph{exactly} by the ansatz for 
the particle densities
\begin{equation}
  \tilde{n}_a(\mathbf{r},t) = \tilde{n}_{0,a}( \mathbf{r} - \mathbf{u}(\mathbf{r},t) ) 
  \, \frac{ \partial r_0 }{ \partial r } 
  \label{equation::I_240} 
\end{equation}
where $\tilde{n}_{0,a}( \mathbf{r} )$ may be any functions of the space coordinates $\mathbf{r}$ 
which are constant in time and which are interpreted as a reference densities. The particle densities 
\eqref{equation::I_240} are expressed in terms of the displacement field $\mathbf{u}(\mathbf{r},t)$. 
While the displacement field appears explicitly in the arguments of the reference densities a further 
implicit dependence arises via the Jacobi determinant \eqref{equation::D_320} which is defined as the 
determinant of the Jacobi matrix \eqref{equation::D_330}. We insert the expressions \eqref{equation::I_240} 
and \eqref{equation::I_230} into the continuity equations \eqref{equation::G_H70} and put these terms 
onto the left-hand side. Then, after some longer calculations we \emph{exactly} obtain
\begin{equation}
  \frac{ \partial \tilde{n}_a }{ \partial t } + \nabla \cdot \tilde{\mathbf{j}}_a = 
  - \, \frac{ \partial r_0 }{ \partial r } \, \nabla_0 \cdot  
  \Bigl[ \tilde{n}_{0,a} \, \Bigl( \frac{ \partial \mathbf{u} }{ \partial t } 
  + ( \mathbf{v} \cdot \nabla ) \, \mathbf{u} - \mathbf{v} \Bigr) \Bigr] 
  = 0 \ .
  \label{equation::I_250}
\end{equation}
On the right-hand side of the first equality sign $\nabla_0$ is the nabla operator which is defined by 
the spatial derivatives with respect to the curvilinear coordinates $\mathbf{r}_0$. It is related to 
the normal nabla operator $\nabla$ by the chain rule which means that the latter operator is multiplied 
by the inverse of the Jacobi matrix \eqref{equation::D_330}. Beyond that we identify 
$\tilde{n}_{0,a} = \tilde{n}_{0,a}( \mathbf{r} - \mathbf{u}(\mathbf{r},t) )$. The continuity equations 
are satisfied if in Eq.\ \eqref{equation::I_250} the last equality sign is true. This requirement 
implies that the expression in the large brackets $( \cdots )$ must be zero so that
\begin{equation}
  \frac{ d \mathbf{u} }{ d t } = \frac{ \partial \mathbf{u} }{ \partial t } 
  + ( \mathbf{v} \cdot \nabla ) \, \mathbf{u} = \mathbf{v} \ .
  \label{equation::I_260}
\end{equation}
This latter equation equals the time-evolution equation for the displacement field in the form 
\eqref{equation::G_E20} where all the transport coefficients \eqref{equation::I_220} are zero so that 
Eq.\ \eqref{equation::G_H60} reduces to $\mathbf{v}_\mathrm{struc} = \mathbf{v}$. Thus, we conclude 
that the particle densities \eqref{equation::I_240} together with the particle current densities 
\eqref{equation::I_230} \emph{exactly} solve the continuity equations \eqref{equation::G_H70}.

According to Eq.\ \eqref{equation::D_150} the particle densities are combined into the mass density. 
Thus, in analogy to Eq.\ \eqref{equation::I_240} we may write the mass density as
\begin{equation}
  \tilde{\rho}(\mathbf{r},t) = \tilde{\rho}_0( \mathbf{r} - \mathbf{u}(\mathbf{r},t) ) 
  \, \frac{ \partial r_0 }{ \partial r } 
  \label{equation::I_270} 
\end{equation}
where $\tilde{\rho}_0( \mathbf{r}_0 )$ is the reference mass density. The related mass current 
density $\tilde{\mathbf{j}}(\mathbf{r},t)$ is defined in Eq.\ \eqref{equation::D_170} which 
because of Galilean invariance equals the momentum density. As a consequence, the mass density 
\eqref{equation::I_270} and the related mass density current exactly solve a continuity equation, 
too.

Now, we turn to the continuity equation for the momentum density which is given by 
Eq.\ \eqref{equation::G_H80} together with the momentum current density \eqref{equation::G_I10}. 
As usual in continuum mechanics we may subtract a continuity equation for the mass density multiplied 
by a velocity. Then, we obtain the \emph{Euler} equation
\begin{equation}
  \tilde{\rho} \, \frac{ d \mathbf{v} }{ d t } = \tilde{\rho} \, \Bigl[ \frac{ \partial \mathbf{v} }{ \partial t } 
  + ( \mathbf{v} \cdot \nabla ) \, \mathbf{v} \Bigr] = \tilde{\mathbf{f}} 
  \label{equation::I_280}
\end{equation}
which may be interpreted as the equation of motion for the continuum mechanics. On the right-hand 
side the force density $\tilde{\mathbf{f}} = \tilde{f}^i \, \mathbf{e}_i$ is given by the vector 
components
\begin{equation}
  \tilde{f}^i = \partial_k [ \sigma_\mathrm{tot}^{ik} + \Lambda^{ik,jl} \, \partial_l v_j 
  - \tilde{\Pi}_\mathrm{fluc}^{ik} ] + \tilde{\rho} \, g^i \, .
  \label{equation::I_290}
\end{equation}
The term in square brackets results from minus the divergence of the momentum current density 
\eqref{equation::G_I10} where the first or kinetic term is omitted. The last term is an optional 
contribution in order to include gravitational effects where $\mathbf{g} = - \nabla \phi$ is 
the gravitational acceleration expressed in terms of the Newton gravitational potential 
$\phi = \phi(\mathbf{r})$.

Following Eqs.\ \eqref{equation::G_A80}-\eqref{equation::G_B00} the reversible contributions 
may be combined into the total stress tensor
\begin{equation}
  \sigma_\mathrm{tot}^{ik} = - p_\mathrm{tot} \, \delta^{ik} + \sigma^{\prime ik} 
  \label{equation::I_300}
\end{equation}
which itself may be decomposed into a scalar contribution described by the total pressure 
$p_\mathrm{tot}$ and an irreducible tensor contribution described by the irreducible or 
traceless stress tensor $\sigma^{\prime ik}$. In linear response we may write
\begin{equation}
  \sigma_\mathrm{tot}^{ik} = \sigma_{0,\mathrm{tot}}^{ik} + C_\mathrm{tot}^{ik,jl} \, u_{jl}
  \label{equation::I_310}
\end{equation}
where $C_\mathrm{tot}^{ik,jl}$ are some total elastic constants and where $u_{jl}$ is the 
Lagrange strain tensor. For polycrystalline or amorphous materials which on macroscopic scales 
imply isotropic symmetry we may write 
\begin{equation}
  C_\mathrm{tot}^{ik,jl} = \lambda \, \delta^{ik} \delta^{jl} 
  + \mu \, [ \delta^{ij} \delta^{kl} + \delta^{il} \delta^{kj} ] 
  \label{equation::I_320}
\end{equation}
in terms of the Lam\'e coefficients $\lambda$ and $\mu$. In this way we recover the elasticity theory 
for the deformation of continuum materials as described in the standard text books \cite{LL07}.

Once again we summarize. For a normal crystal the diffusion of particles is prohibited so that the 
related coefficients \eqref{equation::I_220} are all zero. The continuity equations for the particle 
densities can be solved exactly. The particle densities $\tilde{n}_a(\mathbf{r},t)$ can be expressed 
exactly in terms of the displacement field $\mathbf{u}(\mathbf{r},t)$. As a consequence the particle 
densities can be eliminated completely and exactly in favor of the displacement field. At the end 
there remain two equations for the two relevant variables $\mathbf{u}(\mathbf{r},t)$ and 
$\mathbf{v}(\mathbf{r},t)$. First we have the time-evolution equation for the displacement field 
\eqref{equation::I_260} which describes the motion of the lattice structure strictly coupled to the 
motion of the material. Second there is the Euler equation \eqref{equation::I_280} together with 
the force densities \eqref{equation::I_290} which represents the equation of motion for the continuum 
mechanics.

For simplicity we have restricted the discussions and the derivations to the isothermal case where 
the time-evolution equations of Sec.\ \ref{section::07} were used. However, an extension to the 
general case starting with the time-evolution equations of Sec.\ \ref{section::08} is straight forward. 
In the latter case there are some more off-diagonal transport coefficients which are related to particle 
diffusion. In Eqs.\ \eqref{equation::H_290} and \eqref{equation::H_320} the coefficients $G_a^{kl}$ 
and $\xi^{kp}$ imply explicit and implicit particle diffusion, respectively, which both are caused by 
a temperature gradient. Due to the Onsager symmetry the two coefficients additionally imply an entropy 
current in Eq.\ \eqref{equation::H_310}.

In the ideal limit of no point defects and no diffusion the two additional transport coefficients 
are zero as well so that beyond Eq.\ \eqref{equation::I_220} we have $G_a^{kl} = 0$ and $\xi^{kp} = 0$. 
As a consequence, in the general case for non-constant temperatures we may repeat the above derivations 
starting with Eqs.\ \eqref{equation::I_230} and \eqref{equation::I_240}. As a result we obtain the same 
two time-evolution equations for the displacement field $\mathbf{u}(\mathbf{r},t)$ and for the velocity 
field $\mathbf{v}(\mathbf{r},t)$ which are Eqs.\ \eqref{equation::I_260} and \eqref{equation::I_280}. 
Beyond that we must consider the time-evolution equation for the energy density \eqref{equation::H_210} 
or alternatively for the entropy density \eqref{equation::H_220}. In this way for the general case we 
obtain an additional time-evolution equation which includes the transport of heat and the production 
of entropy.

\subsection{Cartesian and curvilinear coordinates}
\label{section::09e}
For the description of the deformations we extensively use the methods of General Relativity and of 
Riemannian Geometry \cite{LL02}. More precisely, we use the tensor formalism with upper and lower 
indices. While the deformation of a crystal can not cause a curved space there will be curvilinear 
coordinates. However, there is a big difference. In nonlinear elasticity theory we have \emph{two} 
coordinate systems which must be handled in parallel. First, we have the laboratory frame with Cartesian 
coordinates. Second we have a coordinate system which is fixed to the lattice structure of the crystal 
and which becomes \emph{curvilinear} when strain is applied leading to deformations. We denote the 
coordinates by $r^i$ and by $r_0^p$, respectively. Thus, in the tensor formalism we have two types 
of indices. First, there are Cartesian indices which are denoted by the letters $i,j,k,l,\ldots$. 
Second, there are curvilinear indices which are denoted by the letters $p,q,r,s,\ldots$. The indices 
are written as \emph{lower} indices for \emph{covariant} transformations and as \emph{upper} indices 
for \emph{contravariant} transformations.

The appropriate measures for the local strain and the local deformation are the local transformation 
matrices between the two coordinate systems
\begin{eqnarray}
  e_i^p(\mathbf{r}) &=& \frac{ \partial r_0^p }{ \partial r^i } = \delta_i^p - \partial_i u^p(\mathbf{r}) \, ,
  \label{equation::I_330} \\
  e_p^i(\mathbf{r}) &=& \frac{ \partial r^i }{ \partial r_0^p } = \delta_p^i + \partial_{0,p} u^i(\mathbf{r}) \, .
  \label{equation::I_340} 
\end{eqnarray}
Alternative and even better measures are the metric tensor and the inverse metric tensor of the 
curvilinear coordinates which are defined by
\begin{eqnarray}
  g_{0,pq}(\mathbf{r}) &=& e_p^i(\mathbf{r}) \, e_q^j(\mathbf{r}) \, \delta_{ij} 
  = \delta_{pq} + 2 \, u_{pq}(\mathbf{r}) \, ,
  \label{equation::I_350} \\
  g_0^{pq}(\mathbf{r}) &=& e_i^p(\mathbf{r}) \, e_j^q(\mathbf{r}) \, \delta^{ij} 
  = \delta^{pq} - 2 \, u^{pq}(\mathbf{r}) \, . \hspace{8mm}
  \label{equation::I_360} 
\end{eqnarray}
While the transformation matrices describe deformations and rotations, the metric tensors are symmetric 
in the indices and describe deformations only. Right to the last equality signs shifted by a constant 
tensor and optionally multiplied by a factor $2$ or changed by the sign we have defined the commonly 
used measures for strains and deformations. These are
\begin{eqnarray}
  \partial_i u^p(\mathbf{r}) &=& \mbox{spatial linear strain tensor} \, ,
  \label{equation::I_370} \\
  \partial_{0,p} u^i(\mathbf{r}) &=& \mbox{material linear strain tensor} \, ,
  \label{equation::I_380} \\
  u_{pq}(\mathbf{r}) &=& \mbox{Lagrange strain tensor} \, ,
  \label{equation::I_390} \\
  u^{pq}(\mathbf{r}) &=& \mbox{Grabert-Michel strain tensor} \, , \hspace{8mm}
  \label{equation::I_400} 
\end{eqnarray}
respectively. These strain tensors occur conjugate to the stress tensors defined in 
Eqs.\ \eqref{equation::F_840}-\eqref{equation::F_880}. The conjugate relation between strain and stress 
is clearly seen in the differential thermodynamic relations which are given in different variants by 
Eqs.\ \eqref{equation::F_640}, \eqref{equation::F_720}, and \eqref{equation::F_800}. Here, an important 
point is that the total differentials of the strain tensors must exist and that the transformation 
formulas have the forms
\begin{eqnarray}
  du_{pq}(\mathbf{r}) &=& e_p^i(\mathbf{r}) \, [ \partial_{0,q} du^j(\mathbf{r}) \, \delta_{ij} ] \nonumber \\
  &=& g_{0,pr}(\mathbf{r}) \, g_{0,qs}(\mathbf{r}) \, du^{rs}(\mathbf{r}) \, ,
  \label{equation::I_410} \\
  du^{pq}(\mathbf{r}) &=& e_i^p(\mathbf{r}) \, [ \partial_j du^q(\mathbf{r}) \, \delta^{ij} ] \nonumber \\
  &=& g_0^{pr}(\mathbf{r}) \, g_0^{qs}(\mathbf{r}) \, du_{rs}(\mathbf{r}) 
  \label{equation::I_420} 
\end{eqnarray}
which are analogous to the transformation formulas of the stress tensors \eqref{equation::F_740}, 
\eqref{equation::F_750}, \eqref{equation::F_780}, \eqref{equation::F_810}, and \eqref{equation::F_820}.

However, there are only \emph{four} strain tensors defined in Eqs.\ \eqref{equation::I_370}-\eqref{equation::I_400} 
while there are \emph{five} stress tensors defined in Eqs.\ \eqref{equation::F_840}-\eqref{equation::F_880}. 
We are missing a strain tensor $u_{ij}(\mathbf{r}) = u^{ij}(\mathbf{r})$ with \emph{two} Cartesian indices 
which is conjugate to the Cauchy stress tensor $\sigma^{ij}(\mathbf{r}) = \sigma_{ij}(\mathbf{r})$ defined 
in Eq.\ \eqref{equation::F_860}. The reason for this absence is that Cartesian indices are related to straight 
and orthogonal coordinates which can not describe deformations. An alternative explanation is the fact that 
we can not construct a \emph{total differential} $du_{ij}(\mathbf{r}) = du^{ij}(\mathbf{r})$ by transformation 
formulas similar like Eqs.\ \eqref{equation::I_410} and \eqref{equation::I_420}.

An important result are the necessary conditions for the global thermodynamic equilibrium \eqref{equation::F_660} 
which here we write once again so that
\begin{equation}
  \partial_k \, \sigma_p^{\ k}(\mathbf{r}) = 0 \, , \qquad 
  \mathbf{v}(\mathbf{r}) = \mathbf{v} \, , \qquad
  \mu_a(\mathbf{r}) = \mu_a \, .
  \label{equation::I_430} 
\end{equation}
If the crystal does not move then the velocity is zero so that the second condition is trivial. Within the 
crystal we have diffusion of the particles which equilibrates the system. For this process the first 
condition for the stress tensor rules the adjustment of the lattice structure while the third condition 
for the chemical potentials rules the magnitudes of the particle densities. 

In the first condition of \eqref{equation::I_430} the details of the indices of the stress tensor are 
important. This stress tensor which is named in Eq.\ \eqref{equation::F_870} has one lower curvilinear 
index $p$ and one upper Cartesian index $k$. Since $\partial_k$ represents the partial space derivatives 
with respect to the Cartesian coordinates the upper index of the stress tensor $k$ must be Cartesian, too. 
However, the lower curvilinear index $p$ is nontrivial. The same stress tensor appears also in the final 
time-evolution equations \eqref{equation::G_H70}-\eqref{equation::G_H90} where 
Eqs.\ \eqref{equation::G_I00}-\eqref{equation::G_I20} are inserted for the right-hand sides. First, in 
the particle current densities \eqref{equation::G_I00} it occurs in the third term with offdiagonal 
coefficients $K_a^{kp}$. Second, in the velocity of the lattice structure \eqref{equation::G_I20} it 
occurs in the third term with transport coefficients $\zeta^{pq}$. Thus, the time-evolution equations 
support once again that $\sigma_p^{\ k}(\mathbf{r})$ is the proper stress tensor which must be used in 
the formulation of the nonlinear elasticity theory.

We may ask the question whether we may use alternatively the second Piola stress tensor 
$\sigma^{pq}(\mathbf{r})$ with two upper curvilinear indices defined in Eq.\ \eqref{equation::F_840}. 
In this case we must insert the transformation formula
\begin{equation}
  \sigma_p^{\ k}(\mathbf{r}) = g_{0,pr}(\mathbf{r}) \, e_s^k(\mathbf{r}) \, \sigma^{rs}(\mathbf{r})
  \, .
  \label{equation::I_440} 
\end{equation}
Applying the partial space derivative $\partial_k$ there will appear derivatives of the metric tensor 
and of the transformation matrix which result in contributions involving \emph{Christoffel symbols}. In 
this way, the formulas become more complicated. For this reason, while in our treatment of the nonlinear 
elasticity theory we use the transformation matrices and the metric tensors with pleasure to describe 
strains and deformations we have avoided to use Christoffel symbols and covariant derivatives.

For the nonlinear elasticity theory the advantage of the tensor formalism as used in General Relativity 
\cite{LL02} is the possibility that the different coordinates, strain tensors, and stress tensors can 
be classified in a very efficient and transparent way. The notation of the formulas with upper and lower 
indices distinguished between Cartesian and curvilinear coordinates is very clear. The several different 
thermodynamic relations and the transformation formulas are easy to understand.

The situation is very different in the conventional literature about elasticity theory as e.g.\ the book of 
Marsden and Hughes \cite{Ma83} and the book of Truesdell and Noll \cite{TN92}. Here, a matrix notation with 
no indices is used where the different strain and stress tensors are defined with respect to the particular 
physical situation considered. While the physical outcomes are the same, the mathematical notation is 
much more difficult to oversee. Thus, we recommend to use the tensor formalism of General Relativity 
and Riemannian Geometry for the Cartesian and the curvilinear coordinate system in order to obtain a much 
simpler and more transparent mathematical formalism.

\subsection{Elastic constants}
\label{section::09f}
The first stage of the continuum mechanics of deformed crystals is the theory of the static phenomena 
in the thermodynamic equilibrium. In our case it is a conventional density-functional theory which starts 
with the microscopic Helmholtz free energy $F[T,\mathbf{j},n]$. We then derive a macroscopic free energy 
$F[T,\mathbf{u},\tilde{\mathbf{j}},\tilde{n}]$ by a constraint minimization procedure where the macroscopic 
relevant variables are kept fixed. In this way we obtain the free energy as the integral of the free energy 
\emph{density} $f(T,\partial \mathbf{u},\tilde{\mathbf{j}},\tilde{n})$ over the whole space which is 
defined in Eq.\ \eqref{equation::E_520}. The free energy becomes local in space so that the theory is 
extended to nonequilibrium by the concept of the local thermodynamic equilibrium. Thus, the relevant 
physical properties of the crystal system are described by the local free energy density 
$f(T,\partial \mathbf{u},\tilde{\mathbf{j}},\tilde{n})$. 

In order to proceed for our theory of continuum mechanics we need the free energy density 
$f(T,\partial \mathbf{u},\tilde{\mathbf{j}},\tilde{n})$ as an explicit function of the relevant 
variables. The dependence on the momentum density $\mathbf{j}$ follows exactly from Galilean 
invariance. According to Eq.\ \eqref{equation::E_510} the \emph{kinetic energy} is separated which 
is given explicitly and exactly by Eq.\ \eqref{equation::E_500}. The remaining free energy density 
$f(T,\partial \mathbf{u},\tilde{n})$ describes the local internal properties of the crystal. 
We may try to calculate this latter function within first principles from the underlying microscopic 
theory of many interacting particles. This, however, is not possible in practice. Alternatively, 
we may construct a model function with some free parameters and adjust these parameters so that 
the predictions of the theory agree with experimental data. In this way, we obtain a 
\emph{phenomenological} theory for the static properties of the local thermodynamic equilibrium.

In case of small deformations of the crystal and of small deviations from the thermodynamic 
equilibrium we may expand the free energy density into a Taylor series up to second order. From 
the first order terms we obtain the stress tensor and the chemical potentials of the thermodynamic 
equilibrium. From the second order terms we obtain the \emph{elastic constants} which describe the 
deviations of the stress tensor and the chemical potentials from the equilibrium values in linear 
response. A large fraction of our paper is devoted to calculate these elastic constants. At the 
end, for the Taylor expansion we have obtained Eq.\ \eqref{equation::F_260} which here we write 
once again in the alternative form
\begin{equation}
  \begin{split}
    f(T,\mathsf{u},\tilde{n}) = \,& f_0(T,\tilde{n}_0) + \sigma_0^{ij} \, u_{ij} 
    + \sum_a \mu_{0,a} \, [ \tilde{n}_a - \tilde{n}_{0,a} ] \\
    &+ \frac{ 1 }{ 2 } \, C_n^{ij,kl} \, u_{ij} \, u_{kl} 
    - \sum_a \mu_a^{ij} \, \frac{ [ \tilde{n}_a - \tilde{n}_{0,a} ] }{ \tilde{n}_{0,a} } \, u_{ij} \\
    &+ \frac{ 1 }{ 2 } \sum_{ab} \nu_{ab} \, \frac{ [ \tilde{n}_a - \tilde{n}_{0,a} ] }{ \tilde{n}_{0,a} }
    \, \frac{ [ \tilde{n}_b - \tilde{n}_{0,b} ] }{ \tilde{n}_{0,b} } 
  \end{split}
  \label{equation::I_450} 
\end{equation}
where $u_{ij}$ is the Lagrange strain tensor with two \emph{lower} spatial indices. As a consequence 
in this alternative formula the elastic constants have \emph{upper} spatial indices. The elastic 
constants for the deformations are given by 
\begin{equation}
  C_n^{ij,kl} = C^{ij,kl} + \Delta C_0^{ij,kl} \, .
  \label{equation::I_460} 
\end{equation}
We use the subscript $n$ in order to indicate that these are the elastic constants for deformations 
at constant particle densities $\tilde{n}_a$. This fact arises directly from the Taylor expansion 
\eqref{equation::I_450}. In Eq.\ \eqref{equation::I_460} the first term $C^{ij,kl} = C_{ij,kl}$ 
represents the main elastic constants defined in Eq.\ \eqref{equation::F_130}. The second term 
$\Delta C_0^{ij,kl}$ is defined in Eq.\ \eqref{equation::F_330}. It is needed as a correction to 
compensate the nonlinear terms of the Lagrange strain tensor \eqref{equation::F_050} in the first-order 
Taylor expansion term. Further elastic constants are $\mu_a^{ij}$ for the coupling between deformation 
and change of the particle densities and $\nu_{ab}$ for the change of the particle densities only.

In this subsection we assume that the deviations from the equilibrium are small so that the strain 
tensor $u_{ij}$ is small and hence the metric tensor of the deformed coordinate system 
$g_{0,ij} = \delta_{ij} + 2 \, u_{ij} \approx \delta_{ij}$ is nearly Cartesian. As a consequence, 
all indices $i,j,k,l,\ldots$ are Cartesian. We may transform to curvilinear indices $p,q,r,s,\ldots$
by using a transformation formula of the Lagrange strain tensor from $u_{ij}$ to $u_{pq}$ which 
follows from Eqs.\ \eqref{equation::F_050}, \eqref{equation::F_480}, and \eqref{equation::F_540}. 
However, curvilinear indices bring more complications only but no additional physical insights. 
For this reason, in this subsection we only consider Cartesian indices.

In the general case the deformations and the changes of the particle densities are independent from 
each other which means that the Lagrange strain tensor $u_{ij}$ and the macroscopic particle densities 
$\tilde{n}_a$ are independent variables. Now, we consider the special case of a \emph{normal crystal} 
where the strong repulsive interactions of the particles imply that there is \emph{exactly one particle} 
at each lattice site. In this case the macroscopic particle densities are not independent. Rather, we 
have proved that the formula \eqref{equation::I_240} provides an exact solution of the continuity 
equations. Here, we rewrite this formula as
\begin{equation}
  \tilde{n}_a = \tilde{n}_{0,a} \, \frac{ \partial r_0 }{ \partial r } 
  = \tilde{n}_{0,a} \, g_0^{-1/2} \, .
  \label{equation::I_470} 
\end{equation}
On the right-hand side of the second equality sign we have replaced the Jacobi matrix by the determinant 
of the metric tensor which is defined by
\begin{equation}
  g_0 = \det\{ g_{0,ij} \} = \det\{ \delta_{ij} + 2 \, u_{ij} \} \, .
  \label{equation::I_480} 
\end{equation}
The relation between the Jacobi determinants and the determinants of the metric tensors can be inferred 
from Eqs.\ \eqref{equation::I_330}-\eqref{equation::I_360} if we calculate the determinants of all these 
quantities. In this way we justify the second equality sign in Eq.\ \eqref{equation::I_470}.

In order to proceed we need the Taylor expansion with respect to the Lagrange strain tensor $u_{ij}$ up 
to second order. First, we explicitly evaluate the determinant \eqref{equation::I_480}. Then, we expand 
$g_0^{-1/2}$ up to second order. Finally, we obtain the relative particle density changes
\begin{equation}
  \frac{ \tilde{n}_a - \tilde{n}_{0,a} }{ \tilde{n}_{0,a} } = - \delta^{ij} \, u_{ij} 
  + \frac{ 1 }{ 2 } \, [ \delta^{ij} \delta^{kl} + \delta^{ik} \delta^{jl} + \delta^{il} \delta^{jk} ] 
  \, u_{ij} u_{kl} \, .
  \label{equation::I_490} 
\end{equation}
Now, we insert this result into the Taylor expansion of the free energy \eqref{equation::I_450}. In the 
first-order term we must keep the nonlinear contribution of Eq.\ \eqref{equation::I_490} while in the 
second-order term the linear contribution is sufficient. Thus, we obtain the free energy density
\begin{equation}
  f(T,\mathsf{u}) = f_0(T) + \sigma_1^{ij} \, u_{ij} + \frac{ 1 }{ 2 } \, C_c^{ij,kl} \, u_{ij} \, u_{kl} \, .
  \label{equation::I_500} 
\end{equation}
Here, we define the new stress tensor
\begin{equation}
  \sigma_1^{ij} = \sigma_0^{ij} - \Bigl( \sum_a \mu_{0,a} \, \tilde{n}_{0,a} \Bigr) \, \delta^{ij} 
  \label{equation::I_510} 
\end{equation}
and the new elastic constants
\begin{equation}
  \begin{split}
    C_c^{ij,kl} = \,& C^{ij,kl} + \sum_a ( \mu_a^{ij} \, \delta^{kl} + \delta^{ij} \, \mu_a^{kl} ) 
    + \sum_{ab} \nu_{ab} \, \delta^{ij} \delta^{kl} \\ 
    &+ \Delta C_1^{ij,kl} 
  \end{split}
  \label{equation::I_520} 
\end{equation}
where
\begin{equation}
  \begin{split}
    \Delta C_1^{ij,kl} = \,& \Delta C_0^{ij,kl} \\
    &+ \Bigl( \sum_a \mu_{0,a} \, \tilde{n}_{0,a} \Bigr) 
    \, [ \delta^{ij} \delta^{kl} + \delta^{ik} \delta^{jl} + \delta^{il} \delta^{jk} ] \, .
  \end{split}
  \label{equation::I_530} 
\end{equation}
The new elastic constants $C_c^{ij,kl}$ have the subscript $c$ which indicates that the vacancy densities 
$c_a$ defined in Eq.\ \eqref{equation::D_780} are kept constant. More precisely we have $c_a = 0$. This 
fact is a consequence of our special case where we consider a perfect crystal with no point defects so 
that the particle densities are given by the formula \eqref{equation::I_470}.

Eq.\ \eqref{equation::I_520} must be compared with Eq.\ \eqref{equation::I_460} in order to understand 
the definition of the new correction \eqref{equation::I_530}. Inserting Eq.\ \eqref{equation::F_330} 
for the old correction we obtain the new correction 
\begin{equation}
  \begin{split}
    \Delta C_1^{ij,kl} &= - \frac{ 1 }{ 2 } \, [ \sigma_1^{ik} \, \delta^{jl} + \sigma_1^{jl} \, \delta^{ik} 
    + \sigma_1^{jk} \, \delta^{il} + \sigma_1^{il} \, \delta^{jk} \\ 
    &\hspace{10.5mm} + \sigma_1^{ij} \, \delta^{kl} + \sigma_1^{kl} \, \delta^{ij} ] 
  \end{split}
  \label{equation::I_540} 
\end{equation}
where the new stress tensor $\sigma_1^{ij}$ is defined in Eq.\ \eqref{equation::I_510}. We note that 
the old correction \eqref{equation::F_330} and the new correction \eqref{equation::I_540} have the same 
structure where only the old stress tensor $\sigma_0^{ij}$ is replaced by the new stress tensor 
$\sigma_1^{ij}$. Moreover, we note that the definition of the new stress tensor \eqref{equation::I_510} 
may be compared with the necessary conditions \eqref{equation::E_660} for the global thermodynamic 
equilibrium of a normal crystal.

As a further alternative we consider the free energy \emph{per mass} $\bar{f}(T,\mathsf{u})$ which is 
calculated from the free energy \emph{density} $f(T,\mathsf{u})$ by the formula
\begin{equation}
  \bar{f}(T,\mathsf{u}) = f(T,\mathsf{u}) / \tilde{\rho} \, .
  \label{equation::I_550} 
\end{equation}
For the macroscopic mass density $\tilde{\rho}$ we have a formula analogous to 
Eq.\ \eqref{equation::I_470}. Again, we expand with respect to the Lagrange strain tensor $u_{ij}$ up 
to second order. Thus, we obtain
\begin{equation}
  \begin{split}
    \tilde{\rho} &= \tilde{\rho}_0 \, \frac{ \partial r_0 }{ \partial r } = \tilde{\rho}_0 \, g_0^{-1/2} \\ 
    &= \tilde{\rho}_0 \, \Bigl[ 1 - \delta^{ij} \, u_{ij} 
    + \frac{ 1 }{ 2 } \, [ \delta^{ij} \delta^{kl} + \delta^{ik} \delta^{jl} + \delta^{il} \delta^{jk} ] 
    \, u_{ij} u_{kl} \Bigr] 
  \, .
  \end{split}
  \label{equation::I_560} 
\end{equation}
Now, we insert the free energy density \eqref{equation::I_500} and the macroscopic mass density 
\eqref{equation::I_560} into the above formula \eqref{equation::I_550}. We expand with respect to 
the Lagrange strain tensor up to second order. Then, we obtain the Taylor expansion series for the 
free energy per mass
\begin{equation}
  \bar{f}(T,\mathsf{u}) = \frac{ 1 }{ \tilde{\rho}_0 } \, \Bigl[ f_0(T) + \sigma_2^{ij} \, u_{ij} 
  + \frac{ 1 }{ 2 } \, \bar{C}_c^{ij,kl} \, u_{ij} \, u_{kl} \Bigr] \, .
  \label{equation::I_570} 
\end{equation}
Here, we define a further new strain tensor
\begin{equation}
  \sigma_2^{ij} = \sigma_1^{ij} + f_0 \, \delta^{ij} 
   = \sigma_0^{ij} + \Bigl( f_0 - \sum_a \mu_{0,a} \, \tilde{n}_{0,a} \Bigr) \, \delta^{ij} 
  \label{equation::I_580} 
\end{equation}
and further new elastic constants 
\begin{equation}
  \begin{split}
    \bar{C}_c^{ij,kl} = \,& C^{ij,kl} + \sum_a ( \mu_a^{ij} \, \delta^{kl} + \delta^{ij} \, \mu_a^{kl} ) 
    + \sum_{ab} \nu_{ab} \, \delta^{ij} \delta^{kl} \\ 
    &+ \Delta C_2^{ij,kl} 
  \end{split}
  \label{equation::I_590} 
\end{equation}
where now the corrections are
\begin{equation}
  \begin{split}
    \Delta C_2^{ij,kl} = \,& \Delta C_1^{ij,kl} + \sigma_2^{ij} \, \delta^{kl} + \delta^{ij} \, \sigma_2^{kl} \\
    &- f_0 \, [ \delta^{ij} \delta^{kl} + \delta^{ik} \delta^{jl} + \delta^{il} \delta^{jk} ] \, .
  \end{split}
  \label{equation::I_600} 
\end{equation}
Eq.\ \eqref{equation::I_590} must be compared with Eq.\ \eqref{equation::I_520} in order to understand 
the definition of the further new correction \eqref{equation::I_600}. Inserting Eq.\ \eqref{equation::I_540} 
into Eq.\ \eqref{equation::I_600} we obtain an explicit formula for the further new correction which reads
\begin{equation}
  \begin{split}
    \Delta C_2^{ij,kl} &= - \frac{ 1 }{ 2 } \, [ \sigma_2^{ik} \, \delta^{jl} + \sigma_2^{jl} \, \delta^{ik} 
    + \sigma_2^{jk} \, \delta^{il} + \sigma_2^{il} \, \delta^{jk} \\ 
    &\hspace{10.5mm} - \sigma_2^{ij} \, \delta^{kl} - \sigma_2^{kl} \, \delta^{ij} ] \, .
  \end{split}
  \label{equation::I_610} 
\end{equation}
We have derived three different explicit formulas for the corrections which are given by 
Eqs.\ \eqref{equation::F_330}, \eqref{equation::I_540}, and \eqref{equation::I_610}, respectively. All three 
formulas have a similar structure where in the last formula two signs have changed. On the other hand, the 
formulas differ by the stress tensor which is involved. An interesting result is that in each case all 
nontrivial contributions of the nonlinear terms come together into a respective single stress tensor. We note 
that we have found three different stress tensors which are related to each other by Eqs.\ \eqref{equation::I_510} 
and \eqref{equation::I_580}.

The inspection of the Taylor expansion \eqref{equation::I_570} together with the formulas \eqref{equation::I_590} 
and \eqref{equation::I_610} for the elastic constants implies that nontrivial physical effects are capsuled 
inside the stress tensor $\sigma_2^{ij}$. For this reason, we must consider this stress tensor in more detail.
We know that the density of the grand canonical thermodynamic potential is defined by the Legendre transformation 
\eqref{equation::G_910} which here we rewrite with no kinetic term and with a subscript $0$ for the thermal 
equilibrium as
\begin{equation}
  \omega_0 = f_0 - \sum_a \mu_{0,a} \, \tilde{n}_{0,a} = - p_0 \, . 
  \label{equation::I_620} 
\end{equation}
In this way, we identify the \emph{thermodynamic} pressure $p_0$ so that we may rewrite the stress tensor 
$\sigma_2^{ij}$ defined in Eq.\ \eqref{equation::I_580} as
\begin{equation}
  \sigma_2^{ij} = \sigma_0^{ij} - p_0 \, \delta^{ij} \, .
  \label{equation::I_630} 
\end{equation}
When in Eq.\ \eqref{equation::G_A70} we have discussed the reversible part of the momentum current density 
we have identified the \emph{total} stress tensor given by Eq.\ \eqref{equation::G_A80}. If we compare 
our recent result \eqref{equation::I_630} with the formulas \eqref{equation::G_A80}-\eqref{equation::G_B00}
then we identify 
\begin{equation}
  \sigma_2^{ij} = \sigma_{0,\mathrm{tot}}^{ij} = - p_{0,\mathrm{tot}} \, \delta^{ij} + \sigma_0^{\prime ij} \, .
  \label{equation::I_640} 
\end{equation}
Thus, we conclude: The stress tensor $\sigma_2^{ij}$ which is the coefficient of the linear term in 
the free energy per mass \eqref{equation::I_570} is the total stress tensor of the thermal equilibrium 
$\sigma_{0,\mathrm{tot}}^{ij}$. Beyond that, we may split the total stress tensor into a scalar and 
an irreducible contribution. In this way we define the total pressure $p_{0,\mathrm{tot}}$ and explain 
the last equality sign in Eq.\ \eqref{equation::I_640}.

We may calculate the first derivative of the free energy per mass \eqref{equation::I_570}. Then, we 
obtain a linear response equation for the total stress tensor which reads 
\begin{equation}
  \sigma_\mathrm{tot}^{ij} = \tilde{\rho}_0 \, \frac{ \partial \bar{f} }{ \partial u_{ij} } 
  = \sigma_{0,\mathrm{tot}}^{ij} + \bar{C}_c^{ij,kl} \, u_{kl} \, .
  \label{equation::I_650} 
\end{equation}
Thus, the elastic constants $\bar{C}_c^{ij,kl}$ defined in Eq.\ \eqref{equation::I_590} may be 
interpreted as the linear-response coefficients for the total stress tensor with respect to the 
Lagrange strain tensor.

There is a physical argument to explain why in the free energy \emph{per mass} $\bar{f}(T,\mathsf{u})$ 
the linear coefficient $\sigma_2^{ij}$ is the total stress tensor $\sigma_{0,\mathrm{tot}}^{ij}$. In 
an experiment we may consider a normal crystal of finite size. This crystal consists of certain numbers 
of particles $N_a$ of species $a$ which are constant. Consequently, the total mass $M = \sum_a m_a N_a$ 
has a certain fixed value. On the other hand, when we deform the crystal by a strain $u_{ij}$ then the 
volume changes. Thus, we must consider the free energy $F[T,\mathsf{u}]$ of a varying volume where 
always the same particles and the same mass are inside the volume. Hence, we may divide the free energy 
by the total mass $M$. In this way, we obtain the free energy per mass 
$\bar{f}(T,\mathsf{u}) = f(T,\mathsf{u}) / \tilde{\rho} = F[T,\mathsf{u}] / M$ as the correct function 
which we must consider. We note that the same total stress tensor appears also in the reversible 
part of the momentum current density \eqref{equation::G_A70} when the kinetic part is separated. 
This result may be explained by the fact that the momentum current density is a part of the continuity 
equation which represents the conservation of momentum on a local scale.

In a recent publication Lin \emph{et al.}\ \cite{Li21} calculated the direct correlation function and 
the elastic constants for a crystal of interacting spherical hard-core particles where the lattice 
has a face-centered cubic structure. A density-functional approach was applied using the White-Bear-II 
density functional \cite{Ta00,Ro10}. For the general elastic constants the formulas of Walz and Fuchs 
\cite{WF10} were used which are our Eqs.\ \eqref{equation::E_280}-\eqref{equation::E_300} without the 
correction terms. From these results the special elastic constants $\bar{C}_c^{ij,kl}$ were calculated 
which in our paper are defined in Eq.\ \eqref{equation::I_590}. Lin \emph{et al.}\ \cite{Li21} used their 
Eq.\ (9) which defines the elastic constants in Voigt notation with only two indices. This Eq.\ (9) was 
stated but not derived. In our present paper we provide a derivation of this formula by the above 
calculations. In our formula \eqref{equation::I_590} the first line is equivalent to all contributions 
of Eq.\ (9) except the last terms which involve the pressure $p$. Thus, there remains to show that our 
correction term defined in Eq.\ \eqref{equation::I_610} together with Eq.\ \eqref{equation::I_640} provides 
these pressure terms. In a crystal with a face-centered cubic lattice there is an isotropic symmetry which 
implies that in thermal equilibrium the irreducible stress tensor is zero so that $\sigma_0^{\prime ij} = 0$. 
Thus, we have $\sigma_2^{ij} = \sigma_{0,\mathrm{tot}}^{ij} = - p_{0,\mathrm{tot}} \, \delta^{ij}$. Hence, 
from Eq.\ \eqref{equation::I_610} we obtain the \emph{isotropic} correction term
\begin{equation}
  \Delta C_2^{ij,kl} = p_{0,\mathrm{tot}} \, [ \delta^{ik} \, \delta^{jl} 
  + \delta^{jk} \, \delta^{il} - \delta^{ij} \, \delta^{kl} ] \, .
  \label{equation::I_660} 
\end{equation}
Turning to Voigt notation with two indices this term exactly provides the pressure terms in Eq.\ (9) 
of Lin \emph{et al.}\ \cite{Li21} so that this latter equation is proven.

Lin \emph{et al.}\ \cite{Li21} presented further calculations of the elastic constants including a 
numerical simulation of interacting particles. The result is shown in their Table I. The first 
column shows the results of the formulas of Walz and Fuchs \cite{WF10}. The second column shows 
the results obtained by direct minimization of the free energy functional and extracting the elastic 
constants by explicit numerical differentiation according to the Taylor expansion \eqref{equation::I_570}. 
Our formulas \eqref{equation::E_280}-\eqref{equation::E_300} \emph{including} the correction terms 
should exactly provide these results because our correction terms were calculated by minimizing the 
free energy in linear order. The third column shows the results of the numerical simulations. One 
finds agreement between all three results. However, the second column is closer to the numerical 
simulations than the first column. Thus, we conclude that our correction terms should provide an 
improvement for the elastic constants.

\subsection{Transport coefficients}
\label{section::09g}
The second stage of the continuum mechanics of deformed crystals is the theory of the dynamic 
phenomena in the nonequilibrium. Here, we have derived the time-evolution equations for the relevant 
variables which on the right-hand sides have three terms, a reversible, a dissipative, and a fluctuating 
term. Depending on the version of the theory the time-evolution equations are given by 
Eq.\ \eqref{equation::G_B80}, \eqref{equation::H_180}, or \eqref{equation::H_190}. In these equations 
there are some parameters which describe the explicit physical properties of the dynamical processes. 
The parameters are called \emph{transport coefficients} and are provided by the Onsager matrix which 
describes the strength of the dissipative and of the fluctuating terms. 

For convenience and completeness we consider the general case of Sec.\ \ref{section::08} where we choose 
the time-evolution equation in the form \eqref{equation::H_190}. In this case the nonzero parts of the 
Onsager matrix are given by the transport coefficients for particle diffusion $D_{ab}^{kl}$, the offdiagonal 
coefficients $G_a^{kl}$ and $K_a^{kp}$, the viscosity tensor $\Lambda^{ik,jl}$, the heat conductivity tensor 
$\alpha^{kl}$, the further offdiagonal coefficients $\xi^{kp}$, and the coefficients $\zeta^{pq}$ for the 
diffusive motion of the material relative to the crystal lattice. These transport coefficients are found 
in the formulas for the dissipative contributions of the current densities and in the correlations of the 
stochastic forces. They are the essential parameters for the time evolution of the dynamic phenomena. 
In the isothermal case of Sec.\ \ref{section::07} where the time-evolution equation is given by 
Eq.\ \eqref{equation::G_B80} the nonzero parts of the Onsager matrix are given by a subset of the above 
mentioned transport coefficients where three of them are not present. For this reason the latter case 
need not be considered separately but is included in the general case.

In a first step we calculate the nonzero parts of the Onsager matrix in a \emph{global} thermodynamic 
equilibrium so that we obtain the transport coefficients in \emph{linear response}. For this purpose 
we use the special Mori scalar product 
\begin{equation}
  ( \, \hat{Y}(\mathbf{r}_1,t_1) \mid \hat{Y}(\mathbf{r}_2,t_2) \, )_\mathrm{eq} = 
  \mathrm{Tr} \{ \hat{\varrho}_\mathrm{eq} 
  \, [ \hat{Y}(\mathbf{r}_1,t_1) ]^* \, \hat{Y}(\mathbf{r}_2,t_2) \} 
  \label{equation::I_670}
\end{equation}
which is defined with a phase-space distribution function $\hat{\varrho}_\mathrm{eq} = \varrho_\mathrm{eq}(\Gamma)$ 
in global thermodynamic equilibrium. We combine the definition of the integral kernel \eqref{equation::G_600} 
with the definition of the Onsager matrix \eqref{equation::G_B10} or \eqref{equation::G_B40}. In the case 
of conserved quantities we must apply Eqs.\ \eqref{equation::G_B30} to remove the spatial derivatives 
from the transport coefficients. Furthermore, in the case of the displacement field we may apply the 
Liouville operator $\mathsf{L}$ explicitly onto Eq.\ \eqref{equation::G_370} so that we obtain
\begin{equation}
  \begin{split}
    \mathsf{L} \, \hat{u}^k(\mathbf{r}_1) = \,&- i \sum_{lnb} \int d^dr_2 
    \, \mathcal{N}^{kl} \, w( \mathbf{r}_1 - \mathbf{r}_2 ) \\
    &\times [ \partial_l n_{0,b}(\mathbf{r}_2) ] \, [ \partial_n \hat{j}_b^n(\mathbf{r}_2) ] \, .
  \end{split}
  \label{equation::I_680}
\end{equation}

As a result, for the respective transport coefficients we obtain the following defining equations 
which read
\begin{widetext}
\begin{eqnarray}
  ( \, \mathsf{Q}(t^\prime) \hat{\tilde{j}}_b^l(\mathbf{r}^\prime) \mid \mathsf{U}(t^\prime,t)
  \mid \mathsf{Q}(t) \hat{\tilde{j}}_a^k(\mathbf{r}) \, )_\mathrm{eq}
  &=& 2 \, k_B T \, D_{ab}^{kl} \, w( \mathbf{r} - \mathbf{r}^\prime ) \, \delta( t - t^\prime ) \, ,
  \label{equation::I_690} \\
  ( \, \mathsf{Q}(t^\prime) \hat{q}^l(\mathbf{r}^\prime) \mid \mathsf{U}(t^\prime,t)
  \mid \mathsf{Q}(t) \hat{\tilde{j}}_a^k(\mathbf{r}) \, )_\mathrm{eq}
  &=& 2 \, k_B T \, G_a^{kl} \, w( \mathbf{r} - \mathbf{r}^\prime ) \, \delta( t - t^\prime ) \, ,
  \label{equation::I_700} \\
  i \ ( \, \mathsf{Q}(t^\prime) \mathsf{L} \hat{u}^i(\mathbf{r}^\prime) \mid \mathsf{U}(t^\prime,t)
  \mid \mathsf{Q}(t) \hat{\tilde{j}}_a^k(\mathbf{r}) \, )_\mathrm{eq}
  &=& 2 \, k_B T \, K_a^{ki} \, w( \mathbf{r} - \mathbf{r}^\prime ) \, \delta( t - t^\prime ) \, ,
  \label{equation::I_710} \\
  ( \, \mathsf{Q}(t^\prime) \hat{\tilde{\Pi}}^{jl}(\mathbf{r}^\prime) \mid \mathsf{U}(t^\prime,t)
  \mid \mathsf{Q}(t) \hat{\tilde{\Pi}}^{ik}(\mathbf{r}) \, )_\mathrm{eq}
  &=& 2 \, k_B T \, \Lambda^{ik,jl} \, w( \mathbf{r} - \mathbf{r}^\prime ) \, \delta( t - t^\prime ) \, ,
  \label{equation::I_720} \\
  ( \, \mathsf{Q}(t^\prime) \hat{q}^l(\mathbf{r}^\prime) \mid \mathsf{U}(t^\prime,t)
  \mid \mathsf{Q}(t) \hat{q}^k(\mathbf{r}) \, )_\mathrm{eq}
  &=& 2 \, k_B T \, \alpha^{kl} \, w( \mathbf{r} - \mathbf{r}^\prime ) \, \delta( t - t^\prime ) \, ,
  \label{equation::I_730} \\
  i \ ( \, \mathsf{Q}(t^\prime) \mathsf{L} \hat{u}^i(\mathbf{r}^\prime) \mid \mathsf{U}(t^\prime,t)
  \mid \mathsf{Q}(t) \hat{q}^k(\mathbf{r}) \, )_\mathrm{eq}
  &=& 2 \, k_B T \, \xi^{ki} \, w( \mathbf{r} - \mathbf{r}^\prime ) \, \delta( t - t^\prime ) \, ,
  \label{equation::I_740} \\
  ( \, \mathsf{Q}(t^\prime) \mathsf{L} \hat{u}^j(\mathbf{r}^\prime) \mid \mathsf{U}(t^\prime,t)
  \mid \mathsf{Q}(t) \mathsf{L} \hat{u}^i(\mathbf{r}) \, )_\mathrm{eq}
  &=& 2 \, k_B T \, \zeta^{ij} \, w( \mathbf{r} - \mathbf{r}^\prime ) \, \delta( t - t^\prime ) \, .
  \label{equation::I_750} 
\end{eqnarray}
On the left-hand sides the correlation functions depend only on the differences of the space and time 
variables. Because of the projection operators $\mathsf{Q}(t)$ and $\mathsf{Q}(t^\prime)$ for the fast 
irrelevant variables these correlations functions decay rather rapidly on short space and time scales. 
We assume that the projection operators imply a complete and perfect separation of the space and time 
scales so that the short scales are below the resolution of our macroscopic theory. For this reason, 
on the right-hand sides we have used the coarse-graining function for minimum spatial scales and the 
delta function for effectively zero time scales.

The above equations can be inverted if we integrate over the \emph{relative} space and time coordinates 
which we denote by $\Delta \mathbf{r} = \mathbf{r} - \mathbf{r}^\prime$ and $\Delta t = t - t^\prime$. 
Thus, we obtain the explicit formulas for the transport coefficients 
\begin{eqnarray}
  D_{ab}^{kl} &=& \frac{ 1 }{ 2 \, k_B T } \, \int d^d(\Delta r) \int_{-\infty}^{+\infty} d(\Delta t) 
  \ ( \, \mathsf{Q}(t^\prime) \hat{\tilde{j}}_b^l(\mathbf{r}^\prime) \mid \mathsf{U}(t^\prime,t)
  \mid \mathsf{Q}(t) \hat{\tilde{j}}_a^k(\mathbf{r}) \, )_\mathrm{eq} \, ,
  \label{equation::I_760} \\
  G_a^{kl} &=& \frac{ 1 }{ 2 \, k_B T } \, \int d^d(\Delta r) \int_{-\infty}^{+\infty} d(\Delta t) 
  \ ( \, \mathsf{Q}(t^\prime) \hat{q}^l(\mathbf{r}^\prime) \mid \mathsf{U}(t^\prime,t)
  \mid \mathsf{Q}(t) \hat{\tilde{j}}_a^k(\mathbf{r}) \, )_\mathrm{eq} \, ,
  \label{equation::I_770} \\
  K_a^{ki} &=& \frac{ i }{ 2 \, k_B T } \, \int d^d(\Delta r) \int_{-\infty}^{+\infty} d(\Delta t) 
  \ ( \, \mathsf{Q}(t^\prime) \mathsf{L} \hat{u}^i(\mathbf{r}^\prime) \mid \mathsf{U}(t^\prime,t)
  \mid \mathsf{Q}(t) \hat{\tilde{j}}_a^k(\mathbf{r}) \, )_\mathrm{eq} \, ,
  \label{equation::I_780} \\
  \Lambda^{ik,jl} &=& \frac{ 1 }{ 2 \, k_B T } \, \int d^d(\Delta r) \int_{-\infty}^{+\infty} d(\Delta t) 
  \ ( \, \mathsf{Q}(t^\prime) \hat{\tilde{\Pi}}^{jl}(\mathbf{r}^\prime) \mid \mathsf{U}(t^\prime,t)
  \mid \mathsf{Q}(t) \hat{\tilde{\Pi}}^{ik}(\mathbf{r}) \, )_\mathrm{eq} \, ,
  \label{equation::I_790} \\
  \alpha^{kl} &=& \frac{ 1 }{ 2 \, k_B T } \, \int d^d(\Delta r) \int_{-\infty}^{+\infty} d(\Delta t) 
  \ ( \, \mathsf{Q}(t^\prime) \hat{q}^l(\mathbf{r}^\prime) \mid \mathsf{U}(t^\prime,t)
  \mid \mathsf{Q}(t) \hat{q}^k(\mathbf{r}) \, )_\mathrm{eq} \, ,
  \label{equation::I_800} \\
  \xi^{ki} &=& \frac{ i }{ 2 \, k_B T } \, \int d^d(\Delta r) \int_{-\infty}^{+\infty} d(\Delta t) 
  \ ( \, \mathsf{Q}(t^\prime) \mathsf{L} \hat{u}^i(\mathbf{r}^\prime) \mid \mathsf{U}(t^\prime,t)
  \mid \mathsf{Q}(t) \hat{q}^k(\mathbf{r}) \, )_\mathrm{eq} \, ,
  \label{equation::I_810} \\
  \zeta^{ij} &=& \frac{ 1 }{ 2 \, k_B T } \, \int d^d(\Delta r) \int_{-\infty}^{+\infty} d(\Delta t) 
  \ ( \, \mathsf{Q}(t^\prime) \mathsf{L} \hat{u}^j(\mathbf{r}^\prime) \mid \mathsf{U}(t^\prime,t)
  \mid \mathsf{Q}(t) \mathsf{L} \hat{u}^i(\mathbf{r}) \, )_\mathrm{eq} \, .
  \label{equation::I_820} 
\end{eqnarray}
\end{widetext}
In general the results for the transport coefficients will depend on the average coordinates 
$\mathbf{R} = ( \mathbf{r} + \mathbf{r}^\prime ) / 2$ and the average times $T = ( t + t^\prime ) / 2$. 
However, in global thermodynamic equilibrium for an undeformed crystal we have translation invariance 
in the space and time directions. Hence, the correlation functions depend only on the relative coordinates 
so that the transport coefficients are constant.

In the above calculation we assume that the laboratory coordinates $\mathbf{r}$ and the intermediate 
coordinates $\mathbf{r}_1$ are the same. Thus, all spatial indices in the above quantities are Cartesian. 
We may transform to a non-Cartesian coordinate system with coordinates $\mathbf{r}_0$. Since in a global 
thermodynamic equilibrium any deformations are homogeneous in space in the present case these latter 
coordinates are never curvilinear but just linear. In the above formulas for the transport coefficients 
we have the displacement field $\hat{\mathbf{u}}$ which must be transformed. We note that 
$\mathsf{L} \hat{\mathbf{u}}$ must be treated as the differential $d\hat{\mathbf{u}}$. As a consequence, 
we must apply the simple transformation formula \eqref{equation::F_490}. Thus, for the related transport 
coefficients we obtain the transformation formulas
\begin{equation}
  K_a^{kp} = e_i^p \, K_a^{ki} \, , \qquad
  \xi^{kp} = e_i^p \, \xi^{ki} \, , \qquad
  \zeta^{pq} = e_i^p \, e_j^q \, \zeta^{ij} \, .
  \label{equation::I_830} 
\end{equation}

Now, we turn to the next step of our calculation. Via the constraints a global thermodynamic equilibrium 
can be parameterized by the averages of the relevant variables. These are the particle densities 
$\tilde{n}_a$, the momentum density $\mathbf{j}$, and the strain tensor $\partial_k u^i$. Furthermore, 
there is the constant temperature $T$. Thus, if we repeat the calculation many times for all possible 
states in a global thermodynamic equilibrium then we obtain the transport coefficients as functions of 
the constant temperature and of the relevant variables similar like the free energy density 
\eqref{equation::E_510}. While in the isothermal case of Sec.\ \ref{section::07} the temperature $T$ 
is a constant external parameter in the general case of Sec.\ \ref{section::08} the entropy density $s$ 
is the appropriate variable. Hence, we must perform a Legendre transformation from $T$ to $s$. As a 
consequence, in our general case where the time-evolution is defined by Eq.\ \eqref{equation::H_190} the 
transport coefficients depend on the variables of the internal energy density \eqref{equation::H_080}. 
Thus, we obtain
\begin{eqnarray}
  D_{ab}^{kl} &=& D_{ab}^{kl}(s,\partial \mathbf{u},\tilde{\mathbf{j}},\tilde{n}) \, ,
  \label{equation::I_840} \\
  G_a^{kl} &=& G_a^{kl}(s,\partial \mathbf{u},\tilde{\mathbf{j}},\tilde{n}) \, ,
  \label{equation::I_850} \\
  K_a^{kp} &=& K_a^{kp}(s,\partial \mathbf{u},\tilde{\mathbf{j}},\tilde{n}) \, ,
  \label{equation::I_860} \\
  \Lambda^{ik,jl} &=& \Lambda^{ik,jl}(s,\partial \mathbf{u},\tilde{\mathbf{j}},\tilde{n}) \, ,
  \label{equation::I_870} \\
  \alpha^{kl} &=& \alpha^{kl}(s,\partial \mathbf{u},\tilde{\mathbf{j}},\tilde{n}) \, ,
  \label{equation::I_880} \\
  \xi^{kp} &=& \xi^{kp}(s,\partial \mathbf{u},\tilde{\mathbf{j}},\tilde{n}) \, ,
  \label{equation::I_890} \\
  \zeta^{pq} &=& \zeta^{pq}(s,\partial \mathbf{u},\tilde{\mathbf{j}},\tilde{n}) \, .
  \label{equation::I_900} 
\end{eqnarray}
Since the Onsager matrix and hence the transport coefficients control the strengths of the 
dissipations and of the fluctuations, the dependence on the relevant variables implies that our 
theory provides \emph{nonlinear dissipation} and \emph{nonlinear fluctuations}.

In Eqs.\ \eqref{equation::I_690}-\eqref{equation::I_750} or equivalently in 
Eqs.\ \eqref{equation::I_760}-\eqref{equation::I_820} the transport coefficients are defined as Mori 
scalar products of local quantities. Since they have a \emph{tilde} and a \emph{hat} on top, these 
quantities are \emph{coarse grained} and depend on the \emph{microscopic} variables of the phase 
space, the coordinates and momenta $\Gamma = ( \mathbf{r}_{ai},\mathbf{p}_{ai} )$. Most of the local 
quantities are the current densities of the conserved quantities which are well defined. Furthermore, 
some of the Mori scalar products are calculated with the displacement field which has a hat on top 
and which is well defined by the microscopic formula \eqref{equation::G_370}. As a consequence, most 
of the transport coefficients are well defined. 

However, there remain the three transport coefficients $G_a^{kl}$, $\alpha^{kl}$, and $\xi^{kp}$ which 
on a first sight are not well defined because they involve the entropy current $\hat{\mathbf{q}}(\mathbf{r})$ 
in the Mori scalar product. The reason is that the entropy density $s(\mathbf{r})$ and the entropy 
current density $\mathbf{q}(\mathbf{r})$ are quantities of the local thermodynamics and hence defined 
only macroscopically. As a consequence, the microscopic forms $\hat{s}(\mathbf{r}) = s(\mathbf{r},\Gamma)$ 
and $\hat{\mathbf{q}}(\mathbf{r}) = \mathbf{q}(\mathbf{r},\Gamma)$ do not exist so that a hat may not be 
put on top of these quantities.

Nevertheless, there is an elegant way out. We may alternatively calculate the three transport 
coefficients $G_{E,a}^{kl}$, $\alpha_E^{kl}$, and $\xi_E^{kp}$ where the energy current density 
$\hat{\tilde{\mathbf{j}}}_E(\mathbf{r})$ is inserted into the Mori scalar product. From the differential 
thermodynamic relation of the energy density \eqref{equation::H_090} we infer a thermodynamic relation 
for the current densities which reads
\begin{equation}
  \hat{\tilde{j}}_E^l = T \, \hat{q}^l + \sigma_p^{\ k} \, \hat{J}_k^{pl} + v_k \, \hat{\tilde{\Pi}}^{kl} 
  + \sum_a \mu_a \, \hat{\tilde{j}}_a^l \, .
  \label{equation::I_910} 
\end{equation}
Here we put tildes and hats on top of the current densities so that this thermodynamic relation is 
used for coarse grained microscopic current densities. From the continuity equation of the linear 
strain tensor \eqref{equation::G_I30} and the special form \eqref{equation::G_I40} we infer the 
microscopic three-index current 
$\hat{J}_k^{pl}(\mathbf{r}) = - i \, \mathsf{L} \hat{u}^p(\mathbf{r}) \, \delta_k^{\ l}$. In the next 
step, we use the thermodynamic relation \eqref{equation::I_910} in order to calculate the Mori scalar 
products. Thus, as a result we obtain thermodynamic relations for the transport coefficients which read 
\begin{eqnarray}
  G_{E,a}^{kl} &=& G_a^{kl} \, T + K_a^{kq} \, \sigma_q^{\ l} + \sum_b D_{ab}^{kl} \, \mu_b \, ,
  \label{equation::I_920} \\
  \alpha_E^{kl} &=& T \, \alpha^{kl} \, T + T \, \xi^{kq} \, \sigma_q^{\ l} 
  + \sum_b T \, G_b^{lk} \, \mu_b \nonumber \\ 
  &&+ \sigma_p^{\ k} \, \xi^{lp} \, T + \sigma_p^{\ k} \, \zeta^{pq} \, \sigma_q^{\ l} 
  + \sum_b \sigma_p^{\ k} \, K_b^{lp} \, \mu_b \nonumber \\ 
  &&+ \sum_a \mu_a \, G_a^{kl} \, T + \sum_a \mu_a \, K_a^{kq} \, \sigma_q^{\ l} 
  + \sum_{ab} \mu_a \, D_{ab}^{kl} \, \mu_b \nonumber \\ 
  &&+ v_i \, \Lambda^{ik,jl} \, v_j \, ,
  \label{equation::I_930} \\
  \xi_E^{kp} &=& T \, \xi^{kp} + \sigma_q^{\ k} \, \zeta^{qp} + \sum_a \mu_a \, K_a^{kp} \, .
  \label{equation::I_940} 
\end{eqnarray}
The transport coefficients $G_{E,a}^{kl}$, $\alpha_E^{kl}$, and $\xi_E^{kp}$ calculated with the 
energy current density are well defined. We may interpret the thermodynamic relations 
\eqref{equation::I_920}-\eqref{equation::I_940} as linear equations for $G_a^{kl}$, $\alpha^{kl}$, 
and $\xi^{kp}$ and resolve with respect to these latter coefficients. In this way, we obtain well 
defined results also for the transport coefficients calculated with the entropy current density.

The above calculation is a procedure which must be proven and justified. For this purpose we consider 
the time-evolution equation in the GENERIC form \eqref{equation::H_180}. Here, the energy density 
$\tilde{\varepsilon}(\mathbf{r})$ is a natural variable so that the transport coefficients 
$G_{E,a}^{kl}$, $\alpha_E^{kl}$, and $\xi_E^{kp}$ are involved in the Onsager matrix. Then, we 
transform to the time-evolution equation in the alternative form \eqref{equation::H_190} where 
the entropy density $s(\mathbf{r})$ is the related variable and where the transport coefficients 
$G_a^{kl}$, $\alpha^{kl}$, and $\xi^{kp}$ are elements of the Onsager matrix. The transformation 
has been established for a conventional liquid and described in Sec.\ 4.4 of our previous publication 
\cite{Ha16}. Here, in this work we must extend the transformation to an elastic crystal where the 
displacement field arises as a further variable. The extension is straight forward. As a result we recover 
the thermodynamic relations for the transport coefficients \eqref{equation::I_920}-\eqref{equation::I_940} 
so that the above procedure is proven and justified.

In the final step we apply the concept of the local thermodynamic equilibrium. We consider a 
nonequilibrium state and a special point $P$ in the space time with the coordinate $\mathbf{r}_P$ 
and the time $t_P$. Then, we consider a small epsilon surroundings as defined in 
Eq.\ \eqref{equation::I_200}. We assume that within this epsilon surroundings the nonequilibrium 
state is approximately an equilibrium so that we may approximate this nonequilibrium locally 
by a \emph{tangential} global equilibrium. Thus, we may use our transport coefficients which 
we have calculated above with our global-equilibrium formulas. Eventually, the relevant 
variables which parametrize the equilibrium state become space and time dependent so that 
$\tilde{n}_a = \tilde{n}_a(\mathbf{r},t)$, $\tilde{\mathbf{j}} = \tilde{\mathbf{j}}(\mathbf{r},t)$, 
$s = s(\mathbf{r},t)$, and $\partial_k u^i = \partial_k u^i(\mathbf{r},t)$. Thus, via the formulas 
\eqref{equation::I_840}-\eqref{equation::I_900} all the transport coefficients become 
\emph{implicitly} space and time dependent.

Our explicit results \eqref{equation::I_760}-\eqref{equation::I_820} may be interpreted as 
\emph{Green-Kubo} formulas. The first formula for the transport coefficients of particle diffusion 
$D_{ab}^{kl}$ and the fourth formula for the viscosity tensor $\Lambda^{ik,jl}$ are well known 
since a long time. The last three formulas for the coefficients of heat transport and diffusive 
lattice motion $\alpha^{kl}$, $\xi^{ki}$, $\zeta^{ij}$ have been derived first by Szamel \cite{Sz97} 
and later by Miserez \cite{Mi21}. In phenomenological theories the latter three coefficients have 
been introduced much earlier by Martin, Parodi, and Pershan \cite{MPP72} and by Fleming and Cohen 
\cite{FC76}. 

Within an alternative microscopic approach the transport coefficients have been obtained also by 
Mabillard and Gaspard \cite{MG21}. Even though these authors do not use the projection operators 
explicitly, in their Green-Kubo formulas (3.34)-(3.39) the global currents (3.27)-(3.30) are used 
where effectively all those terms are subtracted which in our case are subtracted by our projection 
operator $\mathsf{Q}(t)$. Thus, for the transport coefficients the formulas of Mabillard and Gaspard 
(3.34)-(3.36) and (3.39) are equivalent to our formulas \eqref{equation::I_800}-\eqref{equation::I_820} 
and \eqref{equation::I_790}, respectively. Beyond that, Mabillard and Gaspard define two more coefficients 
in their Eqs.\ (3.37) and (3.38). These coefficients are offdiagonal elements of the antisymmetric 
matrix \eqref{equation::G_E50}. They imply \emph{reactive} contributions which we have discussed 
in Subsecs.\ \ref{section::07e} and \ref{section::08f} but which we have not considered later on.

In principle the microscopic formulas defined by the Mori scalar product are always the same. 
If we compare our explicit formulas with those of other authors we may find differences in the 
normalization factors which are due to different definitions and notations. Furthermore, some of 
the coefficients depend on whether they are calculated with the energy current or with the entropy 
current. Here, different authors make different choices. Nevertheless, both versions of the 
transport coefficients are related to each other by the thermodynamic relations 
\eqref{equation::I_920}-\eqref{equation::I_940}.

However, there is an important detail which causes a real and important difference. 
The formulas depend on the particular choice and definition of the projection operators 
$\mathsf{P}(t)$ and $\mathsf{Q}(t)$. Here, approximations may occur. If we use our projection 
operators defined in Eqs.\ \eqref{equation::G_290}, \eqref{equation::H_170} and 
\eqref{equation::G_490} together with the integral kernels \eqref{equation::D_060} and 
\eqref{equation::G_360} then we claim that our formulas for the transport coefficients 
\eqref{equation::I_760}-\eqref{equation::I_820} are exact in the macroscopic limit of continuum 
mechanics.

\section{Conclusions}
\label{section::10}
Starting from a general microscopic theory of many interacting particles we derive the macroscopic 
time-evolution equations of the continuum mechanics of solid crystals. We use the step by step procedure 
which in a previous paper \cite{Ha16} we have developed to derive the hydrodynamic equations of a simple 
liquid. Here, we have extended the approach to obtain the nonlinear elasticity theory including fluctuation 
effects. For this purpose two main extensions must be done. First, we implement a coarse-graining procedure 
for all densities of the conserved quantities. In a solid crystal the densities have a periodic lattice 
structure on microscopic scales which must be smeared out in the macroscopic theory of continuum mechanics. 
Second, we define a displacement field in terms of the microscopic densities which describes the translations, 
rotations, and deformations of the crystal on a local level.

As results we obtain a density-functional theory for the local thermodynamic equilibrium and time-evolution 
equations for the relevant variables. This theory may be interpreted as a \emph{hybrid} dynamic density-functional 
theory which treats the microscopic degrees of freedom by a constraint minimization of the free-energy functional 
and the macroscopic degrees of freedom by dynamic time-evolution equations. Eventually, we derive the dynamic 
equations of elasticity which include all the nonlinear effects and the fluctuations in terms of nonlinear 
Gaussian stochastic forces. We find the time-evolution equations which previously have been derived by Grabert 
and Michel \cite{GM83} as a phenomenological theory by general physical principles applied to macroscopic scales. 
Our theory also agrees with the results of Temmen \emph{et al.}\ \cite{Te00}.

We have subjected our theory to several consistency checks. Thus, we have found that our time-evolution 
equations can be written as Langevin equations with nonlinear Gaussian fluctuations which can be transformed 
into a Fokker-Planck equation. We proved that all conditions are satisfied so that this Fokker-Planck equation 
has a simple Boltzmann like distribution function as an equilibrium solution. The \emph{sufficient} conditions 
were stated in our previous paper \cite{Ha16} which require that on the right-hand side of the time-evolution 
equations all terms are written as divergences of current densities. For the coarse-grained densities of the 
conserved quantities these conditions are always satisfied since the time-evolution equations are continuity 
equations. However, a special problem is the displacement field. Nevertheless, the problem can be solved or 
circumvented. If we take the spatial derivative then we obtain the linear strain tensor as an alternative 
relevant variable. The related time-evolution equation has a spatial derivative on both sides so that the 
right-hand side can be written formally as the divergence of a current density. In this way, the sufficient 
conditions are satisfied by all relevant variables of our nonlinear elasticity theory.

In order to describe the strong deformations we have applied the concepts of General Relativity and Riemannian 
Geometry by using the tensor formalism with upper an lower indices. In this way we have found a simple 
classification of the many different strain and stress tensors which are commonly used in the elasticity 
theory. On the other hand, we have derived explicit formulas for the elastic constants which extend the 
previous results of Walz and Fuchs \cite{WF10} by correction terms. Since these correction terms are calculated 
by minimizing the free energy and solving related linear-response equations our formulas are \emph{exact} 
results for any given free-energy functional. Furthermore, we have found explicit formulas for the transport 
coefficients. Part of them, the coefficients for the heat or entropy transport, for the diffusive motion of the 
crystal lattice relative to the material, and for the related offdiagonal effects have been derived previously 
by Szamel \cite{Sz97} who states that the formulas are exact. Up to normalization factors and up to a different 
notation we obtain the same formulas. Moreover, we confirm that these formulas are exact in the limit of 
continuum mechanics for large space and time scales if the projection operators are chosen correctly.

Our time-evolution equations include the diffusive motion of the particles in a natural way. The reason is 
that the coarse-grained particle densities are included within the set of relevant variables where we have 
explicit continuity equations for these particle densities. For this reason, our theory is well suited for 
the cluster crystals which explicitly allows the diffusion of particles. Beyond that our theory includes 
crystals of several different species of particles where on the other hand the phenomenological theories of 
Grabert and Michel \cite{GM83} and of Temmen \emph{et al.}\ \cite{Te00} are restricted to crystals of only 
one species of particles. In this regard our theory is more general.

In a normal crystal the diffusive motion of particles is possible only via point defects which are vacant 
lattice sites and interstitial particles. However, these point defects are strongly suppressed by energy 
barriers which are much larger than the thermal energies of the particles. Both their densities and their 
motions by hopping and diffusion are strongly reduced. As a consequence, nearly no diffusion of the particles 
is present. In this case we have solved the continuity equations exactly and explicitly in order to eliminate 
the particle densities in favor of the displacement field. As a result there remain only two time-evolution 
equations. The first equation for the displacement field describes the macroscopic motion of the material 
fixed to the crystal lattice. The second equation for the velocity field is the Euler equation which 
represents the equation of motion in continuum mechanics. In this way we recover the conventional elasticity 
theory of normal crystals which is well known from the text books \cite{LL07,CL95}.

In order to stay within the framework of conventional density-functional theory in the first and main part 
of the paper we have assumed that the temperature is constant and that heat transport and warming by friction 
are neglected. In a later section we have released these restrictions and considered the general case. Thus, 
we finally obtain the complete time-evolution equations for nonlinear elasticity. However, an alternative 
more general and directer approach is possible by starting with the entropy functional. Here, an alternative 
maximization procedure can be implemented which uses the constraints that the coarse-grained densities, 
including the energy density, and the displacement field are kept constant. Then, the full GENERIC formalism 
in its original form \cite{GO97A,GO97B,Ot05} can be applied in order to obtain the complete and most general 
time-evolution equations. Unfortunately, explicit formulas for the entropy functional are not available so 
that explicit calculations are not possible in this latter approach.

\acknowledgments
The author would like to thank Profs.\ M.\ Fuchs, M.\ Oettel, G.\ Kahl and Drs.\ F.\ Miserez, J.\ M.\ H\"aring, 
S.\ Ganguly, S.-C.\ Lin, G.\ P.\ Shrivastav for many discussions about the topic. 
This work is supported by the Deutsche Forschungsgemeinschaft (DFG) through a D-A-CH grant FU 309/11-1 
and OE 285/5-1 and by the Austrian Funding Agency (FWF) under grant number I3846-N36.



\bibliography{nonlinear_elasticity}
\end{document}